\documentclass[12pt]{article}
\pdfoutput=1

\usepackage{draft}
\usepackage{hyperref}
\usepackage{graphicx,color,subfig}
\usepackage[nosort]{cite}
\usepackage{mciteplus}
\usepackage{skak}
\usepackage{empheq}
\usepackage{anyfontsize}
\usepackage{dsfont}
\usepackage{float}
\usepackage{tabu}
\usepackage{multirow}
\usepackage{hhline}

\usepackage{tikz}
\usetikzlibrary{cd}
\tikzcdset{column sep/normal=0cm}
\DeclareFontFamily{OT1}{pzc}{}
\DeclareFontShape{OT1}{pzc}{m}{it}{<-> s * [1.10] pzcmi7t}{}
\DeclareMathAlphabet{\mathpzc}{OT1}{pzc}{m}{it}

\def\be#1\ee{\begin{align}#1\end{align}}

\begin{document}

\unitlength = .8mm

\begin{titlepage}

\preprint{ CALT-TH 2017-067 \\ PUPT-2546 }

\begin{center}

\hfill \\
\hfill \\
\vskip 1cm

\title{Topological Defect Lines and Renormalization Group Flows in Two Dimensions}

\author{Chi-Ming Chang,$^{\textsymknight}$ Ying-Hsuan Lin,$^{\textsymbishop}$ Shu-Heng Shao,$^{\textsympawn}$ Yifan Wang,$^{\textsymking}$ Xi Yin$^\textsymrook$}

\address{
$^{\textsymknight}$Center for Quantum Mathematics and Physics (QMAP), University of California,
\\
Davis, CA 95616, USA
\\
$^\textsymbishop$Walter Burke Institute for Theoretical Physics, California Institute of Technology,  Pasadena, CA 91125, USA
\\
$^\textsympawn$School of Natural Sciences, Institute for Advanced Study, Princeton, NJ 08540, USA
\\
$^\textsymking$Joseph Henry Laboratories, Princeton University, Princeton, NJ 08544, USA
\\
$^\textsymrook$Jefferson Physical Laboratory, Harvard University, 
Cambridge, MA 02138 USA}

\email{wychang@ucdavis.edu, yhlin@caltech.edu, shuhengshao@gmail.com, yifanw@princeton.edu,
xiyin@fas.harvard.edu}
\end{center}

\abstract{We consider topological defect lines (TDLs) in two-dimensional conformal field theories. Generalizing and encompassing both global symmetries and Verlinde lines, TDLs together with their attached defect operators provide models of fusion categories without braiding. We study the crossing relations of TDLs, discuss their relation to the 't Hooft anomaly, and use them to constrain renormalization group flows to either conformal critical points or topological quantum field theories (TQFTs). We show that if certain non-invertible TDLs are preserved along a RG flow, then the vacuum cannot be a non-degenerate gapped state.  
For various massive flows, we determine the infrared TQFTs completely from the consideration of TDLs together with modular invariance.}

\vfill

\end{titlepage}

\eject

\begingroup
\hypersetup{linkcolor=black}
\tableofcontents
\endgroup

\section{Introduction}
Two-dimensional conformal field theories (CFTs) are usually defined in terms of the data of local ``bulk" point-like operators, namely the spectrum of Virasoro primaries and their structure constants, subject to the associativity of operator product expansion (OPE) and modular invariance. It is well known that there are extended objects, or ``defects", such as boundary conditions \cite{Cardy:1986gw, Verlinde:1988sn,Cardy:1989vyr,Cardy} and line defects/interfaces \cite{Oshikawa:1996ww, Oshikawa:1996dj, Petkova:2000ip,Fuchs:2002cm, Fuchs:2003id, Fuchs:2004dz, Fuchs:2004xi, Frohlich:2004ef, Frohlich:2006ch, Quella:2006de,Fuchs:2007vk,Fuchs:2007tx, Bachas:2007td, Kong:2008ci, Petkova:2009pe,2012CMaPh.313..351K, Carqueville:2012dk,Brunner:2013xna, Davydov:2013lma, Kong:2013gca, Petkova:2013yoa,Bischoff:2014xea,Bhardwaj:2017xup}, in the CFT that can be characterized in terms of the response of bulk local operators in the presence of the defect, but typically obey strong notions of locality that do not obviously follow from those of the bulk local operators.\footnote{See \cite{Hauru:2015abi,Aasen:2016dop,Bridgeman:2017etx,Bal:2018wbw} for the lattice realization of some topological line defects and their constraints on infrared phases.}

A basic example is a global symmetry element $g$, which by definition is a linear transformation on the bulk local operators that preserve their OPEs: the action of $g$ on a bulk local operator may be viewed as the contraction of a loop of a topological defect line (TDL) on the bulk local operator \cite{Frohlich:2006ch,Davydov:2010rm,Kapustin:2014gua,Gaiotto:2014kfa}. A TDL that corresponds to a   global symmetry will be referred to as an {\it invertible line} in this paper, for the reason that such a line is associated to a symmetry action and therefore its inverse must exist.\footnote{Here the inverse ${\cal L}^{-1}$ of a line $\cal L$ is defined such that their fusion relation is ${\cal L} {\cal L}^{-1}={\cal L}^{-1} {\cal L}=1$. For a more general TDL $\cal L$ (such as the $N$ line in the critical Ising model), its inverse might not exist.
} 
In all known examples of global symmetries in a CFT, the corresponding invertible lines are subject to a strong locality property, namely, that it can end on defect operators, which obey an extended set of OPEs.\footnote{In the absence of an 't Hooft anomaly, such defect operators are related to orbifold twisted sector states upon a symmetry-invariant projection, but the existence of the defect operator Hilbert space is more general and applies to global symmetries with an 't Hooft anomaly as well.} 
  In the case of a  continuous global symmetry such as $U(1)$,  Noether's theorem states  that there must be an associated conserved spin-one current $j_\mu$.  The contour integral of this conserved current $e^{i\theta \int ds^\mu j_\mu}$ then defines a family of  invertible TDLs labeled by $\theta\in S^1$, where the topological property follows from the conservation equation. 
 In the case of discrete global symmetry such as $\mathbb{Z}_N$, the existence of the associated invertible TDLs with the above-mentioned properties can be  thought of as a discrete version of Noether's theorem. It has nontrivial implications on the action of $g$ on the bulk local operators that do not obviously follow from the standard axioms on the bulk local operators.

Interestingly, there are TDLs that do not correspond to any global symmetries, and they are ubiquitous in $2d$ CFTs \cite{Verlinde:1988sn,Petkova:2000ip, Frohlich:2006ch, Petkova:2009pe, Petkova:2013yoa}. This is possible because the general TDLs need not obey group-like fusion relations, but instead form a (semi)ring (generally non-commutative) under fusion. Though invariant under isotopy transformations (by definition), a general TDL cannot simply be reconnected within the same homological class; rather, it obeys nontrivial crossing relations under the splitting/joining operation, and consequently, the action of a general TDL on bulk local operators by contraction need not preserve the OPE as a global symmetry action would. 

A special class of TDLs (not necessarily invertible) are known as Verlinde lines \cite{Verlinde:1988sn,Petkova:2000ip,Drukker:2010jp,Gaiotto:2014lma}. They exist in rational CFTs defined by diagonal modular invariants. The fusion ring generated by the Verlinde lines in an RCFT is formally identical to that of the representations of the chiral vertex algebra (although the physical interpretation of the fusion of Verlinde lines is entirely different from the OPEs). In particular, such a fusion ring is commutative and admits braiding, which is {\it not} the case for the most general system of TDLs. The fusion of general TDLs may not be commutative (as is the case for nonabelian global symmetry), and even when they are commutative, they may not admit braiding.

The structure of fusion and crossing relations of the Verlinde lines is captured by what is known as a \textit{modular tensor category} \cite{Moore:1988qv,Moore:1989vd,TuraevV.G2010QIoK,KITAEV20062}, which requires the crossing relations to obey the pentagon identity, and braiding relations to further obey the hexagon identity. The general TDLs of interest in this paper are models of a more general mathematical structure known as {\it fusion category}	 \cite{etingof2016tensor,10.2307/20159926} (at least when there are finitely many simple lines), which still requires the crossing relations to obey the pentagon identity, but does not require braiding. In a sense, fusion categories modeled by TDLs unify and generalize the notions of both nonabelian symmetry groups and modular tensor category (modeled by Verlinde lines). See \cite{Levin:2005aa,Bhardwaj:2017xup} for physicists' expositions on this subject, and \cite{Bhardwaj:2017xup} in particular for the relation to gauging and the 't Hooft anomaly. The goal of this paper is to explore the possible types of TDLs realized in unitarity, compact $2d$ CFTs, and their implications on renormalization group (RG) flows by a generalization of the 't Hooft anomaly matching \cite{tHooft:1980xss,Kapustin:2014zva}.

We will begin by describing a set of physically motivated defining properties of TDLs in Section~\ref{sec:definingproperties}, and discuss their relations to the notion of fusion category in Section~\ref{sec:fusioncategory}. In the context of global symmetry groups corresponding to group-like categories, we will discuss 't Hooft anomalies, orbifolds, and discrete torsion \cite{Vafa:1986wx,Brunner:2014lua} in relation to invertible TDLs in Section~\ref{sec:globalsymmetry}.

In Section~\ref{sec:TDLsRCFTs}, we discuss TDLs in rational CFTs that are generally not invertible. In particular, we review Verlinde lines in Section~\ref{sec:verlindelines}, and describe the explicit crossing relations of Verlinde lines in diagonal Virasoro minimal models. Next, in Section~\ref{sec:generaltdls}, we will discuss examples of TDLs in rational CFTs that are neither Verlinde lines nor invertible lines. The first example is a set of TDLs in the three-state Potts model that preserves the Virasoro algebra but not the $W_3$ algebra, found in \cite{Petkova:2000ip}. The second example is given by the topological Wilson lines in WZW and coset models, generalizing the construction of \cite{Bachas:2004sy}. The third example is a set of TDLs in the non-diagonal $SU(2)_{10}$ WZW model of $E_6$ type (or in the $(A_{10}, E_6)$ minimal model), which realizes the so-called ${1\over 2}E_6$ fusion category \cite{Hagge:2007aa,Ostrik:2013aa}. This fusion category consists of just three simple lines, has commutative fusion relation, and yet does not admit braiding \cite{Hagge:2007aa}.

The primary interest of this paper is to explore the constraints of TDLs on the dynamics of QFTs when the fusion and crossing relations of TDLs are known. A consequence of these relations of TDLs is the restriction on the spin content of defect operators at the end of the TDLs. Typically, only specific fractional spins are allowed for the defect operators at the end of a given type of TDLs. This is the subject of Section~\ref{sec:spinselect}.

When certain TDLs are preserved along an RG flow, say the ones that commute with
the relevant deformation of the UV CFT, these TDLs will survive in the IR.  The IR TDLs obey the same fusion and crossing relations as in the UV.  Physically, this follows from the topological property of the TDL as there is no intrinsic scale associated to it.  More rigorously, the consistent solutions of the fusion and crossing relations are discrete and therefore cannot be deformed continuously under RG flows.   This is known as the Ocneanu rigidity in category theory \cite{10.2307/20159926}.
 This basic observation has interesting implications on an RG flow to a massive phase, where the IR dynamics is described by a topological quantum field theory (TQFT) \cite{Moore:2006dw}.
 The TDLs of the TQFT inherited from the UV CFT will constrain and often allow us to completely determine the fully extended TQFT \cite{doi:10.1063/1.531236,lurie2009,Freed:2009qp,2011arXiv1112.1000S}, \textit{i.e.} with all lines and defect operators included. 
In particular, we show that if a TDL $\cal L$ with non-integral vacuum expectation value $\la {\mathcal L}\ra$ (defined in Section~\ref{Sec:Vac}) is preserved along the flow, then the vacuum cannot be a non-degenerate gapped state.  
 We will analyze various explicit RG flows in Section~\ref{sec:rgflow}, for the tricritical Ising model deformed by the second energy operator $\varepsilon'$, the tricritical Ising model deformed by the second spin operator $\sigma'$, and the $(A_{10},E_6)$ minimal model deformed by $\phi_{2,1}$. In addition, we consider UV CFTs realizing twisted siblings of the ${\rm Rep}(S_3)$ fusion category, and rule out the possibility of flowing to IR TQFTs with unique vacuum.

More generally, it is {\it a priori} not obvious whether a fusion category of TDLs can always be realized by some TQFT, as the latter requires the construction of defect operators and is subject to modular invariance, neither of which is directly captured by the fusion category structure. 
The analogous question for the 't Hooft anomaly of a global symmetry has recently been answered in \cite{Witten:2016cio,Wang:2017loc,Tachikawa:2017gyf}.  There, it was shown that in general spacetime dimensions, given an 't Hooft anomaly of a finite group specified by the group cohomology, there always exists a TQFT realizing this anomaly. 
It would be interesting to either prove this statement or find a counter example for the more general fusion category beyond invertible lines associated with global symmetries.  

Finally, we will describe a class of TDLs in potentially irrational unitary compact CFTs, as RG fixed points of coupled minimal models, in Section~\ref{sec:coupledminmodels}. We conclude with some future prospectives in Section~\ref{sec:conclusion}. Some further details of the H-junction crossing kernels and explicit solutions to the pentagon identities are given in Appendices \ref{CKbasis} and \ref{sec:pentasolutions}.

\section{Definition and properties of topological defect lines}
\label{sec:definingproperties}

\begin{figure}[H]
\centering
\begin{minipage}{0.2\textwidth}
\includegraphics[width=1\textwidth]{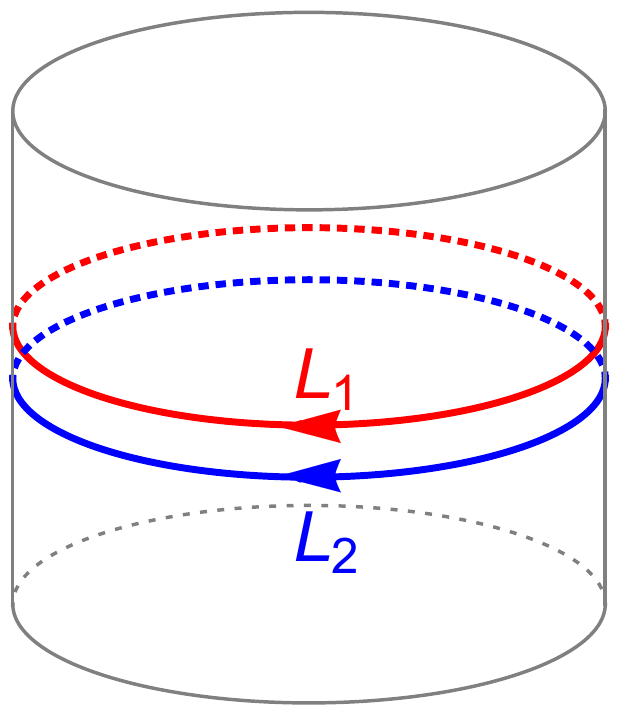}
\end{minipage}%
\begin{minipage}{0.1\textwidth}\begin{eqnarray*} ~=  \\ \end{eqnarray*}
\end{minipage}%
\begin{minipage}{0.2\textwidth}
\includegraphics[width=1\textwidth]{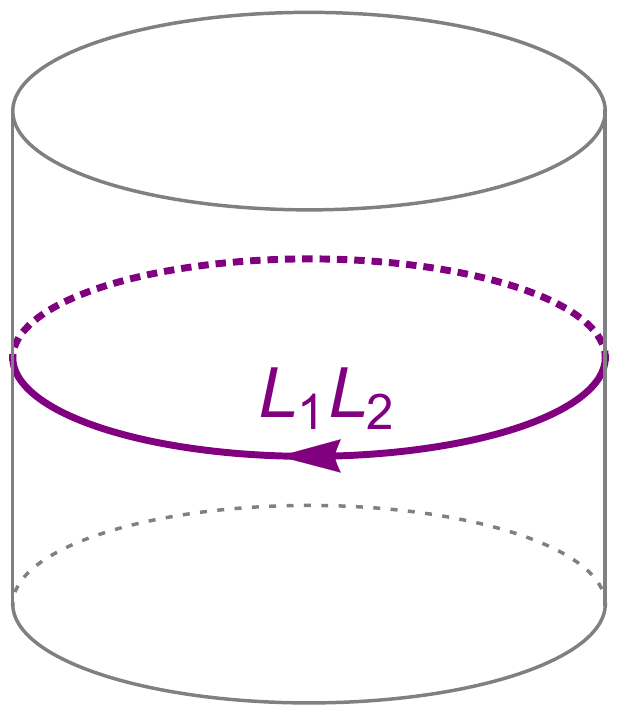}
\end{minipage}%
\caption{Fusion of a pair of TDLs ${\cal L}_1$ and ${\cal L}_2$ wrapping the spatial loop on the cylinder.}
\label{fig:fuse}
\end{figure}

Topological defect lines (TDLs) comprise a special class of extended objects, that are defined along an oriented path, in a two-dimensional quantum field theory. To motivate the formal definitions, let us begin by considering topological defect loops on a cylinder, extended along the compact direction. Such a TDL $\cal L$ can be regarded as a linear operator $\widehat {\cal L}$ acting on the Hilbert space on a circle, that commutes with the left and right Virasoro algebras. This last property implies that the exact location and shape of $\cal L$ is irrelevant. As depicted in Figure~\ref{fig:fuse}, the composition of linear operators $\widehat{\cal L}$ can be understood as the fusion of TDLs, a notion we presently define. Primary of TDLs are those associated to  global symmetries \cite{Kapustin:2014gua,Gaiotto:2014kfa} -- which we will call {\it invertible} defect lines (see Section~\ref{Sec:Homological}) -- acting by symmetry transformations, such that the set of $\widehat {\cal L}$ form a representation of the symmetry group.

The set of TDLs are equipped with an algebraic structure -- the {\it fusion ring} -- comprised of two operations: direct sum ${+}$ and fusion $\triangleright$.\footnote{More precisely, these binary operations define a semiring which can be canonically extended to a ring.} Direct sum is associative and commutative, and fusion is associative but not necessarily commutative. Moreover, fusion is distributive with respect to direct sum, $({\cal L}_1 {+} {\cal L}_2) \triangleright {\cal L}_3 = ({\cal L}_1 \triangleright {\cal L}_3) {+} ({\cal L}_2 \triangleright {\cal L}_3)$. 
 For the class of TDLs we will be investigating, every one of them has a unique decomposition into a direct sum of  {\it simple} objects (precise definition given later), which cannot be decomposed further. There is an identity object $I$ among the simple objects, such that ${\cal L} \triangleright I = I \triangleright {\cal L} = {\cal L}$. For every TDL $\cal L$, there exists an orientation reversed TDL $\overline{\cal L}$. Under direct sum and fusion, $\overline{\cal L}_1 {+} \overline{\cal L}_2 = \overline{{\cal L}_1 {+} {\cal L}_2}$, and $\overline{\cal L}_2 \triangleright \overline{\cal L}_1 = \overline{{\cal L}_1 \triangleright {\cal L}_2}$. In the following, the fusion operator will often be omitted, with the fused TDL ${\cal L}_1 \triangleright {\cal L}_2$ simply abbreviated as ${{\cal L}_1 {\cal L}_2}$.

\begin{figure}[H]
\centering
\begin{minipage}{0.4\textwidth}
\includegraphics[width=1\textwidth]{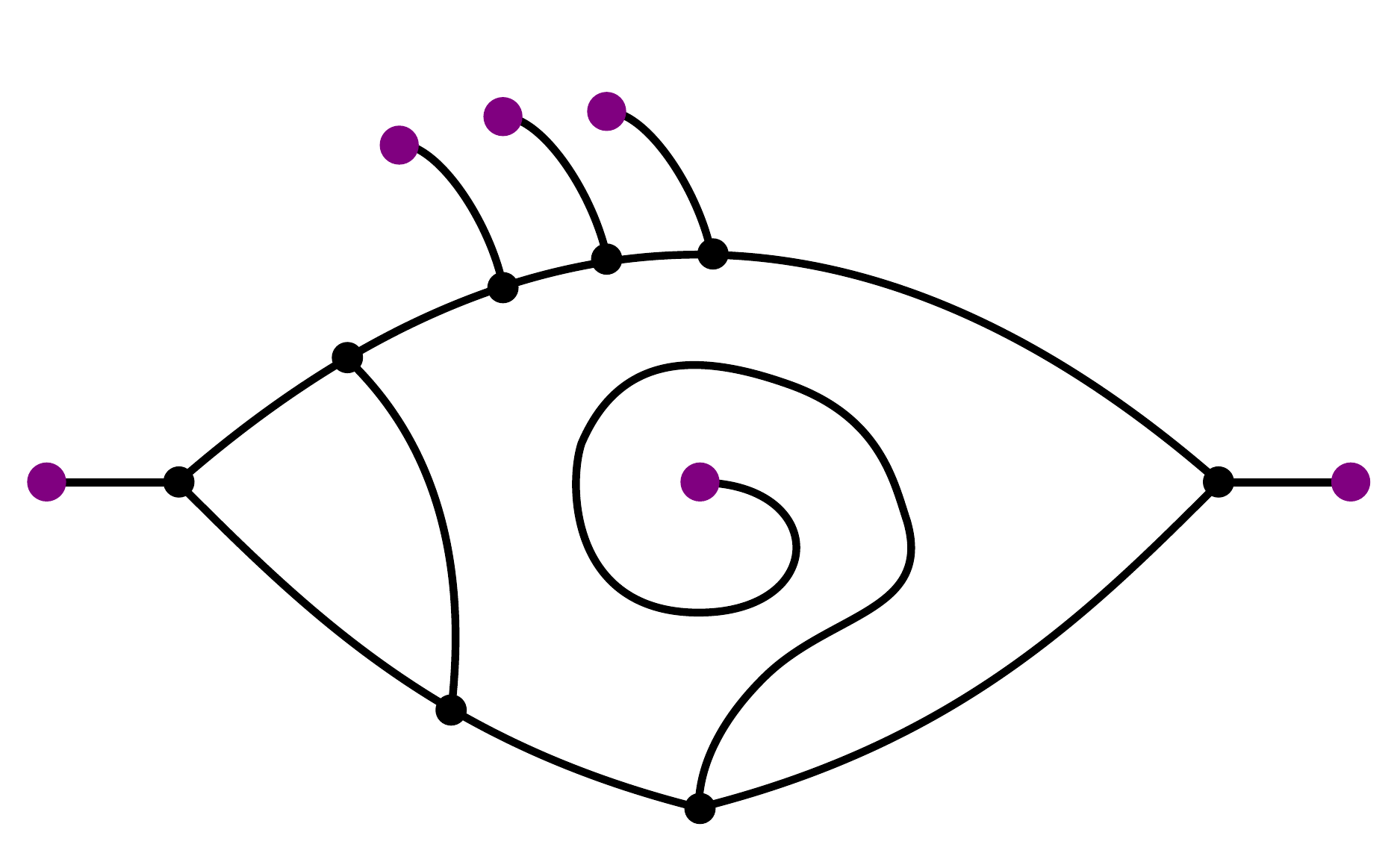}
\end{minipage}%
\caption{An admissible configuration of TDLs with endpoints (purple dots), joined by T-junctions (black dots).
}
\label{fig:TDLfish}
\end{figure}

To construct more general correlation functions with TDL insertions, such as the one involving the TDL configuration shown in Figure~\ref{fig:TDLfish}, the fusion ring alone is insufficient. More specifically, TDLs can end on points or join at junctions, and these points and junctions must be equipped with additional structures. The topological nature of TDLs means that observables only depend on the homotopy class of the TDL configuration, and this property may be formulated more precisely in terms of isotopy invariance. We now define these structures and properties, and discuss several corollaries. Many of these structures have already been defined and explored in both the mathematics and physics literature, such as in the context of modular tensor category \cite{Moore:1988qv,Moore:1989vd,TuraevV.G2010QIoK,KITAEV20062}, and in the works of \cite{10.2307/20159926,Levin:2005aa,Mueger:2008aa,Carqueville:2012dk,Brunner:2013xna,Bischoff:2014xea,etingof2016tensor,Bhardwaj:2017xup}. We hope to first recast them in a language natural for quantum field theory, and then derive various new consequences.

\subsection{Defining properties}
\label{Sec:Properties}

We formulate the defining properties of TDLs in CFTs, and comment on the generalization to non-conformal theories in Section~\ref{Sec:Nonconformal}.

\noindent{\bf 1. (Isotopy invariance)} On a flat surface, all physical observables (including in particular the correlation functional defined in Property~4) are invariant under continuous deformations of TDLs that are ambient isotopies of the graph embedding, and preserves the positions of endpoints and junctions as well as the angles of the TDLs coming out of endpoints and junctions. This is the key property that distinguishes TDLs from conformal or more general defect lines. It follows from isotopy invariance that TDLs commute with the stress-energy tensor.

\noindent{\bf 2. (Defect operator)} A TDL ${\cal L}$ comes with a space ${\cal H}_{\cal L}$ of possible point-like defect operators at its end, from which the TDL is outgoing in our convention (ingoing is given by the orientation reversal ${\cal H}_{\overline{\cal L}}$). By the state/operator mapping, ${\cal H}_{\cal L}$ is the Hilbert space of theory on a circle with a single ${\cal L}$-defect point (future-oriented). For the trivial TDL $I$, ${\cal H}_I$ is the same as the Hilbert space of bulk local operators in theory. 

TDLs can join at point-like junctions, and we adopt the convention that every line is outgoing. Each junction comes with an ordering (as opposed to cyclic ordering) of the lines attached to the junction.\footnote{Equivalently, we can simply mark the last line ${\cal L}_k$ entering a $k$-way junction to specify the ordering of the lines ${\cal L}_1,\cdots,{\cal L}_k$ that meet at the vertex of an embedded graph. The physical motivation for such an ordering prescription comes from the microscopic description of the junction (say in a lattice model), where the lines entering the junction are {\it a priori} distinguished even if they are of the same type. We will see later that this is essential for allowing for TDLs that correspond to global symmetries with 't Hooft anomalies.
}
A $k$-way junction is equipped with a junction Hilbert space ${\cal H}_{{\cal L}_1,{\cal L}_2,\cdots,{\cal L}_k}$ of possible defect operators at the junction. Under cyclic permutations of ${\cal L}_1,{\cal L}_2,\cdots,{\cal L}_k$, the junction Hilbert spaces are isomorphic under possibly nontrivial cyclic permutation maps.

It also follows from isotopy invariance that {\bf (i)} the Hilbert space ${\cal H}_{{\cal L}_1,{\cal L}_2,\cdots,{\cal L}_k}$ of defect operators at a $k$-way junction is a representation of the holomorphic and anti-holomorphic Virasoro algebras, though the states generally have non-integer spins, and {\bf (ii)} the contraction of a TDL loop encircling a local (bulk or defect) primary operator produces a local primary of the same conformal weight. We will discuss this in more detail later in this section.

\noindent{\bf 3. (Junction vector)} In a CFT with a unique vacuum, the junction vector space $V_{{\cal L}_1,{\cal L}_2,\cdots,{\cal L}_k}$ is the space of weight-$(0,0)$ states in ${\cal H}_{{\cal L}_1,{\cal L}_2,\cdots,{\cal L}_k}$, which may be zero, one, or more than one-dimensional. Cyclically permuted junction vector spaces (such as $V_{{\cal L}_1 ,{\cal L}_2, {\cal L}_3}$ and $V_{{\cal L}_2, {\cal L}_3 ,{\cal L}_1}$) are isomorphic via a cyclic permutation map (that may act nontrivially even when ${\cal L}_i$ are of the same type).

A junction associated to a junction vector is call a topological junction. For TDL configurations with topological junctions, the isotopy invariance is extended to ambient isotopy that need not preserve the positions of topological junctions and the angles of the TDLs coming out of topological junctions. We refer to a three-way topological junction as a T-junction, and a four-way topological junction as an X-junction. In the rest of this paper, we take all the $k$-way junctions for $k>1$ to be topological. This restriction is without the loss of generality, since by the locality property introduced in the later part of this section, any TDL configuration can be written equivalently as a sum of TDL configurations with only topological junctions.

In this paper, we will use TDLs and their junction vector spaces to constrain various CFTs and  TQFTs.  We will restrict ourselves to TQFTs that arise at the end of massive RG flows from CFTs. Typically, the space of topological defect operators at a junction in the IR TQFT becomes larger than in the UV CFT.\footnote{For instance, the bulk Hilbert space, regarded as the Hilbert space at a junction of trivial TDLs, often develops degenerate vacua in the IR.}
For such a TQFT, we define the junction vector space $V_{{\cal L}_1,{\cal L}_2,\cdots,{\cal L}_k}$ as the subspace of weight-$(0,0)$ topological defect operators that are inherited from those in the UV CFT.\footnote{The complete structure of defect operators in general TQFTs is rather rich and can be captured by a pivotal 2-category \cite{Davydov:2011kb,Carqueville:2016nqk}.
}

\begin{figure}[H]
\centering
\begin{minipage}{0.2\textwidth}
\includegraphics[width=1\textwidth]{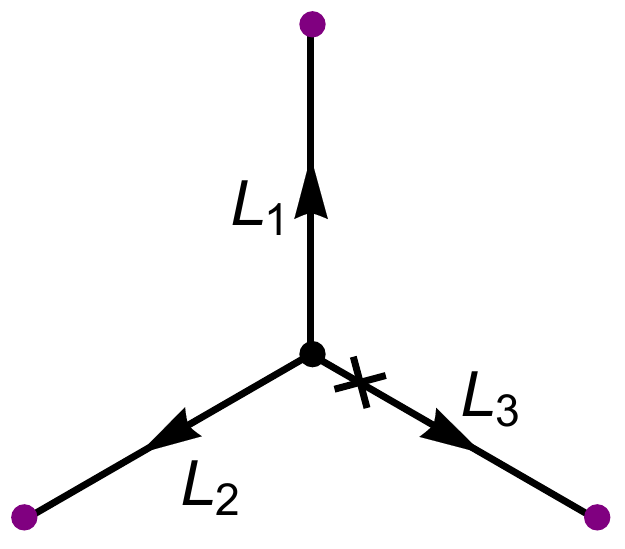}
\end{minipage}%
\caption{A correlation functional (on the plane), where the TDLs are joined by a T-junction (black dot) with the order of lines specified (last leg marked by the ``$\times$"), and ending on defect operators (purple dots). It is a linear function on the junction vector space $V_{{\cal L}_1,{\cal L}_2,{\cal L}_3}$.}
\label{fig:sky}
\end{figure}

\noindent{\bf 4. (Correlation functional)} An admissible configuration of TDLs on an oriented surface is an embedded oriented graph possibly with endpoints and junctions. We associate to each endpoint a defect operator and each junction (with an ordering of lines) a junction vector. The correlation functional of a TDL configuration, along with a given set of defect operators and bulk local operators, is a multi-linear complex-valued function on the tensor product of junction vector spaces. An example is depicted in Figure~\ref{fig:sky}.

The isotopy invariance of a correlation functional can be extended to curved surfaces, but with an important subtlety -- the isotopy anomaly: under deformation of a TDL, 
the correlation functional may acquire a phase that is proportional to the integral of the curvature over the region swept by the deformation. If the TDL ${\cal L}$ is of a different type from its orientation reversal, $\overline{\cal L}$, then the isotopy anomaly can be absorbed by a finite local counter term on the TDL that involves the extrinsic curvature. If ${\cal L}$ is a TDL of the same type as $\overline{\cal L}$, consistency with unitarity and modular invariance sometimes requires a non-vanishing isotopy anomaly on a curved surface, which also introduces an orientation-reversal anomaly. These anomalies are explored and discussed in Section~\ref{isoanomaly}.

\noindent{\bf 5. (Direct sum)} Given two TDLs ${\cal L}_1$ and ${\cal L}_2$, there exists a direct sum TDL ${\cal L}_1+{\cal L}_2$, such that ${\cal H}_{{\cal L}_1+{\cal L}_2}= {\cal H}_{{\cal L}_1}\oplus {\cal H}_{{\cal L}_2}$. Furthermore, junction vector spaces and correlation functionals are additive with respect to the direct sum of TDLs. A TDL ${\cal L}$ is called {\it simple} if the junction vector space $V_{{\cal L},\overline{\cal L}}$  is one-dimensional. It follows that a simple TDL cannot be further written as a positive sum of other simple TDLs, by additivity of the junction vector space. A TDL is called semi-simple if it is a direct sum of finitely many simple TDLs.


We further introduce two notions for the set of TDLs in a theory. {\it Semi-simiplicity}: every TDL in theory is semi-simple. {\it Finiteness}: the number of types of simple TDLs in theory is finite. 
In this paper, we assume semi-simplicity, but not finiteness, even though the latter is typically assumed in the literature on fusion category.\footnote{The finiteness condition clearly fails for TDLs associated to continuous global symmetries.}

\noindent{\bf 6. (Conjugation)} The two-point function of a pair of defect operators in ${\cal H}_{\cal L}$ and ${\cal H}_{\overline{\cal L}}$ connected by a straight $\cal L$ gives a bilinear map $h: {\cal H}_{{\cal L}}\times {\cal H}_{\overline{\cal L}}\to \bC$. There is an antiunitary conjugation map $\iota:{\cal H}_{\cal L}\to{\cal H}_{\overline{\cal L}}$, such that $h$ is related to the inner product $\la \, , \ra$ on ${\cal H}_{\cal L}$ by $h(v_1,\iota(v_2)) = \la v_1,v_2\ra$ for $v_1,v_2\in {\cal H}_{\cal L}$. We define the conjugation map from ${\cal H}_{{\cal L}_1,{\cal L}_2,\cdots,{\cal L}_k}$ to ${\cal H}_{\overline{\cal L}_k,\overline{\cal L}_{k-1},\cdots,\overline{\cal L}_1}$ in a similar fashion. Acting the conjugation map on all defect operators in a correlation function, further combined with a parity action on their locations, $z_i \mapsto \bar z_i$, is equivalent to complex conjugation.

\begin{figure}[H]
\centering
\begin{minipage}{0.25\textwidth}
\includegraphics[width=1\textwidth]{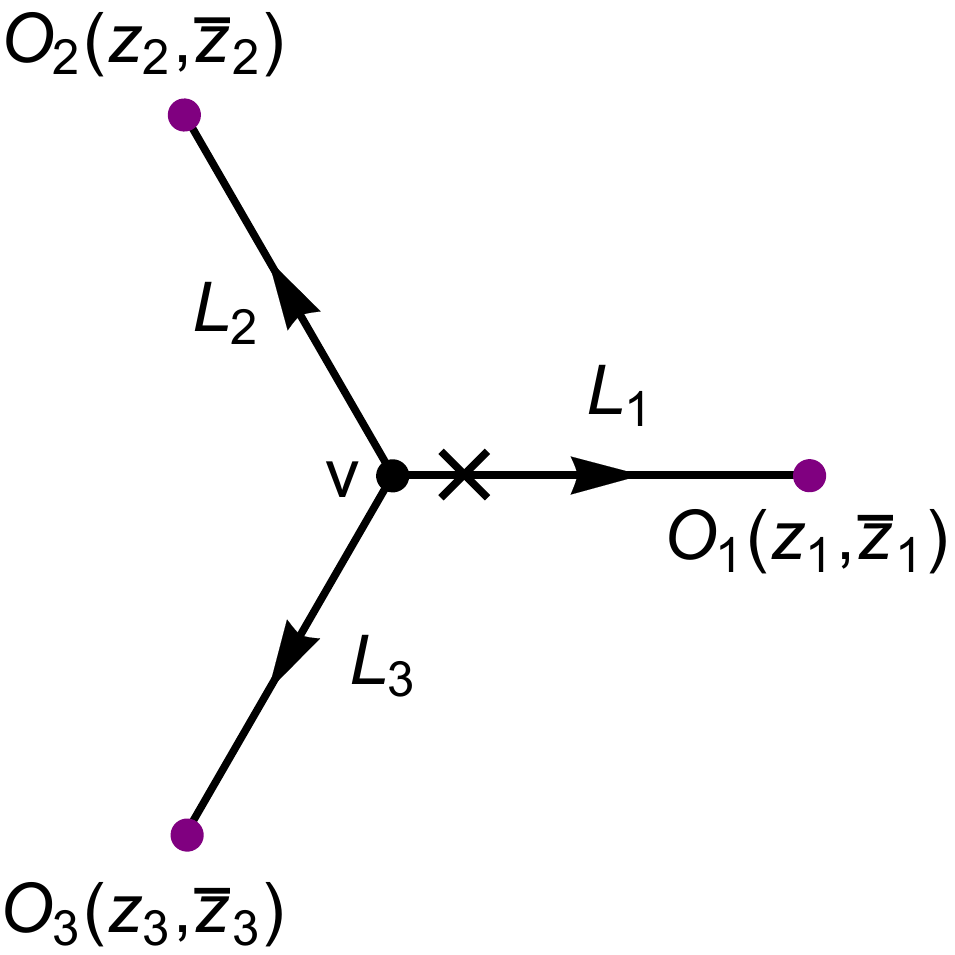}
\end{minipage}%
\hspace{-.2in}
\begin{minipage}{0.15\textwidth}\begin{eqnarray*}  = \\ \end{eqnarray*}
\end{minipage}%
\hspace{-.2in}
$\left( \quad
\begin{minipage}{0.25\textwidth}
\includegraphics[width=1\textwidth]{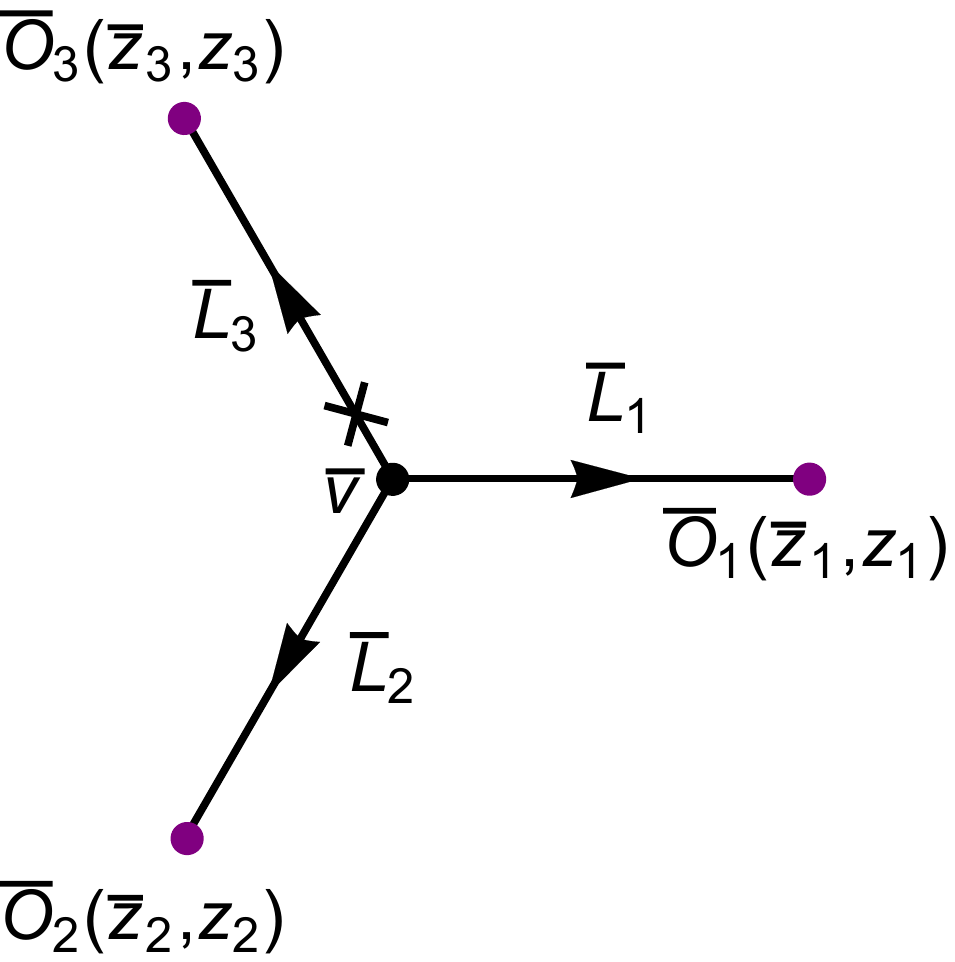}
\end{minipage}%
\right)^*$
\caption{Conjugation map in a correlation function. Here, $O_{i}$ with $i=1,2,3$ denote defect operators in ${\cal H}_{{\cal L}_{i}}$, and $\overline O_{i}\equiv \iota(O_{i})$ denote their conjugates, which are defect operators in ${\cal H}_{\overline{\cal L}_{i}}$. Similarly, $v$ is a junction vector in $V_{{\cal L}_1,{\cal L}_2,{\cal L}_3}$, and $\overline v \equiv \iota(v) \in V_{\overline{\cal L}_3,\overline{\cal L}_2,\overline{\cal L}_1}$.
}
\label{fig:iota}
\end{figure}

\begin{figure}[H]
\centering
\begin{minipage}{0.375\textwidth}
\includegraphics[width=1\textwidth]{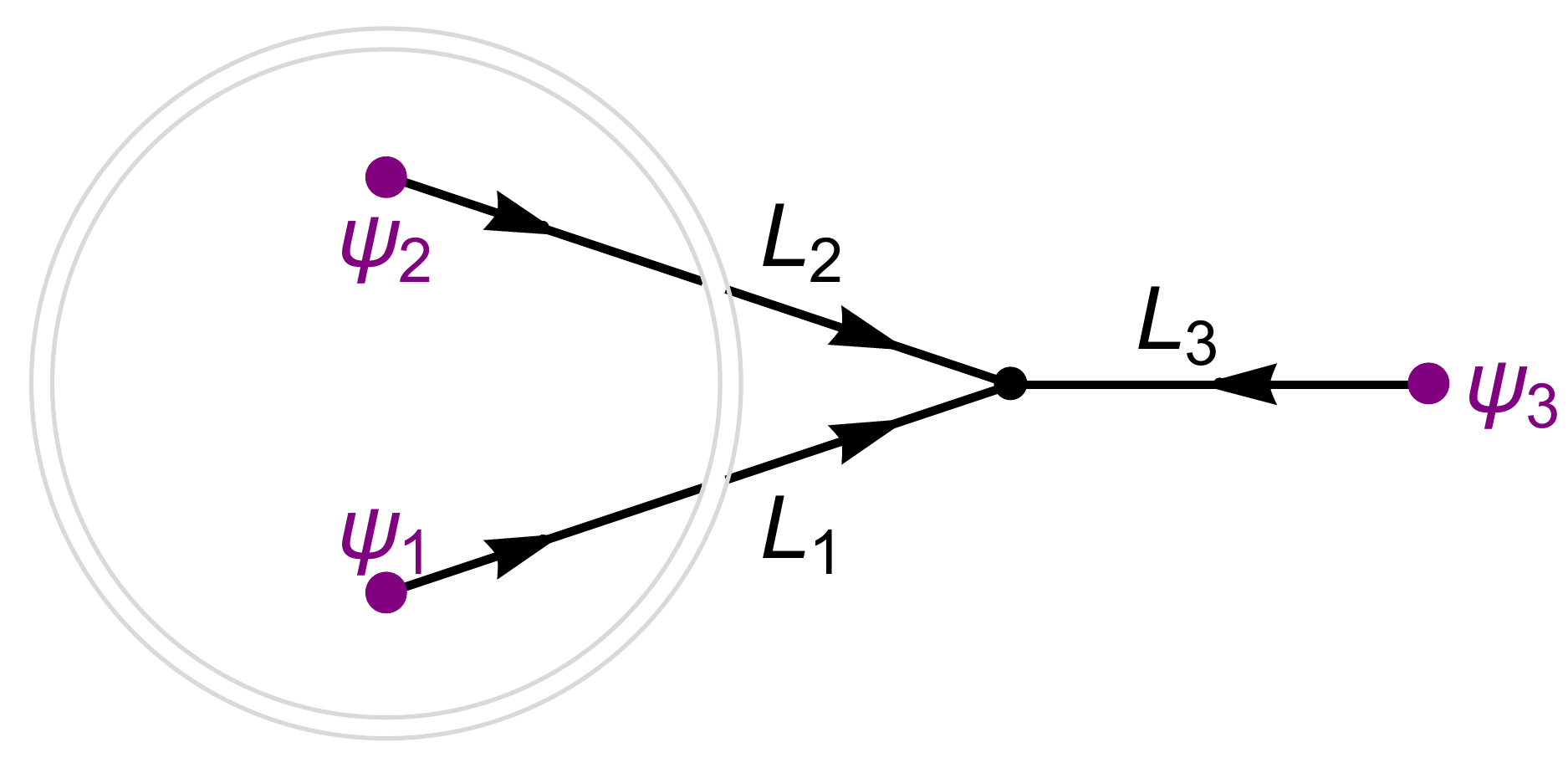}
\end{minipage}%
\hspace{.5in}
\begin{minipage}{0.4\textwidth}
\includegraphics[width=1\textwidth]{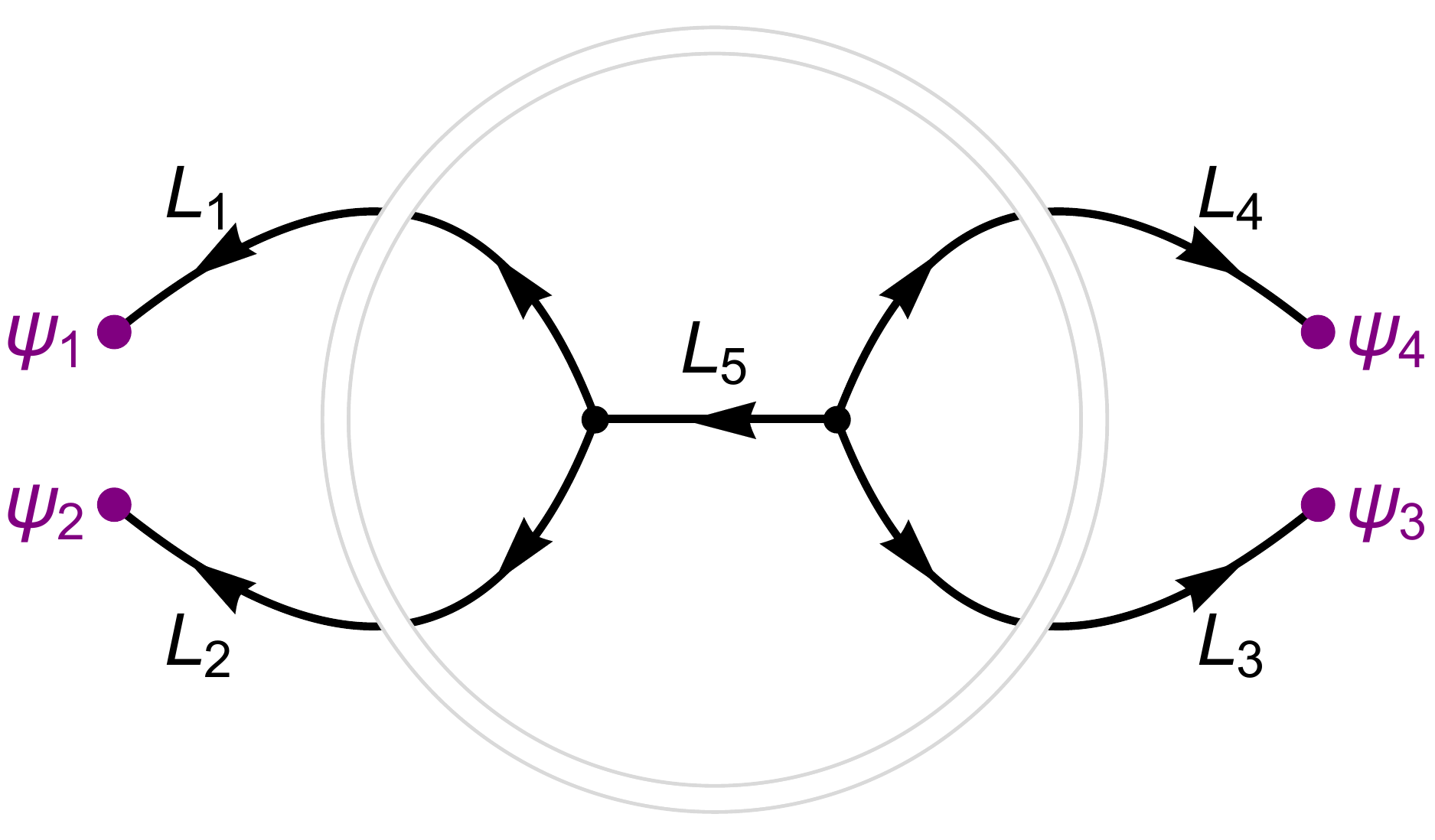}
\end{minipage}%
\caption{{\bf Left:} Cutting a TDL graph along the gray circle and inserting a complete basis of states in ${\cal H}_{{\cal L}_1, {\cal L}_2}$ is equivalent to replacing the defect operators $\Psi_1$ and $\Psi_2$ by their OPE, which is a defect operator in ${\cal H}_{{\cal L}_1, {\cal L}_2}$. {\bf Right:} A similar cut-and-insert procedure, where the defect operators inside the circle are all junction vectors. The 
graph inside the smaller
gray circle implements the map $V_{{\cal L}_1, {\cal L}_2, \overline{\cal L}_5} \otimes V_{{\cal L}_3, {\cal L}_4, {\cal L}_5} \to V_{{\cal L}_1, {\cal L}_2, {\cal L}_3, {\cal L}_4}$.}
\label{fig:cutting}
\end{figure}

\noindent{\bf 7. (Locality)} A TDL configuration on a Riemann surface is equivalent to one obtained by cutting the TDLs transversely along a circle and inserting a complete orthonormal basis of operators in ${\cal H}_{{\cal L}_1{\cal L}_2\cdots{\cal L}_k}$, where ${\cal L}_1,\cdots,{\cal L}_k$ are the TDLs that are cut along the circle. In particular, the locality property encompasses the notion of OPEs between operators in ${\cal H}_{{\cal L}_1}$ and ${\cal H}_{{\cal L}_2}$. See Figure~\ref{fig:cutting} for an illustration.

When only junction vectors are present inside the cut, the insertion of states in ${\cal H}_{{\cal L}_1,{\cal L}_2,\cdots,{\cal L}_k}$ reduces to the insertion of junction vectors in $V_{{\cal L}_1,{\cal L}_2,\cdots,{\cal L}_k}$. In particular, in this case, the TDL graph on a disc with  ${\cal L}_1,\cdots,{\cal L}_k$ crossing the boundary of the disc gives a multi-linear map from the tensor product of junction vector spaces associated with the graph to $V_{{\cal L}_1,{\cal L}_2,\cdots,{\cal L}_k}$. Such maps on junction vector spaces are represented by gray circles, as illustrated on the right of Figure~\ref{fig:cutting}.

\noindent {\bf 8. (Partial fusion)} A pair of TDLs ${\cal L}_1$ and ${\cal L}_2$ wrapping the compact direction on a cylinder fuses to a single (not necessarily simple) TDL ${\cal L}_1{\cal L}_2$ when there is no other TDL or local operator inserted between them, as shown in Figure~\ref{fig:fuse}. Fusion endows the set of TDLs with a ring structure. The defect Hilbert space ${\cal H}_{{\cal L}_1,\cdots,{\cal L}_i,{\cal L}_{i+1},\cdots,{\cal L}_k}$ of a $k$-way junction is isomorphic to the defect Hilbert space ${\cal H}_{{\cal L}_1,\cdots,({\cal L}_i{\cal L}_{i+1}),\cdots,{\cal L}_k}$ of a $(k-1)$-way junction under the fusion between ${\cal L}_i$ and ${\cal L}_{i+1}$.

\begin{figure}[H]
\centering
\begin{minipage}{0.2\textwidth}
\includegraphics[width=1\textwidth]{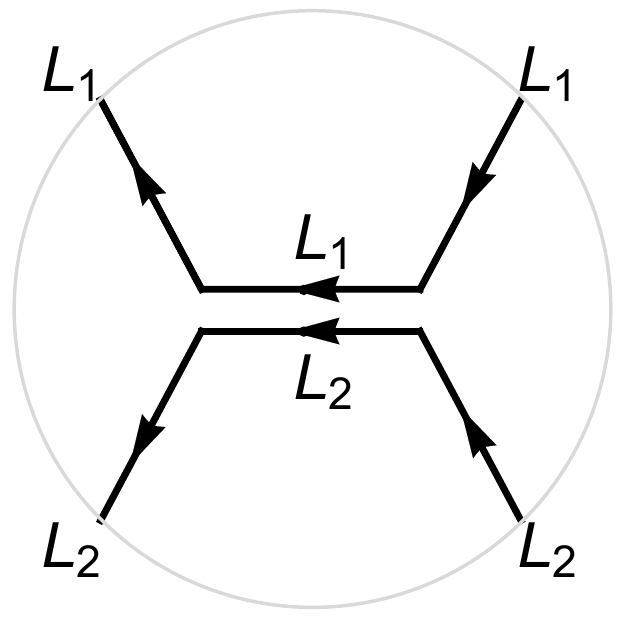}
\end{minipage}%
\begin{minipage}{0.11\textwidth}\begin{eqnarray*} ~~= ~ \sum_i \\  \end{eqnarray*}
\end{minipage}%
\begin{minipage}{0.2\textwidth}
\includegraphics[width=1\textwidth]{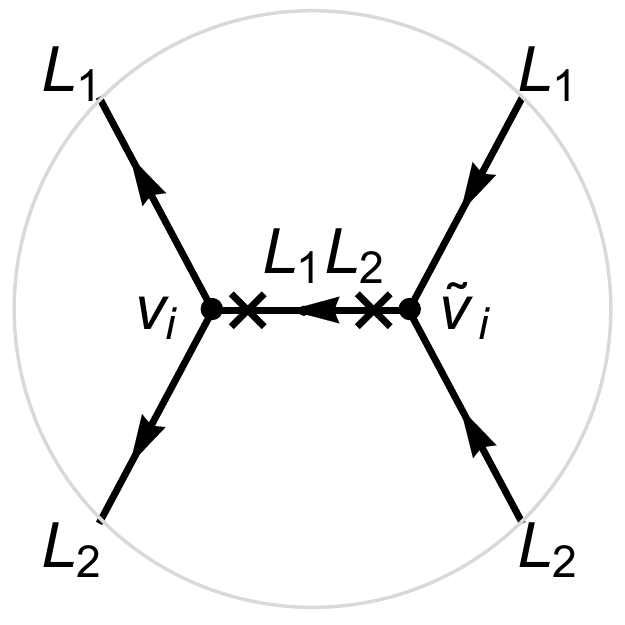}
\end{minipage}
\caption{Partial fusion of a pair of TDLs ${\cal L}_1$ and ${\cal L}_2$.}
\label{fig:partialFusion}
\end{figure}

On a local patch, a pair of TDLs ${\cal L}_1$ and ${\cal L}_2$ can be partially fused to a TDL ${\cal L}_1{\cal L}_2$, as shown in Figure~\ref{fig:partialFusion}, with a set of junction vectors $v_i \in V_{{\cal L}_1,{\cal L}_2,\overline{{\cal L}_1{\cal L}_2}}$ and $\widetilde v_i \in V_{\overline {\cal L}_2,\overline{\cal L}_1, ({\cal L}_1{\cal L}_2)}$ inserted at the T-junctions. Moreover, $\sum_i v_i \otimes \widetilde v_i$ is uniquely determined in the partial fusion.

\begin{figure}[H]
\centering
\begin{minipage}{0.2\textwidth}
\includegraphics[width=1\textwidth]{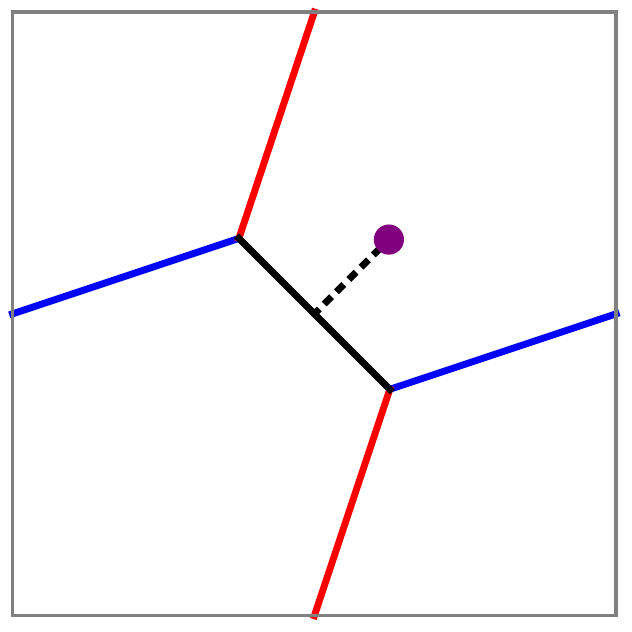}
\end{minipage}%
\begin{minipage}{0.15\textwidth}\begin{eqnarray*}~~\longrightarrow^{\!\!\!\!\!\!\!\!\! S}~ \\ \end{eqnarray*}
\end{minipage}%
\begin{minipage}{0.2\textwidth}
\includegraphics[width=1\textwidth]{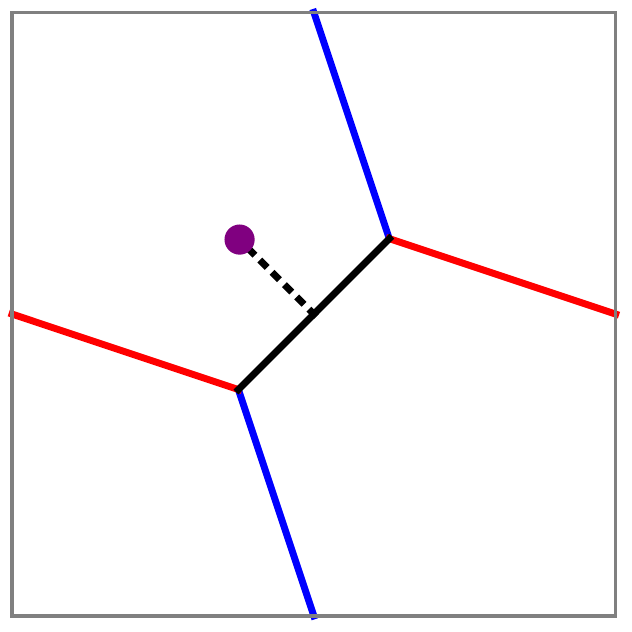}
\end{minipage}%
\caption{The modular $S$ transform of the torus one-point function of a defect operator (purple dot) attached to a TDL graph (consisting of red, blue, and dotted lines, representing TDLs of different types).}
\label{fig:torusmod}
\end{figure}

\noindent{\bf 9. (Modular covariance)} The torus one-point functional of a primary defect operator $\Psi\in{\cal H}_{\cal L}$ attached to a TDL graph $\Gamma$ transforms covariantly under the modular group ${\rm PSL}(2,\mathbb{Z})$. Namely, under the modular $T$ transformation that sends the torus modulus $\tau\to \tau+1$, $\Gamma$ is mapped to a new graph $\Gamma^T$ (with the same set of junction vector spaces) attached to the same defect operator $\Psi$, while preserving the torus correlation functional
\ie
\left\la \Psi(\Gamma^T)\right\ra_{\tau+1,\bar\tau+1} = \left\la \Psi(\Gamma)\right\ra_{\tau,\bar\tau}.
\fe
Under the modular $S$ transformation (as shown in Figure~\ref{fig:torusmod}), $\tau\to -1/\tau$, $\Gamma$ is mapped to $\Gamma^S$ attached to the same $\Psi$, with
\ie
\left\la \Psi(\Gamma^S)\right\ra_{-1/\tau,-1/\bar\tau} =e^{{\pi i\over 2}(h-\tilde h)} (-i\tau)^h (i\bar\tau)^{\tilde h}\left\la \Psi(\Gamma)\right\ra_{\tau,\bar\tau},
\fe
where $(h,\tilde h)$ are the conformal weights of $\Psi$.

We will argue below that the modular covariance of a graph with TDLs on a general punctured Riemann surface follows from the modular covariance of torus one-point functions and the crossing invariance of sphere four-point functions of defect operators (connected via H-junctions), generalizing the results of  \cite{Moore:1988qv,Moore:1989vd,Sonoda:1988mf,Sonoda:1988fq,Bakalov2000}. Here the punctures include both local and defect operators.
The notion of modular invariance in the presence of TDLs/defect operators is such that the partition function on a general punctured Riemann surface can be unambiguously computed by any choice of pairs-of-pants decomposition.   
The boundaries of the pairs-of-paints will generally intersect TDLs, and each pair-of-pants  corresponds to a state in the tensor product of three defect Hilbert spaces.  
The partition function is then given by the contractions of these states in the defect Hilbert spaces.  
Now to show that such a definition of the partition function with defect operators is unambiguous, one simply needs to know that two different pairs-of-paints decompositions give the same answer.  
Once we know that two different decompositions can be related by a unique sequence of simple moves, {\it i.e.}\  the sphere four-point crossing and torus one-point S-transform, it suffices to know that the state on the boundary of a four-punctured sphere or a one-punctured torus in the presence of TDLs is invariant under the simple move. These follow from two of our axioms, namely the H-junction crossing and torus one-point modular covariance.  
We emphasize that in our argument, the simple moves do not depend on the location of TDL junction operators. This is because we are free to slide a loop joining two pair-of-pants along the surface past any topological junctions. 

\subsection{Corollaries}

Let us now derive a number of important corollaries of the defining properties of TDLs. Again, we restrict to CFTs and comment on the non-conformal case at the end.

\subsubsection{H-junction crossing relation} 

By the locality property, an H-junction involving four external TDLs ${\cal L}_1,\cdots,{\cal L}_4$ and an internal TDL ${\cal L}_5$ is a bilinear map
\ie
H_{{\cal L}_2,{\cal L}_3}^{{\cal L}_1,{\cal L}_4}({\cal L}_5)\equiv
\begin{minipage}{0.2\textwidth}
\includegraphics[width=1\textwidth]{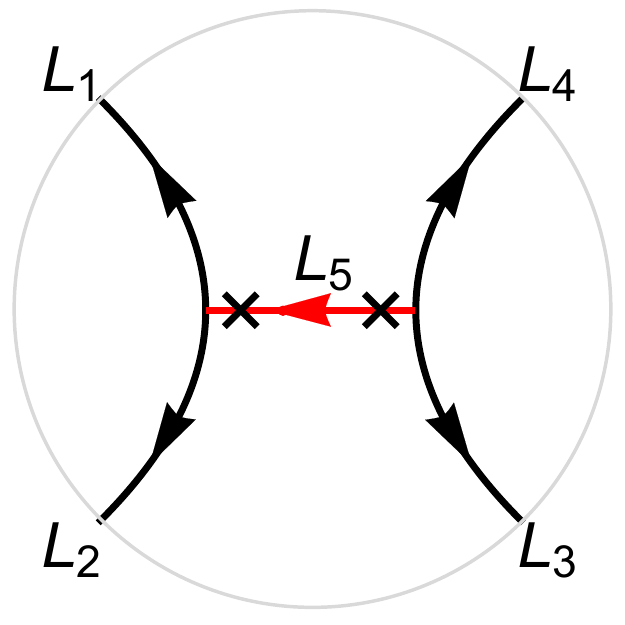}
\end{minipage}%
~:~V_{{\cal L}_1,{\cal L}_2,\overline{\cal L}_5}\otimes V_{{ {\cal L}_3,{\cal L}_4,{\cal L}_5}} \to V_{{\cal L}_1,{\cal L}_2,{\cal L}_3,{\cal L}_4}.
\fe 
Given four simple TDLs ${\cal L}_1,\cdots,{\cal L}_4$, the direct sum of all possible H-junctions gives a map
\ie
H_{{\cal L}_2,{\cal L}_3}^{{\cal L}_1,{\cal L}_4}\equiv\bigoplus_{{\rm simple}~{\cal L}_5} H_{{\cal L}_2,{\cal L}_3}^{{\cal L}_1,{\cal L}_4}({\cal L}_5) ~:~ \bigoplus_{{\rm simple}~{\cal L}_5}V_{{\cal L}_1,{\cal L}_2,\overline{\cal L}_5}\otimes V_{{ {\cal L}_3,{\cal L}_4,{\cal L}_5}} \to V_{{\cal L}_1,{\cal L}_2,{\cal L}_3,{\cal L}_4}.
\fe

\begin{figure}[H]
\centering
\hspace{-.4in}
\begin{minipage}{0.25\textwidth}
\includegraphics[width=1\textwidth]{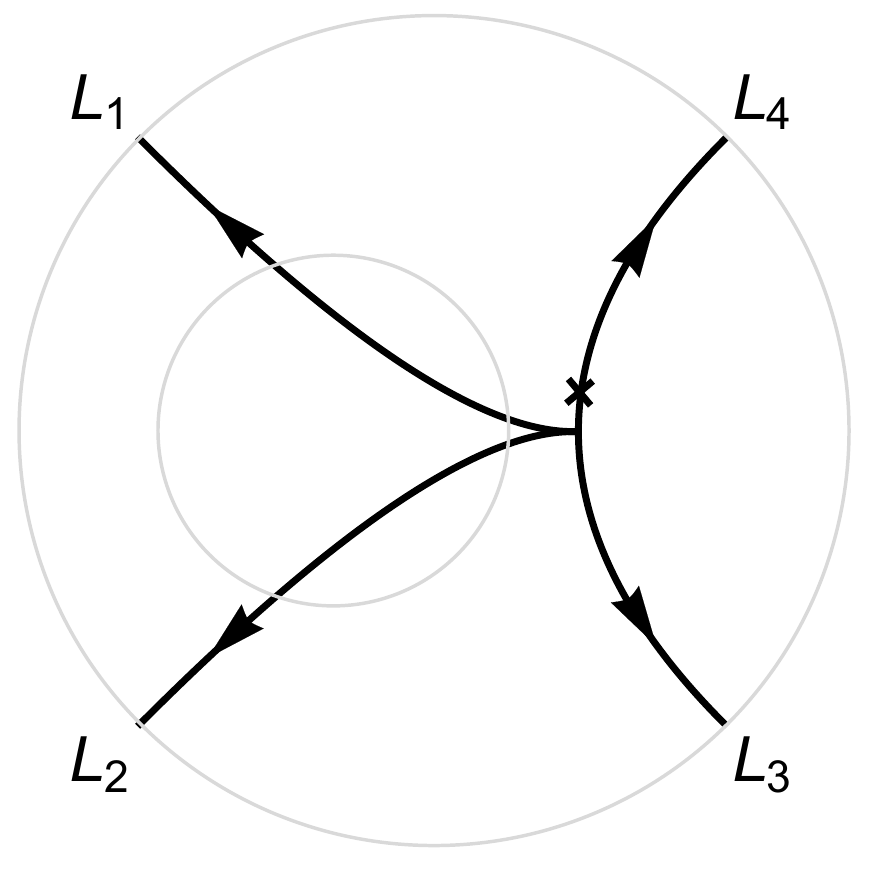}
\end{minipage}%
\begin{minipage}{0.09\textwidth}\begin{eqnarray*}~= \sum_i\\ \end{eqnarray*}
\end{minipage}%
\begin{minipage}{0.25\textwidth}
\includegraphics[width=1\textwidth]{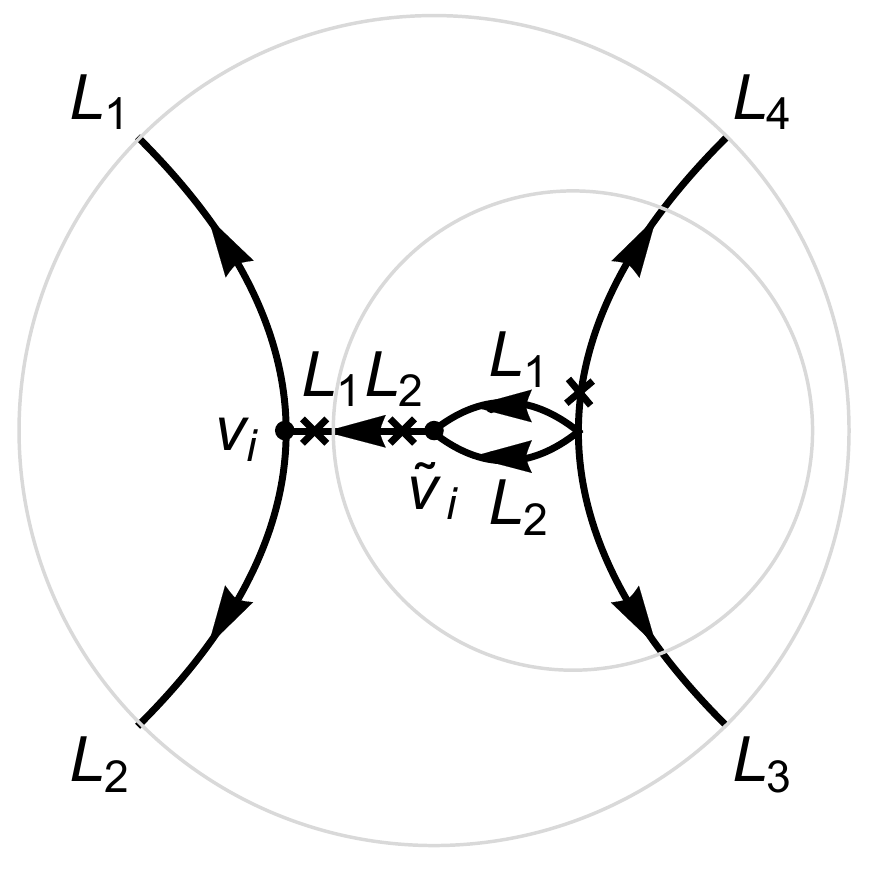}
\end{minipage}%
\begin{minipage}{0.05\textwidth}\begin{eqnarray*}~= \\ \end{eqnarray*}
\end{minipage}%
\begin{minipage}{0.25\textwidth}
\includegraphics[width=1\textwidth]{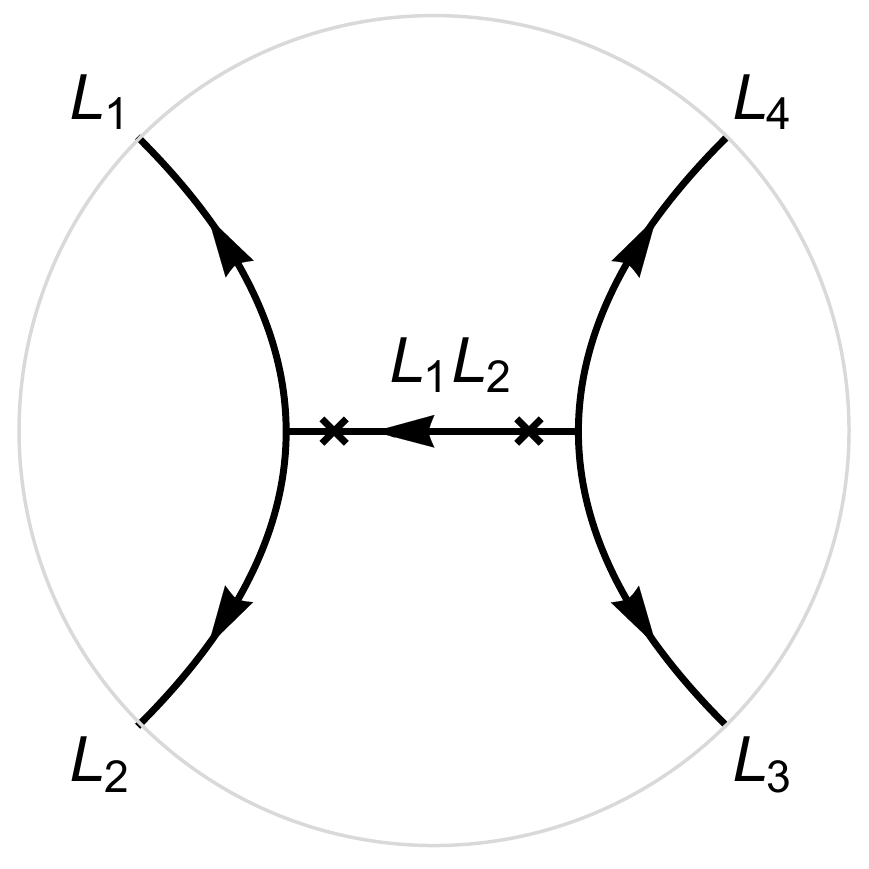}
\end{minipage}%
\begin{minipage}{0.05\textwidth}\begin{eqnarray*}~\circ \overline H_{{\cal L}_2,{\cal L}_3}^{{\cal L}_1,{\cal L}_4} \\ \end{eqnarray*}
\end{minipage}%
\caption{The inverse map $\overline H_{{\cal L}_2,{\cal L}_3}^{{\cal L}_1,{\cal L}_4}$. By the locality property, each graph inside each gray circle represents a multilinear map from the unspecified junction vectors inside the graph to the junction vector space of the lines intersecting the circle. In particular, the graph inside the smaller gray circle in the second graph represents a function that we denote by $f_{\widetilde v_i} : V_{{\cal L}_1,{\cal L}_2,{\cal L}_3,{\cal L}_4}\to V_{{\cal L}_3,{\cal L}_4,({\cal L}_1{\cal L}_2)}$. Then the inverse map is $\overline H_{{\cal L}_2,{\cal L}_3}^{{\cal L}_1,{\cal L}_4} \equiv \sum_iv_i \otimes f_{\widetilde  v_i} : V_{{\cal L}_1,{\cal L}_2,{\cal L}_3,{\cal L}_4}\to V_{{\cal L}_1,{\cal L}_2,\overline{{\cal L}_1{\cal L}_2}}\otimes  V_{{\cal L}_3,{\cal L}_4,({\cal L}_1{\cal L}_2)}$.}
\label{fig:InvH}
\end{figure}

An inverse map $\overline H_{{\cal L}_2,{\cal L}_3}^{{\cal L}_1,{\cal L}_4}$ can be constructed by combining partial fusion and locality, as illustrated in Figure~\ref{fig:InvH}. By the assumption of semi-simplicity, ${\cal L}_1{\cal L}_2$ is a finite sum of simple TDLs. Let ${\cal L}_6$ be a simple line in this sum, and define $\overline H_{{\cal L}_2,{\cal L}_3}^{{\cal L}_1,{\cal L}_4}({\cal L}_6)$ as the projection of $\overline H_{{\cal L}_2,{\cal L}_3}^{{\cal L}_1,{\cal L}_4}$ to the subspace ${V_{{\cal L}_1,{\cal L}_2,\overline{\cal L}_6} \otimes V_{{ {\cal L}_3,{\cal L}_4,{\cal L}_6}}}$.

\begin{figure}[H]
\centering
\begin{minipage}{0.18\textwidth}
\includegraphics[width=1.1\textwidth]{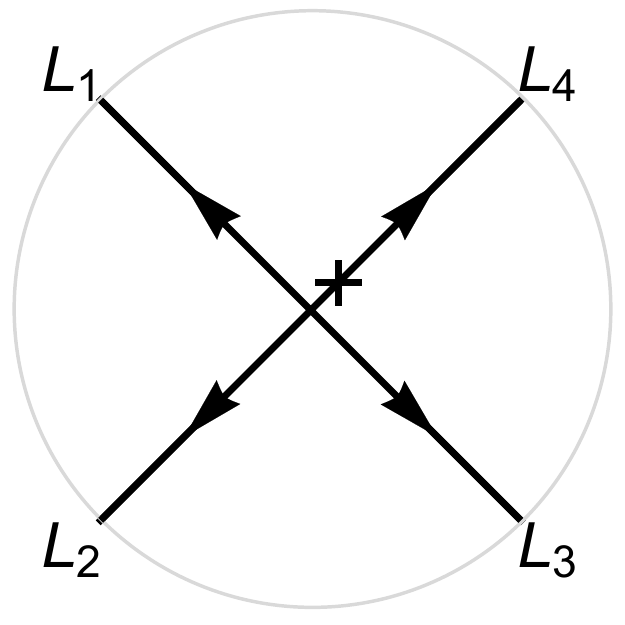}
\end{minipage}%
\begin{minipage}{0.1\textwidth}\begin{eqnarray*} \quad \quad=~\\  \end{eqnarray*}
\end{minipage}%
\begin{minipage}{0.18\textwidth}
\includegraphics[width=1.1\textwidth]{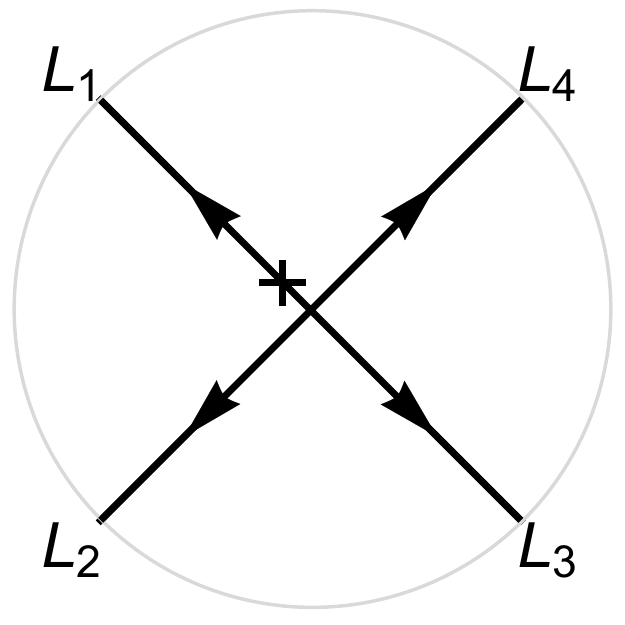}
\end{minipage}
\begin{minipage}{0.2\textwidth}
\begin{eqnarray*}
\circ~ C_{{\cal L}_1,{\cal L}_2,{\cal L}_3,{\cal L}_4}
\\
\end{eqnarray*}
\end{minipage}
\caption{The cyclic permutation map on an X-junction.}
\label{fig:Xperm}
\end{figure}

To formulate the crossing kernel, it is useful to define the \textit{cyclic permutation map}\footnote{This is not to be confused with a rotation. In particular, its action on the junction vectors is in general nontrivial, whereas a rotation acts trivially since junction vectors are topological operators of weight (0,0).
}
\ie
C_{{\cal L}_1,{\cal L}_2,\cdots,{\cal L}_k} ~:~ V_{{\cal L}_1,{\cal L}_2,\cdots,{\cal L}_k}\to V_{{\cal L}_2,{\cal L}_3,\cdots,{\cal L}_k,{\cal L}_1}
\fe
on a $k$-way junction vector spaces, as in Figure~\ref{fig:Xperm} for the case of $k=4$.

\begin{figure}[H]
\centering
\begin{minipage}{0.2\textwidth}
\includegraphics[width=1\textwidth]{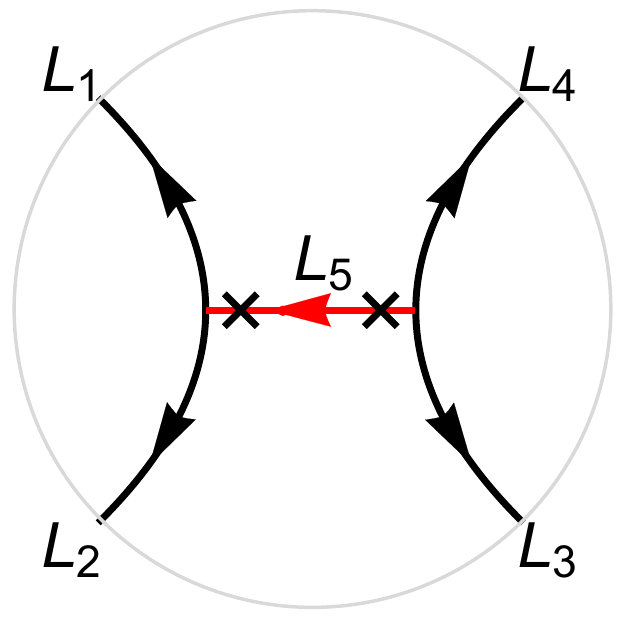}
\end{minipage}%
\begin{minipage}{0.15\textwidth}\begin{eqnarray*} ~\,=\,\sum_{{\rm simple}~{\cal L}_6}\\  \end{eqnarray*}
\end{minipage}%
\begin{minipage}{0.2\textwidth}
\includegraphics[width=1\textwidth]{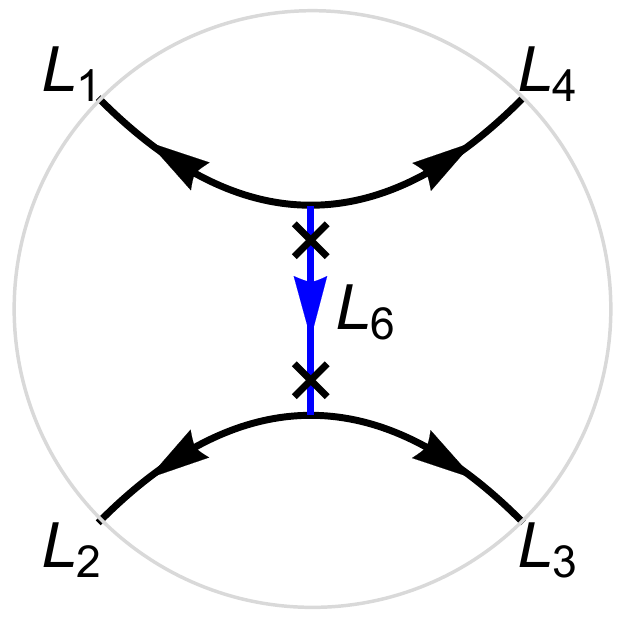}
\end{minipage}
\begin{minipage}{0.2\textwidth}
\begin{eqnarray*}
~\circ~ K_{{\cal L}_2,{\cal L}_3}^{{\cal L}_1,{\cal L}_4}({\cal L}_5,{\cal L}_6)
\\
\end{eqnarray*}
\end{minipage}
\caption{The H-junction crossing relation. Note that the crossing kernel $K_{{\cal L}_2,{\cal L}_3}^{{\cal L}_1,{\cal L}_4}({\cal L}_5, {\cal L}_6)$ is defined such that the internal line is marked as the last leg on both T-junctions.}
\label{fig:Hcross}
\end{figure}

The composition of the map $\overline H_{{\cal L}_3,{\cal L}_4}^{{\cal L}_2,{\cal L}_1}({\cal L}_6)$, the cyclic permutation map $C_{{\cal L}_1,{\cal L}_2,{\cal L}_3,{\cal L}_4}$, and the map $H_{{\cal L}_2,{\cal L}_3}^{{\cal L}_1,{\cal L}_4}({\cal L}_5)$ leads to an overall map
\ie\label{hcrossk}
K_{{\cal L}_2,{\cal L}_3}^{{\cal L}_1,{\cal L}_4}({\cal L}_5,{\cal L}_6) &\equiv \overline H_{{\cal L}_3,{\cal L}_4}^{{\cal L}_2,{\cal L}_1}({\cal L}_6)\circ C_{{\cal L}_1,{\cal L}_2,{\cal L}_3,{\cal L}_4}\circ H_{{\cal L}_2,{\cal L}_3}^{{\cal L}_1,{\cal L}_4}({\cal L}_5)
\\
&~:~V_{{\cal L}_1,{\cal L}_2,\overline{\cal L}_5}\otimes V_{{ {\cal L}_3,{\cal L}_4,{\cal L}_5}} \to V_{{\cal L}_2,{\cal L}_3,\overline{\cal L}_6}\otimes V_{{ {\cal L}_4,{\cal L}_1,{\cal L}_6}},
\fe
where ${\cal L}_5$ and ${\cal L}_6$ are any pair of simple TDLs. We refer to such a linear relation as an {\it H-junction crossing relation}, depicted in Figure~\ref{fig:Hcross}, and $K_{{\cal L}_2,{\cal L}_3}^{{\cal L}_1,{\cal L}_4}({\cal L}_5,{\cal L}_6)$ as the {\it H-junction crossing kernels} that relate H-junctions in the $12 \to 34$ channel to those in the $23 \to 41$ channel.

\begin{figure}[H]
\centering
\begin{minipage}{0.23\textwidth}
\includegraphics[width=1\textwidth]{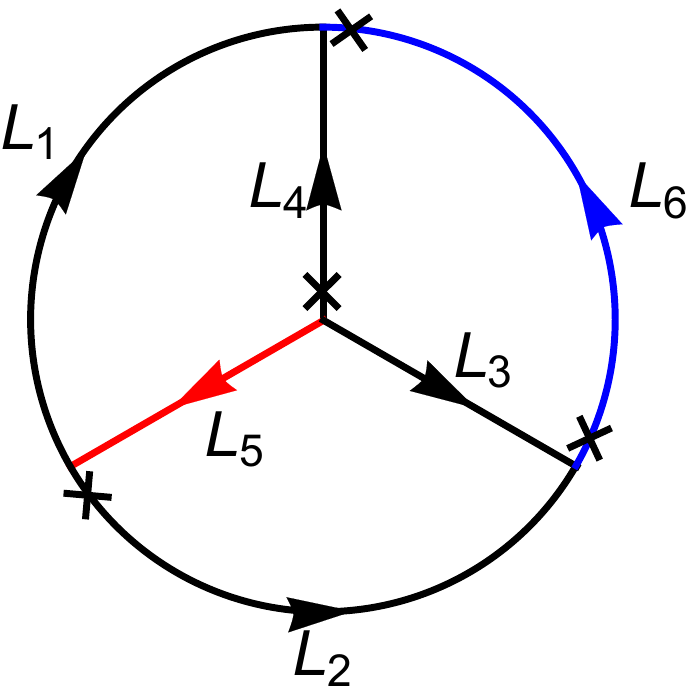}
\end{minipage}%
\caption{The H-junction crossing kernel $K_{{\cal L}_2,{\cal L}_3}^{{\cal L}_1,{\cal L}_4}({\cal L}_5, {\cal L}_6)$ represented as the correlation functional of the Mercedes graph.}
\label{fig:Mercedes}
\end{figure}

The map $K_{{\cal L}_2,{\cal L}_3}^{{\cal L}_1,{\cal L}_4}({\cal L}_5,{\cal L}_6)$ 
regarded as a multi-linear complex-valued function on $V_{\overline{\cal L}_5,{\cal L}_1,{\cal L}_2}\otimes V_{{{\cal L}_5}, {\cal L}_3,{\cal L}_4} \otimes V_{\overline{\cal L}_3,\overline{\cal L}_2,{\cal L}_6}\otimes V_{\overline{\cal L}_1,\overline {\cal L}_4,\overline{\cal L}_6}$, is the same as the correlation functional of the Mercedes TDL graph in Figure~\ref{fig:Mercedes}.

\subsubsection{Pentagon identity}

\begin{figure}[H]
\centering
\begin{minipage}{0.18\textwidth}
\includegraphics[width=1.1\textwidth]{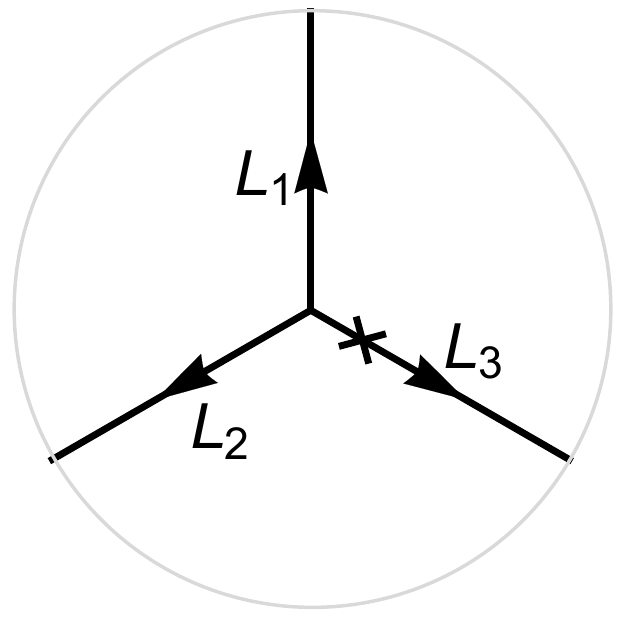}
\end{minipage}%
\begin{minipage}{0.1\textwidth}\begin{eqnarray*} \quad \quad=~\\  \end{eqnarray*}
\end{minipage}%
\begin{minipage}{0.18\textwidth}
\includegraphics[width=1.1\textwidth]{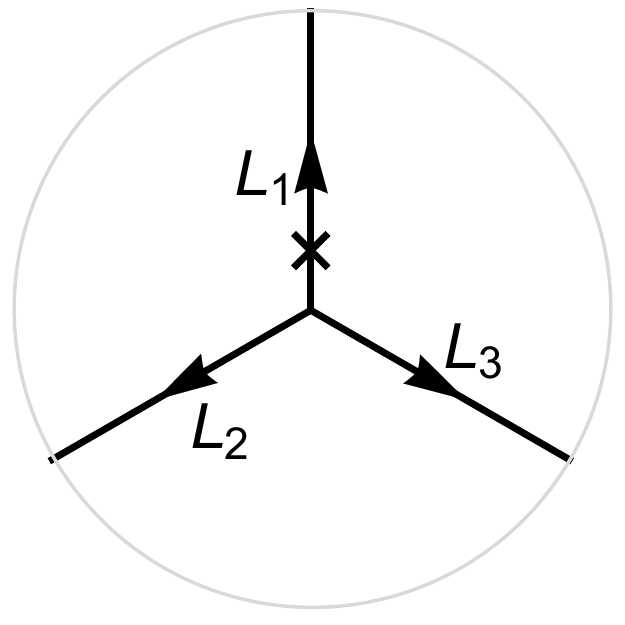}
\end{minipage}
\begin{minipage}{0.2\textwidth}
\begin{eqnarray*}
\circ~ C_{{\cal L}_1,{\cal L}_2,{\cal L}_3}
\\
\end{eqnarray*}
\end{minipage}
\caption{The cyclic permutation map on a T-junction.}
\label{fig:cyclic}
\end{figure}

\begin{figure}[H]
\centering
\begin{minipage}{0.18\textwidth}
\includegraphics[width=1.1\textwidth]{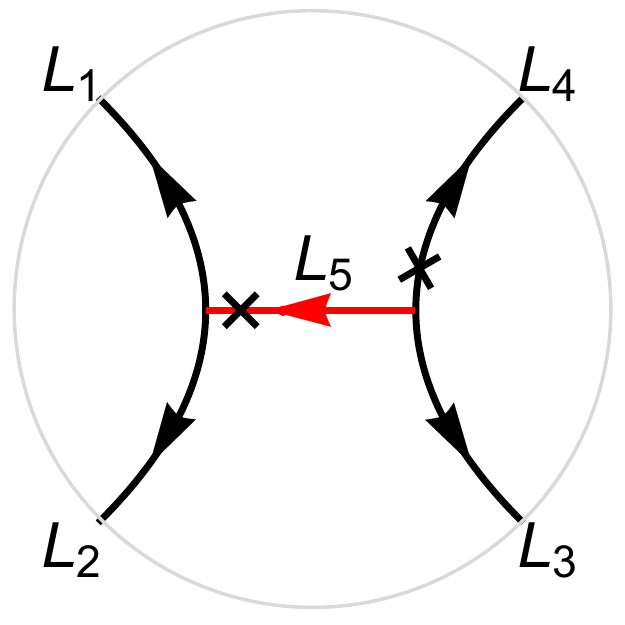}
\end{minipage}%
\begin{minipage}{0.2\textwidth}\begin{eqnarray*} \quad \quad=~\sum_{{\rm simple}~{\cal L}_6}~\\  \end{eqnarray*}
\end{minipage}%
\begin{minipage}{0.18\textwidth}
\includegraphics[width=1.1\textwidth]{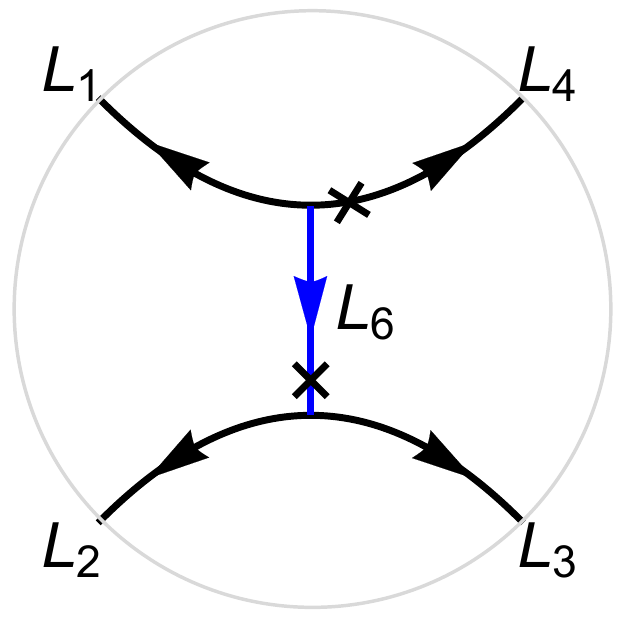}
\end{minipage}
\begin{minipage}{0.2\textwidth}
\begin{eqnarray*}
\quad \circ~ {\widetilde K}_{{\cal L}_2,{\cal L}_3}^{{\cal L}_1,{\cal L}_4}({\cal L}_5,{\cal L}_6)
\\
\end{eqnarray*}
\end{minipage}
\caption{The permuted crossing kernel $\widetilde K$. Note that the last legs at the T-junctions are marked differently in comparison to the crossing kernel $K$ defined in Figure~\ref{fig:Hcross}.}
\label{fig:Khat}
\end{figure}

It follows from the locality property applied to a TDL graph on the disc with five external lines crossing the boundary that the H-junction crossing relations must obey the pentagon identity. Using the cyclic permutation map
\ie
\label{CyclicPerm}
C_{{\cal L}_1,{\cal L}_2,{\cal L}_3} ~:~ V_{{\cal L}_1,{\cal L}_2,{\cal L}_3}\to V_{{\cal L}_2,{\cal L}_3,{\cal L}_1}
\fe
on T-junction vector spaces as in Figure~\ref{fig:cyclic}, the permuted H-junction crossing kernels $\widetilde K$ is related to $K$ by 
\ie\label{permK}
{\widetilde K}_{{\cal L}_2,{\cal L}_3}^{{\cal L}_1,{\cal L}_4}({\cal L}_5,{\cal L}_6) &\equiv C_{{\cal L}_4,{\cal L}_1,{\cal L}_6}\circ K_{{\cal L}_2,{\cal L}_3}^{{\cal L}_1,{\cal L}_4}({\cal L}_5,{\cal L}_6)\circ C_{{\cal L}_5,{\cal L}_3, {\cal L}_4}
\\
&~:~V_{{\cal L}_1,{\cal L}_2,\overline{\cal L}_5}\otimes V_{{{\cal L}_5} ,{\cal L}_3,{\cal L}_4} \to V_{{\cal L}_2,{\cal L}_3,\overline{\cal L}_6}\otimes V_{{\cal L}_1,{{\cal L}_6}, {\cal L}_4},
\fe
such that they appear in the H-junction crossing relations as depicted in Figure~\ref{fig:Khat}.

As illustrated by the following commuting diagram,
\\

\noindent
\begin{tikzcd}
& V_{j_1,j_2,\overline j}\otimes V_{j,j_4 ,j_3}\otimes V_{\overline j_3, k_4 ,k_1 }  
\arrow[rd, "{{\widetilde K}_{j_4,k_4}^{j,k_1}(\overline j_3,k_2)}"]
\arrow[ld,"{{\widetilde K}_{j_2,j_4}^{j_1,j_3}(j,j')}" ]
& & &
\\
V_{j_2,j_4, \overline j'}\otimes V_{j_1,j',j_3  }\otimes V_{\overline j_3 ,k_4 ,k_1 }  
\arrow[dd,"{{\widetilde K}_{j',k_4}^{j_1,k_1}(\overline j_3,\overline k_3)}"]
& &V_{j_1,j_2,\overline j}\otimes V_{j_4, k_4, \overline k_2}\otimes V_{j,k_2 ,k_1  } 
\arrow[dd,"{{\widetilde K}_{j_2,k_2}^{j_1,k_1}(j,\overline k_3)}"] 
\\
\\
V_{j_2,j_4 ,\overline j'}\otimes V_{j' ,k_4 , k_3}\otimes V_{j_1,\overline k_3,k_1  }
\arrow[rr,"{{\widetilde K}_{j_4,k_4}^{j_2,k_3}(j',k_2)}"]
&& V_{j_2 ,k_2  ,k_3}\otimes V_{j_4, k_4, \overline k_2} \otimes V_{ j_1,\overline k_3, k_1}
\end{tikzcd}
\\

\noindent
the permuted crossing kernels $\widetilde K$ satisfy the pentagon identity
\ie\label{eqn:pentagon}
& {\widetilde K}_{{\cal L}_{j_2},{\cal L}_{k_2}}^{{\cal L}_{j_1},{\cal L}_{k_1}}({\cal L}_{j},\overline{\cal L}_{k_3})\circ {\widetilde K}_{{\cal L}_{j_4},{\cal L}_{k_4}}^{{\cal L}_{j},{\cal L}_{k_1}}(\overline{\cal L}_{j_3},{\cal L}_{k_2})
\\
&= \sum_{j'}{\widetilde K}_{{\cal L}_{j_4},{\cal L}_{k_4}}^{{\cal L}_{j_2},{\cal L}_{k_3}}({\cal L}_{j'},{\cal L}_{k_2})\circ {\widetilde K}_{{\cal L}_{j'},{\cal L}_{k_4}}^{{\cal L}_{j_1},{\cal L}_{k_1}}(\overline{\cal L}_{j_3},\overline{\cal L}_{k_3})\circ {\widetilde K}_{{\cal L}_{j_2},{\cal L}_{j_4}}^{{\cal L}_{j_1},{\cal L}_{j_3}}({\cal L}_{j},{\cal L}_{j'}).
\fe
In the above diagram, we abbreviated ${\widetilde K}_{{\cal L}_{j_2},{\cal L}_{j_3}}^{{\cal L}_{j_1},{\cal L}_{j_4}}({\cal L}_{j_5},{\cal L}_{j_6})$ by ${\widetilde K}_{j_2, j_3}^{j_1, j_4}(j_5,j_6)$, and $V_{{\cal L}_{j_1},{\cal L}_{j_2},{\cal L}_{j_3}}$ by $V_{j_1,j_2,j_3}$.

The admissible crossing relations are classified by the solutions to the pentagon identity. The solutions are rigid in the sense that they admit no continuous deformation, modulo the gauge transformations corresponding to the change of basis vectors in each junction vector space (see Appendix~\ref{CKbasis}). This is a proven property in category theory known as Ocneanu rigidity \cite{10.2307/20159926}. Since the solution set is discrete, a solution cannot  change continuously under the RG flow, and hence the crossing relations should match between the UV and the IR theories.  This is a key property that will be used later to constrain various RG flows.

\subsubsection{Action on bulk local operators and defect operators}\label{sec:action} 

\begin{figure}[H]
\centering
\begin{minipage}{0.05\textwidth}~
\end{minipage}%
\begin{minipage}{0.14\textwidth}
\includegraphics[width=1\textwidth]{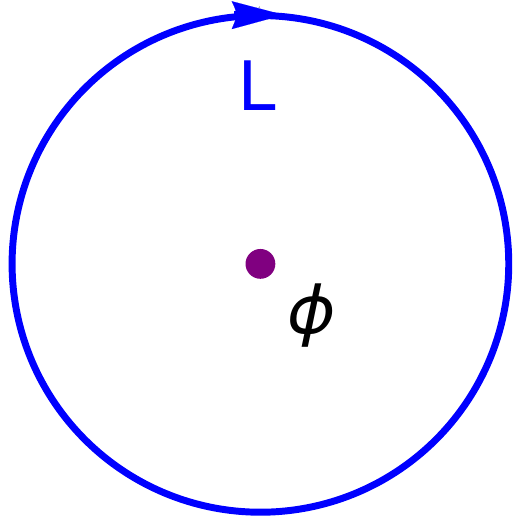}
\end{minipage}%
\begin{minipage}{0.01\textwidth}\begin{eqnarray*}~=~ \widehat{\cal L}\cdot \phi \\ \end{eqnarray*}
\end{minipage}%
\begin{minipage}{0.2\textwidth}~
\end{minipage}%
\begin{minipage}{0.3\textwidth}
\includegraphics[width=1\textwidth]{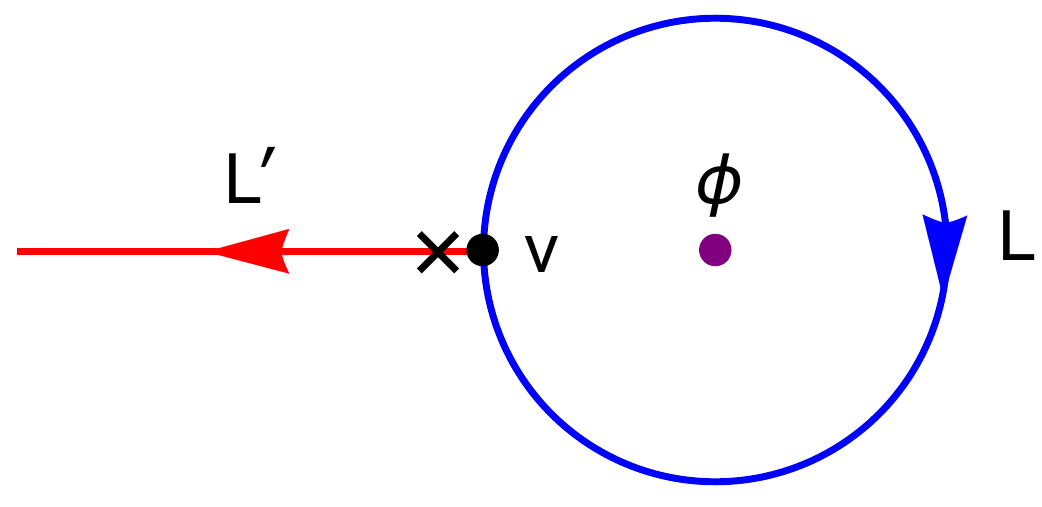}
\end{minipage}%
\begin{minipage}{0.07\textwidth}\begin{eqnarray*}~= \\ \end{eqnarray*}
\end{minipage}%
\begin{minipage}{0.3\textwidth}
\includegraphics[width=1\textwidth]{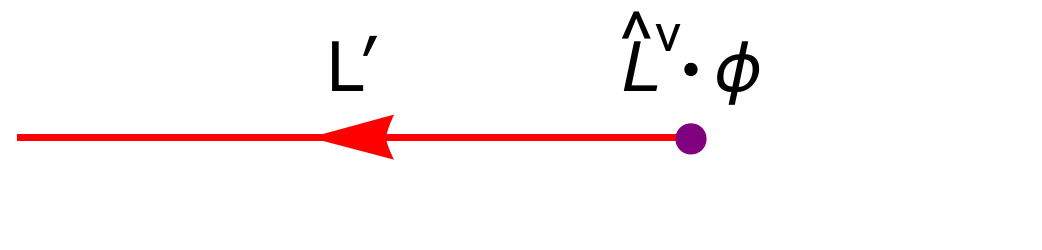}
\end{minipage}%
\caption{{\bf Left:} Contracting a TDL ${\cal L}$ loop on a bulk local operator $\phi$ produces another bulk local operator $(\widehat{\cal L} \cdot \phi)$ of the same weight. {\bf Right:} ``Lassoing" the bulk local operator $\phi$ with an ${\cal L}$ loop attached to the ${\cal L}'$ line produces a defect operator in ${\cal H}_{{\cal L}'}$.}
\label{fig:Lact}
\end{figure}

\begin{figure}[H]
\centering
\begin{minipage}{0.3\textwidth}
\includegraphics[width=1\textwidth]{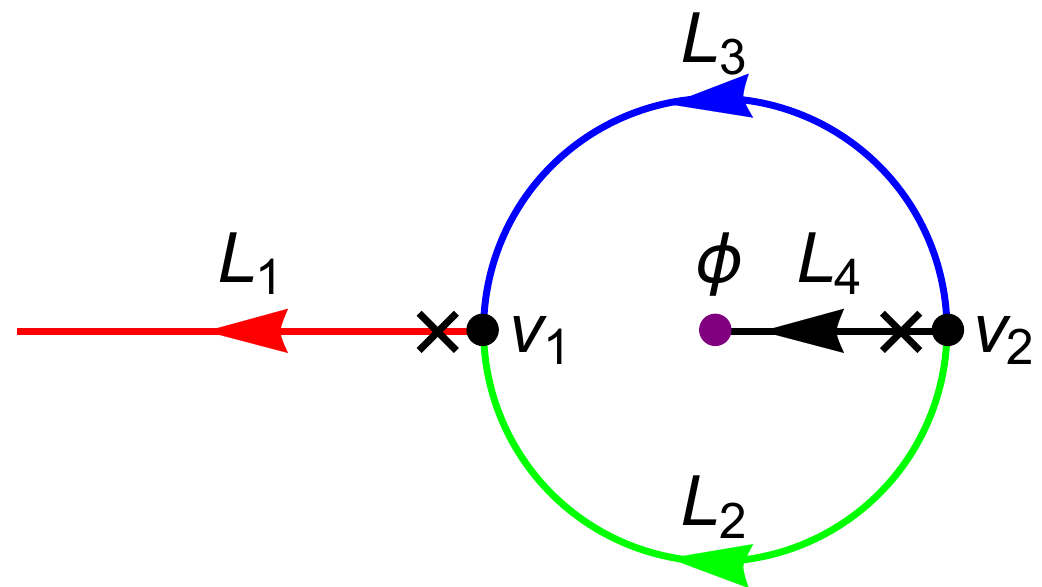}
\end{minipage}%
\caption{ ``Lassoing" the defect operator $\phi$ in ${\cal H}_{{\cal L}_4}$ by the TDLs ${\cal L}_2$ and ${\cal L}_3$ joined by T-junctions with ${\cal L}_1$ and ${\cal L}_4$ produces a defect operator in ${\cal H}_{{\cal L}_1}$.}
\label{fig:Lact2}
\end{figure}

\begin{figure}[H]
\centering
\begin{minipage}{0.2\textwidth}
\includegraphics[width=1\textwidth]{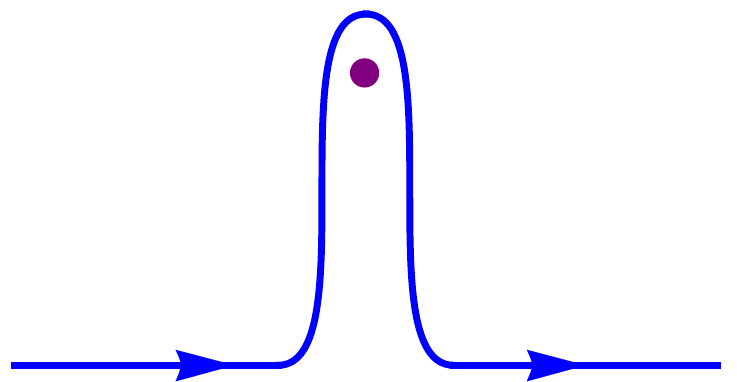}
\end{minipage}%
\begin{minipage}{0.05\textwidth}\begin{eqnarray*}~=~ \\ \end{eqnarray*}
\end{minipage}%
\begin{minipage}{0.2\textwidth}
\includegraphics[width=1\textwidth]{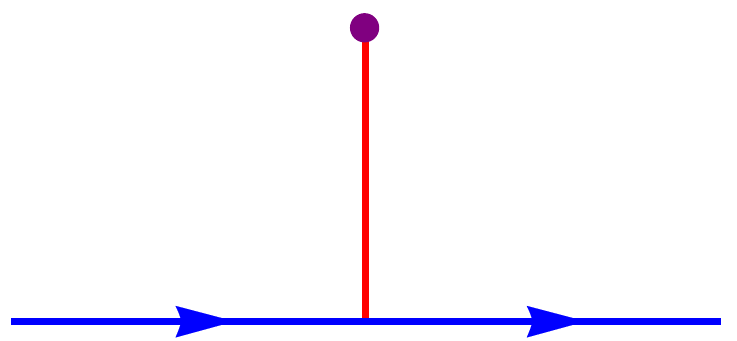}
\end{minipage}%
\caption{Moving a TDL ${\cal L}$ (blue) past a bulk local operator leaves behind another (possibly non-simple) TDL (red) attached to a defect operator and a T-junction (with a specified junction vector).}
\label{fig:LactM}
\end{figure}

Given a TDL ${\cal L}$, let $\widehat{\cal L}:{\cal H} \rightarrow {\cal H}$ be the linear operator on the bulk Hilbert space ${\cal H}$ of the CFT on the cylinder defined by wrapping an ${\cal L}$ loop around the spatial circle. Assuming no isotopy anomaly, we can think of $\widehat{\cal L}$ as an operation on a bulk local operator $\phi$ by contracting an ${\cal L}$ loop encircling $\phi$.\footnote{In the presence of isotopy anomaly, $\widehat{\cal L}$ differs from the left of Figure~\ref{fig:Lact} by a phase, a scenario which we ignore for the moment, and return to in  Section~\ref{isoanomaly}.} 
 As already mentioned, the isotopy invariance property implies that  $\widehat{\cal L}$ preserves the conformal weight of $\phi$. 
More generally, as shown in Figure~\ref{fig:Lact}, contracting a TDL ${\cal L}$ loop on a bulk local operator $\phi$, where ${\cal L}$ has another TDL ${\cal L}'$ attached to it with a junction vector $v$, produces a defect operator $(\widehat{\cal L}^v \cdot \phi)$ in ${\cal H}_{{\cal L}'}$ of the same conformal weight as $\phi$. We refer to this operation as ``lassoing'' for its resemblance. Note that the information of ${\cal L}'$ is implicit in the notation of $\widehat{\cal L}^v: {\cal H} \rightarrow {\cal H}_{ {\cal L}'}$, since $v \in V_{{\cal L}',\overline{\cal L},{\cal L}}$. A similar lassoing can be performed when the TDL loop circles a defect operator, as in Figure~\ref{fig:Lact2}. Combined with crossing (partial fusion), passing a bulk local operator (or a defect operator) through a TDL $\cal L$ generally introduces a T-junction that ends on a defect operator, as shown in Figure~\ref{fig:LactM}.

\subsubsection{Vacuum expectation value $\la {\cal L} \ra$ and defect Hilbert space ${\cal H}_{\cal L}$} \label{Sec:Vac}
If theory has a unique vacuum, which we denote by $|0\ra$, then we can define
\begin{align}
\la{\cal L}\ra \equiv \la 0|\widehat {\cal L}|0\ra
\end{align}
as the expectation value of an empty ${\cal L}$ loop on a cylinder. $\la {\cal L}\ra$ is a fundamental quantity for a TDL that we will repeatedly make use of.


We first show that $\la {\cal L} \ra \geq 0$ in a unitary theory with a unique ground state. Define the ${\cal L}$-twisted torus character with $\widehat{\cal L}$ acting on the bulk Hilbert space,
\ie
Z^{\cal L}(\tau,\bar\tau) \equiv {\rm tr}\, \widehat{\cal L} \, q^{L_0 - {c\over 24}}\bar q^{\bar L_0 - {\tilde c\over 24}},
\fe
where $q=e^{2\pi i \tau}$. This is related to the torus partition function of the Hilbert space  ${\cal H}_{\cal L}$ by a modular $S$ transformation
\ie\label{SZL}
Z^{\cal L}(-1/\tau,-1/\bar\tau) = Z_{\cal L}(\tau,\bar\tau) \equiv {\rm tr}_{{\cal H}_{\cal L}} q^{L_0 - {c\over 24}}\bar q^{\bar L_0 - {\tilde c\over 24}}\,.
\fe
Let us take $\tau =-\bar \tau = it$ and send $t\to 0$, then $Z^{\cal L}(-1/\tau,-1/\bar \tau)$ only picks up the contribution $\la{\cal L}\ra e^{{\pi c \over 12t}}$ from the one ground state.\footnote{This argument also holds in non-unitary CFTs, where the ground state may not correspond to the identity operator, as long as the degeneracies are non-negative ({\it e.g.}, no ghost). For example, in the Lee-Yang model, a.k.a. the (2,5) minimal model, the limit $t\to 0$ of $Z^{\cal L}(-1/\tau ,-1/\bar \tau)$ is dominated not by the identity operator, but by the primary of weight $(-{1\over5},-{1\over5})$.
}
In this limit, \eqref{SZL} reduces to
\begin{align}
\label{Lge0}
\la {\cal L} \ra    = \lim_{t\to 0}e^{-{\pi c \over 12t}} \text{tr} _{{\cal H}_{\cal L}} e^{-2\pi t ( L_0 +\bar L_0 -{c \over 12})} \ge 0\,.
\end{align}

An important corollary of the locality property is that the set of $\la {\cal L}\ra$ satisfies a system of polynomial equations with {\it positive integer} coefficients, given by the abelianization of the fusion ring. In a unitary, {\it compact} CFT, this leads to the stronger constraint
\ie
\label{Lge1}
\la \cL \ra \geq 1,
\fe
since by compactness we expect $\la \cL\ra$ for all TDLs $\cL$ to be bounded below by a positive number.\footnote{Using the folding trick, one can relate the TDL $\cL$ in a CFT $\cal T$ to a boundary state $\cal B$ in the tensor product CFT $\overline{\cal T}\otimes \cal T$. Furthermore $\la \cL \ra $ maps to the $g$-function of $\cal B$ ($\log g$ being the boundary entropy). When $\overline{\cal T}\otimes \cal T$ is described by a nonlinear sigma model, for instance, $\cal B$ corresponds to a D-brane wrapping a submanifold of the target space whose mass is proportional to the $g$-function. Thus a vanishing $g$ can only occur in a singular sigma model. More generally, we expect the $g$-function, and hence $\la {\cal L}\ra$, to be bounded from below by a positive number in any unitary, {\it compact} CFT.}
Had there been any $\cL$ with $\la \cL\ra<1$, the sequence of TDLs $\cL^n$ with $n \in \mathbb{N}$ would go to zero and hence violate this condition.

Furthermore, since $\la {\cal L}\ra$ satisfies the polynomial equations, it is protected under RG flow. In the case where there are degenerate vacua in the IR, $\la {\cal L}\ra$ is defined as the cylinder expectation value in the vacuum inherited from the UV.

Given the expectation \eqref{Lge1}, we can run the modular invariance argument in reverse to argue that in a unitary, compact CFT with a unique vacuum, the Hilbert space ${\cal H}_{\cal L}$ of defect operators at the end of a TDL $\cal L$ must always be non-empty,
\ie\label{HLnonempty}
{\cal H}_{\cal L}\neq \emptyset.
\fe
Importantly, if we relax the condition of a unique vacuum, then there could be TDLs on which no defect operator can end.  We will encounter such an example in some TQFTs with degenerate vacua in Section~\ref{tritoising}.

Let us also introduce the expectation value of an empty clockwise ${\cal L}$ loop on the plane, which we denote by $R({\cal L})$, and the expectation value of a counterclockwise ${\cal L}$ loop, which we denote by $R(\overline{\cal L})$.  The expectation value on a plane $R({\cal L})$ might not equal to the expectation value on a cylinder $\la {\cal L}\ra $. Their relation will be discussed in Section~\ref{isoanomaly}.

\begin{figure}[H]
\centering
\begin{minipage}{0.2\textwidth}
\includegraphics[width=1\textwidth]{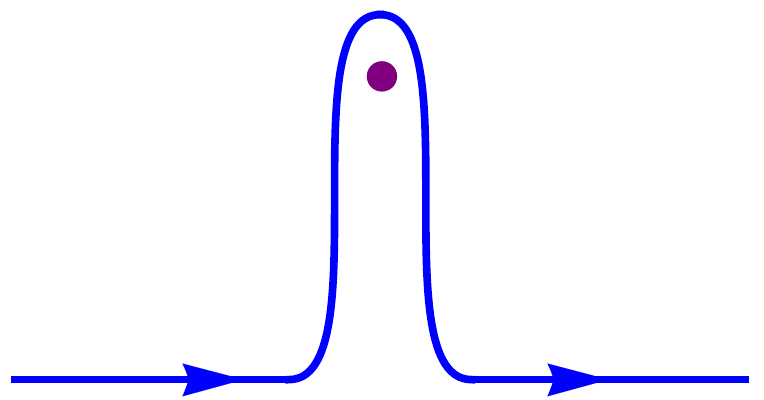}
\end{minipage}%
\begin{minipage}{0.05\textwidth}\begin{eqnarray*}~=~ \\ \end{eqnarray*}
\end{minipage}%
\begin{minipage}{0.2\textwidth}
\includegraphics[width=1\textwidth]{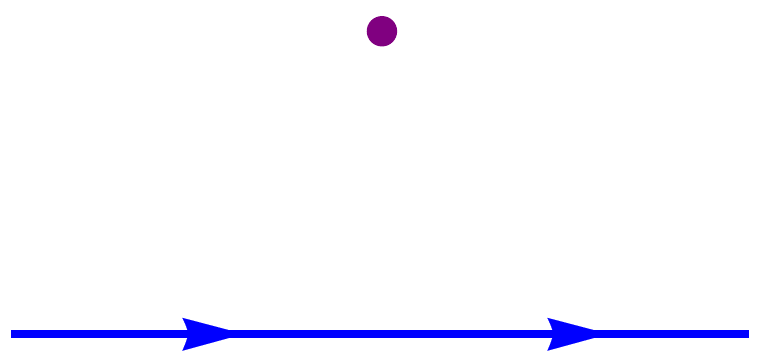}
\end{minipage}%
\begin{minipage}{0.05\textwidth}\begin{eqnarray*}~+~ \\ \end{eqnarray*}
\end{minipage}%
\begin{minipage}{0.2\textwidth}
\includegraphics[width=1\textwidth]{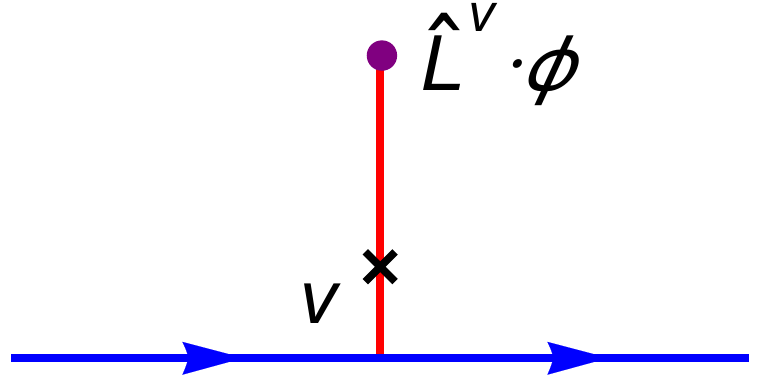}
\end{minipage}%
\caption{Moving a TDL $\cL$ (blue) past a bulk local operator $\phi$ (purple) that satisfies \eqref{com1}. The third diagram captures all the contributions involving non-identity TDLs attached to $\cL$ via a junction vector $v$.}
\label{fig:com}
\end{figure}

\begin{figure}[H]
\centering
\begin{minipage}{0.2\textwidth}
\includegraphics[width=1\textwidth]{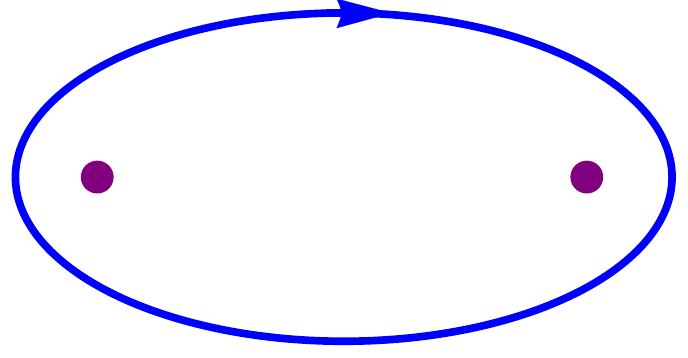}
\end{minipage}%
\begin{minipage}{0.05\textwidth}\begin{eqnarray*}~=~ \\ \end{eqnarray*}
\end{minipage}%
\begin{minipage}{0.2\textwidth}
\includegraphics[width=1\textwidth]{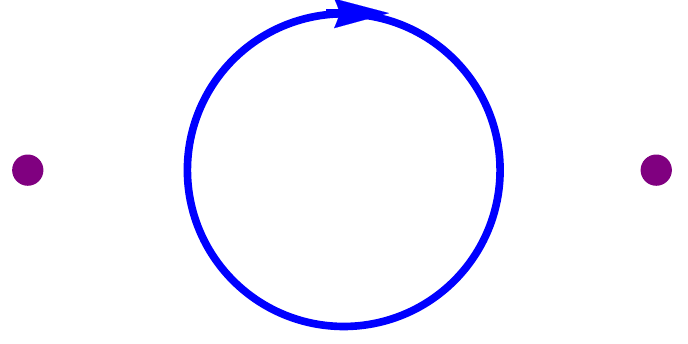}
\end{minipage}%
\begin{minipage}{0.05\textwidth}\begin{eqnarray*}~+~ \\ \end{eqnarray*}
\end{minipage}%
\begin{minipage}{0.22\textwidth}
\includegraphics[width=1\textwidth]{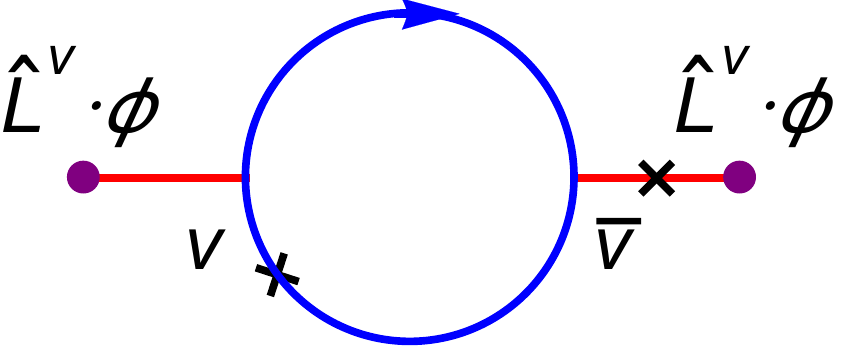}
\end{minipage}%
\caption{Gluing two copies of the diagrams in Figure~\ref{fig:com} by conjugation. Here we have dropped diagrams on the LHS that vanish by the tadpole condition (see next subsection). }
\label{fig:comg}
\end{figure}

A bulk local operator $\phi$ is said to \textit{commute} with a TDL ${\cal L}$ if they commute as operators on the cylinder. In particular, this requires
\begin{align}
\widehat{\cal L}|\phi\ra = \la{\cal L}\ra |\phi\ra\,.
\label{com1}
\end{align}
In unitary theories, \eqref{com1} further implies that $\widehat\cL^v\cdot \phi =0$. To see this, we start with the configuration in Figure~\ref{fig:com}.
Next we glue two copies of the diagrams in Figure~\ref{fig:com} together by summing over the states in ${\cal H}_{\cL,\overline\cL}$ and obtain the relation in Figure~\ref{fig:comg}.
The correlation function of the TDL configurations in Figure~\ref{fig:comg} is such that the first diagram on the RHS saturate the contribution from the LHS. Consequently, the last diagram in  Figure~\ref{fig:comg} must vanish as a two point function of $\widehat \cL^v \cdot \phi$ and its conjugate. By unitarity we conclude  $\widehat\cL^v\cdot \phi =0$. 
In other words, the TDL $\cal L$ does not ``feel" the insertion of such a bulk local operator $\phi$ when they pass through each other (the last diagram in  Figure~\ref{fig:com} vanishes). This is sufficient to ensure that $\phi$ commutes with $\cL$.\footnote{We thank Davide Gaiotto for discussion on this point.}

\subsubsection{Vanishing tadpole} \label{Sec:VanishingTadpole}
Under some mild assumptions that we will specify below, one can show that a TDL configuration containing a tadpole -- that is, a nontrivial simple TDL ${\cal L}'$ ending on a loop ${\cal L}$, as shown on the left of Figure~\ref{fig:Lact}, with the operator $\phi $ chosen to be the identity -- has vanishing correlation functional.  We will refer to this property as the vanishing tadpole property.  

\begin{figure}[H]
\centering
\begin{minipage}{0.25\textwidth}
\includegraphics[width=1\textwidth]{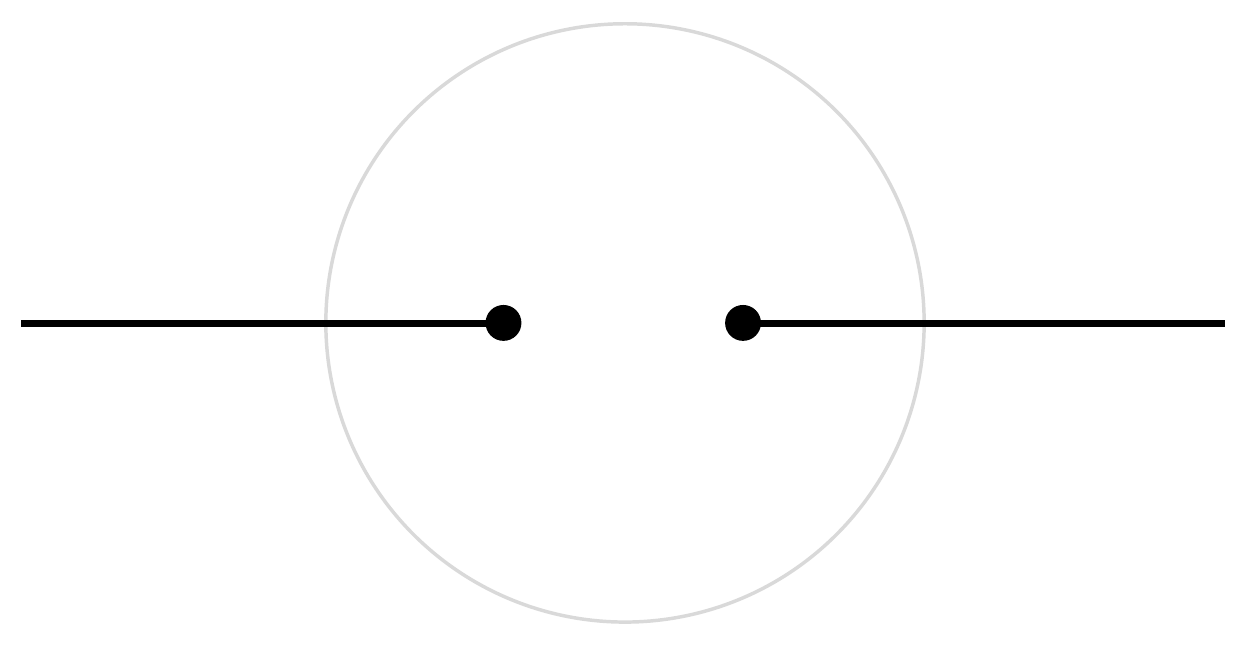}
\end{minipage}%
\begin{minipage}{0.05\textwidth}\begin{eqnarray*}~\propto~ \\ \end{eqnarray*}
\end{minipage}%
\begin{minipage}{0.25\textwidth}
\includegraphics[width=1\textwidth]{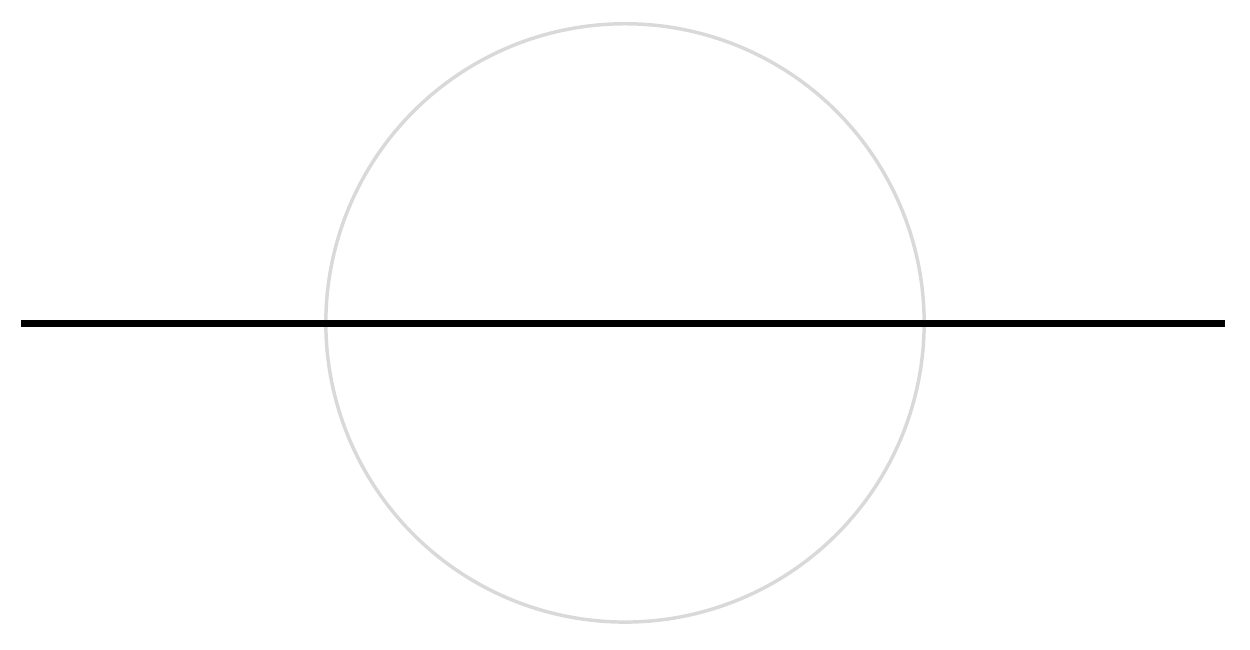}
\end{minipage}%
\caption{The OPE of a pair of weight-$(0,0)$ topological defect operators in $V_{{\cal L}'}$  gives the simple TDL ${\cal L}'$.}
\label{fig:Open}
\end{figure}

In order to argue the vanishing tadpole property, let us first prove a lemma: if the collection of all TDLs acts \textit{faithfully} on the bulk local operators, {\it i.e.}, the only TDL that commutes with all bulk local operators is the trivial TDL, then $V_{{\cal L}'}$ is an empty set. We prove by contradiction. Since $V_{{\cal L}',\overline {\cal L}'}$ is one-dimensional, the OPE of a pair of  defect operators in $V_{{\cal L}'}$ should be proportional to inserting the trivial defect operator in $V_{{\cal L}',\overline {\cal L}'}$, which is equivalent to inserting nothing on ${\cal L}'$ (see Figure~\ref{fig:Open}).
In other words, such an ${\cal L}'$ line can be ``opened up" with the weight-(0,0) defect operators in $V_{{\cal L}'}$ inserted at the two endpoints. It then follows that ${\cal L}'$ commutes with all bulk local operators, and hence must be a trivial line by our assumption of faithfulness. This contradicts with the initial assumption that ${\cal L}'$ is nontrivial.
Note that in a TQFT that arises at the end of an RG flow from a CFT, the Hilbert space ${\cal H}_{{\cal L}',\overline {\cal L}'}$ is generally multi-dimensional, but the junction vector space $V_{{\cal L}',\overline {\cal L}'}$ inherited from the CFT is one-dimensional.

To argue that tadpoles vanish, we use the locality property to cut a disk that contains the tadpole, and replace the TDL configuration by ${\cal L}'$ ending on a particular defect operator in $V_{{\cal L}'}$.
The latter is an empty set  by our lemma. Hence, the vanishing tadpole property holds for a CFT that satisfies the faithfulness assumption.

The faithfulness assumption is typically true in a compact CFT with a unique vacuum, but is generally not preserved under RG flows.  For example, consider the $\mathbb{Z}_2$ line which acts nontrivially in the critical Ising model, and the  RG flow triggered by the energy operator $\varepsilon$. Since $\varepsilon$ commutes with the $\mathbb{Z}_2$ line, the latter is preserved along the entire RG flow.  For one sign of the coupling, the IR theory has a unique vacuum on which the $\mathbb{Z}_2$ acts trivially \cite{Onsager,Yang}, and hence violates the faithfulness assumption in the IR.   
Nonetheless, since the vanishing tadpole property holds true in the UV, it must persist under the RG flow. Therefore, we can still constrain the IR TQFT arising at the end of a massive RG flow from a CFT with vanishing tadpoles. Various concrete examples will be given in Section~\ref{Sec:specificflow}.

\subsubsection{Trivial junctions} \label{Sec:Trivial} 

\begin{figure}[H]
\centering
\begin{minipage}{0.18\textwidth}
\includegraphics[width=1\textwidth]{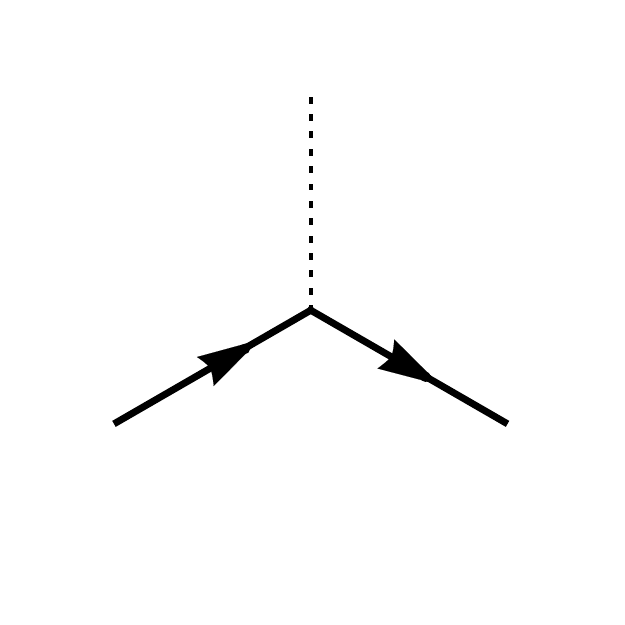}
\end{minipage}%
\begin{minipage}{0.05\textwidth}\begin{eqnarray*}~=~  \\ \end{eqnarray*}
\end{minipage}%
\begin{minipage}{0.18\textwidth}
\includegraphics[width=1\textwidth]{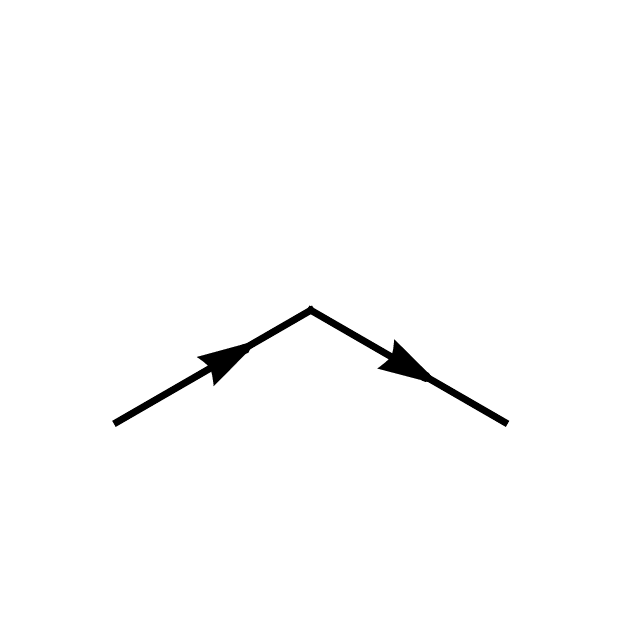}
\end{minipage}%
\caption{Removal of a trivial junction (with identity junction vector).}
\label{fig:trivialjunction}
\end{figure}

The fusion of a simple TDL ${\cal L}$ with its orientation reversal $\overline{\cal L}$ (which may or may not be equivalent to ${\cal L}$) contains the trivial TDL $I$ as a direct summand with multiplicity 1. {\it A priori}, according to the ordering of the legs, there are three ``trivial junction" vector spaces $V_{\overline{\cal L},I,{\cal L}}$, $V_{I,{\cal L},\overline{\cal L}}$, and $V_{{\cal L},\overline{\cal L},I}$, all of which are isomorphic to $\bC$. However, there exist canonical choices for these junction vectors, which we denote by ${1}_{\overline{\cal L}, I ,{\cal L}}$, ${1}_{ I ,{\cal L},\overline{\cal L}}$, and ${1}_{{\cal L},\overline{\cal L},I}$, such that the correlation functional of a TDL graph that contains a trivial junction evaluated on the identity junction vector is equivalent to that of the TDL graph where the trivial junction is forgotten (Figure~\ref{fig:trivialjunction}). It follows from this definition that the identity junction vectors map to themselves under the cyclic permutation maps,
\ie
C_{\overline{\cal L}, I ,{\cal L}}({1}_{\overline{\cal L}, I ,{\cal L}})={1}_{ I ,{\cal L},\overline{\cal L}},\quad C_{ I ,{\cal L},\overline{\cal L}}({1}_{ I ,{\cal L},\overline{\cal L}})={1}_{{\cal L},\overline{\cal L},I},\quad C_{ {\cal L},\overline{\cal L},I}({1}_{ {\cal L},\overline{\cal L},I})={1}_{\overline{\cal L},I,{\cal L}}.
\fe

The two-point functions of the canonical identity junction vectors are
\ie
h({1}_{\overline{\cal L}, I ,{\cal L}},{1}_{\overline{\cal L}, I ,{\cal L}})=R({\cal L}),\quad h({1}_{{\cal L},\overline{\cal L}, I },{1}_{I ,{\cal L},\overline{\cal L}})=R(\overline {\cal L}).
\fe
By Property~6, the antiunitarity of the conjugation map implies that the norm of an identity junction vector is equal to $\sqrt{|R({\cal L})|} = \sqrt{\la {\cal L}\ra}$ (see Section~\ref{isoanomaly} for the relation between $R({\cal L})$ and $\la {\cal L}\ra$).

When evaluating on the identity junction vectors, the crossing kernels with a trivial external line in three of the four positions become
\ie\label{singleID}
&{\widetilde K}_{{\cal L}_2,{\cal L}_3}^{I,{\cal L}_4}({\cal L}_2,\overline{\cal L}_4)~:~{1}_{I,{\cal L}_2,\overline{\cal L}_2}\otimes v_{{{\cal L}_2}, {\cal L}_3,{\cal L}_4} \mapsto v_{{\cal L}_2,{\cal L}_3,{\cal L}_4}\otimes {1}_{I,{\overline{\cal L}_4}, {\cal L}_4},
\\
&{\widetilde K}_{I,{\cal L}_3}^{{\cal L}_1,{\cal L}_4}({\cal L}_1,{\cal L}_3)~:~{1}_{{\cal L}_1,I,\overline{\cal L}_1}\otimes v_{{{\cal L}_1}, {\cal L}_3,{\cal L}_4} \mapsto {1}_{I,{\cal L}_3,\overline{\cal L}_3}\otimes v_{{\cal L}_1,{{\cal L}_3}, {\cal L}_4},
\\
&{\widetilde K}_{{\cal L}_2,I}^{{\cal L}_1,{\cal L}_4}(\overline{\cal L}_4,{\cal L}_2)~:~v_{{\cal L}_1,{\cal L}_2,{\cal L}_4}\otimes {1}_{{\overline{\cal L}_4},I,{\cal L}_4} \mapsto {1}_{{\cal L}_2,I,\overline{\cal L}_2}\otimes v_{{\cal L}_1,{{\cal L}_2} ,{\cal L}_4},
\fe
for arbitrary junction vectors $v_{{\cal L}_i,{\cal L}_j,{\cal L}_k}\in V_{{\cal L}_i,{\cal L}_j,{\cal L}_k}$. Those involving a trivial external line in the fourth position act by
\ie\label{KToC}
&{\widetilde K}_{{\cal L}_2,{\cal L}_3}^{{\cal L}_1,I}(\overline{\cal L}_3,\overline{\cal L}_1)~:~v_{{\cal L}_1,{\cal L}_2,{\cal L}_3}\otimes {1}_{\overline{\cal L}_3, {\cal L}_3,I} \mapsto C_{{\cal L}_1,{\cal L}_2,{\cal L}_3}(v_{{\cal L}_1,{\cal L}_2,{\cal L}_3})\otimes {1}_{{\cal L}_1,\overline{\cal L}_1, I},
\fe
where $C_{{\cal L}_1,{\cal L}_2,{\cal L}_3}$ is the cyclic permutation map \eqref{CyclicPerm}.

The expectation values $R({\cal L})$ and $R(\overline{\cal L})$, defined earlier in this section, can be determined by splitting empty ${\cal L}$ loops using the crossing relations, and applying the vanishing tadpole property. More explicitly, they are related to the crossing kernels by 
\ie\label{eqn:KtoR}
{\widetilde K}_{\overline{\cal L},{\cal L}}^{{\cal L},\overline{\cal L}}(I,I)& ~:~ {1}_{{\cal L},\overline{\cal L},I}\otimes {1}_{I,{\cal L},\overline{\cal L}}\mapsto R({\cal L})^{-1}\times{1}_{\overline{\cal L},{\cal L},I}\otimes {1}_{{\cal L},I, \overline{\cal L}},
\fe
and there is a similar relation between ${\widetilde K}_{{\cal L},\overline{\cal L}}^{\overline{\cal L},{\cal L}}(I,I)$ and $R(\overline{\cal L})$.

\subsubsection{Fusion coefficients} \label{Sec:FusionCoefficients} The fusion coefficients are related to the dimensions of junction vector spaces as
\ie\label{eqn:FRinDim}
{\cal L}_1{\cal L}_2 = \sum_{{\cal L}_i}{\rm dim}(V_{{\cal L}_1,{\cal L}_2,\overline{\cal L}_i}){\cal L}_i,
\fe
where ${\cal L}_i$ are simple TDLs.\footnote{ A simple consequence of \eqref{eqn:FRinDim} is that, if every line is the orientation reversal of itself, then the fusion ring is commutative.
}  
This relation can be derived by applying the H-junction crossing relations and the cyclic permutation maps on the TDL configuration on the LHS of Figure~\ref{fig:fuse}. The detailed derivation is given in Appendix~\ref{sec:FR}. A similar derivation shows that $R({\cal L})$, with the gauge choice defined in Appendix~\ref{CKbasis}, obeys the polynomial equations given by the abelianization of the fusion ring, just as $\la {\cal L} \ra$ does.

\subsubsection{Rotation on defect operators} As already stated, a defect primary operator $\Psi \in {\cal H}_{\cal L}$ of weight $(h,\tilde h)$ generally have non-integer spin $s=h-\tilde h$.
This phase rotation is relevant for the monodromy property of the two-point function of defect operators, as we presently explain.

\begin{figure}[H]
\centering
\begin{minipage}{0.23\textwidth}
\includegraphics[width=1\textwidth]{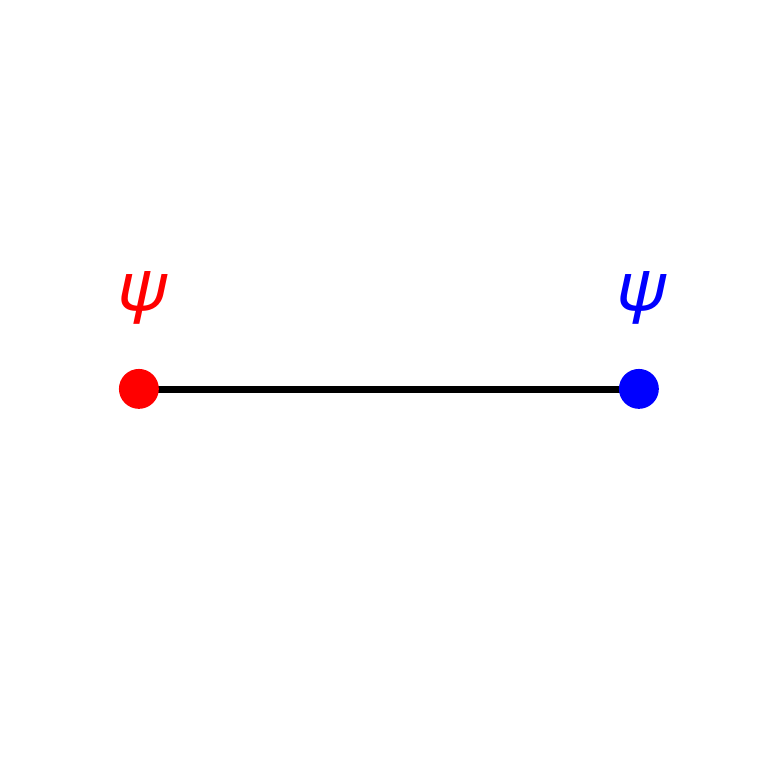}
\end{minipage}%
\begin{minipage}{0.08\textwidth}\begin{eqnarray*}~~\Longrightarrow~~ \\ \end{eqnarray*}
\end{minipage}%
\begin{minipage}{0.23\textwidth}
\includegraphics[width=1\textwidth]{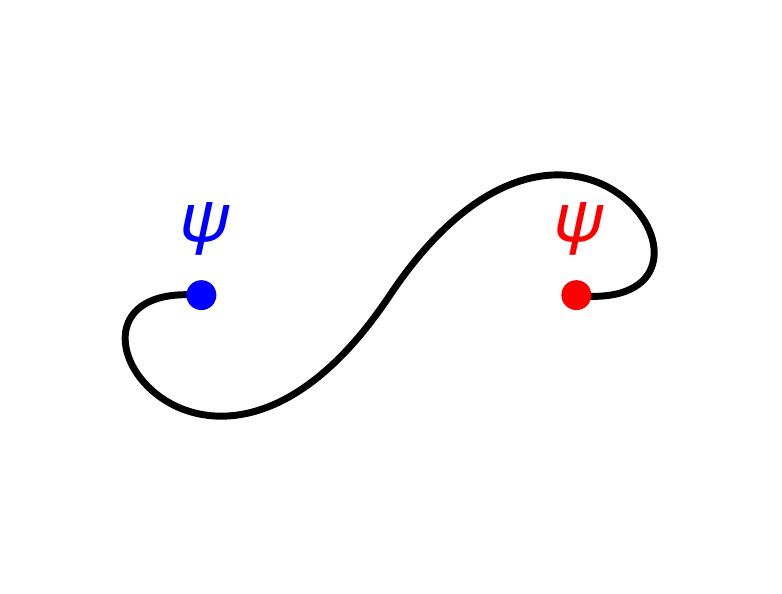}
\end{minipage}%
\begin{minipage}{0.1\textwidth}\begin{eqnarray*}~~\Longrightarrow~~ \\ \end{eqnarray*}
\end{minipage}%
\begin{minipage}{0.23\textwidth}
\includegraphics[width=1\textwidth]{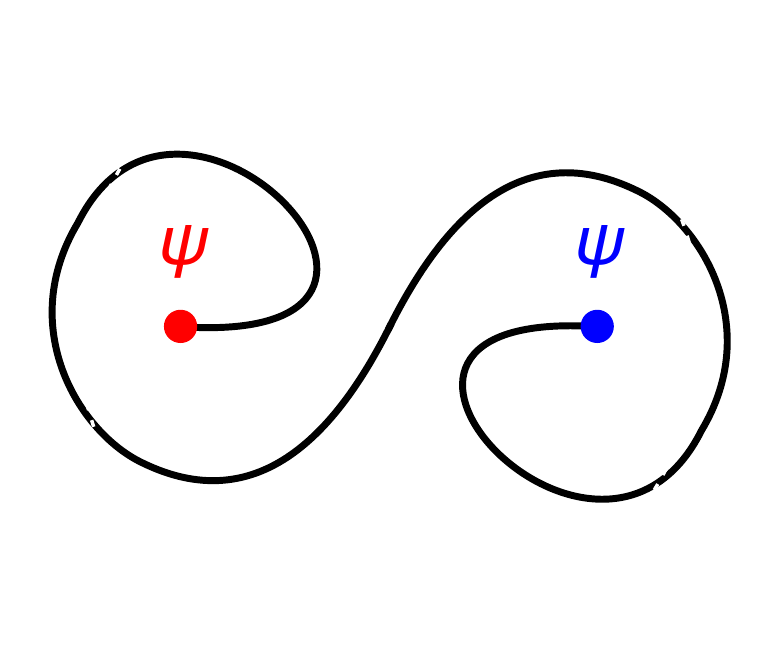}
\end{minipage}%
\caption{Monodromy operation on a two-point function of the defect operator $\Psi \in {\cal H}_{\cal L}$: moving one $\Psi$ around the other.}
\label{fig:twopointmonodromy}
\end{figure}

We may normalize $\Psi$ so that the two-point function of a pair of $\Psi$'s connected by ${\cal L}$ takes the form
\ie
\la \Psi(z_1, \bar z_1) \Psi(z_2, \bar z_2)\ra = z_{12}^{-2h} \bar z_{12}^{-2\tilde h}.
\fe
In order to define the two-point function unambiguously, it is necessary to specify the direction in which the ${\cal L}$ line is attached to $\Psi$ at the ends. The $z_i$ dependence is such that if we translate one of the $\Psi$'s, the direction of the ${\cal L}$ line at its end must remain fixed. For instance, if we bring $z_2$ around $z_1$ in the counterclockwise direction while maintaining the angles of ${\cal L}$ at the ends, as shown in Figure~\ref{fig:twopointmonodromy}, the two-point function picks up a phase $e^{-4\pi i s}$. We can then perform a $2\pi$ rotation on each defect operator, giving another phase factor $e^{4\pi i s}$, after which the TDL configuration returns to the initial one in Figure~\ref{fig:twopointmonodromy}, consistent with the expected monodromy property of the two-point function.

\subsection{Generalization to non-conformal case}
\label{Sec:Nonconformal}

Many of the definitions and properties above admit straightforward generalizations to non-conformal theories. Of particular interest are CFTs coupled to a gauge theory or deformed by a marginal or relevant operator, in which case the TDLs that commute with the gauge current or deformation operator will retain its isotopy invariance property in the gauged or deformed theory (and will still be referred to as TDLs). The major difference in the non-conformal case is the following. Without the state/operator mapping, whenever we spoke of bulk local operators or defect operators in the above, we should replace them by states radially quantized on a circle of some radius. We can speak of correlators of these states as functions of the positions and radii of the circles. The locality property should be understood as the cutting and sewing using the Hilbert space of states on the cut circle. Junction vector spaces are states on a cut circle whose correlation functions are invariant under isotopy that need not preserve the location or size of the circle.
 
The key property that we need is the existence of crossing relations by the modified notion of locality. The crossing kernels are still classified by the solutions to the pentagon identity, and are rigid up to the freedom of choosing junction vectors.
 
These extended definitions and properties will not be explicitly used in non-conformal theories, but only serve to demonstrate that the notions of TDLs and crossing kernels are well-defined along RG flows, and can interpolate between the UV and IR theories. This preservation is a generalization of the 't Hooft ``anomaly'' matching, and will be used in Section~\ref{sec:rgflow} to constrain various RG flows. Until then, we restrict our discussions to CFTs.

\subsection{Isotopy anomaly and orientation-reversal anomaly}
\label{isoanomaly}

The isotopy invariance of a TDL on the plane is equivalent to the statement that the TDL commutes with the stress-energy tensor $T(z)$ and $\widetilde T(\bar z)$. This property extends to TDLs on a curved surface, up to a possible isotopy anomaly due to a contact term in the OPE of the stress tensor with the TDL, of the form $T(x+iy) \sim i \A_{\cal L} \partial_y \delta(y)$, in the presence of a TDL extended along $y\equiv{\rm Im}z =0$ on the plane.\footnote{More precisely, the contact terms for all components of the stress tensor are constrained by conservation to be
$
T_{zz},~T_{\bar z\bar z},~T_{z\bar z} \sim i\A_{\cal L} \partial_y \delta(y).
$
} As a result, on a curved surface, when a TDL ${\cal L}$ is deformed to sweep past a domain $D$, its correlation functional may change by a phase factor
\ie\label{eqn:isoanomaly}
\exp\left[ {i\A_{\cal L}\over 4\pi} \int_D d^2\sigma \sqrt{g} R(g) \right],
\fe
where $g$ is the metric on the surface, and $R$ the scalar curvature, normalized such that $\int d^2\sigma \sqrt{g} R(g) = 8\pi$
 on a unit two-sphere. Note that (\ref{eqn:isoanomaly}) is the only possible form of the isotopy anomaly that is compatible with locality and conformal invariance.\footnote{The formula \eqref{eqn:isoanomaly} is determined by diffeomorphism, conformal invariance, and the contact term of the stress tensor one-point function in the presence of line defect $\cal L$,
 \ie
\vev{\partial_y T(iy){\cal L}}=-{\delta\over \delta g_{zz}(0,y)}{i\A_{\cal L}\over 4\pi}\int_{y=0}dx\sqrt{g}R(g)\vev{{\cal L}}_g\Big|_{g=\delta}=-{i\A_{\cal L}\over 16\pi}\partial_y^2\delta(y)\vev{{\cal L}} \,.
\fe} For the orientation reversed TDL $\overline{\cal L}$, we have $\A_{\overline{\cal L}}=-\A_{\cal L}$.

The isotopy anomaly can also be detected by the phase in the expectation value of an empty clockwise ${\cal L}$ loop on the plane, $R({\cal L})$. Recall that the vacuum expectation value of $\widehat{\cal L}$ on the cylinder, denoted by $\la{\cal L}\ra$, is a positive real number in a unitary theory. $\la {\cal L}\ra$ can also be thought of as the expectation value of an ${\cal L}$ loop on the equator of the sphere. Contracting this loop to near the south pole or the north pole, we have
\ie
R({\cal L}) = e^{i\A_{\cal L}} \la {\cal L}\ra, \quad R(\overline{\cal L})= e^{-i\A_{\cal L}} \la {\cal L}\ra ,
\fe 
where $\A_{\cal L}$ is the isotopy anomaly coefficient in (\ref{eqn:isoanomaly}).

The isotopy anomaly is not entirely physical because it may be absorbed by introducing a finite local counter term on the TDL that is proportional to the extrinsic curvature. That is, we can redefine the TDL by including the factor
\ie\label{extrinsic}
\exp\left[ {i\widetilde\A_{\cal L}\over 2\pi} \int_{\cal L} ds \, K \right],
\fe
where $K$ is the extrinsic curvature of the line, normalized such that the counter-clockwise integral $\int ds \, K$ along the boundary of a flat disc is $2\pi$. While this term with $\widetilde\A_{\cal L} = \A_{\cal L}$ would cancel the isotopy anomaly \eqref{eqn:isoanomaly} by the Gauss-Bonnet theorem (except when ${\cal L} = \overline{\cal L}$ as we will discuss shortly), it would also rotate the junction vectors involving ${\cal L}$ or defect operators at the end of ${\cal L}$ by certain angle dependent phases, and correspondingly the crossing kernels involving ${\cal L}$ undergo a ``gauge" transformation while preserving the pentagon identity. For instance, consider the removal of a trivial junction as in Figure~\ref{fig:trivialjunction}. If we modify the TDL by the extrinsic curvature counter term \eqref{extrinsic}, the right graph in Figure~\ref{fig:trivialjunction} would acquire an extra phase due to the extrinsic curvature at the kink. To maintain the equivalence in Figure~\ref{fig:trivialjunction}, in the left graph, the extra phase would be interpreted as a phase rotation of the trivial junction vector. 

There is one exception to this, namely, when ${\cal L}$ is a TDL of the same type as its orientation reversal $\overline{\cal L}$. In this case, insisting on the equivalence of ${\cal L}$ and $\overline{\cal L}$ on the plane would require $R({\cal L})=R(\overline{\cal L})$, and forbid modifying ${\cal L}$ by an extrinsic curvature term of the form (\ref{extrinsic}). In this case, a nonzero $\A_{\cal L}$ would imply an orientation-reversal anomaly for ${\cal L}$, as correlation functions involving ${\cal L}$ on curved surfaces would differ by a phase depending on the choice of orientation of ${\cal L}$.\footnote{We thank L. Bhardwaj and Y. Tachikawa for correspondence on this point.
}
However, $R({\cal L})=R(\overline{\cal L})$ is still consistent with an isotopy anomaly (\ref{eqn:isoanomaly}) if $\A_{\cal L}$ is a multiple of $\pi$ (rather than $2\pi$), since a small clockwise ${\cal L}$ loop at the north pole of the sphere can be deformed to a small clockwise $\overline{\cal L}$ loop at the south pole of the sphere, differing by the phase $e^{2i\A_{\cal L}}$. This leaves the possibility of $R({\cal L})$ differing from $\la {\cal L}\ra$ by a sign, namely, $\A_{\cal L}=\pi$.\footnote{This particular phase from the isotopy anomaly corresponds to the (second) Frobenius-Schur indicator for the TDL $\cL$ in the fusion category language (see \cite{KITAEV20062} for a discussion).
}
This is necessary for consistency, for instance, if ${\cal L}$ is a $\mathbb{Z}_2$ invertible line with an 't Hooft anomaly, in which case $R({\cal L})=-1$ (as determined by the crossing kernel ${\widetilde K}_{{\cal L},{\cal L}}^{{\cal L},{\cal L}}(I,I)=-1$) while $\la {\cal L}\ra=1$.

\section{Relation to fusion categories}
\label{sec:fusioncategory}

Our definition of TDLs encompasses the structure of a fusion category, at least in the case where the number of simple TDLs is finite. This dictionary between the TDLs and the fusion category, together with various physical applications and consequences, has recently been discussed in great detail in \cite{Bhardwaj:2017xup}. 
  Below we review and elaborate on part of this dictionary that is relevant for our discussion.

An object in the fusion category corresponds to the Hilbert space ${\cal H}_{{\cal L}}$ associated to the endpoint of a TDL ${\cal L}$. A morphism between the objects ${\cal H}_{{\cal L}_1}$ and ${\cal H}_{{\cal L}_2}$ is a weight-(0,0) topological defect operator $m$ between the TDLs ${\cal L}_1$ and ${\cal L}_2$, which gives a linear map between the Hilbert spaces,
\ie
m ~:~ {\cal H}_{{\cal L}_1} \to {\cal H}_{{\cal L}_2}.
\fe
The existence of the trivial line and the additive structure with respect to direct sum are evident. The tensor structure of the fusion category is specified by the junction vector $v\in V_{{\cal L}_1,{\cal L}_2,{\cal L}_3}$, which defines a linear map,
\ie
v ~:~ {\cal H}_{{\cal L}_1}\otimes {\cal H}_{{\cal L}_2}\to {\cal H}_{{\cal L}_3},
\fe
 subject to the H-junction crossing relations which are equivalent to associators in fusion category.  
Simple objects are the Hilbert spaces of defect operators at the end of simple TDLs, and the semi-simplicity of all objects is assumed, that is every TDL can be decomposed as direct sum of finitely many simple ones. The finiteness of simple objects is not a necessary assumption for our purpose, and may not hold for TDLs (such as invertible lines associated with continuous global symmetries).  The number of simple objects is called the rank of a fusion category. The fusion ring is the Grothendieck ring of a fusion category (see \cite{etingof2016tensor}).


In this paper, we prefer to work with a definition of TDLs with H-junction crossing relations that are independent of the angles at the junctions, which leads to the possibility of isotopy anomalies. As will be discussed in the next section, it is possible to eliminate the isotopy anomaly by choosing a finite counter term in the definition of the TDL that involves the extrinsic curvature of the TDL; however, one may then need to keep track of the angle dependence in the H-junction crossing relations, which is implicitly allowed in the notion of the associator of a fusion category. 

The dual object corresponds to the orientation reversal of the TDL ${\cal L}$. The (co-)evaluation maps are related to the identity junction vectors ${1}_{\overline{\cal L}, I ,{\cal L}}$, ${1}_{ I ,{\cal L},\overline{\cal L}}$, and ${1}_{{\cal L},\overline{\cal L},I}$. The vacuum expectation value $\la{\cal L}\ra$ is the ``quantum dimension" of the corresponding object. Note however that in our definition of TDLs, the expectation value of an empty ${\cal L}$ loop on the plane may differ from $\la {\cal L}\ra$ by a phase, due to the isotopy anomaly.

We summarize the relations discussed above in Table~\ref{Tab:category}.

\begin{table}[H]
\centering
\begin{tabular}{|c|c|}
\hline
fusion category & TDLs
\\\hline\hline
object & Hilbert space ${\cal H}_{\cal L}$ of defect operators
\\\hline
simple object & ${\cal H}_{\cal L}$ for a simple (indecomposable) $\cal L$
\\\hline
rank&number of simple TDLs
\\\hline
morphism & junction vector in $V_{{\cal L}_1,{\cal L}_2}$
\\
& (as linear map ${\cal H}_{{\cal L}_1} \to {\cal H}_{{\cal L}_2}$).
\\\hline
tensor structure & junction vector spaces $V_{{\cal L}_1,{\cal L}_2,{\cal L}_3}$
\\\hline
(co-)evaluation & identity junction vectors ${1}_{\overline{\cal L}, I ,{\cal L}}$, ${1}_{ I ,{\cal L},\overline{\cal L}}$, ${1}_{{\cal L},\overline{\cal L},I}$
\\\hline
associator & H-junction crossing kernel
\\\hline
dual object & orientation reversal $\overline{\cal L}$
\\\hline
quantum dimension & cylinder vacuum expectation value $\la {\cal L} \ra$
\\\hline
Grothendieck ring & fusion ring
\\\hline
\end{tabular}
\caption{Summary of the relations between fusion category and TDLs.}
\label{Tab:category}
\end{table}

\subsection{On fusion categories of small ranks}
\label{Sec:SmallRanks}

Fusion categories containing only two simple objects were classified in \cite{Ostrik:aa}. In the TDL language, there is only one nontrivial TDL $X$ in addition to the trivial line $I$, with the fusion relation $X^2=I+a X$, where $a=0$ or 1. In the $a=0$ case, $X$ corresponds to a $\mathbb{Z}_2$ global symmetry, while in the $a=1$ case, $X$ is a more general TDL that does not correspond to any global symmetry. In each of these cases, there are two sets of solutions of the H-junction crossing kernels to the pentagon identity.

For $a=0$, the crossing phase ${\widetilde K}^{X,X}_{X,X}(I,I)$ is either 1 or $-1$. As discussed in the previous section, ${\widetilde K}^{X,X}_{X,X}(I,I)=-1$ occurs if $X$ corresponds to a $\mathbb{Z}_2$ symmetry with an 't Hooft anomaly. Note that such a crossing phase also implies that an empty $X$ loop on the plane has expectation value $-1$, and an isotopy anomaly is required for this to be compatible with unitarity.

For $a=1$, the two solutions to the pentagon identity lead to the empty $X$ loop expectation values $R(X) = {1\pm\sqrt{5} \over 2}$ \cite{Moore:1988qv}. Since the isotopy anomaly is a phase, and that there is only a binary choice $\la X \ra = {1\pm\sqrt{5}\over 2}$ to solve the abelianization of the fusing ring, it follows that $\la X \ra = R(X)$. The choice $\la X \ra = {1-\sqrt{5}\over 2}$ is incompatible with unitarity, and is realized by the nontrivial TDL in the Lee-Yang model $M(2,5)$. The case $\la X\ra = {1+\sqrt{5}\over 2}$ occurs in many examples, such as in certain TDLs in the tricritical Ising model, three-state Potts model, and WZW models.

The impossibility of $a\geq 2$ was proven in \cite{Ostrik:aa} by indirect arguments. We formulate the pentagon identity with the gauge condition in Appendix~\ref{CKbasis} using \texttt{Mathematica}, and directly verify in the $a=2$ case (the junction vector space $V_{X,X,X}$ is two-dimensional) that indeed a solution to the pentagon identity does not exist.

Fusion categories containing three simple objects have been classified in \cite{Ostrik:2013aa} (assuming pivotal structure), and their fusion rings are all commutative. Let the two nontrivial TDLs be $X$ and $Y$, the possible fusion relations are

\noindent (i) $X^2=Y$, $Y^2=X$, $XY= 1$. Here, $X$ and $Y$ are the invertible lines associated with a $\mathbb{Z}_3$ global symmetry. There are three solutions to the pentagon identity: one is non-anomalous, while the other two have 't Hooft anomalies (discussed in Section~\ref{tHooftanomaly}).

\noindent (ii) $X^2=I$, $Y^2=I+X$, $XY=Y$. There are two solutions to the pentagon identity. One of them is realized by the TDLs in the critical Ising model, while the other one is realized by the tensor product theory of the critical Ising model and the $SU(2)_1$ WZW model. In the latter case, the $\mathbb{Z}_2$ line $X$ is realized by the product of the $\mathbb{Z}_2$ line in the critical Ising model, and that associated with the center of the left $SU(2)$ symmetry in the $SU(2)_1$ WZW model (see Section~\ref{Sec:Z2TY} for more details). These two fusion categories are in fact the Tambara-Yamagami extensions \cite{TAMBARA1998692} (discussed in more detail in Section~\ref{Sec:DualityTY}) of the (non-anomalous) $\mathbb{Z}_2$ fusion category.

\noindent (iii) $X^2=I$, $Y^2=I+X+Y$, $XY=Y$. This is the representation ring ${R}_{\bC}(S_3)$ of the permutation group $S_3$. There are three solutions to the pentagon identity \cite{Etingof:ab}. One of them gives the ${\rm Rep}(S_3)$ fusion category, which is realized by a subset of TDLs in either the tetracritical Ising model or the $SU(2)_4$ WZW model. The other two solutions are referred to as twisted ${\rm Rep}(S_3)$ fusion categories in this paper.

\noindent (iv) $X^2=I+Y$, $Y^2=1+X+Y$, $XY=X+Y$. This is the representation ring of the integrable highest-weight representations of the affine Lie algebra $\widehat{su(2)}_5$, restricted to integral spins. We denote this fusion ring by ${R}_{\bC}(\widehat{so(3)}_5)$. There are three solutions to the pentagon identity (see Appendix~\ref{app:sl2}). One of them is the ${\rm Rep}(\widehat{so(3)}_5)$ fusion category, which is realized by a subset of TDLs in either the pentacritical Ising model or the $SU(2)_5$ WZW model. The other two categories do not admit unitary realizations.

\noindent (v) $X^2=I$, $Y^2=I+X+2Y$, $XY=Y$. This fusion category, known as ``${1\over 2}E_6$", does not admit braiding \cite{2005math,Hagge:2007aa}. There are four solutions to the pentagon identity (see Appendix~\ref{1/2E6CK}). Two of them are realized by (subsets of) the TDLs in the $(A_{10},E_6)$ minimal model and by the TDLs in the $E_6$ type non-diagonal $SU(2)_{10}$ WZW model. The other two categories do not admit unitary realizations.

Let us comment here, that a basic difference between a fusion category of TDLs and the OPEs of bulk local operators (say in rational CFTs) is that the former does not require the existence of braiding, while the latter does. This is evident for the invertible lines associated with a (discrete) nonabelian global symmetry. We will see later (in the examples of WZW models) that there are more general simple TDLs that admit nonabelian fusion relations.

\begin{table}[H]
\centering
\begin{tabular}{|c|c|c|c|c|c|}
\hline
rank& fusion ring & \# categories & TDL models in CFTs & section & see also
\\
\hline\hline
\multirow{4}{*}{$2$}  &\multirow{2}{*}{$\bZ_2$} & \multirow{2}{*}{2} & Ising & \ref{Sec:Ising} & \\
\hhline{~~~--~}
&& & $SU(2)_1$ WZW & \ref{Sec:CompactBoson} &
\\\hhline{~-----}
&\multirow{2}{*}{Lee-Yang} & \multirow{2}{*}{2} & Lee-Yang & \ref{Sec:LeeYang} & \multirow{2}{*}{\ref{Sec:LeeYangAlgebra}} \\
\hhline{~~~--~}
&& & tricritical Ising & \ref{Sec:Trising} &
\\\hline\hline
\multirow{9}{*}{3}  &\multirow{2}{*}{$\bZ_3$} & \multirow{2}{*}{3} & three-state Potts & \ref{potts} & \\
\hhline{~~~--~}
&& & $SU(2)_1$ WZW & \ref{Sec:CompactBoson} &
\\\hhline{~-----}
&$\bZ_2$ TY & 2 & Ising ($\otimes SU(2)_1$) & \ref{Sec:Ising}, \ref{Sec:DualityTY} & \ref{Sec:Z2TY}
\\\hhline{~-----}
&\multirow{3}{*}{${R}_{\bC}(S_3)$} & \multirow{3}{*}{3} & tetracritical Ising & \ref{Sec:Tetra} &
\\\hhline{~~~--~}
& & & $SU(2)_4$ WZW & \ref{Sec:WZW} & \ref{repsthree}
\\\hhline{~~~--~}
&& & $\bZ_2$ orbifold of $S_3$ & &
\\\hhline{~-----}
&\multirow{2}{*}{${R}_{\bC}(\widehat{so(3)}_5)$} & \multirow{2}{*}{3} & pentacritical Ising & \ref{Sec:Penta} & \\
\hhline{~~~--~}
& && $SU(2)_5$ WZW & \ref{Sec:WZW} &
\\\hhline{~-----}
&\multirow{2}{*}{${1\over2} E_6$} & \multirow{2}{*}{4} & $(A_{10}, E_6)$ minimal model & \ref{Sec:hE6} & \multirow{2}{*}{\ref{esixselect}} \\
\hhline{~~~--~}
&& & non-diagonal $SU(2)_{10}$ WZW & \ref{Sec:hE6} &
\\\hline\hline
4&$\bZ_3$ TY & 2 & three-state Potts ($\otimes SU(2)_1$) & \ref{potts}, \ref{Sec:DualityTY} & \ref{Sec:Z3TY}
\\\hline
\end{tabular}
\caption{Fusion rings of small ranks, the corresponding number of fusion categories (solutions to the pentagon identity), and some examples of CFTs in which the fusion categories are realized by (a subset of) the TDLs. TY stands for Tambara-Yamagami. Note that not all the categories counted in the third column are realized in the models given in the fourth column.}
\end{table}

\section{Invertible lines, 't Hooft anomalies, and orbifolds}
\label{sec:globalsymmetry}

This section is devoted to a discussion of the most familiar class of topological lines -- the invertible lines associated with global symmetries. Familiar concepts such as 't Hooft anomalies and orbifolds are recast in the formalism of TDLs, and allow us to derive selection rules for the spin content of defects operators at the end of these TDLs.

\subsection{Global symmetries and invertible lines}
\label{Sec:Homological}

The simplest class of simple TDLs are the ones associated with global symmetries, which we refer to as \textit{invertible lines}. For every symmetry group element $g$, there is an invertible line ${\cal L}_g$ such that $\widehat{\cal L}_g$ acts on the states/bulk local operators of the CFT according to the action of $g$ itself, namely $\widehat{\cal L}_g |\phi\ra = \widehat g|\phi\ra$ \cite{Frohlich:2006ch,Davydov:2010rm,Kapustin:2014gua,Gaiotto:2014kfa}. The invariance of the vacuum implies that $\la {\cal L}_g \ra=1$. In a unitary, compact CFT, a typical simple TDL ${\cal L}$ that is not an invertible line has $\la {\cal L}\ra>1$.

The fusion relation of the invertible lines takes the same form as the group multiplication, namely ${\cal L}_g {\cal L}_h = {\cal L}_{gh}$. The T-junction vector space $V_{{\cal L}_{g_1},{\cal L}_{g_2},{\cal L}_{g_3}}$ is one-dimensional if $g_1g_2g_3=1$, and trivial otherwise. The orientation reversal $\overline{\cal L}_g$ is a TDL of the same type as ${\cal L}_{g^{-1}}$.  The identity junction vector in $V_{{\cal L}_g,\overline{\cal L}_g,I}$ has unit norm. Note that for general $g_1, g_2, g_3=(g_1g_2)^{-1}$, there need not be a canonical choice for a (unit norm) T-junction vector in $V_{{\cal L}_{g_1},{\cal L}_{g_2},{\cal L}_{g_3}}$.

\subsection{'t Hooft anomaly}
\label{tHooftanomaly}

The 't Hooft anomalies for global symmetries/invertible lines may be classified by phases in the H-junction crossing relation that solve the pentagon identity \cite{Dijkgraaf:1989pz}. Namely, given four symmetry group elements $g_1,g_2,g_3,g_4$ with $g_1g_2g_3g_4=1$, and choices of T-junctions vectors with unit norm in $V_{{\cal L}_{g_1}, {\cal L}_{g_2},\overline{\cal L}_{g_1g_2}}$, $V_{{\cal L}_{g_3}, {\cal L}_{g_4},\overline{\cal L}_{g_3g_4}}$, $V_{{\cal L}_{g_2}, {\cal L}_{g_3},\overline{\cal L}_{g_2g_3}}$, $V_{{\cal L}_{g_4}, {\cal L}_{g_1},\overline{\cal L}_{g_4g_1}}$, the crossing kernels ${\widetilde K}^{{\cal L}_{g_1},{\cal L}_{g_4}}_{{\cal L}_{g_2},{\cal L}_{g_3}}({\cal L}_{g_1g_2},{\cal L}_{g_2g_3})$ may be nontrivial phases.  Let us define
\ie
e^{i\theta(g_1,g_2,g_3)} = {\widetilde K}^{{\cal L}_{g_1},\overline{\cal L}_{g_1g_2g_3}}_{{\cal L}_{g_2},{\cal L}_{g_3}}({\cal L}_{g_1g_2},{\cal L}_{g_2g_3}),
\fe
which can be viewed as a 3-cochain $C^3(G,U(1))$, {\it i.e.},
\ie
\theta: G^3 \to U(1).
\fe
The pentagon identity can be written as
\ie
\theta(g_1,g_2,g_3g_4)+ \theta(g_1g_2,g_3,g_4)=\theta(g_2,g_3,g_4)+ \theta(g_1,g_2g_3,g_4)+ \theta(g_1,g_2,g_3),
\fe
which is precisely the cocycle condition on the cochain. A phase rotation of the junction vector in $V_{{\cal L}_{g_1}, {\cal L}_{g_2},\overline{\cal L}_{g_1g_2}}$ by $e^{i\varphi(g_1,g_2)}$ results in a shift of $\theta(g_1,g_2,g_3)$ by the coboundary
\ie
\delta\theta(g_1,g_2,g_3) = \varphi(g_2,g_3) + \varphi(g_1,g_2g_3) - \varphi(g_1g_2,g_3) - \varphi(g_1,g_2).
\fe
We see that the inequivalent crossing phases precisely correspond to the group cohomology $H^3(G,U(1))$. The non-anomalous case corresponds to the trivial element of $H^3(G,U(1))$, which has a representative where all the phases vanish.

\subsection{Orbifolds}
\label{sec:orbifolds}

If a CFT has a global symmetry group $G$ that is free of an 't Hooft anomaly, we can construct a $G$-orbifold CFT whose space of bulk local operators is the $G$-invariant projection of $\bigoplus_{g\in G} {\cal H}_{{\cal L}_g}$.\footnote{The generalization of the orbifold construction beyond invertible topological defects has been proposed in \cite{Frohlich:2009gb}.} This construction requires defining 4-way junctions of ${\cal L}_{g_1}, {\cal L}_{g_2}, \overline{\cal L}_{g_1},\overline{\cal L}_{g_2}$, each of which can be split into an H-junction in two different ways. While the absence of an 't Hooft anomaly means that the two ways of splitting are equivalent, there is the freedom to rotate the set of T-junction vectors in $V_{{\cal L}_{g_1}, {\cal L}_{g_2},\overline{\cal L}_{g_1g_2}}$ a phase $e^{i\A(g_1, g_2)}$, such that
\ie
\A(g_2, g_3) + \A(g_1, g_2 g_3) - \A(g_1g_2, g_3) - \A(g_1, g_2)=0.
\fe
Namely, $\A(g_1, g_2)$ is a cocycle in $C^2(G,U(1))$. Furthermore, a change in the phase of the trivial junctions by
\ie
\delta\A(g_1, g_2) = \B(g_2) - \B(g_1g_2) + \B(g_1)
\fe
does not affect the orbifold partition function. Thus, inequivalent orbifolds by $G$ are classified by $H^2(G,U(1))$, known as the discrete torsion \cite{Vafa:1986wx, Brunner:2014lua}.

Next, we can ask which TDLs of the CFT survive or give rise to TDLs in the orbifolded theory. A potential candidate is a TDL $W$ (which is typically not an invertible lines) that commutes with $G$, in the sense of fusion product, {\it i.e.}, $W{\cal L}_g = {\cal L}_g W$ for all $g\in G$. If $W$ does not commute with $G$, we can consider the sum of its $G$-orbits
\ie
W^G = \sum_{[g]\in G/G^{W}} {\cal L}_g {W}\overline{\cal L}_{g},
\fe
where $G^{W}$ is the commutant of ${W}$ in $G$ and $G/G^W$ denotes the set of right cosets.  The answer does not depend on the  choice of the representative $g$ in the summand for $[g]$.  
Note that ${W}^G$ is non-simple in the original CFT but could give rise to a simple TDL in the orbifold theory.

The $G$-invariance of ${W}$ does not guarantee that ${W}$ defines a TDL in the orbifold CFT. For example, if there is a nontrivial crossing phase ${\widetilde K}^{{\cal L}_g,{W}}_{{W},\overline{\cal L}_g}$, then ${W}$ would {\it a priori} be ill-defined in the orbifold theory. However, in the next section, we discuss one important exception.

\subsubsection{Duality defects and Tambara-Yamagami categories}
\label{Sec:DualityTY}

Suppose $g$ is a $\mathbb{Z}_n$ group generator that is free of an 't Hooft anomaly, and suppose there is an unoriented TDL ${\cal N}$ that commutes with $g$ and obeys the  following fusion relation
\ie\label{nln}
&\mathcal{N} \mathcal{L}_{g'} =\mathcal{L}_{g'} \mathcal{N} = \mathcal{N}\,,\quad \forall\, g'\in\mathbb{Z}_n,\\
&{\cal N}^2 = \sum_{g'\in\mathbb{Z}_n}{\cal L}_{g'}.
\fe
Note that while ${\cal L}_{g^k}$ are invertible, $\cal N$ is \textit{not} because ${\cal N}^2$ contains the trivial line $I$ but also other lines.  
The fusion relation also implies that $\widehat{\cal N}$ annihilates any state that transforms with nontrivial $\bZ_n$ charge. Choosing the basis junction vectors in $V_{{\cal N},{\cal N},{\cal L}_{g^m}}$ to have the same norm $n^{1\over 4}$ for all $m$, and applying the H-junction crossing relation to ${\cal N}$ loops connected by ${\cal L}_{g^m}$ lines, we can determine the crossing kernels to be
\ie\label{TYcrossing}
{\widetilde K}^{{\cal L}_{g^m},\,{\cal N}}_{{\cal N},{\cal L}_{g^{k}}}({\cal N},{\cal N}) = \omega^{mk},
\quad
{\widetilde K}^{\cal N,\,N}_{\cal N,\,N}({\cal L}_{g^m},{\cal L}_{g^k}) = \pm {1\over \sqrt{n}} \omega^{mk},
\fe
where $\omega = e^{2\pi i \over n}$. The crossing kernels involving only the $\bZ_n$ group elements are trivial, since the symmetry is non-anomalous.

\begin{figure}[H]
\centering
\begin{minipage}{0.2\textwidth}
\includegraphics[width=1\textwidth]{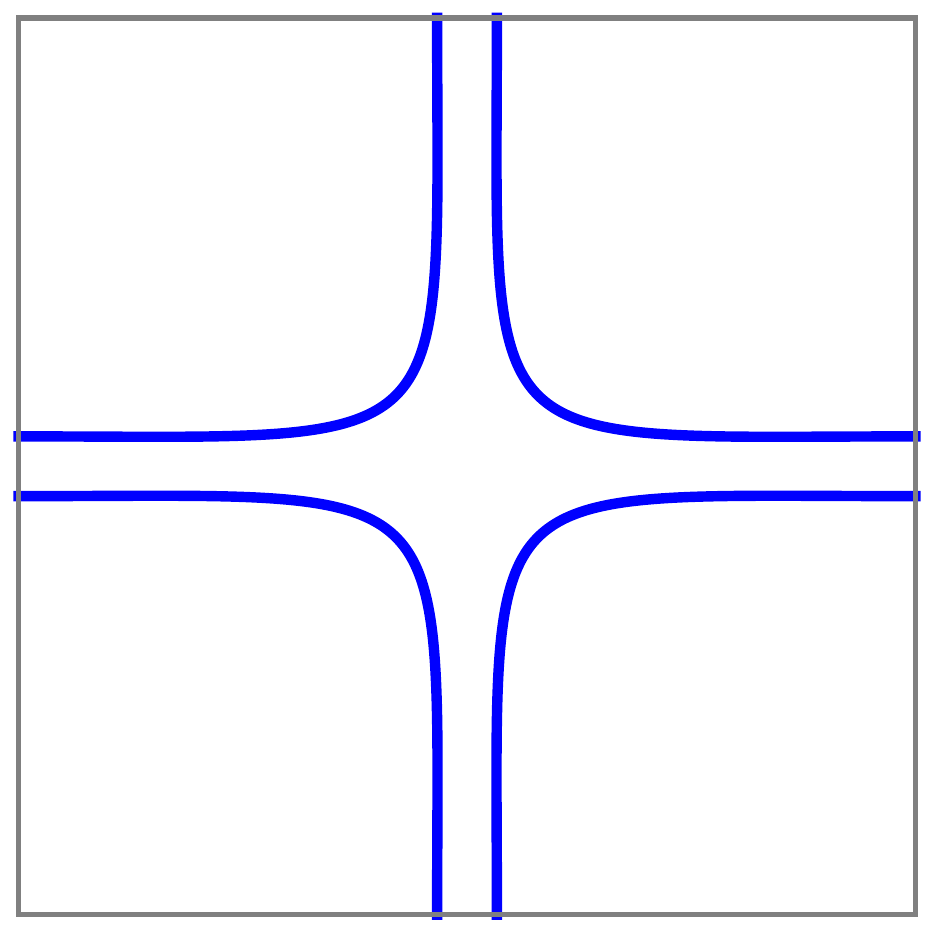}
\end{minipage}%
\begin{minipage}{0.2\textwidth}\begin{eqnarray*}\quad\quad=\sum_{m,k=0}^{n-1} {1\over n}\quad\\ \end{eqnarray*}
\end{minipage}%
\begin{minipage}{0.2\textwidth}
\includegraphics[width=1\textwidth]{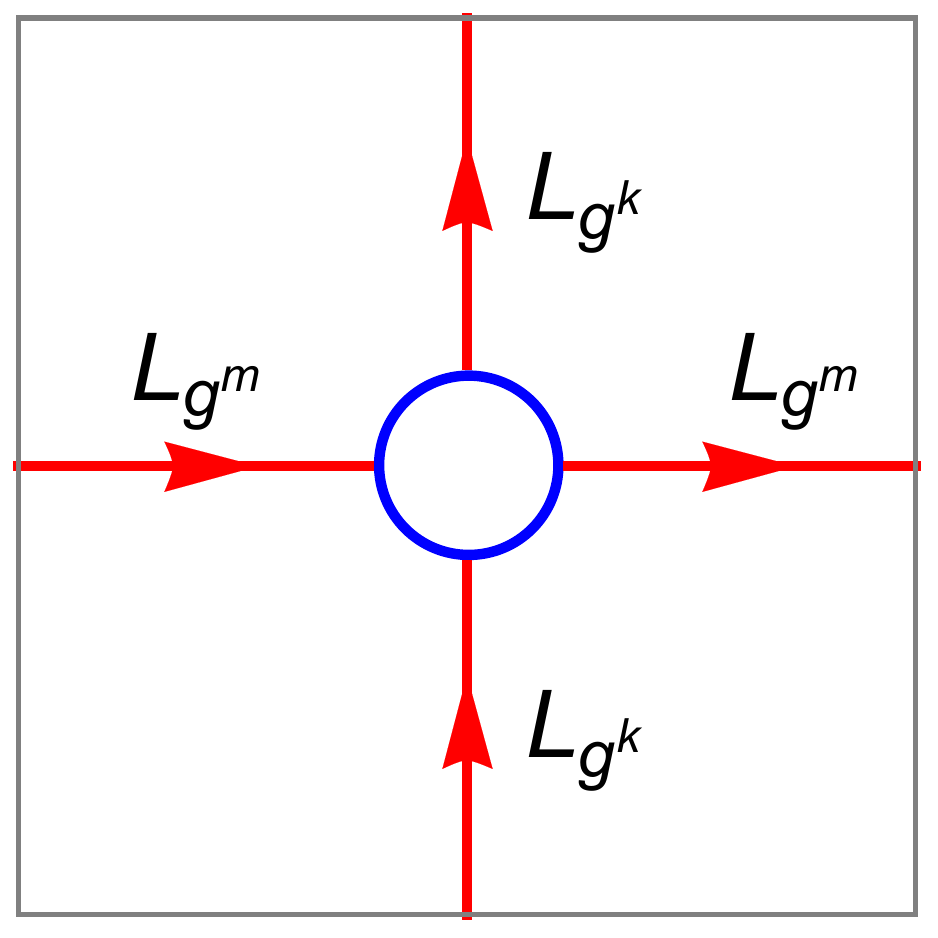}
\end{minipage}%
\caption{Fusing an 8-way junction of ${\cal N}$ (blue) lines into a 4-way intersection of invertible (red) lines.
}
\label{fig:N8way}
\end{figure}

In fact, the TDL $\cal N$ described above is a duality defect relating a $2d$ CFT $\cal T$ with a non-anomalous $\mathbb{Z}_n$ symmetry to its orbifold ${\cal T}/\mathbb{Z}_n$. It generalizes the well-known order-disorder duality (Kramers-Wannier duality) in the critical Ising model \cite{PhysRev.60.252}.
Consider the 8-way junction of ${\cal N}$ as shown in Figure~\ref{fig:N8way}, where the ${\cal N}$ lines may be fused pairwise by applying the H-junction crossing relations \eqref{TYcrossing} to produce a 4-way intersection of the TDL $\sum_{m=0}^{n-1} {\cal L}_{g^m}$. This relation can be used to show that the CFT is isomorphic to its orbifold by the $\mathbb{Z}_n$. For example, on the LHS, if we put this 8-way junction on a torus, the network of TDLs can be shrunk to a contractible circle of $\cal N$, giving the torus partition function of $\cal T$ times the expectation value of $\cal N$ on the torus.  
On the RHS, summing over all the $\mathbb{Z}_n$  invertible lines gives the torus partition function of the orbifold theory ${\cal T}/\mathbb{Z}_n$, again times the expectation value of $\cal N$ on the torus. The equivalence of these two networks of TDLs on the torus then proves the equality between the torus partition functions of $\cal T $ and ${\cal T}/\mathbb{Z}_n$. In the next section, we will see examples of ${\cal N}$, such as the TDLs $N$ in the critical and tricritical Ising models, which are isomorphic to their respective $\mathbb{Z}_2$ orbifolds, as well as the TDLs $N$ and $N'$ in the three-state Potts model, which is isomorphic to its own $\mathbb{Z}_3$ orbifold.

The construction described above is a special case of a Tambara-Yamagami category \cite{TAMBARA1998692}, which is a fusion category whose simple objects are the invertible lines associated with an abelian group (taken to be $\mathbb{Z}_n$ above), plus an additional TDL ${\cal N}$ obeying the fusion relations \eqref{nln}, such that the crossing kernels depend on a choice of the symmetric non-degenerate bicharacter of the abelian group, and a choice of sign. In the $\mathbb{Z}_n$ case, there is a unique symmetric non-degenerate bicharacter because $H^2(\mathbb{Z}_n,U(1))=1$. Our conclusion above can then be rephrased as follows: a $2d$ CFT $\mathcal{T}$ with a non-anomalous abelian finite group global symmetry $G$ is isomorphic to its $G$-orbifold theory $\mathcal{T}/G$, if $\mathcal{T}$ contains a Tambara-Yamagami extension of the $G$ fusion category \cite{Frohlich:2006ch}. The choice of the bicharacter of $G$ in defining the Tambara-Yamagami extension physically corresponds to a choice of the discrete torsion in orbifolding.

\subsection{Cyclic permutation map and spin selection rule}
\label{sec:ZnSSR}

\begin{figure}[H]
\centering
\begin{minipage}{0.2\textwidth}
\includegraphics[width=1\textwidth]{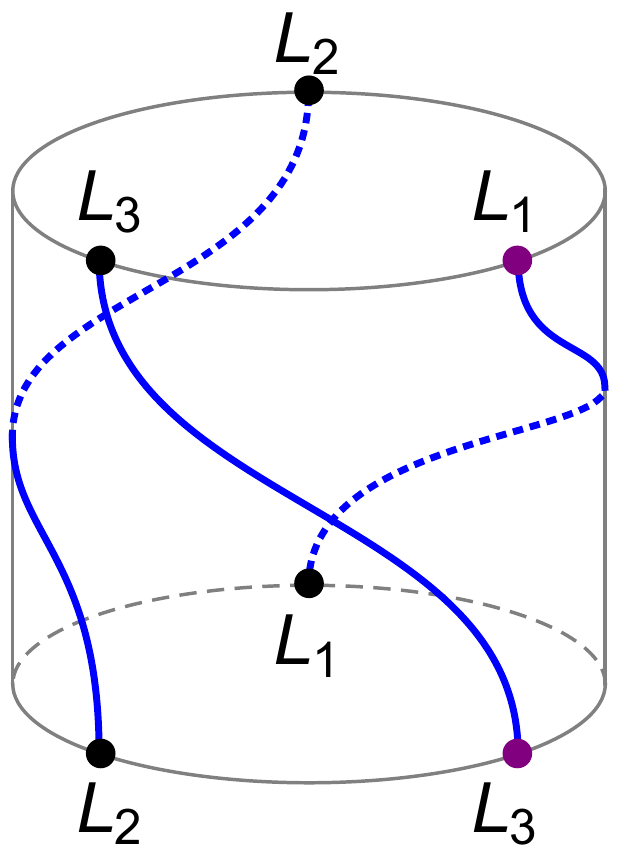}
\end{minipage}%
\caption{The cyclic permutation map from $V_{{\cal L}_1,{\cal L}_2,{\cal L}_3}$ to $V_{{\cal L}_2,{\cal L}_3,{\cal L}_1}$. The ordering of the defect points is such that the purple dot is the last one.}
\label{fig:perm}
\end{figure}

Given an order-$n$ element $g$ of the symmetry group, {\it i.e.}, $g^n=1$, and the corresponding invertible line ${\cal L}_g$, we can deduce a spin selection rule on the defect operators in $V_{{\cal L}_g}$, as follows. On the $n$-way junction vector space $V_{{\cal L}_g,{\cal L}_g,\cdots, {\cal L}_g}$, we can define the cyclic permutation map $\widehat C\in{\rm Aut}(V_{{\cal L}_g,{\cal L}_g,\cdots, {\cal L}_g})$ by the TDL graph on the cylinder that connects the $i$-th ${\cal L}_g$ defect point to the $(i+1)$-th, $i=1,\cdots,n$, as shown in Figure~\ref{fig:perm}, and restricted to 
the junction vector space $V_{{\cal L}_g,{\cal L}_g,\cdots, {\cal L}_g}$. Note that $\widehat C^n$ is the identity map, and thus $\widehat C$ acts by a phase that is an $n$-th root of unity.

\begin{figure}[H]
\centering
\begin{tikzcd}
\begin{minipage}{0.22\textwidth}
\includegraphics[width=1\textwidth]{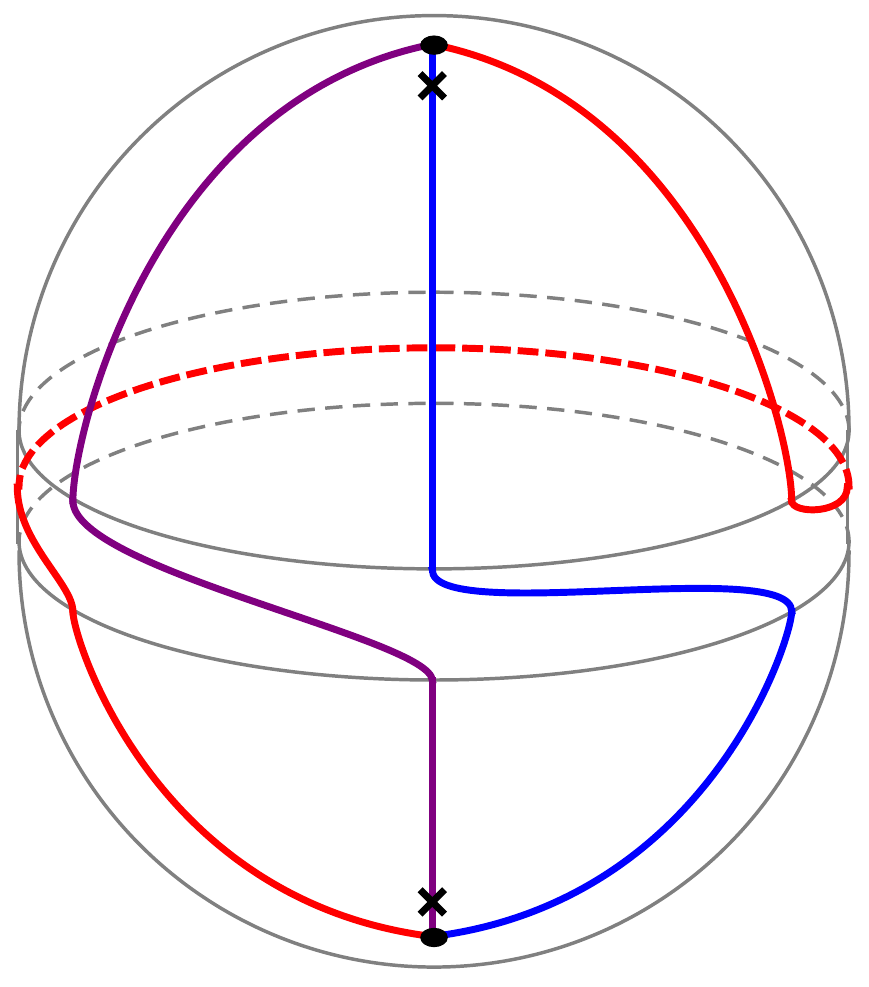}
\end{minipage}%
\begin{minipage}{0.05\textwidth}~
\end{minipage}%
\begin{minipage}{0.12\textwidth}\begin{eqnarray*}~~{\rm isotopy} \\ \longrightarrow ~~ \\ \\ \end{eqnarray*}
\end{minipage}%
\begin{minipage}{0.05\textwidth}~
\end{minipage}%
\begin{minipage}{0.22\textwidth}
\includegraphics[width=1\textwidth]{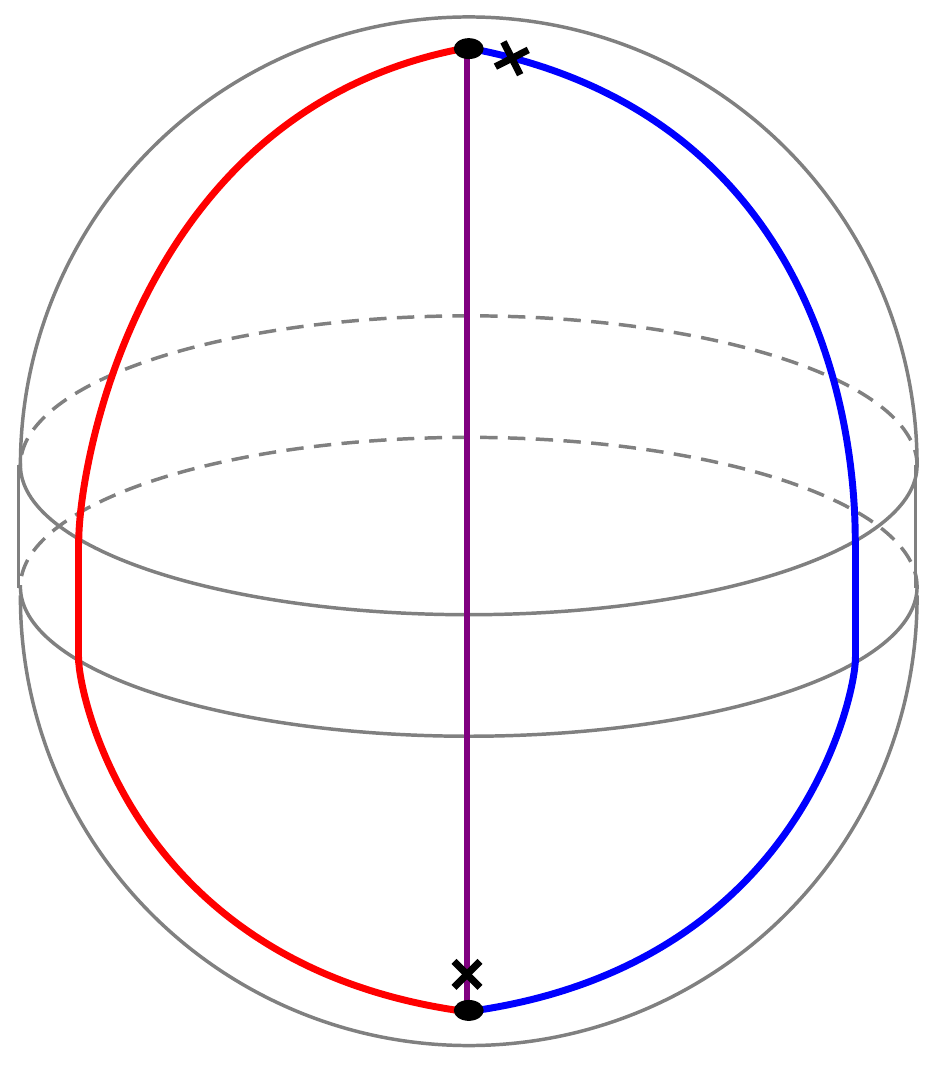}
\end{minipage}%
\end{tikzcd}
\caption{The cyclic permutation map on the cylinder (left) is related to the cyclic permutation map on the junction vector (right) by an isotropy transformation of the TDLs, sweeping through the upper hemisphere.}
\label{fig:permiso}
\end{figure}

In the presence of an isotopy anomaly, the cyclic permutation map $\widehat C$ defined on the cylinder differs from the cyclic permutation map $C$ on a junction of ${\cal L}_g$ lines by the phase $e^{i\A_g}$. This can be seen either by applying a set of H-junction crossing relations, or by deforming the TDLs on a hemisphere as in Figure~\ref{fig:permiso}. While both the isotopy anomaly and the cyclic permutation map $C$ are {\it a priori} dependent on the choice of the extrinsic curvature counter term (\ref{extrinsic}) in the definition of the TDLs, the cyclic permutation map $\widehat C$ on the cylinder is free of such ambiguity.

\begin{figure}[H]
\centering
\begin{minipage}{0.16\textwidth}
\includegraphics[width=1\textwidth]{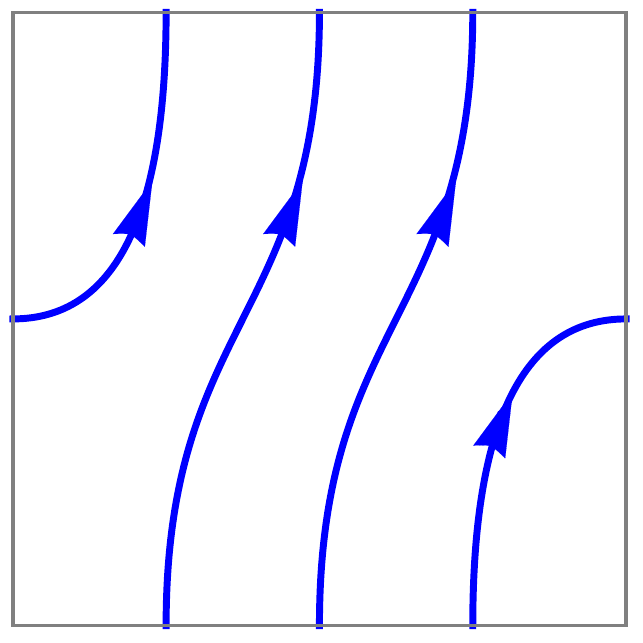}
\end{minipage}%
\begin{minipage}{0.12\textwidth}\begin{eqnarray*}~~{\rm crossing} \\ \longrightarrow ~~ \\ \\ \end{eqnarray*}
\end{minipage}%
\begin{minipage}{0.16\textwidth}
\includegraphics[width=1\textwidth]{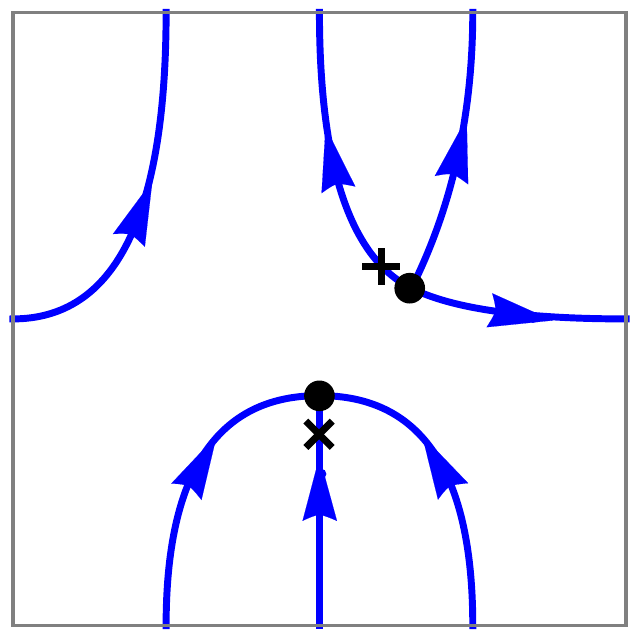}
\end{minipage}%
\begin{minipage}{0.12\textwidth}\begin{eqnarray*}~~{\rm isotopy} \\ =~~ ~ \\ \\ \end{eqnarray*}
\end{minipage}%
\begin{minipage}{0.16\textwidth}
\includegraphics[width=1\textwidth]{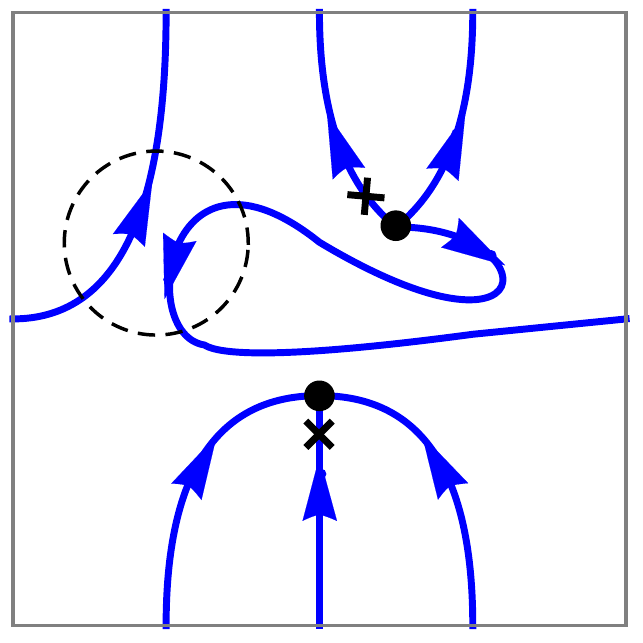}
\end{minipage}%
\begin{minipage}{0.12\textwidth}\begin{eqnarray*}~~{\widetilde K}^{{\cal L}_g, \overline{\cal L}_g}_{\overline{\cal L}_g,{\cal L}_g} \\ ~\longrightarrow ~ \\ \\ \end{eqnarray*}
\end{minipage}%
\begin{minipage}{0.16\textwidth}
\includegraphics[width=1\textwidth]{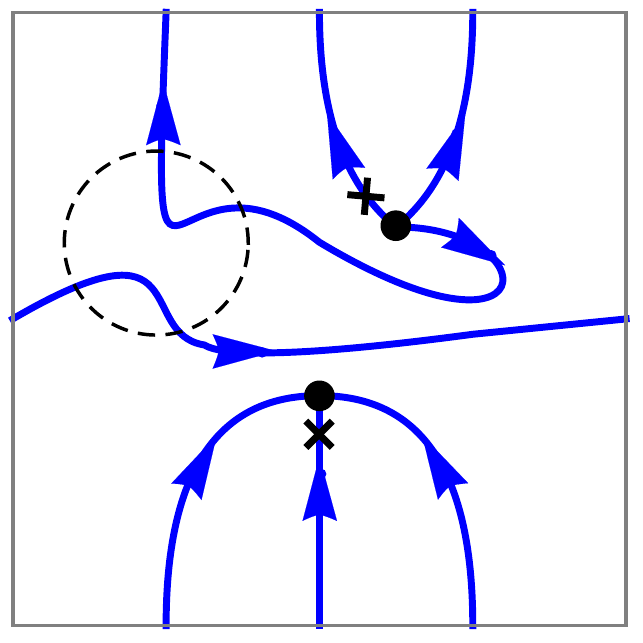}
\end{minipage}%
\\
\begin{minipage}{0.05\textwidth}\begin{eqnarray*} \\ =\quad  \\ \\ \end{eqnarray*}
\end{minipage}%
\begin{minipage}{0.16\textwidth}
\includegraphics[width=1\textwidth]{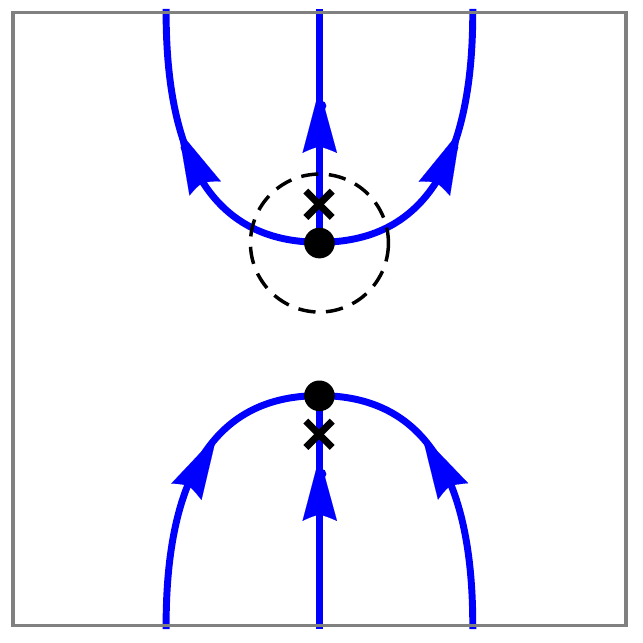}
\end{minipage}%
\begin{minipage}{0.17\textwidth}\begin{eqnarray*}~~{\rm permutation} \\ ~\longrightarrow \quad ~~ \\ \\ \end{eqnarray*}
\end{minipage}%
\begin{minipage}{0.16\textwidth}
\includegraphics[width=1\textwidth]{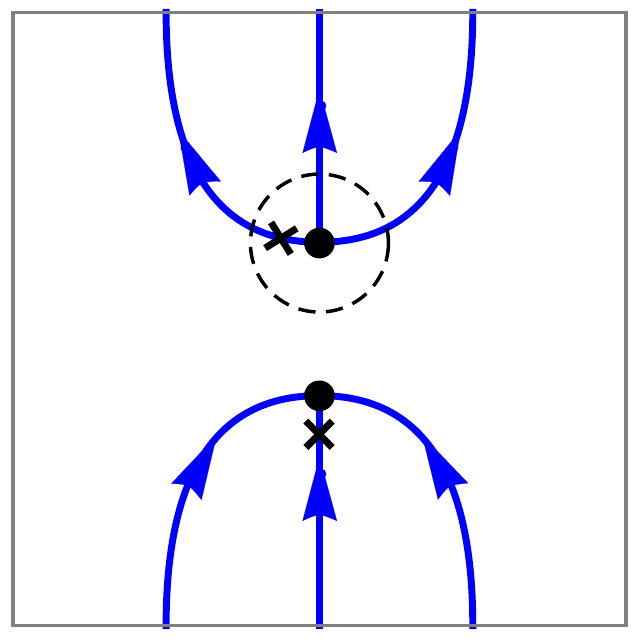}
\end{minipage}%
\begin{minipage}{0.14\textwidth}\begin{eqnarray*}~~{\rm crossing}^{-1} \\ \longrightarrow \; \quad  \\ \\ \end{eqnarray*}
\end{minipage}%
\begin{minipage}{0.16\textwidth}
\includegraphics[width=1\textwidth]{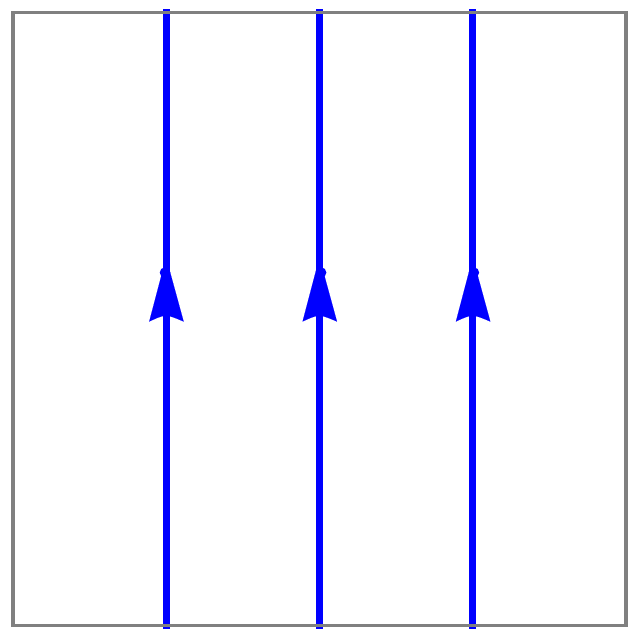}
\end{minipage}%
\caption{Relating the cyclic permutation map $\widehat C$ on the cylinder to the H-junction crossing kernels. The crossing transformation in the first and last steps are precisely the inverse of one another, with the specified ordering of legs specified by the ``$\times$''. Note that in the first and the last step, a certain crossing kernel involving trivial junction and its inverse are applied, respectively. In going to the second line, the horizontal loop is removed because it acts trivially on the bulk ground state.
}
\label{fig:perminv}
\end{figure}

There is a direct relation between the cylinder cyclic permutation map $\widehat C$ and the H-junction crossing kernels, that can be derived through a sequence of crossing and isotopy transformations (recall from Section~\ref{Sec:Trivial} that trivial junctions admit canonical junction vectors where the cyclic permutation map is trivial, and hence the ordering of lines is immaterial), as illustrated in Figure~\ref{fig:perminv} for the $\mathbb{Z}_3$ case. Given a $\mathbb{Z}_3$ element $g$, the cylinder cyclic permutation map $\widehat C$ acts on $V_{{\cal L}_g,{\cal L}_g,{\cal L}_g}$ by the product of the crossing phase ${\widetilde K}^{{\cal L}_g,\overline{\cal L}_g}_{\overline{\cal L}_g,{\cal L}_g}(I,I)$ with the cyclic permutation map $C_{{\cal L}_g,{\cal L}_g,{\cal L}_g}$ on the T-junction, the latter being equivalent to the crossing kernel ${\widetilde K}^{{\cal L}_g,I}_{{\cal L}_g,{\cal L}_g}(I,I)$ by \eqref{KToC}. In other words, the phase arising from the isotopy anomaly of Figure~\ref{fig:permiso} is precisely ${\widetilde K}^{{\cal L}_g,\overline{\cal L}_g}_{\overline{\cal L}_g,{\cal L}_g}(I,I)$. One can indeed verify that gauge-equivalent solutions to the pentagon identity belonging to the same cohomology class in $H^3(\mathbb{Z}_3, U(1))$ give rise to the same phase for $\widehat C$.

\begin{figure}[H]
\centering
\begin{minipage}{0.15\textwidth}
\includegraphics[width=1\textwidth]{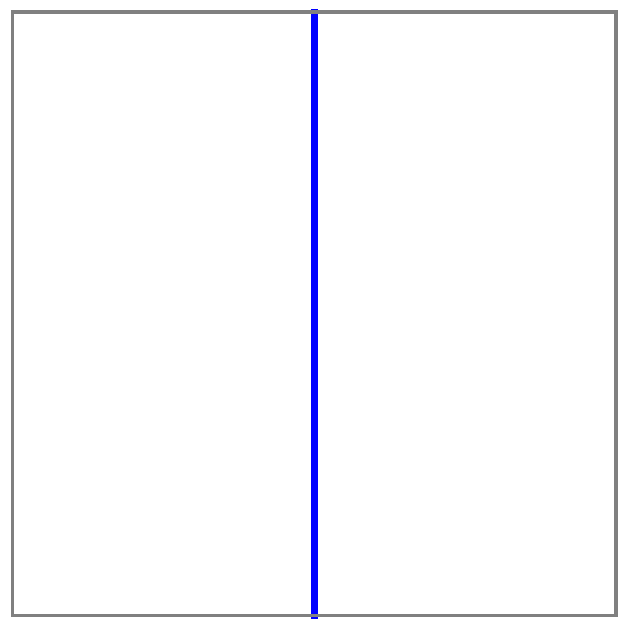}
\end{minipage}%
\begin{minipage}{0.1\textwidth}\begin{eqnarray*}~~\longrightarrow^{\!\!\!\!\!\!\!\!\! T^n}~ \\ \end{eqnarray*}
\end{minipage}%
\begin{minipage}{0.15\textwidth}
\includegraphics[width=1\textwidth]{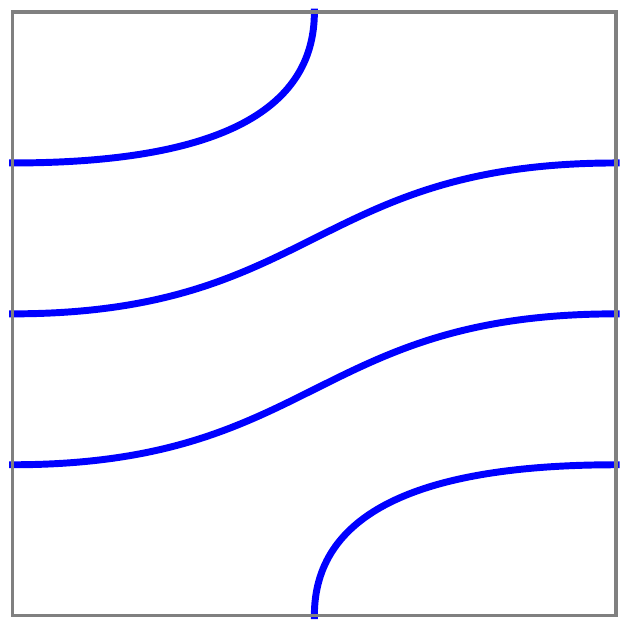}
\end{minipage}%
\begin{minipage}{0.1\textwidth}\begin{eqnarray*}~~\longrightarrow^{\!\!\!\!\!\!\!\!\! S}~ \\ \end{eqnarray*}
\end{minipage}%
\begin{minipage}{0.15\textwidth}
\includegraphics[width=1\textwidth]{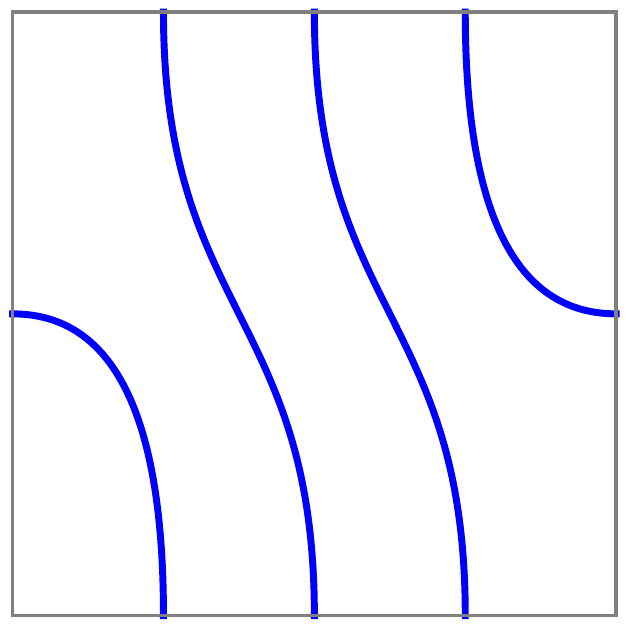}
\end{minipage}%
\caption{Relation between the spins of states in ${\cal H}_{{\cal L}_g}$ and the cyclic permutation map $\widehat C$, for an order $n=3$ element $g$.}
\label{fig:tHooftanomaly}
\end{figure}

We now derive a selection rule on the spin content of the defect Hilbert space ${\cal H}_{{\cal L}_g}$ at the end of the invertible line ${\cal L}_g$. Let $Z_{{\cal L}_g}$ be the partition function of ${\cal H}_{{\cal L}_g}$, {\it i.e.}, the space of defect operators on which ${\cal L}_g$ may end. Suppose $g^n=1$, then consider the modular $T^n$ transformation of $Z_{{\cal L}_g}$. We can bring it back to $Z_{{\cal L}_g}$ by a sequence of H-junction crossing relations, which accumulate to an overall phase $\C$, namely,
\ie\label{tcz}
T^n Z_{{\cal L}_g}(\tau,\bar\tau) = \C Z_{{\cal L}_g}(\tau,\bar\tau).
\fe
One may try to compute $\C$ from the crossing relations, but it is simpler to determine the answer by considering the modular $S$-transformation of (\ref{tcz}), and restricting to the junction vector space $V_{{\cal L}_g, \dotsc, {\cal L}_g}$, as in Figure~\ref{fig:tHooftanomaly}. We find
\ie
\widehat C = \gamma.
\fe
Since $\widehat C$ acts by a phase $e^{2\pi i k \over n}$, where $k$ is an integer that labels a class in $H^3(G, U(1))$ representing the 't Hooft anomaly, we learn that the states in ${\cal H}_{{\cal L}_g}$ have spin
\ie
\label{ZnSpin}
s \in {k\over n^2} + {1\over n}\mathbb{Z}.
\fe
In particular, this is the spin selection rule in ${\cal H}_{{\cal L}_g}$ for a TDL  ${\cal L}_g$ generating the $\bZ_n$ symmetry with an anomaly $[ k ] \in H^3(\bZ_n, U(1))= \bZ_n$ where $[ k] =k\, \, \text{mod} \,\, n$.

This is indeed the case in many known examples of 't Hooft anomalies, such as the chiral $U(1)$ rotation associated with a current algebra at nonzero level.

\subsubsection{Example: chiral $U(1)$ rotation in free compact boson}
\label{Sec:CompactBoson}

Let us consider the explicit example of the chiral $U(1)$ symmetry in the free boson CFT at the self-dual radius $R=1$, generated by the current $j = -2\partial X$, where $X$ is normalized such that\footnote{At a generic radius $R$, the chiral current generates a noncompact $\mathbb{R}$ symmetry, instead of $U(1)$.
}
\ie
X(z,\bar z) X(0) \sim - {1\over 2} \ln|z|^2.
\fe
Let ${\cal L}_\A$ be the invertible line associated with the shift symmetry $X_L \to X_L + \A$. An ${\cal L}_\A$ loop may be written explicitly as
\ie\label{Lalpha}
{\cal L}_\A = :\exp\left[-2\A \oint {dz\over 2\pi i} \partial X(z) \right]:,
\fe
with $:\cdots :$ the standard normal ordering of free fields. This definition is such that the expectation value of an empty ${\cal L}_\A$ loop on the plane is equal to 1, and that there is no isotopy anomaly. At the self-dual radius, the line ${\cal L}_{2\pi}$ is equivalent to the trivial line.  
The ground state in ${\cal H}_{{\cal L}_\A}$ is a defect operator that may be written as
\ie
\Psi_\A(z) = :e^{{i\A\over \pi} X_L(z)}: \, ,
\fe
with an implicit ${\cal L}_\A$ line attached. When $\Psi_\A(z)$ is inserted in a correlator, the location of the ${\cal L}_\A$ line is equivalent to a choice of branch cut in $z$.

The lines $I \equiv {\cal L}_{0}$ and ${\cal L}_{\pm{2\pi\over 3}}$ form the elements of a $\bZ_3$ fusion ring, and realize one of the 't Hooft anomalous solutions to the pentagon identity discussed in Section~\ref{Sec:SmallRanks}. The other 't Hooft anomalous solution can be realized by the analogous construction to \eqref{Lalpha} but with the anti-holomorphic $U(1)$ current. Three $\Psi_{2\pi\over 3}$'s joined by a T-junction of ${\cal L}_{2\pi\over 3}$ can be expanded in terms of bulk local operators, of the form
\ie\label{ooo}
\Psi_{2\pi\over 3}(z_1) \Psi_{2\pi\over 3}(z_2) \Psi_{2\pi\over 3}(z_3) = (z_{12}z_{23}z_{31})^{2\over 9} \Psi_{2\pi}({z_1+z_2+z_3\over 3}) + \cdots,
\fe
where $\Psi_{2\pi}$ has weight $(1,0)$, and the omitted terms involve operators of higher weights. Now taking $z_m = z e^{2\pi i m \over 3}$, the cyclic permutation map on the junction has the same effect as a ${2\pi \over 3}$ rotation on the operator $\Psi_{2\pi}(0)$ appearing on the RHS of the OPE, which produces the phase $e^{2\pi i \over 3}$ since $\Psi_{2\pi}$ is a spin 1 operator. Note that the defect operator $\Psi_{2\pi\over 3}$ has weight $({1 \over 9},0)$, which satisfies the spin selection rule \eqref{ZnSpin} for $n = 3$ and $k=1$.

\begin{figure}[H]
\centering
\begin{minipage}{0.22\textwidth}
\includegraphics[width=1\textwidth]{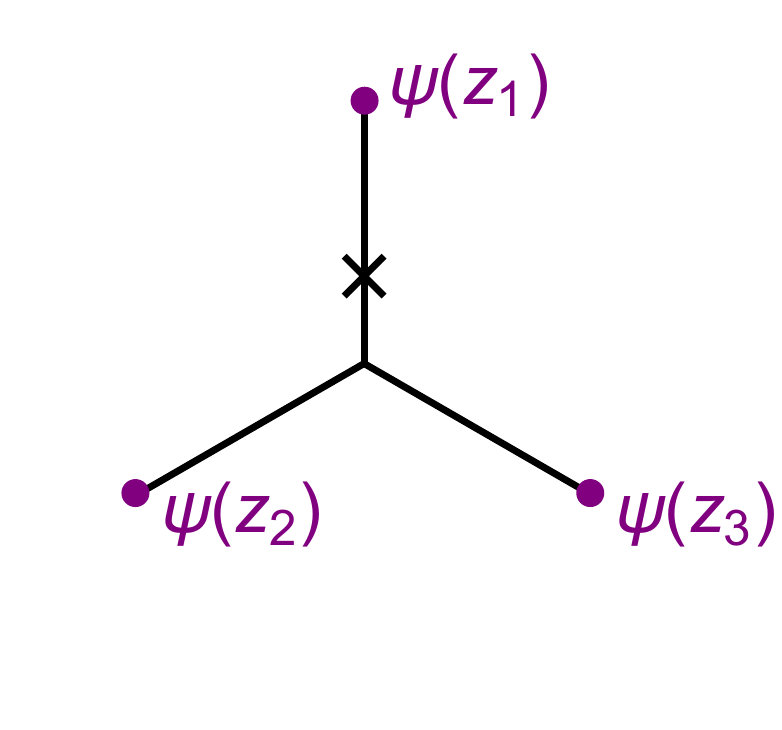}
\end{minipage}%
\begin{minipage}{0.08\textwidth}\begin{eqnarray*}~~\Longrightarrow~~ \\ \\ \end{eqnarray*}
\end{minipage}%
\begin{minipage}{0.22\textwidth}
\includegraphics[width=1\textwidth]{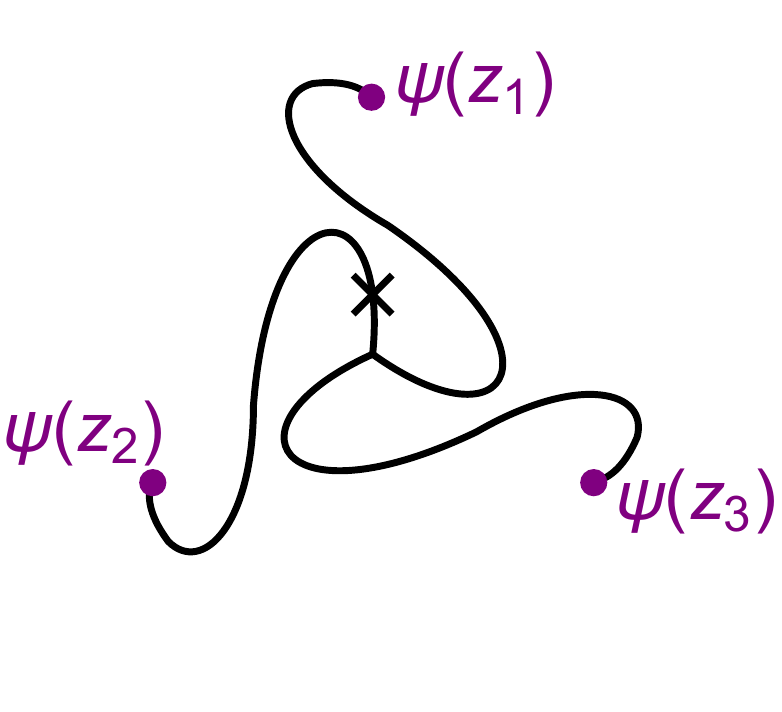}
\end{minipage}%
\begin{minipage}{0.08\textwidth}\begin{eqnarray*}~~\Longrightarrow~~ \\ \\ \end{eqnarray*}
\end{minipage}%
\begin{minipage}{0.22\textwidth}
\includegraphics[width=1\textwidth]{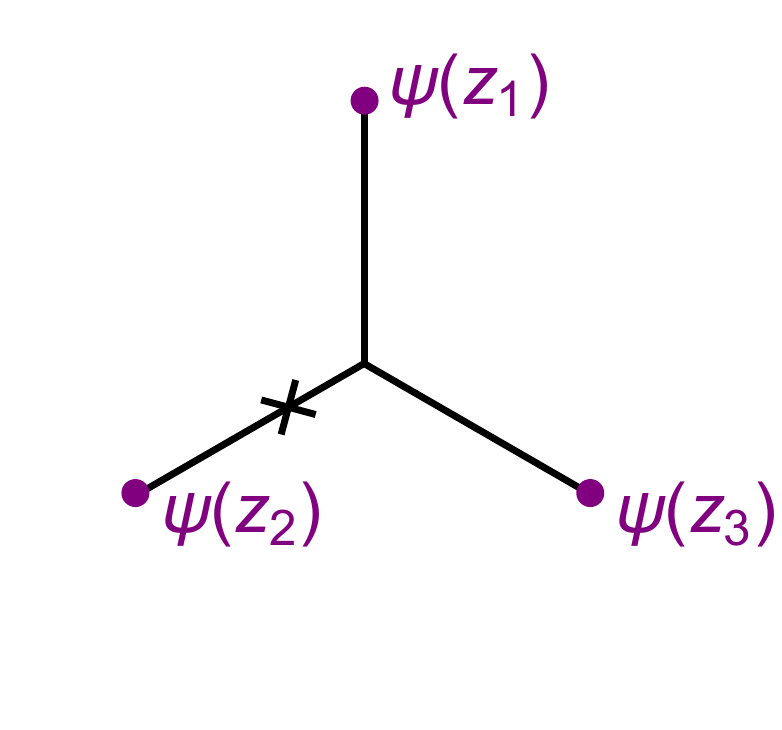}
\end{minipage}%
\caption{Moving the defect operators $\Psi(z_1), \Psi(z_2), \Psi(z_3)$ cyclically and then rotating each $\Psi$ by $2\pi \over 3$ have the same net effect as the cyclic permutation map on the T-junction, which amounts to relabeling the last leg at the junction (marked by ``$\times$").}
\label{fig:psipsipsi}
\end{figure}

Alternatively, we may consider a continuous deformation that moves $z_1$ to $z_2$, $z_2$ to $z_3$, and $z_3$ to $z_1$, while maintaining the junction and the angles at which the ${\cal L}_{2\pi\over 3}$ lines end on $\Psi_{2\pi\over 3}(z_i)$, as in Figure~\ref{fig:psipsipsi}. Under this transformation, the RHS of (\ref{ooo}) picks up a phase $e^{2\pi i {2\over 9}}$ due to the prefactor $(z_{12}z_{23}z_{31})^{2\over 9}$. We can deform this configuration isotopically to the original one transformed by the cyclic permutation map on the T-junction, provided that we also perform a $2\pi \over 3$ rotation on each of the three defect operators, totaling an extra phase $e^{2\pi i \over 9}$. The final net effect is again such that the cyclic permutation map on the T-junction produces the phase $e^{2\pi i \over 3}$.

\section{Topological defect lines in rational CFTs}
\label{sec:TDLsRCFTs}

Let us now discuss TDLs in rational conformal field theories (RCFTs), which are in general not invertible lines, and examine several concrete examples. We begin with the simplest case of diagonal modular invariant theories, where the TDLs that commute with the chiral vertex algebra are completely classified by the Verlinde lines. We then move on to considering TDLs in more general RCFTs, and further discuss their relations to orbifolds and dualities.

\subsection{Verlinde lines in diagonal RCFTs}
\label{sec:verlindelines}

In an RCFT with diagonal modular invariance, there is a simple and explicit construction for a family of TDLs, known as the \textit{Verlinde lines}, that commute with not only the Virasoro algebra but the entire (left and right) chiral vertex algebra of the RCFT \cite{Verlinde:1988sn,Petkova:2000ip}.  
In fact, modular invariance constrains the  Verlinde lines in a diagonal RCFT to be in one-to-one correspondence with the chiral vertex algebra primaries.

We begin with some general discussions of diagonal RCFTs.  The partition function of an RCFT takes the form
\ie
Z(\tau,\bar\tau) = \sum_{i, j} n_{i j} \chi_i(\tau) \bar\chi_{ j}(\bar\tau),
\fe
where $\chi_i(\tau)$ is the character of an irreducible representation of the chiral vertex algebra, labeled by the index $i$. Under the modular $S$ transformation, we have $\chi_i(-1/\tau) = \sum_j S_{ij} \chi_j(\tau)$, where the matrix $S_{ij}$ is unitary and symmetric. The vacuum representation is labeled by $i=0$, and the degeneracies $n_{ij}$ are non-negative integers, with $n_{i j}=\delta_{i j}$ for a diagonal modular invariant theory. The fusion ring takes the form
\ie{}
[i] \times [j] = \sum_k N_{ij}^k [k],
\fe
where the fusion coefficients $N_{ij}^k$ are non-negative integers that obey the Verlinde formula,
\ie
N_{ij}^k = \sum_\ell {S_{i\ell} S_{j\ell} S^*_{k\ell}\over S_{0\ell}}.
\fe

Let us now review the action $\widehat{\cal L}$ of  a Verlinde line on the bulk Hilbert space.  
In a diagonal modular invariant theory, the primaries are simply denoted by $|\phi_k\ra$, in correspondence with irreducible representations of the chiral vertex algebra. There is a one-to-one correspondence between the Verlinde lines and the  primaries of the chiral vertex algebra. A Verlinde line ${\cal L}_k$ corresponding to $\phi_k$ is a TDL with the property
\ie\label{verlinetdl}
\widehat{\cal L}_k |\phi_i\ra = {S_{ki}\over S_{0i}} |\phi_i\ra\,,
\fe
and that it commutes with both the left and the right chiral vertex algebra.  
The action $\widehat{\cal L}$ on the bulk Hilbert space is highly constrained by the requirement that in the $S$ dual channel, the Hilbert space ${\cal H}_{{\cal L}_k}$ of defect operators at the end of ${\cal L}_k$ can be decomposed into the left and right Virasoro multiplets with \textit{non-negative integral} degeneracies.  
Indeed, from the modular $S$ transformation of the torus character, we deduce the partition function of the Hilbert space ${\cal H}_{{\cal L}_k}$ of defect operators at the end of ${\cal L}_k$,
\ie
\label{VerlindeHilbert}
Z_{{\cal L}_k}(\tau,\bar\tau) = \sum_{i,j} N_{ki}^{j} \chi_i(\tau) \bar\chi_j(\bar\tau).
\fe
Namely, ${\cal H}_{{\cal L}_k}$ consists of states in the representation $i$ of the chiral vertex algebra and representation $j$ of the anti-chiral vertex algebra, with degeneracy $N_{ki}^j$, which are in particular non-negative integers. 
For $k\not=0$, states in ${\cal H}_{{\cal L}_k}$ typically have non-integer spins. We will investigate the modular $T$ transformation of $Z_{{\cal L}_k}$ in detail later.

By the Verlinde formula, we see that the fusion ring of the Verlinde lines takes an identical form as the fusion ring of the chiral algebra primary operators in the RCFT,
\ie
{\cal L}_i {\cal L}_j = \sum_k N_{ij}^k {\cal L}_k.
\fe
Interpreted as the fusion relations of TDLs, the RHS is a direct sum of TDLs, where ${\cal L}_k$ appears with multiplicity $N_{ij}^k$. Note that the Verlinde lines always generate a commutative fusion ring.

So far, we have reviewed the action $\widehat{\cal L}_k$ of Verlinde lines on the bulk Hilbert space.  
 The full fusion category of Verlinde lines in a diagonal RCFT, which can be obtained by forgetting the braiding in the modular tensor category, has long been discussed since the work of Moore and Seiberg \cite{Moore:1988qv,Moore:1989yh,Moore:1989vd}. We will review several examples in the rest of this section.

The formula (\ref{verlinetdl}) for the Verlinde lines does not apply to non-diagonal modular invariant theories. There is a straightforward generalization when the non-diagonal theory is defined through an automorphism of the fusion rules  for the bulk local operators  \cite{Petkova:2000ip}. For block-diagonal modular invariant theories, the fusion relations of TDLs can be noncommutative (see, for example, \cite{Fuchs:2007vk} for related discussions).

\subsubsection{Ising model}
\label{Sec:Ising}

The $c={1\over 2}$ critical Ising model has three Virasoro primaries:
\ie
\label{IsingBulk}
1_{0,0},~~\varepsilon_{{1\over2},{1\over2}},~~\sigma_{{1\over16},{1\over16}},
\fe
which are the identity operator, the energy operator, and the spin field, respectively, with their conformal weights indicated in the subscripts.

There are three simple Verlinde lines: the trivial line $I$, the $\mathbb{Z}_2$ invertible line $\eta$, and the $N$ line. 
Together they form a $\mathbb{Z}_2$ Tambara-Yamagami category, which is discussed in Section~\ref{Sec:DualityTY}.  
In particular, the $N$ line  is the  duality defect \cite{Frohlich:2004ef,Frohlich:2006ch,Frohlich:2009gb} for the Kramers-Wannier duality \cite{PhysRev.60.252}. 
These TDLs act on the bulk local primary operators with eigenvalues:
\begin{equation}
\label{isingaction}
\begin{tabu}{ccccc}
 & & 1 & \varepsilon & \sigma
\\
\widehat\eta : & \quad & 1 & 1 & -1
\\
\widehat N : & \quad & \sqrt2 & -\sqrt2 & 0
\end{tabu}
\end{equation}
The fusion relations are given by
\ie
\eta^2 = I,\quad N^2 = I + \eta,\quad  \eta N= N\eta=N,
\fe
from which we see that $\eta$ and $N$ are both unoriented, namely, $\overline\eta=\eta$, $\overline{N}=N$.
The Hilbert spaces of defect operators at the endpoints of $\eta$ and $N$ are determined by the fusion coefficients as in \eqref{VerlindeHilbert}, and are spanned by
\ie{}
\label{IsingDefectSpec}
& {\cal H}_\eta ~:~ \psi_{{1\over 2},0},~~ \tilde\psi_{0,{1\over 2}},~~\mu_{{1\over 16},{1\over 16}},
\\
& {\cal H}_N ~:~ s_{{1\over 16},0},~~\tilde s_{0,{1\over 16}},~~\Lambda_{{1\over 16},{1\over 2}},~~\tilde\Lambda_{{1\over 2},{1\over 16}},
\fe
where we only listed the primaries. Note that the defector operator $\mu$ is the disorder operator in the critical Ising model.

\begin{figure}[H]
\centering
\begin{minipage}{0.15\textwidth}
\includegraphics[width=1\textwidth]{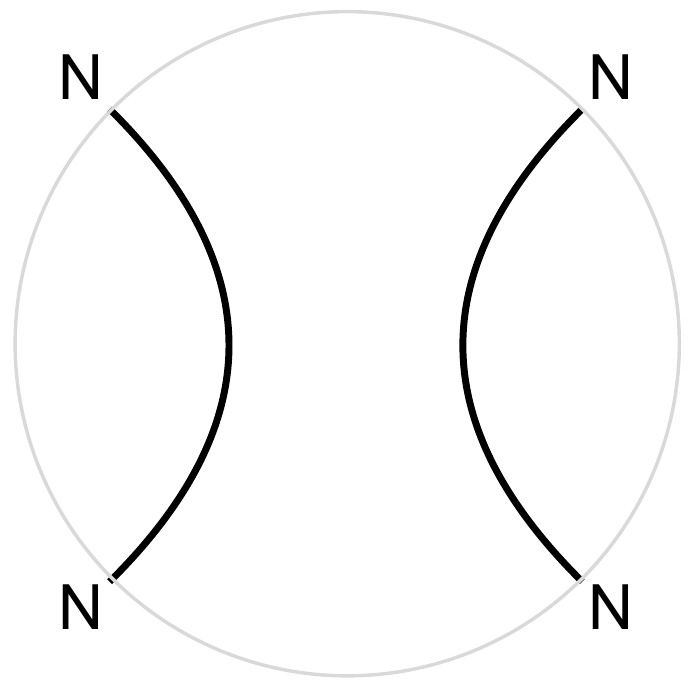}
\end{minipage}%
\begin{minipage}{0.1\textwidth}\begin{eqnarray*}~= {1\over \sqrt{2}} \\ \end{eqnarray*}
\end{minipage}%
\begin{minipage}{0.15\textwidth}
\includegraphics[width=1\textwidth]{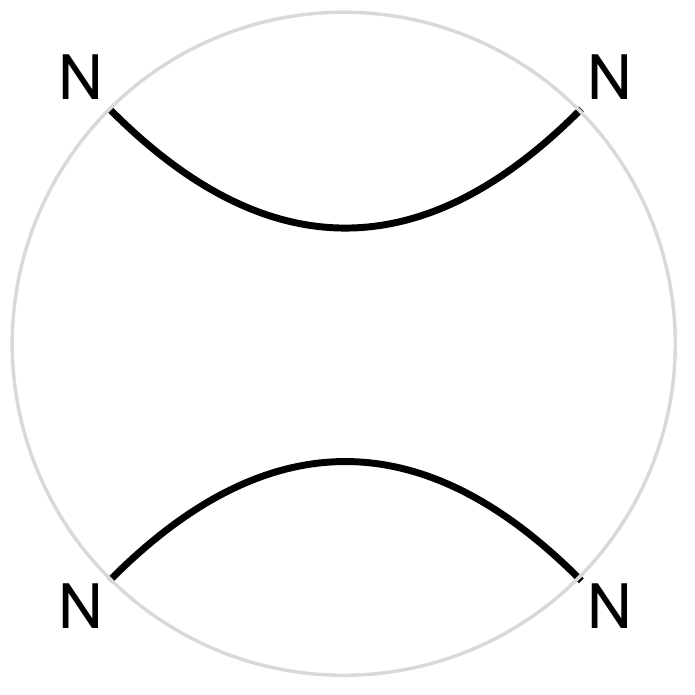}
\end{minipage}%
\begin{minipage}{0.1\textwidth}\begin{eqnarray*}~+ {1\over \sqrt{2}} \\ \end{eqnarray*}
\end{minipage}%
\begin{minipage}{0.15\textwidth}
\includegraphics[width=1\textwidth]{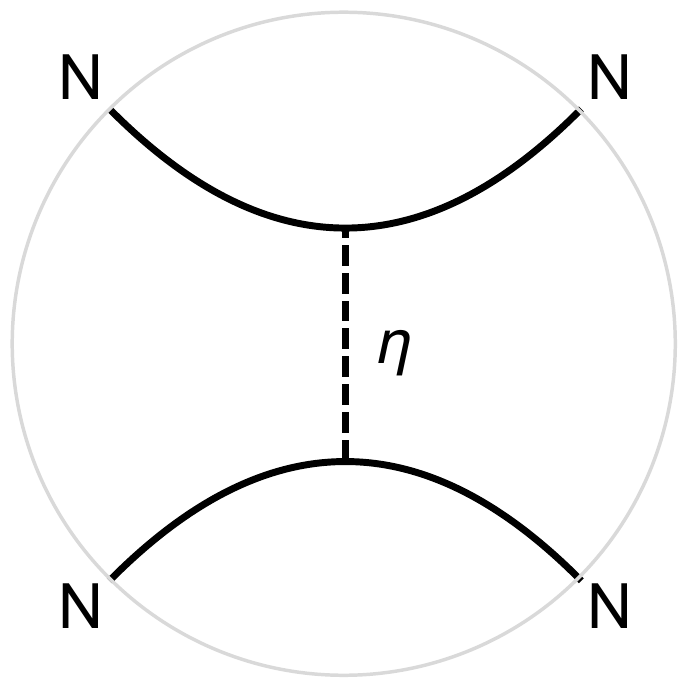}
\end{minipage}%
\\
\bigskip
\begin{minipage}{0.15\textwidth}
\includegraphics[width=1\textwidth]{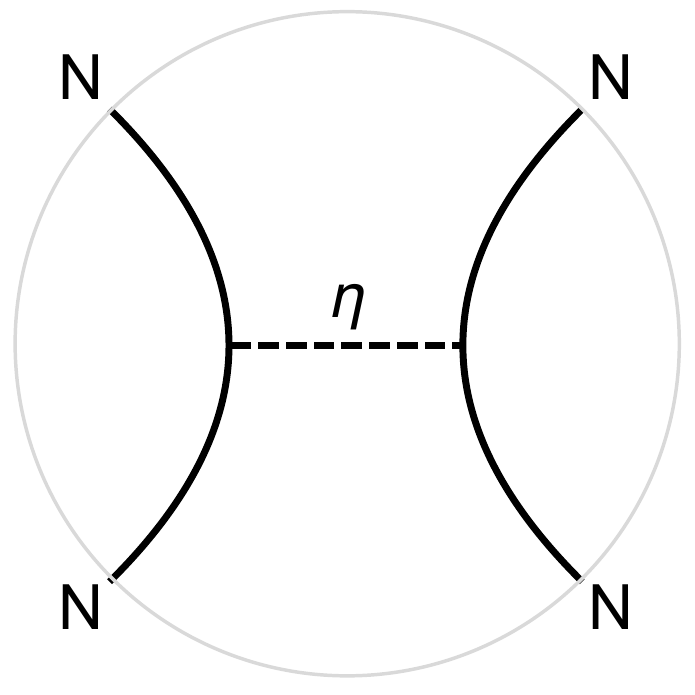}
\end{minipage}%
\begin{minipage}{0.1\textwidth}\begin{eqnarray*}~= {1\over \sqrt{2}} \\ \end{eqnarray*}
\end{minipage}%
\begin{minipage}{0.15\textwidth}
\includegraphics[width=1\textwidth]{figures/HpNNNNI.pdf}
\end{minipage}%
\begin{minipage}{0.1\textwidth}\begin{eqnarray*}~- {1\over \sqrt{2}} \\ \end{eqnarray*}
\end{minipage}%
\begin{minipage}{0.15\textwidth}
\includegraphics[width=1\textwidth]{figures/HpNNNNeta.pdf}
\end{minipage}%
\caption{The H-junction crossing relations involving the  TDL $N$.}
\label{fig:Neumanncross}
\end{figure}

\begin{figure}[H]
\centering
\begin{minipage}{0.15\textwidth}
\includegraphics[width=1\textwidth]{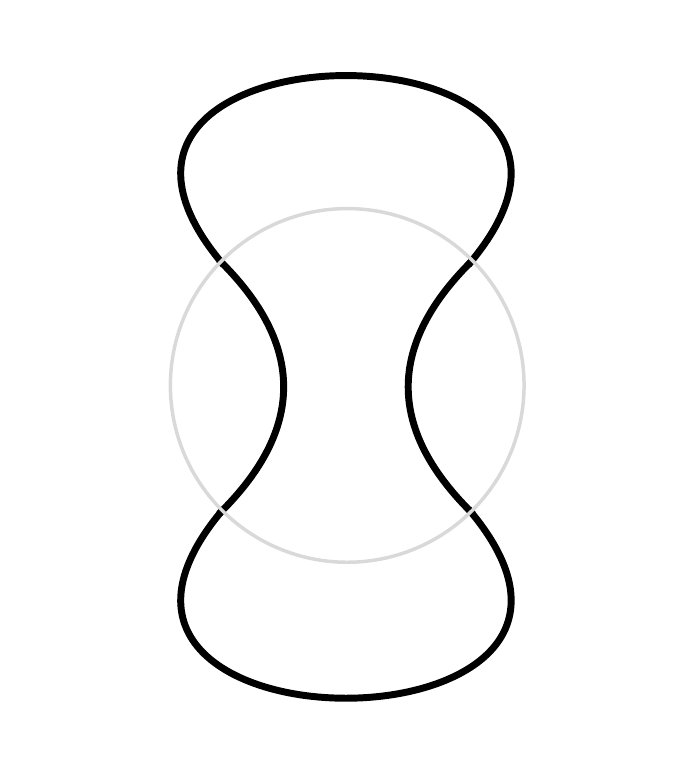}
\end{minipage}%
\begin{minipage}{0.1\textwidth}\begin{eqnarray*}~= {1\over \sqrt{2}} \\ \end{eqnarray*}
\end{minipage}%
\begin{minipage}{0.15\textwidth}
\includegraphics[width=1\textwidth]{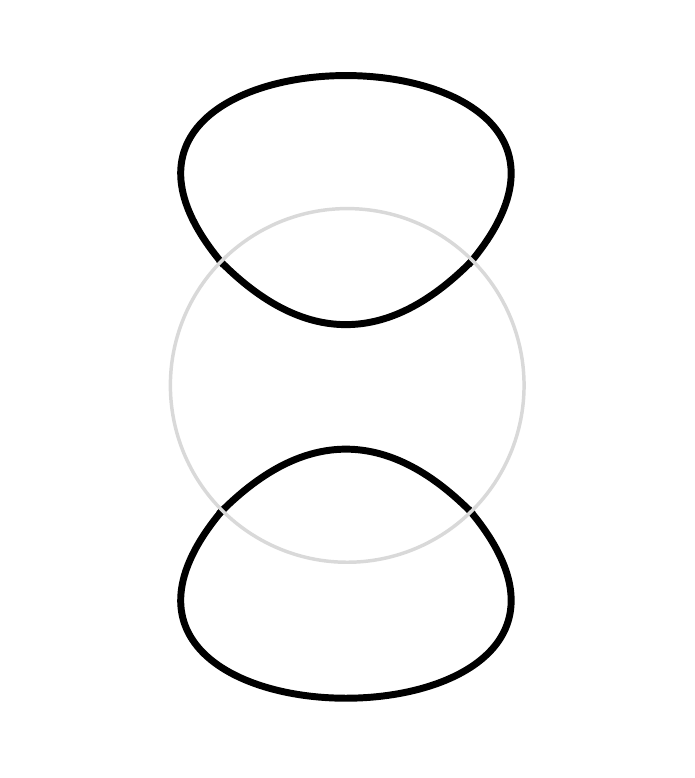}
\end{minipage}%
\begin{minipage}{0.1\textwidth}\begin{eqnarray*}~+ {1\over \sqrt{2}} \\ \end{eqnarray*}
\end{minipage}%
\begin{minipage}{0.15\textwidth}
\includegraphics[width=1\textwidth]{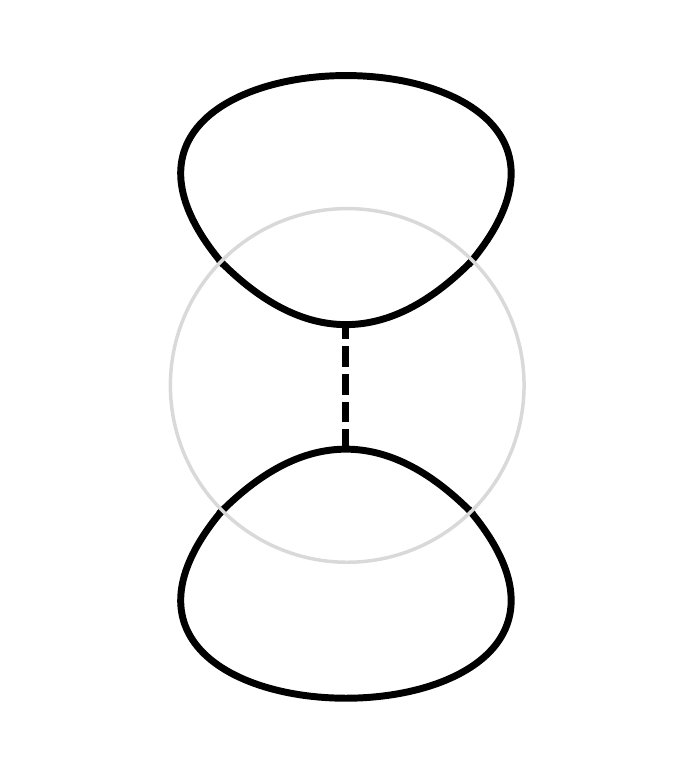}
\end{minipage}%
\quad \quad \quad \quad \begin{minipage}{0.1\textwidth}
\includegraphics[width=1\textwidth]{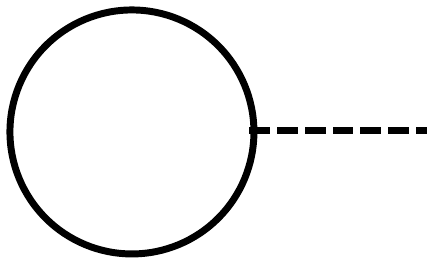}
\end{minipage}$\;=0$%
\\
\begin{minipage}{0.15\textwidth}
\includegraphics[width=1\textwidth]{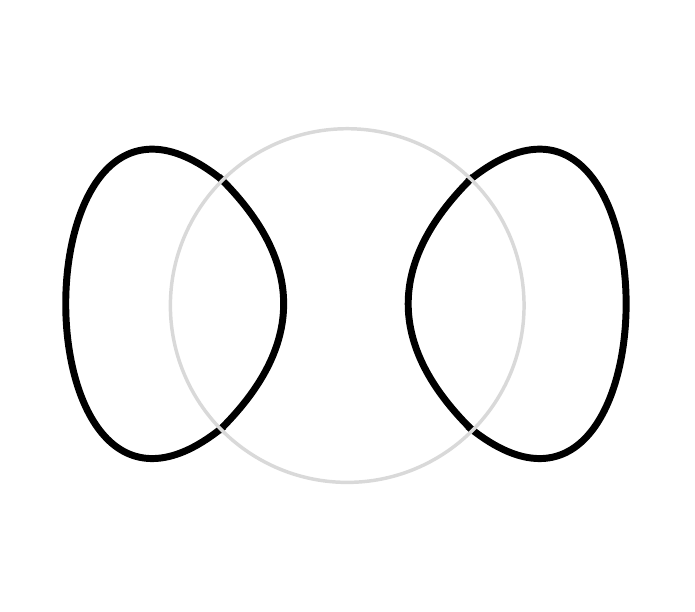}
\end{minipage}%
\begin{minipage}{0.1\textwidth}\begin{eqnarray*}~= {1\over \sqrt{2}} \\ \end{eqnarray*}
\end{minipage}%
\begin{minipage}{0.15\textwidth}
\includegraphics[width=1\textwidth]{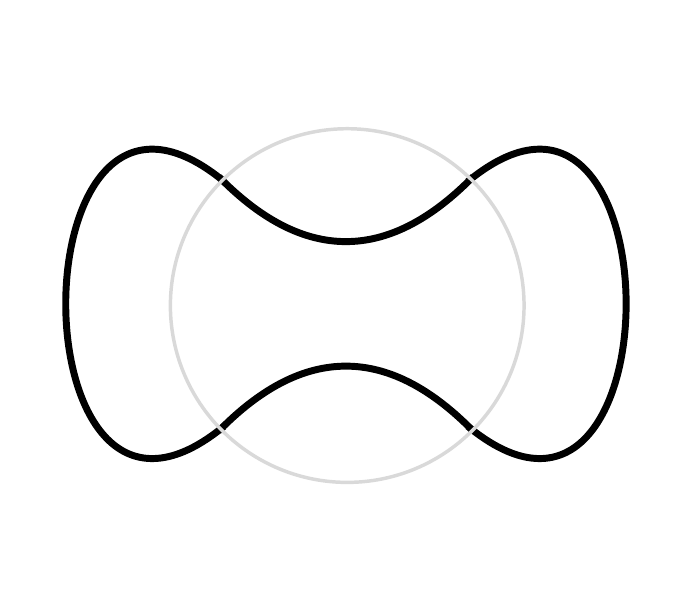}
\end{minipage}%
\begin{minipage}{0.1\textwidth}\begin{eqnarray*}~+ {1\over \sqrt{2}} \\ \end{eqnarray*}
\end{minipage}%
\begin{minipage}{0.15\textwidth}
\includegraphics[width=1\textwidth]{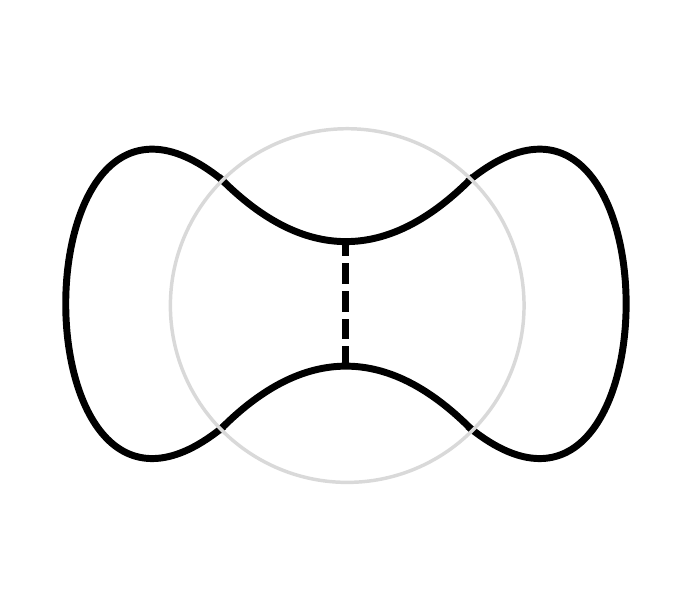}
\end{minipage}%
\quad \quad \quad \quad \begin{minipage}{0.028\textwidth}
\includegraphics[width=1\textwidth]{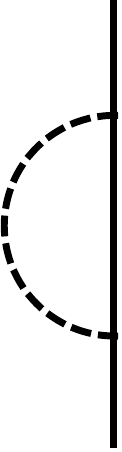}
\end{minipage}$~\; =$%
\begin{minipage}{0.053\textwidth}
\includegraphics[width=1\textwidth]{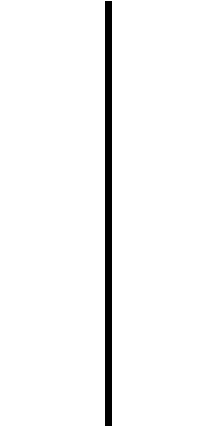}
\end{minipage}
\\
\begin{minipage}{0.15\textwidth}
\includegraphics[width=1\textwidth]{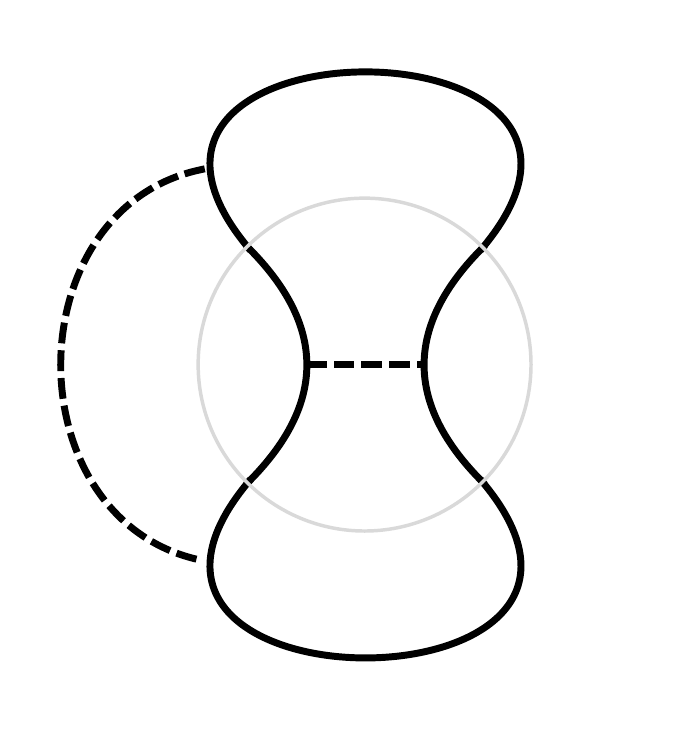}
\end{minipage}%
\begin{minipage}{0.1\textwidth}\begin{eqnarray*}~= {1\over \sqrt{2}} \\ \end{eqnarray*}
\end{minipage}%
\begin{minipage}{0.15\textwidth}
\includegraphics[width=1\textwidth]{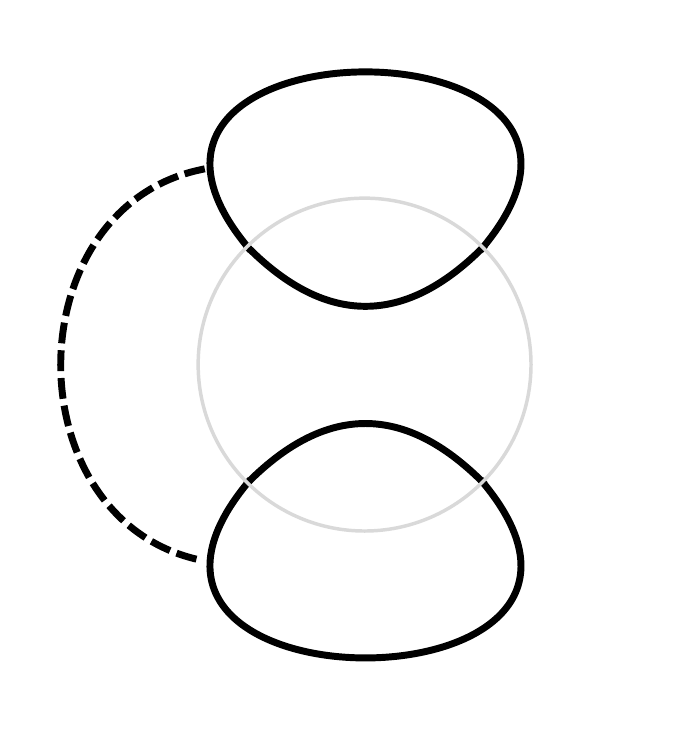}
\end{minipage}%
\begin{minipage}{0.1\textwidth}\begin{eqnarray*}~- {1\over \sqrt{2}} \\ \end{eqnarray*}
\end{minipage}%
\begin{minipage}{0.15\textwidth}
\includegraphics[width=1\textwidth]{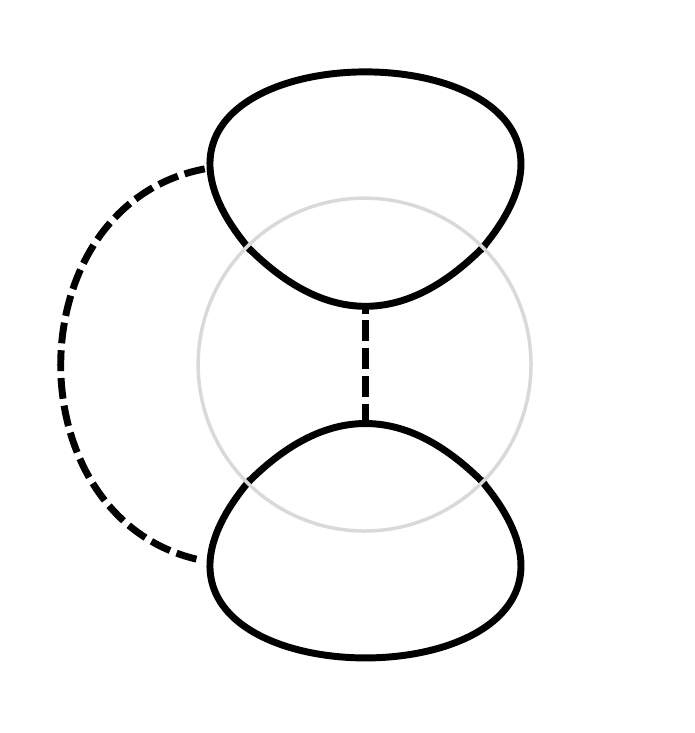}
\end{minipage}%
\quad \quad \quad \begin{minipage}{0.1\textwidth}
\includegraphics[width=1\textwidth]{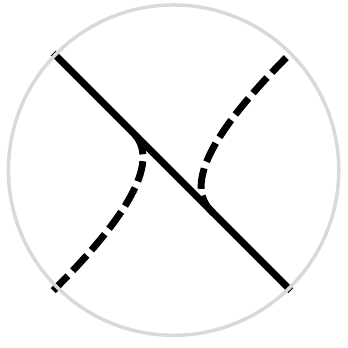}
\end{minipage}$\; = - $%
\begin{minipage}{0.1\textwidth}
\includegraphics[width=1\textwidth]{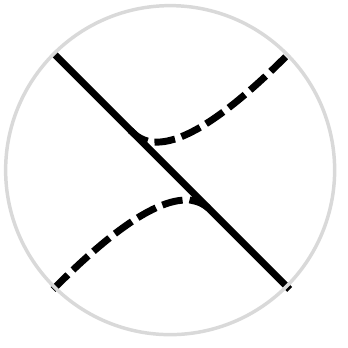}
\end{minipage}
\caption{Some consequences of the H-junction crossing relations and constraints thereof.}
\label{fig:emptycircles}
\end{figure}

The only nontrivial T-junctions involving simple TDLs are the ones corresponding to the junction vector space $V_{N,N,\eta}$ and its cyclic permutations. We fix a basis vector $v \in V_{N,N,\eta}$, along with its permutation images $C_{N,N,\eta}(v) \in V_{N,\eta,N}$, $C_{N,\eta,N}(C_{N,N,\eta}(v)) \in V_{\eta, N,N}$.
We normalize their two-point functions to
\ie
h(v,C_{N,\eta,N}(C_{N,N,\eta}(v)))=h(C_{N,N,\eta}(v),C_{N,N,\eta}(v))=\sqrt{2},
\fe
such their norm is $2^{1\over 4}$, the same as the norm of the identity junction vectors in the trivial junction spaces $V_{N,N,I}$, $V_{N,I,N}$, and $V_{I,N,N}$. In the following, the H-junction crossing kernels will be written with respect to the basis vector $v\in V_{N,N,\eta}$ and the identity junction vector in $V_{N,N,I}$ and $V_{\eta,\eta, I}$. The nontrivial crossing kernels are \cite{Moore:1988qv}
\ie{}
\begin{pmatrix} {\widetilde K}_{N,N}^{N,N}(I, I) & {\widetilde K}_{N,N}^{N,N}(I, \eta) \\ {\widetilde K}_{N,N}^{N,N}(\eta, I) & {\widetilde K}_{N,N}^{N,N}(\eta,\eta) \end{pmatrix} =  \begin{pmatrix} {1\over \sqrt{2}}  & {1\over \sqrt{2}} \\ {1\over \sqrt{2}}  & -{1\over \sqrt{2}} \end{pmatrix},
\quad
{\widetilde K}^{\eta ,N}_{N,\eta}(N,N) = -1,
\fe
The crossing relations and some of their consequences are illustrated in Figure~\ref{fig:Neumanncross} and Figure~\ref{fig:emptycircles}, respectively.

\begin{figure}[H]
\centering
\begin{minipage}{0.2\textwidth}
\includegraphics[width=1\textwidth]{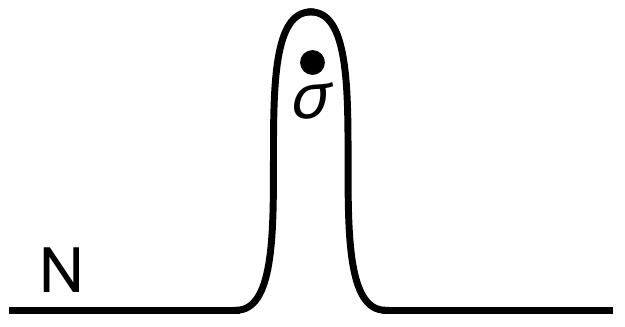}
\end{minipage}%
~~
\begin{minipage}{0.07\textwidth}\begin{eqnarray*}=~{ \alpha \over \sqrt{2}}~~~~~~~~~~~~~~\\ \end{eqnarray*}
\end{minipage}%
~~\begin{minipage}{0.2\textwidth}
\includegraphics[width=1\textwidth]{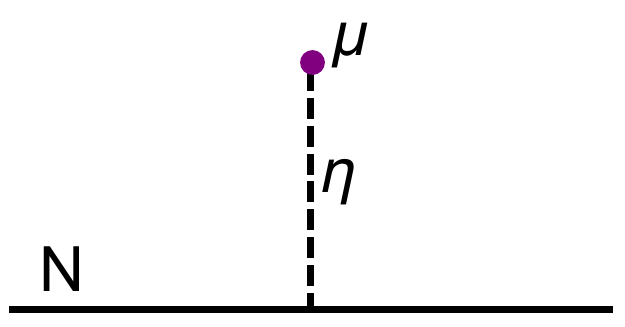}
\end{minipage}%
\caption{In the critical Ising model, moving the TDL  ${ N}$  past the spin field $\sigma$ leaves behind the $\mathbb{Z}_2$ line $\eta$ attached to the  defect operator $\mu\in{\cal H}_\eta$, and a T-junction. $N$ is the order/disorder duality defect that exchanges $\sigma$ with $\mu$.}
\label{fig:KWduality}
\end{figure}

The duality defect $N$ in the critical Ising model is one of the simplest examples of a non-invertible line.  From \eqref{isingaction}, we see that the spin field $\sigma$ has zero eigenvalue under $\widehat{N}$.  
What happens when we deform the TDL $N$ past the spin field $\sigma$ then?  
It turns out that as we do so, the spin field $\sigma$ leaves behind the $\mathbb{Z}_2$ line $\eta$ attached to the defect operator $\mu \in {\cal H}_\eta$, as in Figure~\ref{fig:KWduality}.  
More precisely, the action $\widehat N^v : {\cal H} \rightarrow {\cal H}_\eta$ with $v \in V_{\eta, N,N}$ -- defined by the right figure of Figure~\ref{fig:Lact}, with ${\cal L} = N$, ${\cal L}' = \eta$, and $\phi(x)=\sigma(x)$ -- acts on $\sigma$ as  
\ie\label{Nhatv}
\widehat N^v ~:~
| \sigma \ra &\mapsto  \alpha \, |\mu\ra\,,
\fe
for some coefficient $\alpha$. This is indeed what we expect of the duality defect $N$: it exchanges the spin field $\sigma$ (a bulk local operator) with the disorder operator $\mu$ (a defect operator) \cite{Frohlich:2004ef}. By contrast, the action $\widehat N: {\cal H} \rightarrow {\cal H}$, defined by the left figure of Figure~\ref{fig:Lact}, maps within the bulk Hilbert space, and therefore projects out $\sigma$.

\begin{figure}[H]
\centering
\begin{minipage}{0.15\textwidth}
\includegraphics[width=1\textwidth]{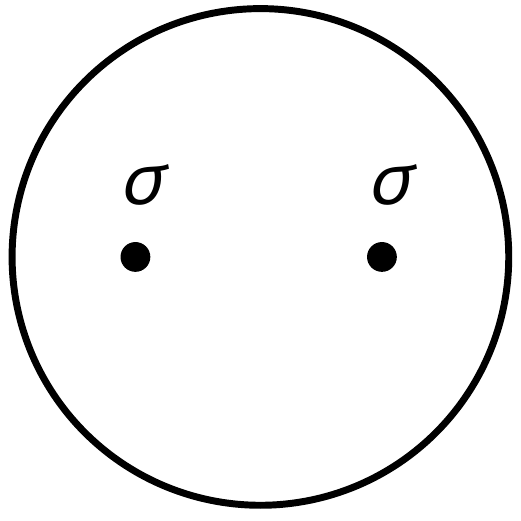}
\end{minipage}%
\begin{minipage}{0.08\textwidth}\begin{eqnarray*}~={1\over\sqrt{2}} ~\\ \end{eqnarray*}
\end{minipage}%
~\begin{minipage}{0.16\textwidth}
\includegraphics[width=1.2\textwidth]{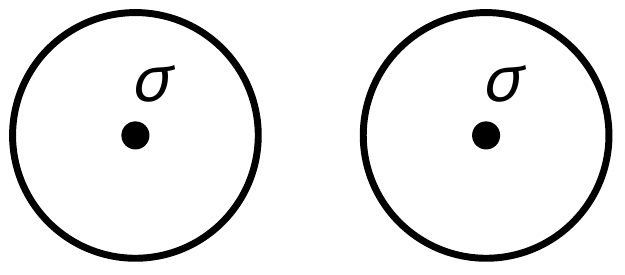}
\end{minipage}%
~\begin{minipage}{0.08\textwidth}\begin{eqnarray*}~~~+~{1\over\sqrt{2}}~~~~~~~\\ \end{eqnarray*}
\end{minipage}%
~~~\begin{minipage}{0.16\textwidth}
\includegraphics[width=1.2\textwidth]{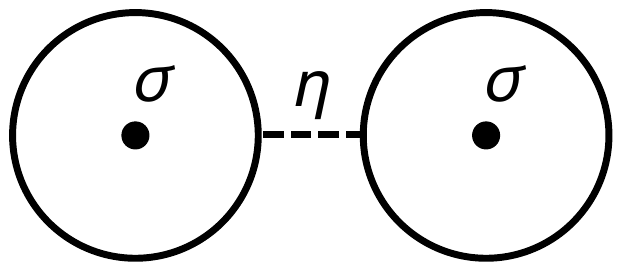}
\end{minipage}%
\caption{Applying partial fusion  to the two-point function of spin fields $\la \sigma(x) \sigma(0)\ra$ circled by an $N$ loop.}
\label{fig:KW2pt}
\end{figure}

Let us prove Figure~\ref{fig:KWduality}, or equivalently, \eqref{Nhatv} and determine the coefficient $\alpha$. Since the TDL $N$ preserves the conformal weights, by inspecting the bulk \eqref{IsingBulk} and defect Hilbert spaces \eqref{IsingDefectSpec}, we see that only the defect operator $\mu \in {\cal H}_\eta$ can potentially be created, as we move $N$ past $\sigma$. The bulk local operator $\sigma$ itself is not created because the eigenvalue of $\widehat N$ on $\sigma$ is 0. It remains to compute the coefficient $\A$ in \eqref{Nhatv}, and show that it is nonzero. We normalize $\sigma(x)$ and $\mu(x)$ such that they have identical two-point functions.  Consider the two-point function of spin fields circled by an $N$ loop, as in Figure~\ref{fig:KW2pt}, and apply partial fusion to a pair of $N$ lines using Figure~\ref{fig:Neumanncross}. The first term on the RHS is zero since $\widehat N|\sigma\ra=0$, while the second term is $1\over\sqrt2$ multiplied by the two-point function of the defect operator $\widehat{N}^v\cdot \sigma\in {\cal H}_\eta$. 
Matching the coefficients on the two sides, we obtain $\alpha=\sqrt{2}$.

\subsubsection*{Correlation functions of defect operators}

\begin{figure}[H]
\centering
\bigskip
\begin{minipage}{0.15\textwidth}
\includegraphics[width=1\textwidth]{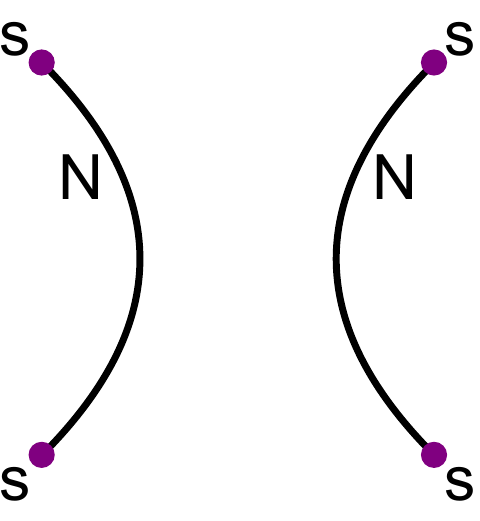}
\end{minipage}%
\begin{minipage}{0.23\textwidth}\begin{eqnarray*}\quad  \\ \end{eqnarray*}
\end{minipage}%
\begin{minipage}{0.15\textwidth}
\includegraphics[width=1\textwidth]{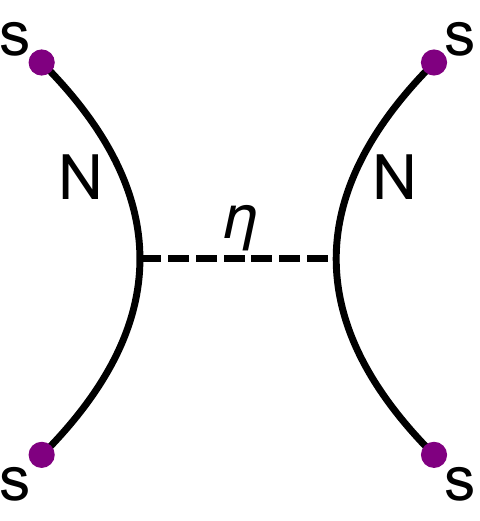}
\end{minipage}%
\caption{The four-point functions of $s(x)$ joined by H-junctions, $\left\la s(x_1) s(x_2) s(x_3) s(x_4) \right\ra_{\rm H}^I$ and $\left\la s(x_1) s(x_2) s(x_3) s(x_4) \right\ra_{\rm H}^\eta$.}
\label{fig:sssscorr}
\end{figure}

Let us consider a few examples of correlation functions of defect operators (at the end of TDLs) in the critical Ising model. First consider the four-point function of the weight-$({1\over 16},0)$ operator $s(x)$ in ${\cal H}_N$, where $s(x_1)$, $s(x_2)$ are connected by an $N$ line, and $s(x_3)$, $s(x_4)$ connected by another $N$ line. The $12\to 34$ channel contains a single Virasoro conformal block, namely, the identity channel block (see the left figure of Figure~\ref{fig:sssscorr}). Thus, we have
\ie\label{trivialchannel}
\left\la s(x_1) s(x_2) s(x_3) s(x_4) \right\ra_{\rm H}^I = x_{13}^{-{1\over 8}} x_{24}^{-{1\over 8}} {\cal F}({1\over 16},{1\over 16},{1\over 16},{1\over 16}; 0; x),
\fe
where $x\equiv {x_{12}x_{34}\over x_{13}x_{24}}$ is the conformally invariant cross ratio, and $s(x)$ is normalized by $\la s(x) s(0)\ra=x^{-{1\over 8}}$ (with the two $s$'s connected by an $N$ line).

Similarly, we can consider the four-point function of $s(x)$ joined by an H-junction where the internal TDL is the $\mathbb{Z}_2$ line $\eta$ rather than the trivial line $I$. In this case, the $12\to 34$ channel contains a single conformal block corresponding to the internal primary $\psi \in {\cal H}_\eta$ (see the right figure of Figure~\ref{fig:sssscorr}). We have
\ie\label{etachannel}
\left\la s(x_1) s(x_2) s(x_3) s(x_4) \right\ra_{\rm H}^\eta = C_{ss\psi}^2 x_{13}^{-{1\over 8}} x_{24}^{-{1\over 8}} {\cal F}({1\over 16},{1\over 16},{1\over 16},{1\over 16}; {1\over 2}; x),
\fe
The conformal blocks appearing in (\ref{trivialchannel}) and (\ref{etachannel}) are
\ie{}
& {\cal F}({1\over 16},{1\over 16},{1\over 16},{1\over 16}; 0; x) = x^{-{1\over 8}}(1-x)^{-{1\over 8}} \sqrt{1+\sqrt{1-x}\over 2},
\\
& {\cal F}({1\over 16},{1\over 16},{1\over 16},{1\over 16}; {1\over 2}; x) =2 x^{-{1\over 8}}(1-x)^{-{1\over 8}} \sqrt{1-\sqrt{1-x}\over 2}.
\fe
Under the crossing transformation that permutes $x_1,x_2,x_3,x_4$ cyclically, $x\to 1-x$, and the blocks transform as
\ie\label{isingblocks}
& {\cal F}({1\over 16},{1\over 16},{1\over 16},{1\over 16}; 0; 1-x) = {1\over \sqrt{2}} {\cal F}({1\over 16},{1\over 16},{1\over 16},{1\over 16}; 0; x) + {1\over 2\sqrt{2}} {\cal F}({1\over 16},{1\over 16},{1\over 16},{1\over 16}; {1\over 2}; x),
\\
& {\cal F}({1\over 16},{1\over 16},{1\over 16},{1\over 16}; {1\over 2}; 1-x) = \sqrt{2} {\cal F}({1\over 16},{1\over 16},{1\over 16},{1\over 16}; 0; x) - {1\over \sqrt{2}} {\cal F}({1\over 16},{1\over 16},{1\over 16},{1\over 16}; {1\over 2}; x).
\fe
On the other hand, it follows from the H-junction crossing relation that,
\ie
\left\la s(x_1) s(x_2) s(x_3) s(x_4) \right\ra_{\rotatebox{90}{\footnotesize H}}^I = {1\over \sqrt{2}}\left\la s(x_1) s(x_2) s(x_3) s(x_4) \right\ra_{\rm H}^I + {1\over \sqrt{2}}\left\la s(x_1) s(x_2) s(x_3) s(x_4) \right\ra_{\rm H}^\eta .
\fe
Writing the LHS as a single conformal block in the $23\to 14$ channel (with the identity being the internal primary), and using the first line of (\ref{isingblocks}), we determine $C_{ss\psi} = {1\over \sqrt{2}}$ (the overall sign can be absorbed into a redefinition of $\psi$).

\begin{figure}[H]
\centering
\bigskip
\begin{minipage}{0.18\textwidth}
\includegraphics[width=1\textwidth]{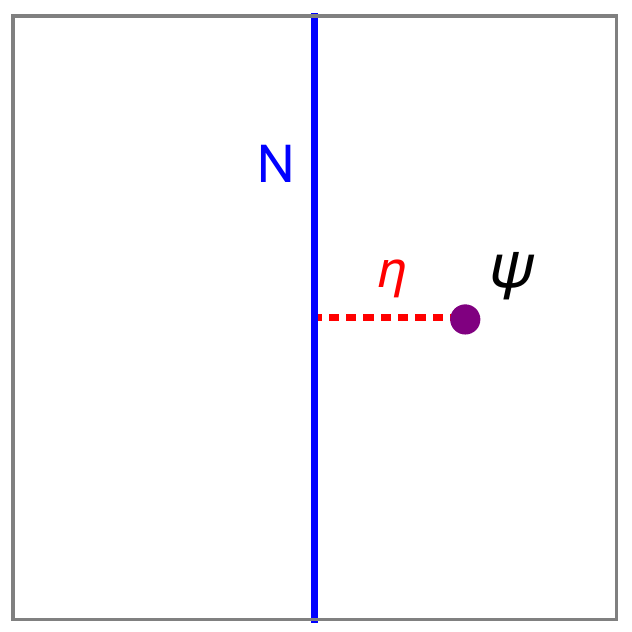}
\end{minipage}%
\begin{minipage}{0.23\textwidth}\begin{eqnarray*}\quad  \\ \end{eqnarray*}
\end{minipage}%
\begin{minipage}{0.18\textwidth}
\includegraphics[width=1\textwidth]{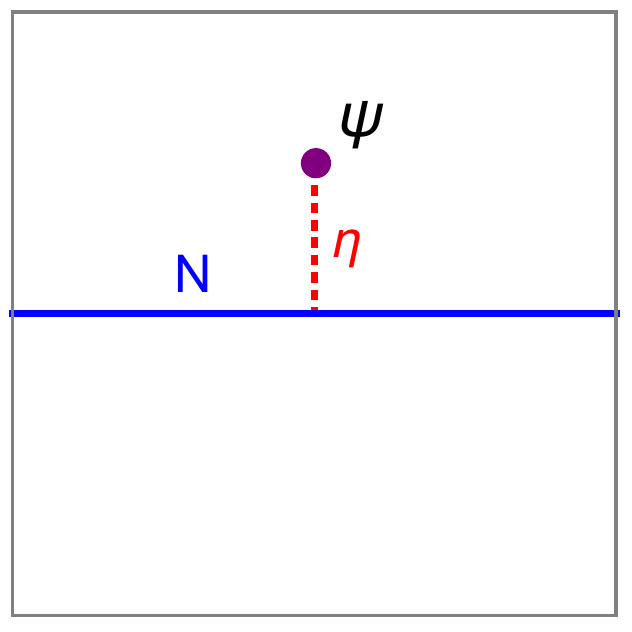}
\end{minipage}%
\caption{The torus one-point function of $\psi$ attached to an $N$ loop via the $NN\eta$ junction, $\la\psi\ra_{T^2,N}$ and $\la\psi\ra_{T^2}^N$.}
\label{fig:psi1pt}
\end{figure}

As another example, consider the torus one-point function of $\psi$ attached to an $N$ loop wrapping the time direction via the $NN\eta$ junction, which we denote by $\la\psi\ra_{T^2,N}$. The analogous torus one-point function with the $N$ loop wrapping the spatial direction, related by the modular $S$ transform, will be denoted by $\la\psi\ra_{T^2}^N$. See Figure~\ref{fig:psi1pt}. It is easy to deduce from the fusion rule that, when cut along a spatial circle, $\la\psi\ra_{T^2,N}$ receives contributions from the conformal families of $s_{{1\over 16},0}$ and $\Lambda_{{1\over 16},{1\over 2}}$ in ${\cal H}_N$, while $\la\psi\ra_{T^2}^N$ receives contribution from the conformal family of the spin field $\sigma_{{1\over 16},{1\over 16}}$ only. From their modular property, together with the structure constant $C_{ss\psi}$ derived above, we can determine\footnote{The expression for $\la\psi\ra_{T^2}^N$ can be derived directly from the OPE coefficient $C_{\sigma\sigma \psi  }$ involving the bulk local operator $\sigma$ and the defect operator $\psi$.}
\ie{}
& \la\psi\ra_{T^2,N} = {1\over \sqrt{2}} \eta(\tau) (\overline\chi_0(\bar\tau) - \overline\chi_{1\over 2}(\bar\tau)),
\quad
\la\psi\ra_{T^2}^N =   \eta(\tau) \overline\chi_{1\over 16}(\bar\tau).
\fe
Here, $\eta(\tau)$ is the Dedekind eta function which represents the $c={1\over 2}$ torus conformal block for the one-point function of a weight-${1\over 2}$ primary, and with a weight-${1\over 16}$ primary running in the loop; and $\chi_h(\tau)$ is the $c={1\over 2}$ (degenerate) Virasoro character associated with a primary of weight $h$.

\subsubsection{Tricritical Ising model}
\label{Sec:Trising}

The $c={7\over 10}$ tricritical Ising model has six Virasoro primaries:
\ie
1_{0,0},~~\varepsilon_{{1\over 10},{1\over 10}},~~\varepsilon'_{{3\over 5},{3\over 5}},~~\varepsilon''_{{3\over 2},{3\over 2}},~~\sigma_{{3\over 80},{3\over 80}},~~\sigma'_{{7\over 16},{7\over 16}}.
\fe
Apart from the trivial line $I$ and the $\mathbb{Z}_2$ invertible line $\eta$, there are four more simple Verlinde lines, $W \equiv {\cal L}_{\phi_{1,3}}$, $\eta W$, $N \equiv {\cal L}_{\phi_{2,1}}$, and $WN$.\footnote{In this paper, we follow the standard convention in \cite{DiFrancesco:1997nk} for $\phi_{r,s}$ with $1\leq r<p',~1\leq s<p$ to label the Virasoro primaries of the minimal model $M(p,p')$.} They act on the bulk local primary operators as
\begin{equation}
\label{triIsingaction}
\begin{tabu}{cccccccc}
 & & 1 & \varepsilon & \varepsilon' & \varepsilon'' & \sigma & \sigma'
\\
\widehat\eta : & \quad & 1 & 1 & 1 & 1 & -1 & -1
\\
\widehat W : & \quad & \zeta & -\zeta^{-1} & -\zeta^{-1} & \zeta & -\zeta^{-1} & \zeta
\\
\widehat N : & \quad & \sqrt2 & -\sqrt2 & \sqrt2 & -\sqrt2 & 0 & 0
\end{tabu}
\end{equation}
where $\zeta \equiv {1+\sqrt{5}\over 2}$ is the golden ratio. Apart from the ones we already stated, some nontrivial fusion relations of the TDLs are
\ie{}
W^2=I+W, \quad  N^2 = I + \eta.
\fe
The defect Hilbert space ${\cal H}_W$ contains 9 primaries with spins
\ie
\label{TrisingWSpins}
{\cal H}_W \ : \ s \in \bZ + \left\{ 0, \pm{2 \over 5} \right\}.
\fe
and the defect Hilbert space of ${\cal H}_N$ contains 8 primaries with spins
\ie
\label{TrisingNSpins}
{\cal H}_N \ : \ s \in {\bZ \over 2} \pm {1 \over 16}.
\fe

Note that $\varepsilon'$ is the only nontrivial primary that commutes with the TDL $N$. Under the RG flow generated by $\varepsilon'$, the tricritical Ising model flows to either the critical Ising model or a massive phase with three degenerate vacua, depending on the sign of the deformation \cite{PhysRevB.30.3908}. In the former case, both $\eta$ and $N$ survive the RG flow, and their fusion relation is preserved. They flow to the $\mathbb{Z}_2$ invertible line $\eta$ and the $N$ line, respectively (denoted by the same symbols), in the critical Ising model.

\subsubsection{Tetracritical Ising model}
\label{Sec:Tetra}

The $c={4\over 5}$ tetracritical Ising model $M(6,5)$ has 10 simple Verlinde lines, which may be labeled by
\ie
I, ~~ C \equiv {\cal L}_{\phi_{1,5}}, ~~ M \equiv {\cal L}_{\phi_{1,3}}, ~~ W \equiv {\cal L}_{\phi_{2,5}}, ~~ MW, ~~ CW, ~~ N, ~~ CN, ~~ WN, ~~ CWN,
\fe
with the fusion relations
\ie
& C^2=I, \quad  M^2=I+M+C, \quad  W^2=I+W, \quad  N^2=I+M, \quad  MN = N+CN.
\fe
In this model, the $C$ and $M$ lines (along with their defect operators) generate the rank-three fusion category ${\rm Rep}(S_3)$, and they act on the bulk local primary operators as
\begin{equation}
\begin{tabu}{cccccccccccc}
 & & 1 & \phi_{1,2} & \phi_{1,3} & \phi_{1,4} & \phi_{1,5} & \phi_{2,1} & \phi_{2,2} & \phi_{2,3} & \phi_{2,4} & \phi_{2,5}
\\
\widehat C : & \quad & 1 & -1 & 1 & -1 & 1 & 1 & -1 & 1 & -1 & 1
\\
\widehat M : & \quad & 2 & 0 & -1 & 0 & 2 & 2 & 0 & -1 & 0 & 2
\end{tabu}
\end{equation}
They are preserved by $\phi_{1,5}$, $\phi_{2,1}$, $\phi_{2,5}$, among which only the weight-$({2\over5}, {2\over5})$ primary $\phi_{2,1}$ is relevant. The relation between these TDLs and those of the $c={4 \over 5}$ three-state Potts model will be discussed in Section~\ref{potts}. The defect Hilbert space ${\cal H}_C$ contains 10 primaries with spins
\ie
\label{TetraCSpins}
{\cal H}_C \ : \ s \in {{\bZ} \over 2}.
\fe
The defect Hilbert space ${\cal H}_M$ contains 18 primaries with spins
\ie
\label{TetraMSpins}
{\cal H}_M \ : \ s \in {\bZ} \pm \left\{ 0, {1 \over 3}, {1 \over 2} \right\}.
\fe
The $W$ line acts on the bulk local primary operators as
\begin{equation}
\begin{tabu}{cccccccccccc}
 & & 1 & \phi_{1,2} & \phi_{1,3} & \phi_{1,4} & \phi_{1,5} & \phi_{2,1} & \phi_{2,2} & \phi_{2,3} & \phi_{2,4} & \phi_{2,5}
\\
\widehat W: & \quad & \zeta & \zeta & \zeta & \zeta & \zeta & -\zeta^{-1} & -\zeta^{-1} & -\zeta^{-1} & -\zeta^{-1} & -\zeta^{-1}
\end{tabu}
\end{equation}
where $\zeta \equiv {1+\sqrt5 \over 2}$ is the golden ratio. The defect Hilbert space ${\cal H}_W$ contains 15 primaries with spins
\ie
\label{TetraWSpins}
{\cal H}_W \ : \ s \in {\bZ} \pm \left\{ 0, {2 \over 5} \right\}.
\fe

\subsubsection{Pentacritical Ising model}
\label{Sec:Penta}

The $c = {6 \over 7}$ pentacritical Ising model $M(7,6)$ has 15 simple Verlinde lines.  Among them, the following three lines
\ie
I, \quad X \equiv {\cal L}_{\phi_{1,5}}, \quad Y \equiv {\cal L}_{\phi_{1,3}}
\fe
form a closed fusion ring with the relations
\ie
X^2 = I + Y, \quad Y^2 = I + X + Y,
\fe
which we recognize as the representation ring ${R}_{\bC}(\widehat{so(3)}_5)$. The defect Hilbert space ${\cal H}_X$ contains 25 primaries with spins
\ie
{\cal H}_X \ : \ s \in \bZ \pm \left\{ 0, {1 \over 7}, {3 \over 7} \right\},
\fe
and the defect Hilbert space ${\cal H}_Y$ contains 30 primaries with spins
\ie
{\cal H}_Y \ : \ s \in \bZ \pm \left\{ 0, {1 \over 7}, {2 \over 7} \right\}.
\fe
It follows from unitarity and the fusion relations that $\la X\ra$ and $\la Y\ra$ are given by the unique positive solution to the quadratic polynomial equations
\ie
1 + \la Y \ra - \la X \ra^2 = 1 + \la X \ra + \la Y \ra -\la Y \ra^2 = 0.
\fe
By \eqref{verlinetdl}, the $X$ and $Y$ lines both commute with the bulk local primaries $\phi_{2,1}$, $\phi_{3,1}$, $\phi_{1,6}$, $\phi_{2,6}$. In particular, the weight-$({3\over8},{3\over8})$ primary $\phi_{2,1}$ generates a relevant deformation, under which the pentacritical Ising model is expected to flow to a TQFT that admits the TDLs $X$ and $Y$.

\subsubsection{Lee-Yang model}
\label{Sec:LeeYang}

Finally, we consider the non-unitary minimal model ${\cal M}(2,5)$ with central charge $c = -{22\over5}$. This theory has two simple Verlinde lines
\ie
I, \quad W \equiv {\cal L}_{\phi_{1,2}},
\fe
which form the fusion ring with the relation
\ie
W^2 = I + W.
\fe
The nontrivial line $W$ has cylinder vacuum expectation value $\la W \ra = - \zeta^{-1}$,
and acts on the bulk local primary operators by
\ie
\widehat W:  ~ (1, \phi_{1,2}) \mapsto ( - \zeta^{-1}, \zeta \phi_{1,2}),
\fe
where $\zeta = {1 + \sqrt5 \over 2}$ is the Golden ratio.  
The crossing kernels are given by \cite{Moore:1988qv},
\ie{}
& 
{\widetilde K}_{W,W}^{W,W}\equiv
\begin{pmatrix} {\widetilde K}_{W,W}^{W,W}(I, I) & {\widetilde K}_{W,W}^{W,W}(I, W) \\ {\widetilde K}_{W,W}^{W,W}(W, I) & {\widetilde K}_{W,W}^{W,W}(W,W) \end{pmatrix} =  \begin{pmatrix} - {\zeta} & -{\zeta} \\ 1  & {\zeta} \end{pmatrix}.
\fe
Some crossing relations are illustrated in Figure~\ref{fig:Wcross} (with $\tilde \zeta = -\zeta^{-1}$). 

From \eqref{VerlindeHilbert},  we see that the defect Hilbert space ${\cal H}_W$ is spanned by three primaries of weights
\ie
\label{LYSpectrum}
(0, -{1\over5}), \quad (-{1\over5},0), \quad (-{1\over5}, -{1\over5}).
\fe

\subsection{More general topological defect lines}
\label{sec:generaltdls}

In this section, we discuss examples of TDLs that are neither invertible lines nor Verlinde lines.

\subsubsection{Three-state Potts model}
\label{potts}

The $c={4\over 5}$ critical three-state Potts model has 12 Virasoro primaries, including eight scalar, two spin-1, and two spin-3 primaries, as listed below:
\ie
1_{0,0},~~\varepsilon_{{2\over 5},{2\over 5}},~~X_{{7\over 5},{7\over 5}},~~Y_{3,3},~~\Phi_{{7\over 5},{2\over 5}},~~\tilde\Phi_{{2\over 5},{7\over 5}},~~\Omega_{3,0},~~\tilde\Omega_{0,3},~~\sigma_{{1\over 15},{1\over 15}},~~\sigma_{{1\over 15},{1\over 15}}^*,~~Z_{{2\over 3},{2\over 3}},~~Z_{{2\over 3},{2\over 3}}^*.
\fe
The three-state Potts model may be regarded as either a non-diagonal Virasoro minimal model, or a diagonal RCFT with respect to the $W_3$ algebra generated by $\Omega_{3,0}$, $\tilde\Omega_{0,3}$, and the Virasoro algebra. The model has an $S_3$ global symmetry generated by an order-3 element $\eta$, and a charge conjugation symmetry $C$. Note that $\eta$ commutes with the $W_3$ algebra, but $C$, which acts on $\Omega_{3,0}$ and $\tilde\Omega_{0,3}$ with a minus sign, does not.

Let us first regard the model as a diagonal RCFT with respect to the $W_3$ algebra, and consider the Verlinde lines. There are six primaries: $1, \varepsilon, \sigma, \sigma^*, Z, Z^*$, and correspondingly six TDLs that commute with the $W_3$. Three of them, $I$, $\eta$, and $\overline\eta = \eta^2$ are the invertible lines for the $\bZ_3$ subgroup of $S_3$ that commutes with $W_3$. The remaining three TDLs, which we denote by $W$, $\eta W$, and $\overline\eta W$, are not invertible. The $W$ line obeys the fusion relation
\ie
W^2 = I+W.
\fe
These are not all the simple TDLs with respect to the Virasoro algebra. Firstly, there is the invertible line $C$, and its fusion product with all six simple TDLs that commute with $W_3$. In addition, there are four more simple TDLs \cite{Petkova:2000ip}, which we denote by $N, N'=CN, WN, WN'$. Note that $N$ and $N'$ are unoriented, namely, $\overline{N}=N$, $\overline{N'}=N'$. They obey the fusion relations
\ie
N^2 = (N')^2= I + \eta + \overline{\eta}.
\fe

The action of the TDLs $\eta$, $W$, and $N$ on the bulk Virasoro primary operators are

\begin{equation}
\begin{tabu}{cccccccccccccc}
 & & 1 & \varepsilon & X & Y & \Phi & \tilde\Phi & \Omega & \tilde\Omega & \sigma & \sigma^* & Z & Z^*
\\
\widehat \eta : & \quad & 1 & 1 & 1 & 1 & 1 & 1 & 1 & 1 & \omega & \omega^2 & \omega & \omega^2
\\
\widehat W : & \quad & \zeta & -\zeta^{-1} & -\zeta^{-1} & \zeta & -\zeta^{-1} & -\zeta^{-1} & \zeta & \zeta & -\zeta^{-1} & -\zeta^{-1} & \zeta & \zeta
\\
\widehat N : & \quad & \sqrt3 & -\sqrt3 & \sqrt3 & -\sqrt3 & \sqrt3 & -\sqrt3 & -\sqrt3 & \sqrt3 & 0 & 0 & 0 & 0
\end{tabu}
\end{equation}
where $\omega \equiv e^{2\pi i \over 3}$ and $\zeta \equiv {1+\sqrt{5}\over 2}$. The $C$ line acts on the first few bulk local primaries by
\begin{equation}
\begin{tabu}{cccccccccccccc}
 & & 1 & \varepsilon & X & Y & \Phi & \tilde\Phi & \Omega & \tilde\Omega
\\
\widehat C : & \quad & 1 & 1 & 1 & 1 & -1 & -1 & -1 & -1
\end{tabu}
\end{equation}
and exchanges $\sigma$ with $\sigma^*$, and $Z$ with $Z^*$. By the modular $S$ transformation, one deduces the spectra of ${\cal H}_{ N}$ and ${\cal H}_{ N'}$, 
\ie
Z_{ N}(\tau,\bar\tau) = (\chi_{1,2} + \chi_{1,4})(\chi_{1,1} + \chi_{1,5}+2\chi_{1,3})^* + (\chi_{2,4}+\chi_{2,2})(\chi_{2,5}+\chi_{2,1}+2\chi_{2,3})^*,
\fe
and $Z_{ N'}(\tau,\bar\tau)=(Z_{ N}(\tau,\bar\tau) )^*$. Here, we observe that the states have spins
\ie
\label{PottsSpins}
{\cal H}_N \ : \ s \in {\bZ \over 2} + \left\{ {1 \over 8}, - {1 \over 24} \right\}, \quad {\cal H}_{N'} \ : \ s \in {\bZ \over 2} + \left\{ - {1 \over 8}, {1 \over 24} \right\}.
\fe
In the next section, we will see that these consistent with the spin selection rule imposed by the fusion ring. Similarly, the spectrum of ${\cal H}_W$ is
\ie
Z_W(\tau,\bar \tau) &= (\chi_{1,1}+\chi_{1,5})(\chi_{2,1}+\chi_{2,5})^*+(\chi_{2,1}+\chi_{2,5})(\chi_{1,1}+\chi_{1,5})^*+|\chi_{2,1}+\chi_{2,5}|^2
\\
& \hspace{.5in} +2(\chi_{2,3}\chi_{1,3}^*+\chi_{1,3}\chi_{2,3}^*)+2|\chi_{2,3}|^2\,,
\fe
whose states have spins
\ie\label{eqn:SSR3PHW}
s\in \bZ+\left\{0,\pm{2\over 5}\right\}.
\fe

The $\mathbb{Z}_2$ orbifold of the three-state Potts model by the charge conjugation symmetry $C$ is isomorphic to the diagonal modular invariant tetracritical Ising model $M(6,5)$. The TDLs $M=\eta + \overline\eta$ and $W$ commute with $C$, and the crossing phases between $M, W$ and $C$ are trivial. They survive the orbifold and give rise to TDLs in the tetracritical Ising model $M(6,5)$. In particular, $M=\eta+\overline\eta$ becomes a simple TDL in $M(6,5)$, and obeys the fusion relation
\ie
M^2 = I + M + \widetilde C,
\fe
where $\widetilde C$ is the invertible line associated with the dual $\widetilde{\mathbb{Z}_2}$ symmetry that assigns $-1$ to the twisted sector states. The fusion product $MW$ is also simple. $N$ and $N'$ give rise to the same simple TDL in $M(6,5)$, which we denote by $\widetilde N$ and obeys the fusion relations
\ie
M\widetilde N = \widetilde N + \widetilde C\widetilde N,\quad \widetilde N^2 =  I+M.
\fe 
Altogether, the fusion among $M,\widetilde C,W,\widetilde N$ generate the 10 simple Verlinde lines of $M(6,5)$, as discussed in Section~\ref{Sec:Tetra}.

\subsubsection{Topological Wilson lines in WZW and coset models}
\label{Sec:WZW}

The WZW model as a diagonal modular invariant theory with respect to the $G_k$ current algebra admits a continuous family of invertible lines associated with the ${(G\times G)/ Z(G)}$ global symmetry, where $Z(G)$ is the center of $G$ that acts axially on $G\times G$.\footnote{Note that in the diagonal $G_k$ WZW model, an axial center symmetry transformation $(g,g^{-1})\in G\times G$ with $g\in Z(G)$ commutes with the current algebra, and acts trivially on all bulk operators.  It follows that there is only one copy of the $Z(G)$ global symmetry present in theory.
}
In addition, there are also the Verlinde lines ${\cal L}_R$ of the current algebra, one for each $G \times G$ representation $(R,R)$ that appears in the spectrum of current algebra primaries. It was observed in \cite{Bachas:2004sy,Alekseev:2007in} that the Verlinde lines of the WZW model coincide with a family of topological Wilson lines that are defined by holomorphic flat connections, of the form
\ie\label{wjline}
W_R(G) = {\rm tr}_R \, P \exp\left[ -{i\over k} t^a \oint dz j^a(z) \right],
\fe
where $t^a$ are the generators of $G$. The regularization used to define $W_R(G)$ is compatible with the isotopy invariance only when the coefficient of the connection is fixed as in (\ref{wjline}) (up to a finite renormalization).

We can generalize (\ref{wjline}) to topological Wilson lines of holomorphic flat connections built out of the currents of a subgroup $H \subset G$, and traced over the representations $R$ of $H$, which we denote by $W_R(H)$. The analogous topological Wilson lines constructed out of the anti-holomorphic currents of $H$ will be denoted by $\overline W_R(H)$. Note that while previously $W_R(G)$ and $\overline{W}_R(G)$ act on WZW primaries in identical ways and are isomorphic to the Verlinde line ${\cal L}_R$, the generalizations $W_R(H)$ and $\overline W_R(H)$ are generally different and do not correspond to Verlinde lines. When $H$ is a $U(1)$ subgroup, $W_R(H)$ and $\overline W_R(H)$ are the invertible lines corresponding to the left and right $H$ symmetries. The set of invertible lines associated to all $U(1)$ subgroups of $G$ generates the left and right $G$ symmetries.

Given a subgroup $K$ of $G$ with $H\subset K\subset G$, and representation $R$ of $K$, the topological Wilson lines $W_R(K)$ and $\overline W_R(K)$ commute with all currents of $H$, and therefore are TDLs of the $G/H$ gauged WZW model. Thus, $W_R(K)$ and $\overline W_R(K)$ flow to TDLs in the $G/H$ coset CFT  \cite{Bachas:2009mc}.

As a simple example, the $SU(2)_3$ WZW model admits three topological Wilson (Verlinde) lines $W_j$, where $j=0,{1\over 2},1,{3\over 2}$ labels the spin of the $SU(2)$ representation. Upon gauging the $U(1)$ subgroup, the gauged WZW model flows to the $SU(2)_3/U(1)$ coset which is isomorphic to the three-state Potts model. Indeed, $W_{3\over 2}$ flows to the $\mathbb{Z}_2$ invertible line $C$, while $W_1$ flows to the $W$ line of the three-state Potts model, with their fusion relations preserved by the RG flow.

More generally, the $SU(2)_k$ WZW model admits topological Wilson lines $W_j$, with $j=0,{1\over 2},\cdots,{k\over 2}$. In particular, $C\equiv W_{k\over 2}$ is an invertible line corresponding to the  $\mathbb{Z}_2$ center symmetry. The $W_j$ lines remain in the $SU(2)_k/U(1)$ coset CFT, and commute with the parafermion algebra. In particular, $SU(2)_k/U(1)$ deformed by the weight-$({k-1\over k},{k-1\over k})$ parafermion bilinear $e^{\pi i \over k}\psi_1\widetilde\psi_1+e^{-{\pi i \over k}}\overline\psi_1\overline{\widetilde\psi}_1$ flows to the $(A_{k},A_{k+1})$ minimal model \cite{Fateev:1991bv}. The $W_j$'s are preserved along the RG flow and become a subset of the Verlinde lines of the  $(A_{k},A_{k+1})$ minimal model.

Next, let us consider the $c={6\over 5}$ coset model ${SU(3)_2\over U(1)\times U(1)}$. The TDLs of the $SU(3)_2$ WZW model that commute with the $U(1)\times U(1)$ current algebra survive the gauging and give the TDLs of the coset CFT. These include the topological Wilson lines $W_{\ell_1,\ell_2}$ associated with the representations $[\ell_1,\ell_2]$, $\ell_1,\ell_2=0,1,2$, $\ell_1+\ell_2\leq 2$, that obey the fusion relations
\ie
& W_{1,0} W_{1,0} = W_{0,1} + W_{2,0},\quad  W_{1,0} W_{0,1} = I + W_{1,1},\quad W_{1,1}^2 = I + W_{1,1},
\\
& W_{2,0} W_{2,0} = W_{0,2},\quad  W_{2,0} W_{0,2} = I,\quad  W_{2,0}W_{1,1}=W_{0,1} .
\fe
In fact, we see that $W_{2,0}, W_{0,2}$ are the invertible lines associated with the $\mathbb{Z}_3$ center symmetry, which commute with $W_{1,1}$. There are also invertible lines of the coset model associated with the $S_3$ Weyl group symmetry. Furthermore, given a subgroup $H\simeq SU(2)\times U(1)$ of $SU(3)$, we have the topological Wilson lines $W_j(H)$ and $\overline W_j(H)$, where $j={1\over 2},1$ labels the spin of an $SU(2)_2$ representation. There are three such subgroups $H_1, H_2, H_3$ that contain a given maximal torus $U(1)\times U(1)$, permuted by the Weyl group action; they lead to the $W_j(H_i)$ and $\overline W_j(H_i)$ lines in the coset CFT. Note that the $j=1$ Wilson lines $W_1(H_i)$ and $\overline W_1(H_i)$ correspond to the $\mathbb{Z}_2$ center symmetry of the $SU(2)$ in $H_i$, acting left and right respectively. While $W_{1\over 2}(H_i)$ obeys the $SU(2)_2$ fusion relation $W_{1\over 2}(H_i) W_{1\over 2}(H_i) = I + W_1(H_i)$, $W_{1\over 2}(H_i) W_{1\over 2}(H_j)$ is a simple TDL for $i\not=j$.

\subsubsection{Models with ${1\over 2}E_6$ fusion ring}
\label{Sec:hE6}

The ``${1\over 2}E_6$" fusion ring contains three simple objects, $I, X, Y$, with the fusion relations
\ie\label{halfesix}
X^2 = I,~~ Y^2=I+X+2Y,~~XY=YX=Y.
\fe
The H-junction crossing kernels that solve the pentagon identity were obtained in~\cite{Hagge:2007aa}. This set of TDLs has the peculiar property that it does not admit braiding, despite the fusion ring being commutative. Note that the junction vector space $V_{Y,Y,Y}$ is two-dimensional, and the cyclic permutation map acts nontrivially on one of the two basis junction vectors in $V_{Y,Y,Y}$.

Consider the non-diagonal $SU(2)_{10}$ WZW model of $E_6$ type \cite{DiFrancesco:1997nk}, whose torus partition function is
\ie
Z = |\chi_0 + \chi_3|^2 + |\chi_{3\over 2} + \chi_{7\over 2}|^2 + |\chi_2 + \chi_5|^2,
\fe
where $\chi_j(\tau)$ is the spin-$j$ affine $SU(2)$ character. Now, consider the TDLs $X$ and $Y$ that preserve the $SU(2)$ current algebra, and act on the $SU(2)$ primaries according to the following twisted partition functions
\ie{}
& Z^X 
=  |\chi_0 + \chi_3|^2 - |\chi_{3\over 2} + \chi_{7\over 2}|^2 + |\chi_2 + \chi_5|^2,
\\
& Z^Y=  ((1+\sqrt{3})\chi_0 + (1-\sqrt{3})\chi_3) (\overline\chi_0 + \overline\chi_3) + ((1-\sqrt{3})\chi_2 + (1+\sqrt{3})\chi_5)(\overline\chi_2 + \overline\chi_5).
\fe
Indeed, they obey the fusion relation (\ref{halfesix}), and in particular the modular $S$ transform of $Z^Y$ gives the partition function of the defect operator Hilbert space ${\cal H}_Y$,
\ie
Z_{{\cal H}_Y} &= (\chi_1 + \chi_2 + \chi_3 + \chi_4)(\bar \chi_0 + \bar \chi_2 + \bar\chi_3 + \bar\chi_5)
\\
& \hspace{1in} + (\chi_{1\over 2} + \chi_{3\over 2} + 2\chi_{5\over 2} + \chi_{7\over 2} + \chi_{9\over 2}) (\bar\chi_{3\over 2} + \bar\chi_{7\over 2}).
\fe
Here, we observe that the spins of states in ${\cal H}_Y$ obey
\ie
\label{spinsec}
s\in  \mathbb{Z} + \left\{ 0, {1\over 6}, {5\over 12},{1\over 2}, {2\over 3}, {3\over 4} \right\}.
\fe

We can construct analogous TDLs obeying the same fusion relations (\ref{halfesix}) in the $(A_{10}, E_6)$ minimal model \cite{DiFrancesco:1997nk}. The torus partition function is
\ie
Z = \sum_{r=1~{\rm step}~2}^{9} |\chi_{r,1} + \chi_{r,7}|^2 + |\chi_{r,4}+\chi_{r,8}|^2 + |\chi_{r,5}+\chi_{r,11}|^2.
\fe
The TDLs $X$ and $Y$ act on the Virasoro primaries according to the twisted characters
\ie{}
Z^X &=  \sum_{r=1~{\rm step}~2}^{9} |\chi_{r,1} + \chi_{r,7}|^2 - |\chi_{r,4}+\chi_{r,8}|^2 + |\chi_{r,5}+\chi_{r,11}|^2,
\\
Z^Y &=  \sum_{r=1~{\rm step}~2}^{9} ((1+\sqrt{3})\overline\chi_{r,1} +(1-\sqrt{3}) \overline\chi_{r,7})(\chi_{r,1} + \chi_{r,7}) 
\\
& \hspace{1in} + ((1-\sqrt{3})\overline\chi_{r,5}+(1+\sqrt{3})\overline\chi_{r,11}) (\chi_{r,5}+\chi_{r,11}) .
\fe
After performing the modular $S$ transform on $Z^Y$, one sees that once again, the spins of states in ${\cal H}_Y$ obey (\ref{spinsec}). In fact, we will see in Section~\ref{esixselect} that such a spin selection rule follows entirely from the crossing relations of the $X$ and $Y$ lines.

Note that in either the non-diagonal $SU(2)_{10}$ WZW model of $E_6$ type or the $(A_{10}, E_6)$ minimal model, there is also a ``conjugate" TDL $\widetilde Y$, whose action on the primaries is that of the parity reversal of $Y$. The $X$ and $\widetilde Y$ lines generate a fusion ring that is identical to that of $X$ and $Y$, but as we will see later, the crossing kernels involving $X$ and $\widetilde Y$ belong to a different solution to the pentagon identity than that of $X$ and $Y$. In particular, the spin selection rule of the states in ${\cal H}_{\widetilde Y}$ is minus the spin content of (\ref{spinsec}).

\section{Crossing kernels and spin selection rules}
\label{sec:spinselect}

For a given TDL ${\cal L}$ in a CFT, we would like to understand the general constraints on the spins of states in ${\cal H}_{\cal L}$, based on the fusion relations involving ${\cal L}$, by considering the modular $T$ transformation property of the two-point function of defect operators in ${\cal H}_{\cal L}$. For some of the fusion rings considered, we provide alternative derivations of the crossing kernels without having to explicitly solve the pentagon identity. We repeatedly use the fact mentioned in Section~\ref{Sec:FusionCoefficients}, that in the gauge choice defined in Appendix~\ref{CKbasis}, $R({\cal L})$ obeys the system of polynomial equations given by the abelianization of the fusion ring.

\subsection{Tambara-Yamagami categories}

As discussed in Section~\ref{Sec:DualityTY}, a Tambara-Yamagami (TY) category is an extension of an abelian invertible fusion category by an additional TDL $\cal N$, whose self-fusion gives a sum over all TDLs in the invertible fusion category \cite{TAMBARA1998692}. By definition, a TY category has trivial cyclic permutation map; hence, we do not need to mark the ordering on the junctions (see Section~\ref{Sec:Trivial}). Its relation to duality defects is discussed in Section~\ref{Sec:DualityTY}. Below, we study the TY categories associated to the abelian finite groups $\bZ_2$ and $\bZ_3$, and derive their spin selection rules.

\subsubsection{$\mathbb{Z}_2$ symmetry}
\label{Sec:Z2TY}

Consider the fusion ring of the $\mathbb{Z}_2$ Tambara-Yamagami category, with the commutative relations
\ie\label{YCZ2fusionrules}
\eta^2=I,\quad {\cal N}^2=1+\eta,\quad \eta {\cal N}={\cal N}.
\fe
There are two solutions of the crossing kernels to the pentagon identity \cite{Moore:1988qv}. In the basis specified in Appendix~\ref{CKbasis}, the nontrivial crossing kernels are\footnote{Here and in the rest of the main text, we abuse the notation by writing $\widetilde {\cal K}$ as $\widetilde K$. Their relation is spelled out in Appendix~\ref{CKbasis}.
}
\ie\label{knn}
& {\widetilde K}_{{\cal N},{\cal N}}^{{\cal N},{\cal N}}\equiv \begin{pmatrix} {\widetilde K}_{{\cal N},{\cal N}}^{{\cal N},{\cal N}}(I, I) & {\widetilde K}_{{\cal N},{\cal N}}^{{\cal N},{\cal N}}(I, \eta) \\ {\widetilde K}_{{\cal N},{\cal N}}^{{\cal N},{\cal N}}(\eta, I) & {\widetilde K}_{{\cal N},{\cal N}}^{{\cal N},{\cal N}}(\eta,\eta) \end{pmatrix} = {\epsilon\over \sqrt{2}} \begin{pmatrix}  1 & 1 \\ 1  & -1 \end{pmatrix},
\quad
{\widetilde K}^{\eta ,{\cal N}}_{{\cal N},\eta}({\cal N},{\cal N}) = -1.
\fe

\begin{figure}[H]
\centering
\begin{minipage}{0.1\textwidth}
\includegraphics[width=1\textwidth]{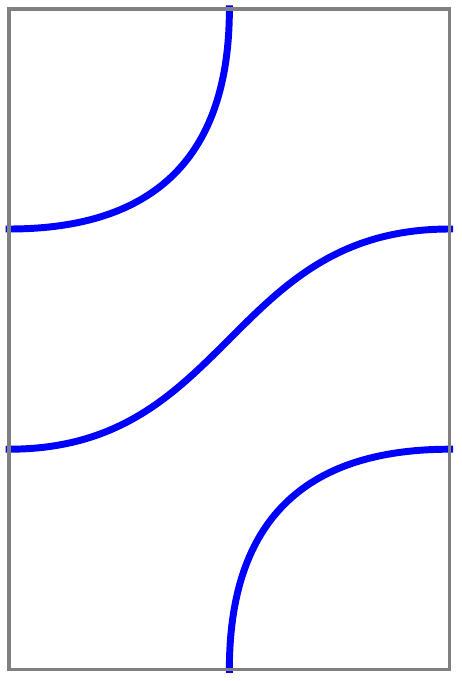}
\end{minipage}%
\begin{minipage}{0.1\textwidth}\begin{eqnarray*}~= {\epsilon\over \sqrt{2}} \\ \end{eqnarray*}
\end{minipage}%
\begin{minipage}{0.1\textwidth}
\includegraphics[width=1\textwidth]{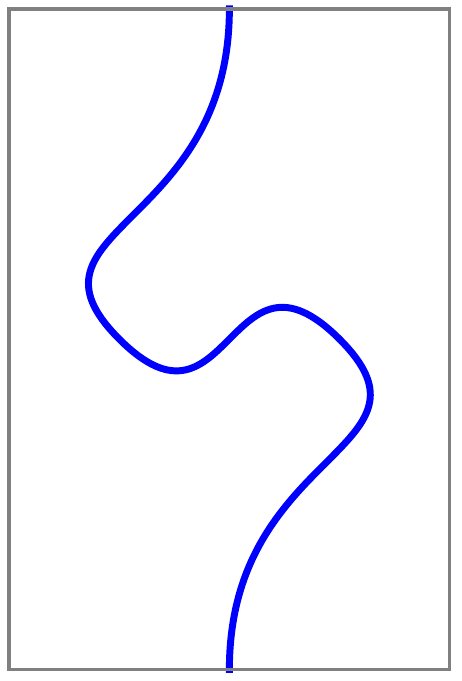}
\end{minipage}%
\begin{minipage}{0.1\textwidth}\begin{eqnarray*}~+ {\epsilon\over \sqrt{2}} \\ \end{eqnarray*}
\end{minipage}%
\begin{minipage}{0.1\textwidth}
\includegraphics[width=1\textwidth]{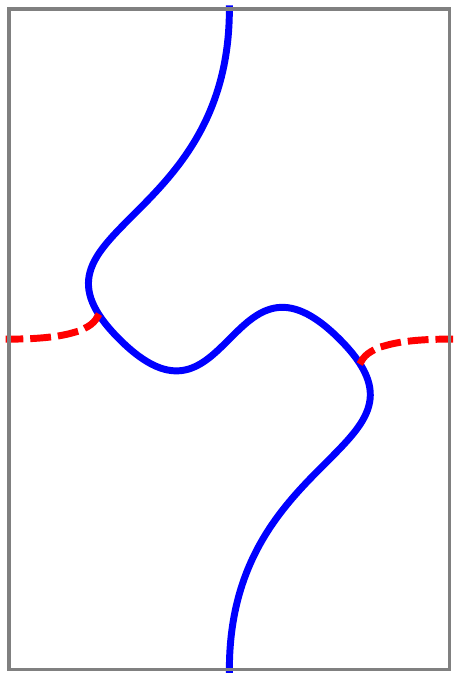}
\end{minipage}%
\caption{Applying the H-junction crossing relations to the $T^2$ transformation of ${\cal N}$ line in time direction.}
\label{fig:spinselectionNIeta}
\end{figure}

\begin{figure}[H]
\centering
\begin{minipage}{0.13\textwidth}
\includegraphics[width=1\textwidth]{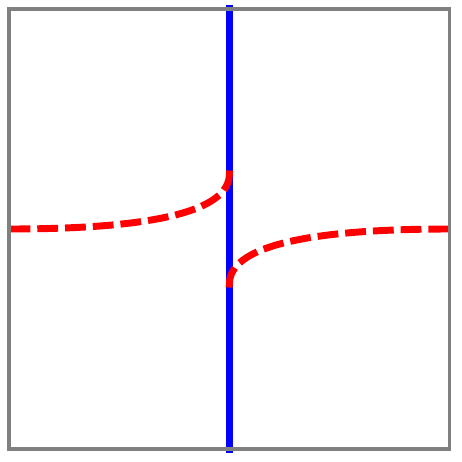}
\end{minipage}\quad \quad \quad \quad \quad \quad \quad 
\begin{minipage}{0.13\textwidth}
\includegraphics[width=1\textwidth]{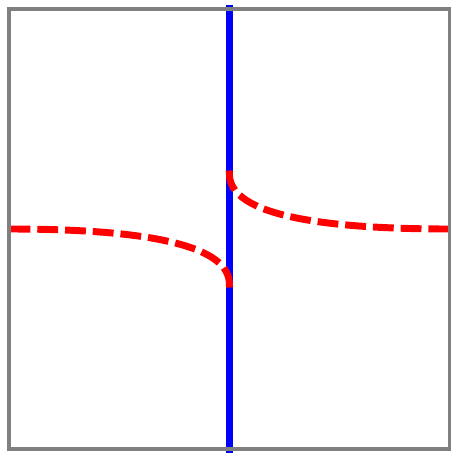}
\end{minipage}%
\caption{The action of $\widehat\eta_-$ (left) and $\widehat\eta_+$ (right) on ${\cal H}_{\cal N}$.}
\label{fig:etaminusplus}
\end{figure}

For any pair of states $|\psi\ra, |\psi'\ra \in {\cal H}_{\cal N}$ with equal conformal weight, let us consider the matrix element of the cylinder propagator
\ie
\la\psi'| q^{L_0 - {c\over 24}} \bar q^{\bar L_0 - {\tilde c\over 24}} |\psi\ra.
\fe
As shown in Figure~\ref{fig:spinselectionNIeta}, we can perform the modular $T^2$ transformation, and then apply the H-junction crossing relation involving ${\widetilde K}_{{\cal N},{\cal N}}^{{\cal N},{\cal N}}$, resulting in
\ie\label{seta}
e^{4\pi i s}\la\psi' |\psi\ra = {\epsilon\over \sqrt{2}}\la\psi'|1+\widehat \eta_-|\psi\ra,
\fe
where $s$ is the spin $h-\tilde h$ of $\psi$, and $\widehat \eta_-$ is the operator acting on ${\cal H}_{\cal N}$ defined by an $\eta$ line wrapping the spatial circle, that is split over the temporal ${\cal N}$ line. We denote by $\widehat\eta_+$ another operator on ${\cal H}_{\cal N}$ defined by a spatial $\eta$ line split over a temporal ${\cal N}$ line, with the opposite ordering of the pair of T-junctions. Both operators $\widehat\eta_-$ and $\widehat\eta_+$, depicted in Figure~\ref{fig:etaminusplus}, are special cases of the more general ``lasso" diagrams shown in Figure~\ref{fig:Lact2}.

The operator $\widehat\eta_-$ is topological in the sense that it commutes with the Virasoro algebra, and thus preserves the weights. Applying the H-junction crossing relation (more specifically, ${\widetilde K}^{\eta ,{\cal N}}_{{\cal N},\eta}({\cal N},{\cal N})=- 1$) to $\la\psi'|\widehat\eta_-^2|\psi\ra$ allows us to determine $\widehat\eta_-^2 = -1$. Now, (\ref{seta}) demands that ${1+\widehat \eta_-\over \sqrt{2}}$ is a phase, but this is precisely the case for $\widehat\eta_-=\pm i$. 
Thus, we learn that the spins of states in ${\cal H}_{\cal N}$ must obey
\ie\label{eqn:SSRising}
s\in {1\over 2}\bZ+\begin{cases}
\pm{1\over 16}\quad&{\rm for}\quad \epsilon=1,
\\
\pm{3\over 16}\quad&{\rm for}\quad \epsilon=-1.
\end{cases}
\fe
Indeed, the spins in the $\epsilon=1$ case are realized in the defect Hilbert space ${\cal H}_N$ of the $N$ line TDL in either the critical Ising model \eqref{IsingDefectSpec} or the tricritical Ising model \eqref{TrisingNSpins}, and we see here that it is a general consequence of the fusion relations of the TDLs. On the other hand, the $\epsilon=-1$ case is realized, for instance, in the tensor product of the critical Ising model (similarly for tricritical Ising) with the $SU(2)_1$ WZW model, which has in particular a $\bZ_2$ invertible line $\cal L_{\pi}$ given in \eqref{Lalpha} (associated to the center of the left $SU(2)$ global symmetry) with an 't Hooft anomaly.
In this case, the identity and $\eta$ lines are realized by the $\bZ_2$ lines of critical Ising model as before, but ${\cal N}$ is instead taken to be the tensor product $N \otimes \cal L_{\pi}$.
The defect Hilbert space factorizes as ${\cal H}_{N \otimes \cal L_{\pi}}={\cal H}_{N}\otimes {\cal H}_{\cal L_{\pi}}$, because the two TDLs belong to two decoupled theories.  The defect Hilbert space ${\cal H}_{ {\cal L}_\pi}$ for the anomalous $\mathbb{Z}_2$ line ${\cal L}_\pi$ obeys the spin selection rule \eqref{ZnSpin} with $k=1$ and $n=2$, which when considered together with the content \eqref{IsingDefectSpec} or \eqref{TrisingNSpins} of ${\cal H}_N$ gives the same spin selection rule for ${\cal H}_{N\otimes {\cal L}_\pi}$ as in \eqref{eqn:SSRising} with $\epsilon=-1$.

\subsubsection{$\bZ_3$ symmetry}
\label{Sec:Z3TY}

Our next example is the $\bZ_3$ Tambara-Yamagami category, with the commutative fusion relations
\ie
& \eta^2=\overline\eta,\quad \bar\eta^2=\eta, \quad \eta\overline\eta=I,
\quad
{\cal N}^2=I+\eta+\overline\eta, \quad \eta {\cal N} = \overline\eta {\cal N}={\cal N}.
\fe
There are two solutions of the crossing kernels to the pentagon identity. In the basis specified in Appendix~\ref{CKbasis}, the nontrivial crossing kernels are
\ie\label{knneta}
& {\widetilde K}_{{\cal N},{\cal N}}^{{\cal N},{\cal N}}\equiv \begin{pmatrix} {\widetilde K}_{{\cal N},{\cal N}}^{{\cal N},{\cal N}}(I, I) & {\widetilde K}_{{\cal N},{\cal N}}^{{\cal N},{\cal N}}(I, \eta)  & {\widetilde K}_{{\cal N},{\cal N}}^{{\cal N},{\cal N}}(I, \overline\eta) \\ {\widetilde K}_{{\cal N},{\cal N}}^{{\cal N},{\cal N}}(\eta, I) & {\widetilde K}_{{\cal N},{\cal N}}^{{\cal N},{\cal N}}(\eta,\eta) & {\widetilde K}_{{\cal N},{\cal N}}^{{\cal N},{\cal N}}(\eta,\overline\eta) \\ {\widetilde K}_{{\cal N},{\cal N}}^{{\cal N},{\cal N}}(\overline\eta, I) & {\widetilde K}_{{\cal N},{\cal N}}^{{\cal N},{\cal N}}(\overline\eta,\eta) & {\widetilde K}_{{\cal N},{\cal N}}^{{\cal N},{\cal N}}(\overline\eta,\overline\eta)  \end{pmatrix} =  {\epsilon\over \sqrt{3}}  \begin{pmatrix}  1 & 1 & 1 \\ 1  & \omega & \omega^2 \\ 1 & \omega^2 & \omega \end{pmatrix},
\\
&{\widetilde K}^{{\cal N},\eta}_{\eta ,{\cal N}}({\cal N},{\cal N}) = {\widetilde K}^{{\cal N},\bar \eta}_{\bar \eta ,{\cal N}}({\cal N},{\cal N})=  {\widetilde K}^{ \eta ,{\cal N}}_{{\cal N},\bar \eta}({\cal N},{\cal N})=\omega,
\\
&{\widetilde K}^{\eta ,{\cal N}}_{{\cal N},\eta}({\cal N},{\cal N}) = {\widetilde K}^{\bar \eta ,{\cal N}}_{{\cal N},\bar \eta}({\cal N},{\cal N})=  {\widetilde K}^{{\cal N},\bar \eta}_{ \eta ,{\cal N}}({\cal N},{\cal N})=\bar \omega,
\fe
where $\omega$ is a third root of unity that is not one, and $\epsilon=\pm 1$. Up to exchanging $\eta$ with $\overline\eta$, we may take $\omega = e^{2\pi i \over 3}$. 

Applying the modular $T^2$ transformation to the cylinder matrix element $\la \psi'|\psi\ra$, with $|\psi\ra, |\psi'\ra\in{\cal H}_{\cal N}$, we deduce that
\ie\label{ssteta}
e^{4\pi i s}\la\psi' |\psi\ra = {\epsilon\over \sqrt{3}}\la\psi'|1+\widehat \eta_- +\widehat{\overline\eta}_- |\psi\ra,
\fe
where the topological operators $\widehat \eta_-$ and $\widehat{\overline\eta}_-$ are defined as spatial $\eta_-$ and $\overline\eta_-$ lines split off a temporal ${\cal N}$ line, acting on ${\cal H}_{\cal N}$. The property that ${\widetilde K}_{\eta,{\cal N}}^{{\cal N},\bar \eta}({\cal N},{\cal N})$ is a third root of unity allows us to deduce that $\widehat\eta_-^3=\widehat{\overline\eta}_-^3=1$, and that $\widehat\eta_-$ commutes with $\widehat{\overline\eta}_-$. Now, suppose 
\ie
\widehat\eta_-|\psi\ra = \omega^a|\psi\ra,\quad \widehat{\overline\eta}_-|\psi\ra = \omega^b|\psi\ra,
\fe
where $a,b=0$, 1, or 2, then (\ref{ssteta}) demands that ${1+\omega^a+\omega^b\over \sqrt{3}}$ must be a phase. This is possible for $(a,b)=(1,0)$, $(2,0)$, $(1,1)$ or $(2,2)$, and correspondingly we have the spin selection rule for the states in ${\cal H}_{\cal N}$,
\ie
\label{TYZ3Spins}
s\in{1\over 2}\bZ \pm \begin{cases}
{1\over 24}, {1\over 8}\quad&{\rm for}\quad \epsilon=1,
\\
{5\over 24}, {1\over 8}\quad&{\rm for}\quad \epsilon=-1.
\end{cases}
\fe
Strikingly, the $\epsilon=1$ case is indeed satisfied by \eqref{PottsSpins} for ${\cal H}_N$ and ${\cal H}_{N'}$ in the three-state Potts model. The $\epsilon=-1$ case can be realized by tensoring these $N$ and $N'$ lines with the $\bZ_2$ line $\cL_{\pi}$ in the $SU(2)_1$ WZW model.

\subsection{Categories with Lee-Yang fusion ring}
\label{Sec:LeeYangAlgebra}

\begin{figure}[H]
\centering
\begin{minipage}{0.1\textwidth}
\includegraphics[width=1\textwidth]{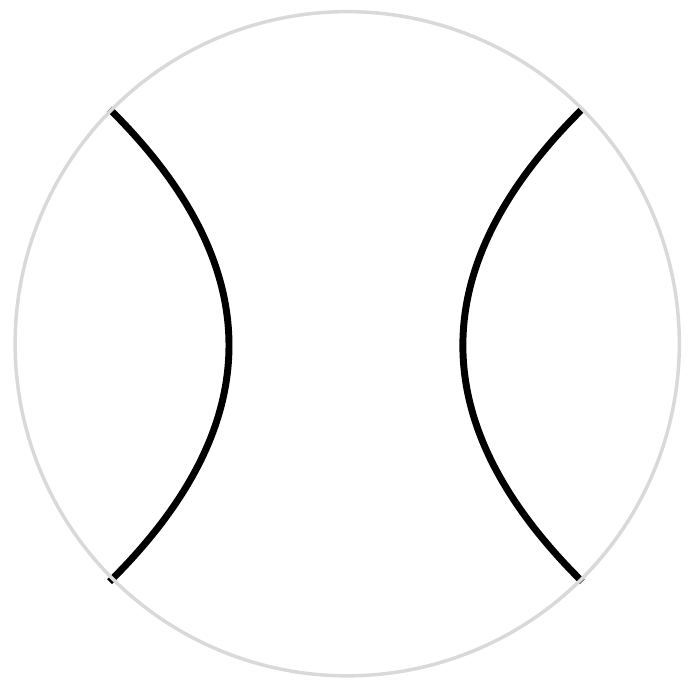}
\end{minipage}%
\begin{minipage}{0.1\textwidth}\begin{eqnarray*}~=\widetilde{\zeta}^{-1} \\ \end{eqnarray*}
\end{minipage}%
\begin{minipage}{0.1\textwidth}
\includegraphics[width=1\textwidth]{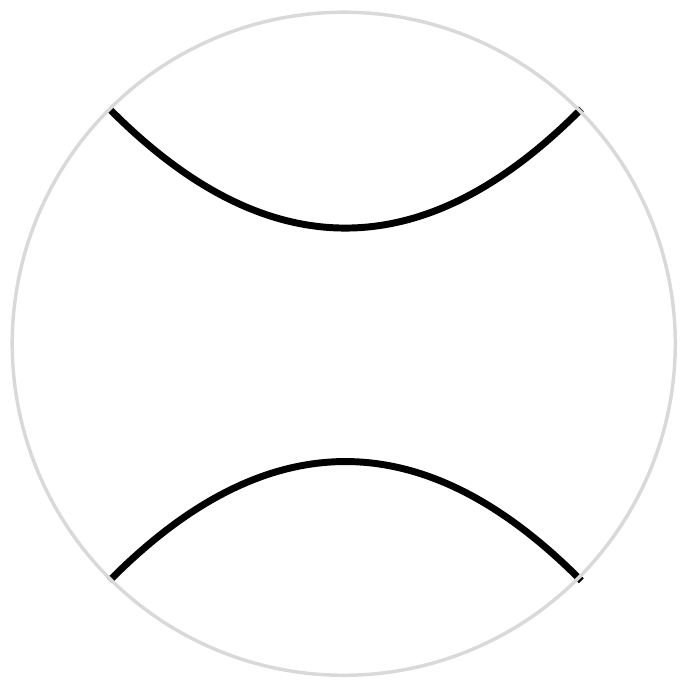}
\end{minipage}%
\begin{minipage}{0.1\textwidth}\begin{eqnarray*}~+ \widetilde{\zeta}^{-1} \\ \end{eqnarray*}
\end{minipage}%
\begin{minipage}{0.1\textwidth}
\includegraphics[width=1\textwidth]{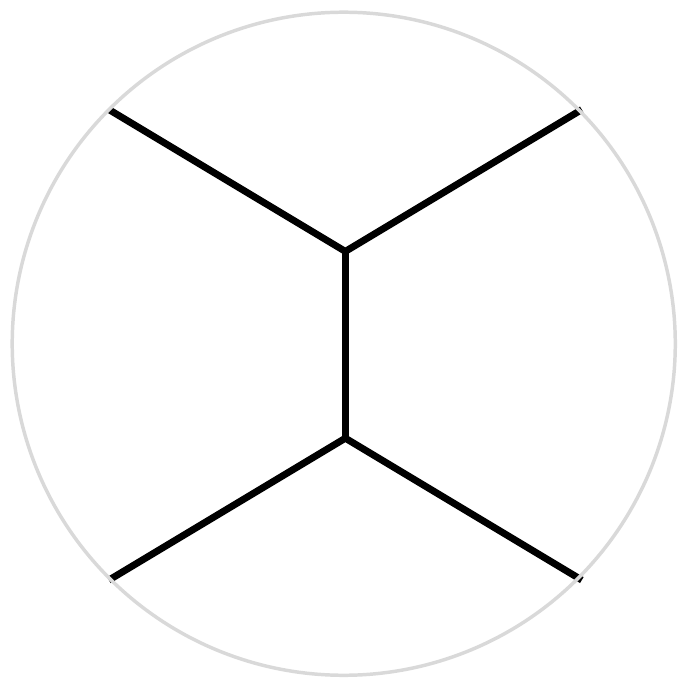}
\end{minipage}%
\\
\bigskip
\begin{minipage}{0.1\textwidth}
\includegraphics[width=1\textwidth]{figures/HpWWWWI.pdf}
\end{minipage}%
\begin{minipage}{0.1\textwidth}\begin{eqnarray*}~= \widetilde{\zeta}^{-1} \\ \end{eqnarray*}
\end{minipage}%
\begin{minipage}{0.1\textwidth}
\includegraphics[width=1\textwidth]{figures/HWWWWI.pdf}
\end{minipage}%
\begin{minipage}{0.1\textwidth}\begin{eqnarray*}~+\widetilde{\zeta}^{-1} \\ \end{eqnarray*}
\end{minipage}%
\begin{minipage}{0.1\textwidth}
\includegraphics[width=1\textwidth]{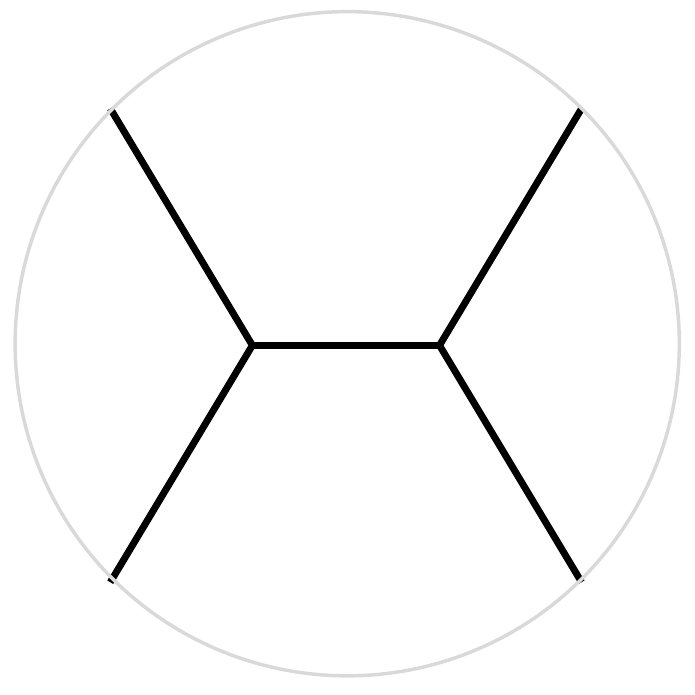}
\end{minipage}%
\\
\bigskip
\begin{minipage}{0.1\textwidth}
\includegraphics[width=1\textwidth]{figures/HpWWWWW.pdf}
\end{minipage}%
\begin{minipage}{0.1\textwidth}\begin{eqnarray*}~= \\ \end{eqnarray*}
\end{minipage}%
\begin{minipage}{0.1\textwidth}
\includegraphics[width=1\textwidth]{figures/HWWWWI.pdf}
\end{minipage}%
\begin{minipage}{0.1\textwidth}\begin{eqnarray*}~ -\widetilde{\zeta}^{-1} \\ \end{eqnarray*}
\end{minipage}%
\begin{minipage}{0.1\textwidth}
\includegraphics[width=1\textwidth]{figures/HWWWWW.pdf}
\end{minipage}%
\\
\bigskip
\begin{minipage}{0.028\textwidth}
\includegraphics[width=1\textwidth]{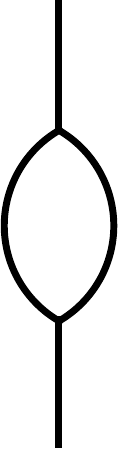}
\end{minipage}$~ = \widetilde{\zeta}\! $%
\begin{minipage}{0.053\textwidth}
\includegraphics[width=1\textwidth]{figures/Nline.pdf}
\end{minipage}\quad \quad \quad \quad ~
\begin{minipage}{0.07\textwidth}
\includegraphics[width=1\textwidth]{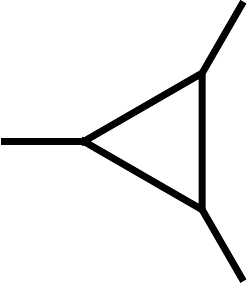}
\end{minipage}$~ = - ~$%
\begin{minipage}{0.07\textwidth}
\includegraphics[width=1\textwidth]{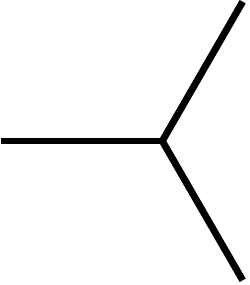}
\end{minipage}
\caption{Some useful crossing relations involving $W$ lines.}
\label{fig:Wcross}
\end{figure}

Consider the Lee-Yang fusion relation,
\ie
W^2 = I + W.
\fe
There are two solutions of the crossing kernels to the pentagon identity \cite{Moore:1988qv}. In the basis specified in Appendix~\ref{CKbasis}, the nontrivial crossing kernels are
\ie{}
& 
{\widetilde K}_{W,W}^{W,W}\equiv
\begin{pmatrix} {\widetilde K}_{W,W}^{W,W}(I, I) & {\widetilde K}_{W,W}^{W,W}(I, W) \\ {\widetilde K}_{W,W}^{W,W}(W, I) & {\widetilde K}_{W,W}^{W,W}(W,W) \end{pmatrix} =  \begin{pmatrix} \widetilde{\zeta}^{-1}  & \widetilde{\zeta}^{-1} \\ 1  & -\widetilde{\zeta}^{-1} \end{pmatrix}.
\label{LeeYangCK}
\fe
Some other useful corollaries of the crossing relations are that, a $W$ bubble on a $W$ line can be collapsed while introducing a factor of $\widetilde{\zeta}$, and a $W$ triangle with a $W$ prong attached to each vertex can be collapsed into a $WWW$ T-junction while introducing a factor of $-1$. These identities are depicted in Figure~\ref{fig:Wcross}.

\begin{figure}[H]
\centering
\begin{minipage}{0.1\textwidth}
\includegraphics[width=1\textwidth]{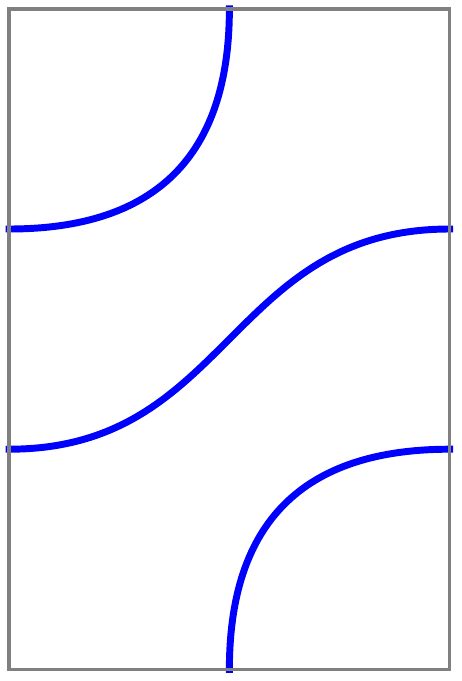}
\end{minipage}%
\begin{minipage}{0.1\textwidth}\begin{eqnarray*}~= \widetilde{\zeta}^{-1} \\ \end{eqnarray*}
\end{minipage}%
\begin{minipage}{0.1\textwidth}
\includegraphics[width=1\textwidth]{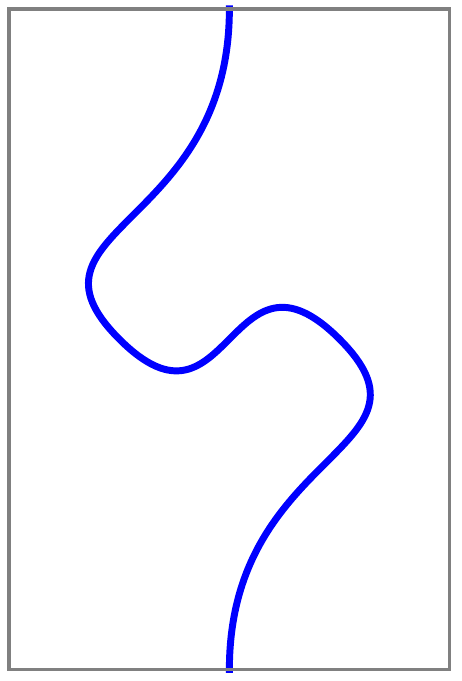}
\end{minipage}%
\begin{minipage}{0.1\textwidth}\begin{eqnarray*}~+ \widetilde{\zeta}^{-1} \\ \end{eqnarray*}
\end{minipage}%
\begin{minipage}{0.1\textwidth}
\includegraphics[width=1\textwidth]{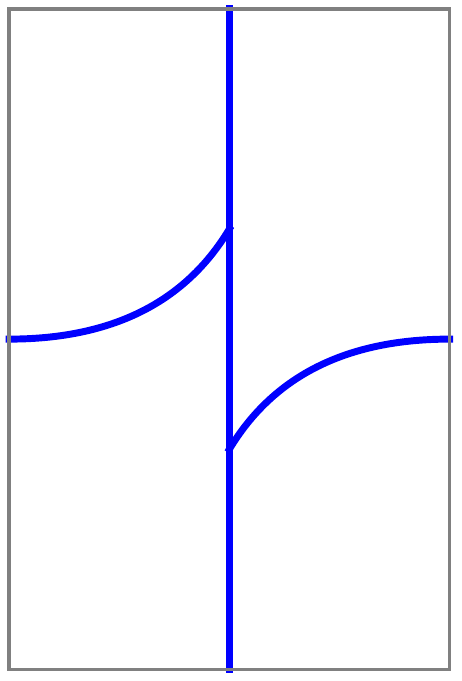}
\end{minipage}%
\\
\bigskip
\begin{minipage}{0.1\textwidth}
\includegraphics[width=1\textwidth]{figures/T2ZWpW.pdf}
\end{minipage}%
\begin{minipage}{0.1\textwidth}\begin{eqnarray*}~=  \\ \end{eqnarray*}
\end{minipage}%
\begin{minipage}{0.1\textwidth}
\includegraphics[width=1\textwidth]{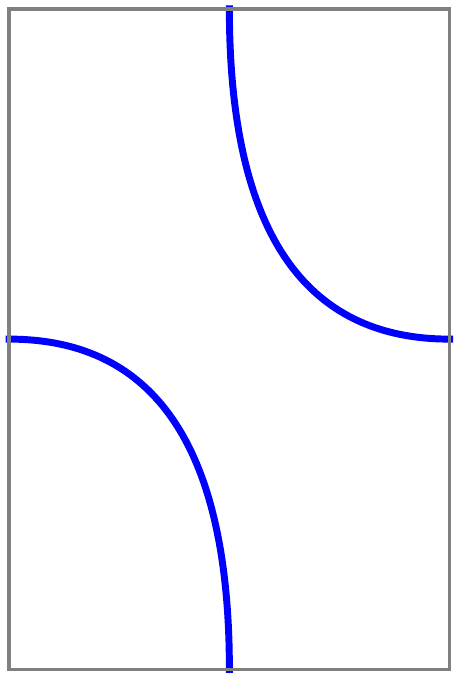}
\end{minipage}%
\begin{minipage}{0.1\textwidth}\begin{eqnarray*}~ -\widetilde{\zeta}^{-1} \\ \end{eqnarray*}
\end{minipage}%
\begin{minipage}{0.1\textwidth}
\includegraphics[width=1\textwidth]{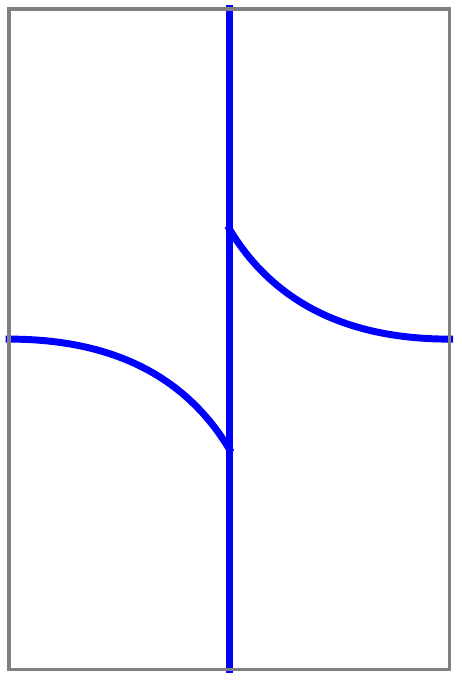}
\end{minipage}%
\caption{Applying the H-junction crossing relations to the $T^2$ transformation of $W$ line in time direction.}
\label{fig:T2ZW}
\end{figure}

Let $|\psi\ra, |\psi'\ra \in {\cal H}_W$. Applying a modular $T^2$ transformation and then the crossing relations to the matrix element $\la\psi'|\psi\ra$ as in Figure~\ref{fig:T2ZW} give the relation
\ie\label{weqa}
e^{4\pi i s} = \widetilde{\zeta}^{-1} + \widetilde{\zeta}^{-1} \widehat W_-= \widetilde{\zeta}^{-1} + e^{-2\pi i s}\widetilde{\zeta}^{-1} - \widetilde{\zeta}^{-2}\widehat W_+,
\fe
where $\widehat W_\pm$ are the operators acting on ${\cal H}_W$ defined by splitting a spatial $W$ line off a temporal $W$ line. Similarly, considering a modular $T^{-2}$ transformation on the matrix element $\la\psi'|\psi\ra$ gives
\ie\label{weqb}
e^{-4\pi i s} = \widetilde{\zeta}^{-1} + \widetilde{\zeta}^{-1} \widehat W_+,
\fe

The solutions to (\ref{weqa}) and (\ref{weqb}) with real $s$ are
\ie\label{spinw}
s\in\bZ+\begin{cases}
0,\pm{2\over 5}\quad&{\rm for}\quad \widetilde{\zeta}={1+\sqrt{5}\over 2},
\\
0,\pm{1\over 5}\quad&{\rm for}\quad \widetilde{\zeta}={1-\sqrt{5}\over 2}.
\end{cases}
\fe
In the $\widetilde{\zeta}={1+\sqrt{5}\over 2}$ case, the spin selection rule \eqref{spinw} is indeed confirmed by the operator content of ${\cal H}_W$ in the tricritical Ising model \eqref{TrisingWSpins}, in the tetracritical Ising model \eqref{TetraWSpins}, or in the three-state Potts model \eqref{eqn:SSR3PHW}. The spins in the defect Hilbert space \eqref{LYSpectrum} of the nontrivial Verlinde line in the Lee-Yang model precisely satisfy the $\widetilde{\zeta} = {1-\sqrt5 \over 2}$ rules.

\subsection{Categories with ${R}_{\bC}(S_3)$ fusion ring}
\label{repsthree}

Consider the fusion ring with the same relations as the decomposition rules for the tensor product of $S_3$ representations,
\ie
&X^2=1, \quad Y^2=I+X+Y, \quad XY=YX=Y.
\fe
There are three solutions of the crossing kernels to the pentagon identity \cite{Etingof:ab}. In the basis specified in Appendix~\ref{CKbasis}, the nontrivial crossing kernels are
\ie\label{kxyrep}
& {\widetilde K}_{Y,Y}^{Y,Y} \equiv \begin{pmatrix} {\widetilde K}_{Y,Y}^{Y,Y}(I, I) & {\widetilde K}_{Y,Y}^{Y,Y}(I, X) & {\widetilde K}_{Y,Y}^{Y,Y}(I, Y)
\\
{\widetilde K}_{Y,Y}^{Y,Y}(X, I) & {\widetilde K}_{Y,Y}^{Y,Y}(X, X) & {\widetilde K}_{YY}^{Y,Y}(X, Y)
\\
{\widetilde K}_{Y,Y}^{Y,Y}(Y, I) & {\widetilde K}_{Y,Y}^{Y,Y}(Y, X) & {\widetilde K}_{Y,Y}^{Y,Y}(Y, Y)
\end{pmatrix} = \begin{pmatrix}
{1\over2} & {1\over2} & {\omega\over2}
\\
{1\over2} & {1\over2} & -{\omega\over2}
\\
1 & -1 & 0
\end{pmatrix},
\\
& {\widetilde K}_{Y,Y}^{Y,I} (Y,Y) = \omega^{-1},
\fe
where $\omega$ is a third root of unity. There is a nontrivial cyclic permutation map $C_{Y,Y,Y}$ given by $C_{Y,Y,Y} = {\widetilde K}_{Y,Y}^{Y,I}(Y,Y) = \omega^{-1}$. Note that all the crossing kernels involving an $X$ external line are trivial.

\begin{figure}[H]
\centering
\begin{minipage}{0.08\textwidth}
\includegraphics[width=1\textwidth]{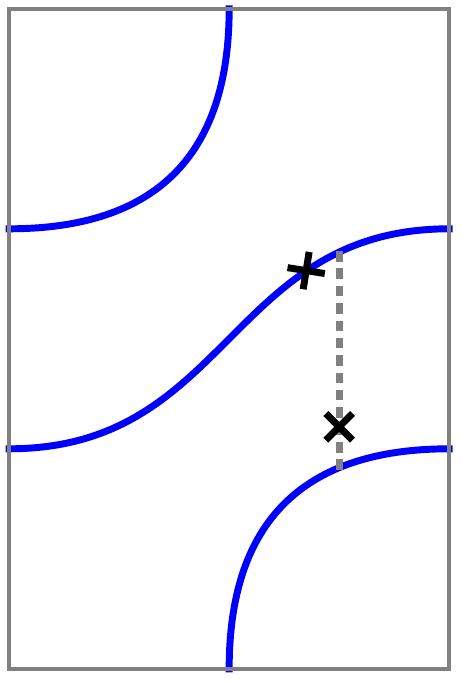}
\end{minipage}%
\begin{minipage}{0.08\textwidth}\begin{eqnarray*}~= {1\over 2} \\ \end{eqnarray*}
\end{minipage}%
\begin{minipage}{0.08\textwidth}
\includegraphics[width=1\textwidth]{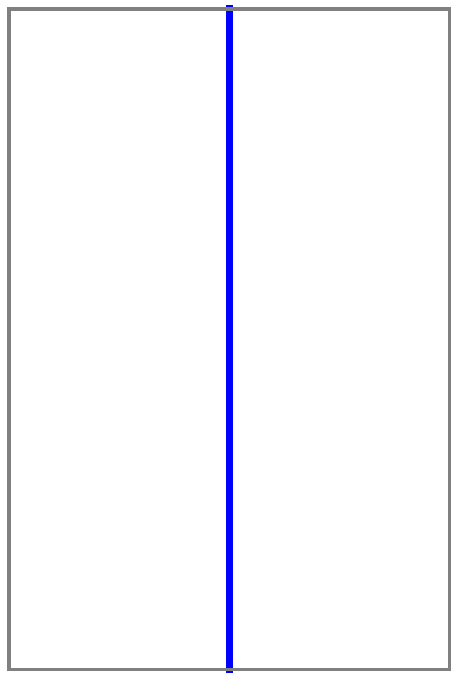}
\end{minipage}%
\begin{minipage}{0.08\textwidth}\begin{eqnarray*}~+ {1\over 2} \\ \end{eqnarray*}
\end{minipage}%
\begin{minipage}{0.08\textwidth}
\includegraphics[width=1\textwidth]{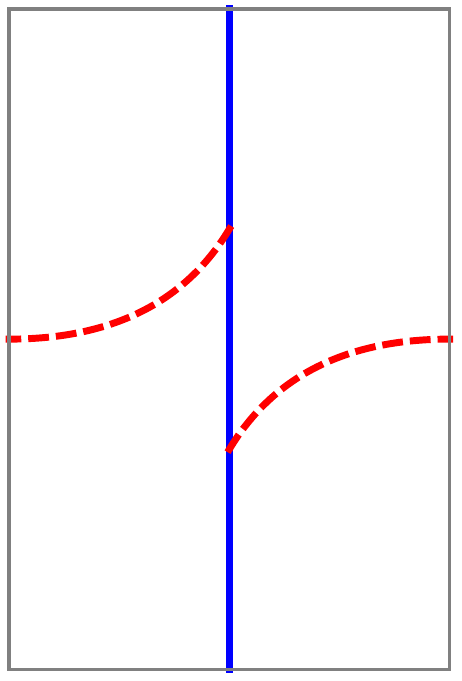}
\end{minipage}%
\begin{minipage}{0.08\textwidth}\begin{eqnarray*}~+ {\omega\over 2} \\ \end{eqnarray*}
\end{minipage}%
\begin{minipage}{0.08\textwidth}
\includegraphics[width=1\textwidth]{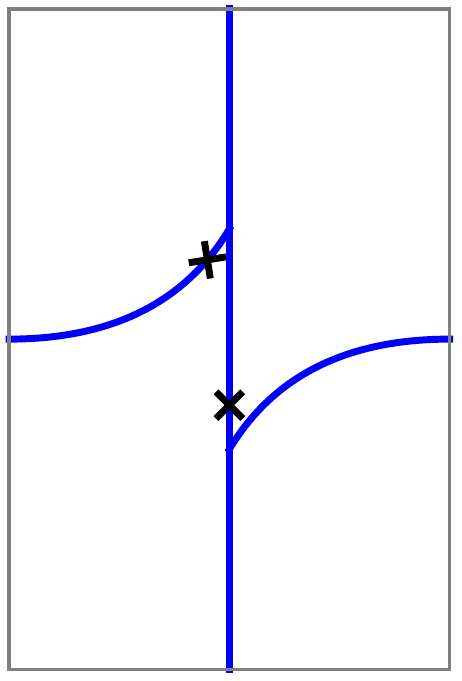}
\end{minipage}%
\\
\bigskip
\begin{minipage}{0.08\textwidth}
\includegraphics[width=1\textwidth]{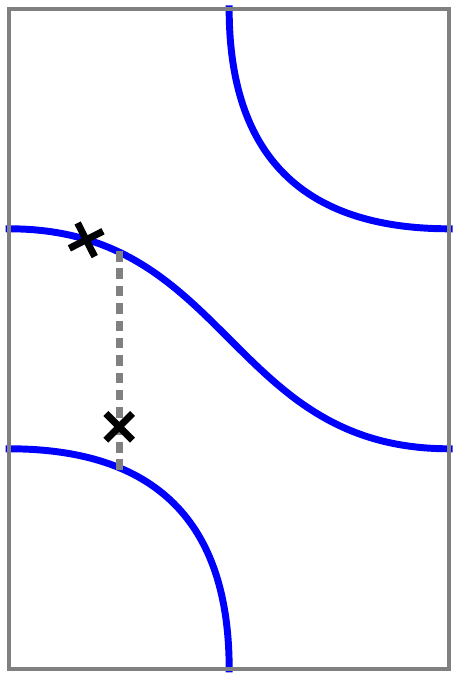}
\end{minipage}%
\begin{minipage}{0.08\textwidth}\begin{eqnarray*}~={1\over 2}  \\ \end{eqnarray*}
\end{minipage}%
\begin{minipage}{0.08\textwidth}
\includegraphics[width=1\textwidth]{figures/RepS3_1.pdf}
\end{minipage}%
\begin{minipage}{0.08\textwidth}\begin{eqnarray*}~ +{1\over 2} \\ \end{eqnarray*}
\end{minipage}%
\begin{minipage}{0.08\textwidth}
\includegraphics[width=1\textwidth]{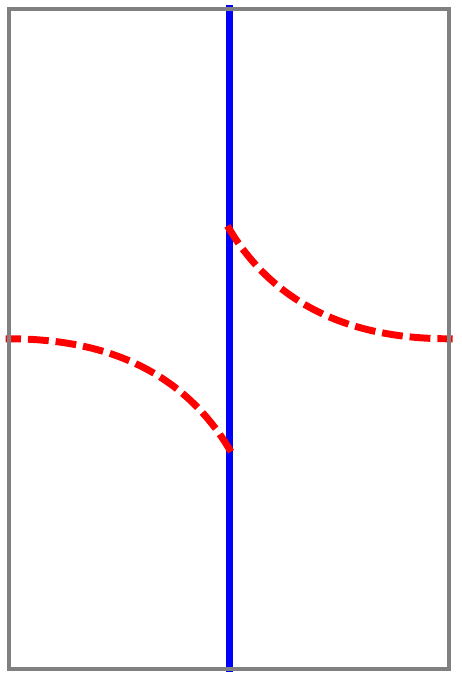}
\end{minipage}%
\begin{minipage}{0.08\textwidth}\begin{eqnarray*}~ +{\omega\over 2} \\ \end{eqnarray*}
\end{minipage}%
\begin{minipage}{0.08\textwidth}
\includegraphics[width=1\textwidth]{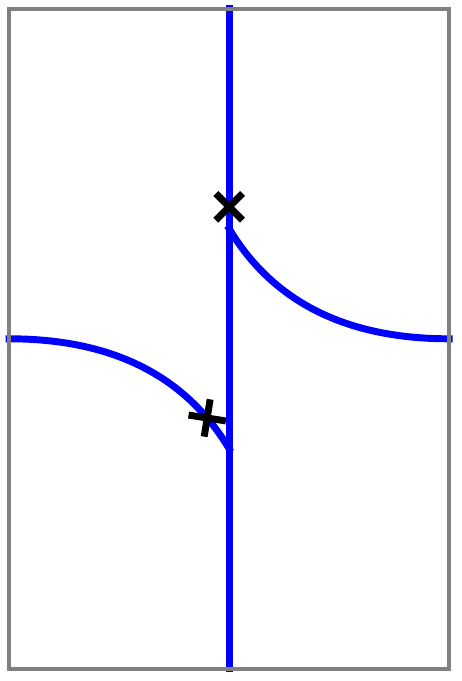}
\end{minipage}%
\\
\bigskip
\begin{minipage}{0.08\textwidth}
\includegraphics[width=1\textwidth]{figures/RepS3_3.pdf}
\end{minipage}%
\begin{minipage}{0.08\textwidth}\begin{eqnarray*}~=  \\ \end{eqnarray*}
\end{minipage}%
\begin{minipage}{0.08\textwidth}
\includegraphics[width=1\textwidth]{figures/TZW.pdf}
\end{minipage}%
\begin{minipage}{0.08\textwidth}\begin{eqnarray*}~ - \\ \end{eqnarray*}
\end{minipage}%
\begin{minipage}{0.08\textwidth}
\includegraphics[width=1\textwidth]{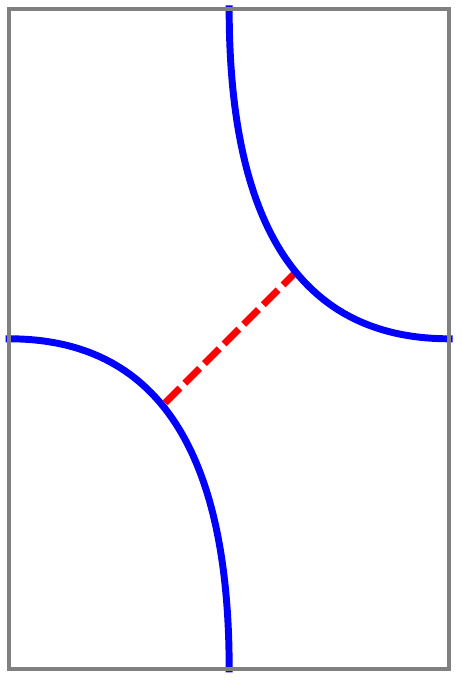}
\end{minipage}%
\\
\bigskip
\begin{minipage}{0.08\textwidth}
\includegraphics[width=1\textwidth]{figures/RepS3_6.pdf}
\end{minipage}%
\begin{minipage}{0.08\textwidth}\begin{eqnarray*}~= \omega \\ \end{eqnarray*}
\end{minipage}%
\begin{minipage}{0.08\textwidth}
\includegraphics[width=1\textwidth]{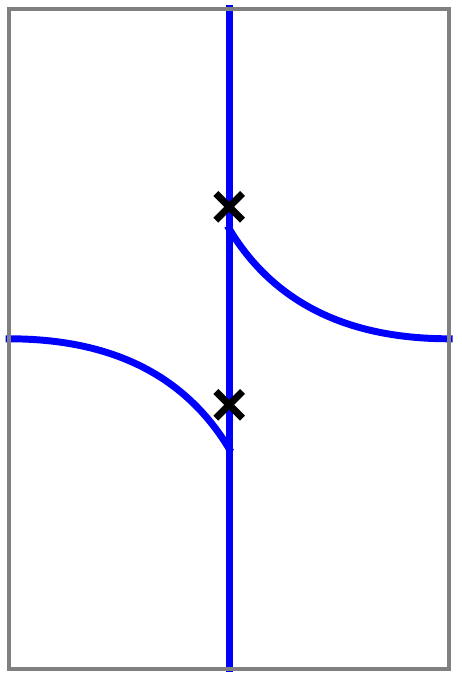}
\end{minipage}%
\begin{minipage}{0.08\textwidth}\begin{eqnarray*}~=  \omega\\ \end{eqnarray*}
\end{minipage}%
\begin{minipage}{0.08\textwidth}
\includegraphics[width=1\textwidth]{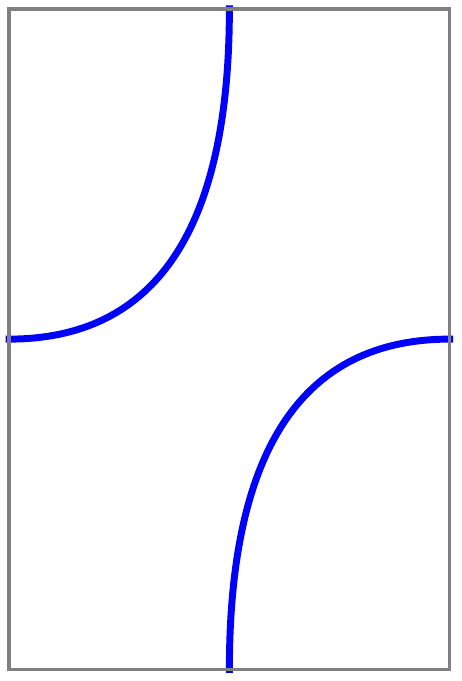}
\end{minipage}%
\begin{minipage}{0.08\textwidth}\begin{eqnarray*}~ - \omega\\ \end{eqnarray*}
\end{minipage}%
\begin{minipage}{0.08\textwidth}
\includegraphics[width=1\textwidth]{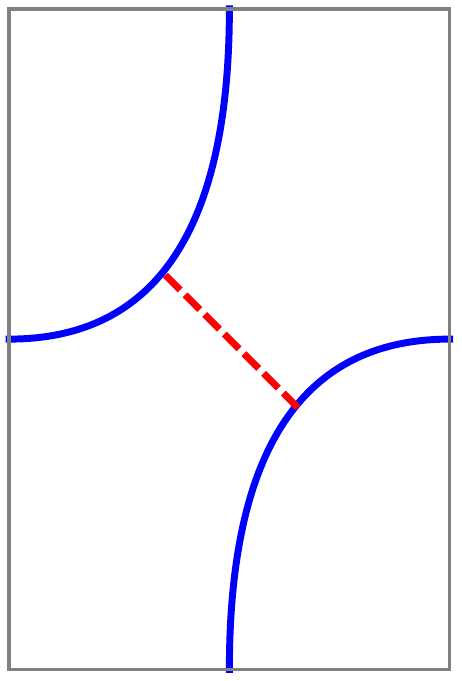}
\end{minipage}%
\caption{Derivation of the spin selection rules for the ${R}_{\bC}(S_3)$ fusion ring. We apply the H-junction crossing relations to the $T^2$ and $T^{-2}$ transformations of a temporal $Y$ line. The blue and red-dashed lines are used to denote the $Y$ and $X$ lines respectively. The black-dotted lines representing the trivial lines, and the crosses marking the ordering at trivial junctions are in fact unnecessary. We include them to make explicit the particular crossing kernels that are applied here. However, it is important to keep track of the marks on the $YYY$ junctions due to the nontrivial cyclic permutation map acting on $V_{Y,Y,Y}$.}
\label{fig:RepS3}
\end{figure}

Let $|\psi\ra, |\psi'\ra \in {\cal H}_Y$. Consideration of modular $T^2$ and $T^{-2}$ transformations on the matrix element $\la\psi'|\psi\ra$ and applying cyclic permutation map and crossing relations as in Figure~\ref{fig:RepS3} gives the equations
\ie
e^{4\pi is} & 
= {1 \over 2} \left[ 1 + \widehat X_- + e^{-2\pi is} \omega \left( 1 - \widehat X_- \right) \right],
\\
e^{-4\pi is} & 
= {1 \over 2} \left[ 1 + \widehat X_- + e^{2\pi is} \omega^2 \left( 1 - \widehat X_-\right) \right]\,,
\fe
where $\widehat X_- :{\cal H}_Y \rightarrow {\cal H}_Y$ is defined on the lower right of Figure~\ref{fig:RepS3}, similar to the $\widehat \eta_-$ operator in Section~\ref{Sec:Z2TY}. 

We find the spin selection rules
\ie
s\in\bZ+\begin{cases}
	0,{1\over 2},{1\over 3},{2\over 3}\quad&{\rm for}\quad \omega=1,
	\\
	0,{1\over 2},{1\over 9},{4\over 9},{7\over 9}\quad&{\rm for}\quad \omega=e^{2\pi i\over 3},
	\\
	0,{1\over 2},{2\over 9},{5\over 9},{8\over 9}\quad&{\rm for}\quad \omega=e^{-{2\pi i\over 3}}.
\end{cases}
\fe
The spins \eqref{TetraMSpins} in the defect Hilbert space ${\cal H}_M$ of the tetracritical Ising model precisely satisfy the $\omega = 1$ rules. 

In the $\omega \neq 1$ cases, which correspond to the twisted ${\rm Rep}(S_3)$ fusion categories, the spin selection rules are realized by the $\bZ_2$ orbifold of CFTs with 't Hooft anomalous $S_3$ symmetry \cite{Bhardwaj:2017xup}. A particular example would be the $\bZ_2$ orbifold of the tensor product of the tetracritical Ising model with $SU(2)_1$ WZW model where the $\bZ_3$ subgroup of the $S_3$ symmetry is taken to be the diagonal combination which has an 't Hooft anomaly.

\subsection{Categories with ${1\over 2} E_6$ fusion ring}
\label{esixselect}

\begin{figure}[H]
\centering
\begin{minipage}{0.1\textwidth}
\includegraphics[width=1\textwidth]{figures/RepS3_0.pdf}
\end{minipage}%
\begin{minipage}{0.16\textwidth}\begin{eqnarray*}~= {\widetilde K}^{Y,Y}_{Y,Y}(I,I) \\ \end{eqnarray*}
\end{minipage}%
\begin{minipage}{0.1\textwidth}
\includegraphics[width=1\textwidth]{figures/RepS3_1.pdf}
\end{minipage}%
\begin{minipage}{0.17\textwidth}\begin{eqnarray*}~+ {\widetilde K}^{Y,Y}_{Y,Y}(I,X) \\ \end{eqnarray*}
\end{minipage}%
\begin{minipage}{0.1\textwidth}
\includegraphics[width=1\textwidth]{figures/RepS3_2.pdf}
\end{minipage}%
\begin{minipage}{0.18\textwidth}\begin{eqnarray*}~+ {\widetilde K}^{Y,Y}_{Y,Y}(I,Y)_{ij} \\ \end{eqnarray*}
\end{minipage}%
\begin{minipage}{0.1\textwidth}
\includegraphics[width=1\textwidth]{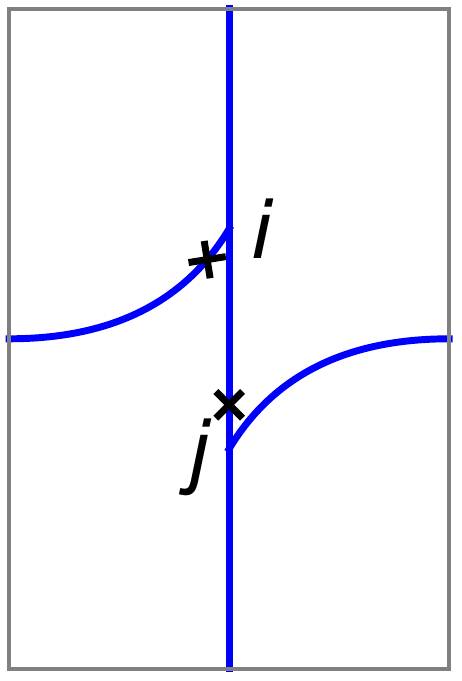}
\end{minipage}%
\\
\bigskip
\begin{minipage}{0.1\textwidth}
\includegraphics[width=1\textwidth]{figures/E6_1.pdf}
\end{minipage}%
\begin{minipage}{0.18\textwidth}\begin{eqnarray*}~={\widetilde K}^{Y,Y}_{Y,Y}(Y,I)_{ji}    \\ \end{eqnarray*}
\end{minipage}%
\begin{minipage}{0.1\textwidth}
\includegraphics[width=1\textwidth]{figures/TZW.pdf}
\end{minipage}%
\begin{minipage}{0.18\textwidth}\begin{eqnarray*}~ +{\widetilde K}^{Y,Y}_{Y,Y}(Y,X)_{ji}  \\ \end{eqnarray*}
\end{minipage}%
\begin{minipage}{0.1\textwidth}
\includegraphics[width=1\textwidth]{figures/RepS3_10.pdf}
\end{minipage}%
\begin{minipage}{0.2\textwidth}\begin{eqnarray*}~ +{\widetilde K}^{Y,Y}_{Y,Y}(Y,Y)_{ji,kl}  \\ \end{eqnarray*}
\end{minipage}%
\begin{minipage}{0.1\textwidth}
\includegraphics[width=1\textwidth]{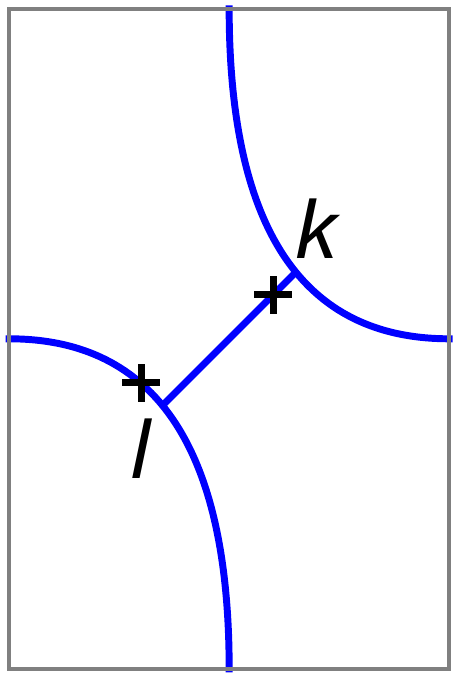}
\end{minipage}%
\caption{Derivation of the spin selection rules for the ${1\over 2} E_6$ fusion ring. We apply the H-junction crossing relations to the $T^2$ and $T^{-2}$ transformations of a temporal $Y$ line. The blue- and red-dashed lines are used to denote the $Y$ and $X$ lines, respectively. Again, the black-dotted line with crosses labeling the (marked) trivial line in the top left diagram merely keeps track of the crossing kernel applied here, and may be omitted. However, it is important to keep track of the marks on the $YYY$ junctions due to the nontriviality of the cyclic permutation map on $V_{Y,Y,Y}$.}
\label{fig:esix}
\end{figure}

For TDLs $X$ and $Y$ that generate the fusion ring of the ${1\over 2}E_6$ fusion category (\ref{halfesix}), there are four possible sets of crossing kernels that solve the pentagon identity, related by Galois group action \cite{Hagge:2007aa}. The explicit formulae of the crossing kernels are summarized in Appendix~\ref{1/2E6CK}. The spin selection rule on states in ${\cal H}_Y$ can be derived from (Figure~\ref{fig:esix})
\ie
&e^{4\pi i s}={\widetilde K}^{Y,Y}_{Y,Y}(I,I)+{\widetilde K}^{Y,Y}_{Y,Y}(I,X) \widehat X_-+{\widetilde K}^{Y,Y}_{Y,Y}(I,Y)_{ij}(\widehat Y_-)_{ij},
\\
&(\widehat Y_-)_{ij}=e^{-2\pi i s}\left[{\widetilde K}^{Y,Y}_{Y,Y}(Y,I)_{ji}+{\widetilde K}^{Y,Y}_{Y,Y}(Y,X)_{ji}\widehat X_-+{\widetilde K}^{Y,Y}_{Y,Y}(Y,Y)_{ji,kl}(\widehat Y_-)_{kl}\right],
\fe
together with the property $\widehat X_-^2=-1$ which follows easily from the crossing relations.  
The subscripts $i,j=1,2$ labels the two basis junction vectors of $V_{Y,Y,Y}$.  The topological operator ${ (\widehat Y_-)}_{ij}$ acting on ${\cal H}_Y$ is defined in the upper right figure of Figure~\ref{fig:esix} with the marked legs specified. 

The resulting selection rule on $s$ corresponding to the four sets of solutions to pentagon identity are
\ie
s\in\bZ+\begin{cases}
0,{1\over 6},{1\over 4}, {1\over 2}, {2\over 3}, {11\over 12}.\quad ~(a)
\\
0,{1\over 4}, {1\over 3}, {1\over 2}, {7\over 12}, {5\over 6},\quad ~(b)
\\
0,{1\over 12}, {1\over 3}, {1\over 2}, {3\over 4}, {5\over 6},\quad ~(c)
\\
0,{1\over 6}, {5\over 12}, {1\over 2}, {2\over 3}, {3\over 4},\quad ~(d)
\end{cases}
\fe
The case $(d)$ is precisely the spin content of ${\cal H}_Y$ in the non-diagonal $SU(2)_{10}$ WZW model of $E_6$ type, and in the $(A_{10}, E_6)$ minimal model. The case $(b)$ is the spin content of ${\cal H}_{\widetilde Y}$ in these models, where $\widetilde Y$ is related to $Y$ by parity. The spin selection rule provides a highly nontrivial check of the existence of the TDLs generating the ${1\over 2}E_6$ fusion ring in these CFTs.

\section{Constraints on RG flows}
\label{sec:rgflow}
 
In this section, we discuss the constraints from TDLs on RG flows. In the case when the RG flow ends in a massive phase, using modular invariance, we derive simple sufficient conditions for degenerate vacua in the IR. Furthermore, for certain massive flows, the IR TQFTs can be completely determined from the consideration of TDLs, together with modular invariance.

We begin with a general discussion on TDLs along  RG flows.  
Recall that in unitary theories, a bulk local operator $\phi$ commutes with a TDL ${\cal L}$ if and only if  $\widehat{\cal L}|\phi\ra = \la{\cal L}\ra |\phi\ra$,  where $\la {\cal L}\ra \equiv \la 0| \widehat {\cal L} |0\ra$ is the expectation value of an empty ${\cal L}$ loop on the cylinder, as defined in Section~\ref{Sec:Vac}. 
If $\phi$ is a relevant conformal primary, then it triggers an RG flow, and all the TDLs that commute with $\phi$ will be preserved along the entire RG flow. The fusion ring, the spin selection rules, and the H-junction crossing relations of the TDLs are also preserved, imposing nontrivial constraints on the IR theory. Furthermore, if the UV CFT has a unique vacuum, the vanishing tadpole property of a nontrivial TDL, as discussed in Section~\ref{Sec:VanishingTadpole}, is expected to hold along the entire RG flow.

The constraints on the Hilbert space ${\cal H}_{\cal L}$ of defect operators at the end of ${\cal L}$ can be readily translated into constraints on the bulk Hilbert space by modular invariance.  
If the spin selection rule on ${\cal H}_{\cal L}$ is such that only non-integer spins are allowed, then ${\cal L}$ can never flow to the trivial line in the IR. This implies the following two possibilities about the IR theory:
\begin{enumerate}
\item  
${\cal H}_{\cal L}$ is non-empty, and obeys the same spin selection rule as in the UV. This can only happen if the flow ends on a nontrivial CFT.
\item ${\cal H}_{\cal L}$ is empty.
Modular invariance then implies that the trace of $\widehat {\cal L}$ over the \textit{bulk} primaries of every conformal weight is also zero, {\it i.e.}, 
\begin{align}
0 = \text{tr}_{{\cal H}_{\cal L}} \,q^{L_0-c/24}\bar q^{\bar L_0-c/24} = \text{tr}\, \widehat{\cal L} \, \tilde q^{L_0-c/24}  {\bar {\tilde q}}^{\bar L_0-c/24},
\end{align}
where $\tilde q$ is the $S$-transform of $q$. This turns out to be a strong constraint on the bulk spectrum in the IR. In particular, there must be degenerate vacua.
\end{enumerate}

A simple application  is the 't Hooft anomaly matching of $\bZ_n$ symmetry. Consider a massive RG flow from a CFT with an anomalous $\bZ_n$ symmetry to an IR TQFT, triggered by a relevant $\bZ_n$ singlet operator. As shown in Section~\ref{sec:ZnSSR}, the Hilbert space $H_{{\cal L}_g}$ for a nontrivial element $g\in \bZ_n$ contains only states with non-integer spins.  Hence, by the above analysis, the Hilbert space $H_{{\cal L}_g}$ of the IR TQFT must be empty. Consequently, there must be degenerate vacua, such that the trace of $\widehat {\cal L}_g$ over the Hilbert space of vacua is zero. 

In addition to constraints on the IR theory from the TDLs that are preserved under the RG flow, we may also learn something from those TDLs that are broken. Generally, if $\phi$ is not charged under any $\mathbb{Z}_2$ symmetry, the RG flows generated by $\phi$ with different signs of the coupling end on different IR theories ${\cal T}_+$ and ${\cal T}_-$. Now, if there is a TDL ${\cal L}'$ that \textit{anticommutes} with $\phi$ in the UV CFT, {\it i.e.},
\begin{align}
\widehat{ \mathcal{L}'} |\phi\ra =  - \la\mathcal{L}'\ra |\phi\ra\,,
\end{align}
then ${\cal L}'$ will survive the RG flow as a topological \textit{interface} between ${\cal T}_+$ and ${\cal T}_-$. Note that $\mathcal{L}'$ is not a TDL in either ${\cal T}_+$ or ${\cal T}_-$ itself because it does not commute with the deformation $\phi$. In particular, the existence of such a topological interface implies that ${\cal T}_+$ and ${\cal T}_-$ must have the same central charge (but can be different CFTs). For instance, if one of the flows ends up in a massive phase, then so must the other, even though the two flows could end up in different TQFTs.

\subsection{A diagnostic for degenerate vacua}
\label{sec:degvacua}

As a first example, let us apply the above strategy to argue the degeneracy of vacua in the IR using TDLs and modular invariance.  
Consider a massive RG flow from a CFT to an IR TQFT, triggered by a relevant operator $\phi$. The CFT need not be unitary, as long as the degeneracies are not negative like in theories with ghosts. If there is a TDL $\mathcal{L}$ that commutes with $\phi$, then ${\cal L}$ is preserved along the entire RG flow. Let us further assume that its loop expectation value $\la \mathcal{L} \ra$ is \textit{not} a non-negative integer. It can then be argued that there must be degenerate vacua in the IR TQFT. 

We prove by contradiction. Suppose that there is a unique vacuum in the IR TQFT. It follows that
\begin{align}
{\rm tr}\, \widehat {\cal L}  =  \la {\cal L}\ra\,,
\end{align}
where here and in the rest of this section, tr denotes a trace over the IR TQFT Hilbert space, unless otherwise specified. Now, by modular invariance, we must also have
\begin{align}
\text{tr} _{{\cal H}_{\cal L}} 1 = \la {\cal L}\ra.
\end{align}
This is a contradiction, because the LHS in the above equation is manifestly a non-negative integer. We have thereby proven the following theorem: if a TDL with loop expectation value $\la {\cal L}\ra$ that is not a non-negative integer is preserved along a massive RG flow, then the IR theory must have degenerate vacua.  

Note that if we have more than one vacuum in the IR, their eigenvalues under $\widehat{\cal L}$ can add up to a non-negative integer, even if each one is not, to be consistent with modular invariance.  As we will see below, this is indeed the case in various massive flows in the minimal models.

This argument can be thought of as a {generalization} of the 't Hooft anomaly matching condition for global symmetry \cite{tHooft:1980xss,Kapustin:2014zva,Gaiotto:2014kfa,Gaiotto:2017yup}: the nontrivial crossing relations of TDLs in the UV have to be captured by certain degrees of freedom in the IR.

\subsection{Constraints on TQFTs in specific flows}
\label{Sec:specificflow}

In this section, we will use TDLs to constrain various massive flows.  For certain flows, one can bootstrap the IR TQFT completely using the data of TDLs that are preserved along the flows.  This can be thought of as a generalized 't Hooft ``anomaly" matching condition, where the IR degrees of freedom are constrained to be consistent with the crossing relations of TDLs inherited from the UV.  We will only consider TQFTs that arise at the endpoints of RG flows from unitary, compact CFTs (with a unique vacuum).

There is one important subtlety to clarify here.  As already mentioned in Section~\ref{Sec:Properties}, we define the junction vector space $V_{{\cal L}_1,\cdots,{\cal L}_k}$ of the IR TQFT as the Hilbert space of weight-(0,0) defect operators that are flowed from the UV CFT. In particular, $V_{{\cal L}_1 , \cdots, {\cal L}_k}$ is a subspace of all the weight-(0,0) operators at the junction in the IR TQFT.  The reason for this restriction is because the TDLs of the UV CFT that are not broken by the flow have crossing kernels that are preserved on these subspaces $V_{{\cal L}_1,\cdots,{\cal L}_k}$. This is analogous to the usual constraint on an RG flow by matching the 't Hooft anomaly of a global symmetry.

In the following, every fusing ring considered is commutative. In such a case, the representations of the TDL actions on the degenerate bulk local operators of a fixed conformal weight can be diagonalized. In this diagonal basis, the set of eigenvalues must solve the polynomial equations given by the abelianization of the fusion ring, just as $\la {\cal L} \ra$ does.  We will always work in such a basis.

\subsubsection{Ising model deformed by $\varepsilon$}
\label{Sec:IsingFlow}

Consider the critical Ising model (Section~\ref{Sec:Ising}) deformed by the energy operator $\varepsilon$.  Depending on the sign of the deformation, we either flow to a TQFT $\mathcal{T}_+$ with only one vacuum, or to $\mathcal{T}_-$ with two vacua \cite{Onsager,Yang}. In the latter case, the $\mathbb{Z}_2$ global symmetry is spontaneously broken by the degenerate vacua. According to the general arguments given above, since the $N$ line anticommutes with the relevant deformation $\varepsilon$, it flows to a topological interface between the two TQFTs $\mathcal{T}_+$ and $\mathcal{T}_-$.  In fact, as discussed in Section~\ref{Sec:DualityTY}, the  $N$ line in the critical Ising model is a duality defect \cite{Frohlich:2004ef} that implements the Kramers-Wannier duality \cite{PhysRev.60.252}.\footnote{See \cite{Kapustin:2014gua} for discussion on a subtlety with the Kramers-Wannier duality on general Riemann surfaces.}

\subsubsection{Tricritical Ising model deformed by $\sigma'$}
\label{lytqft}

Let us now consider a simple example where a non-invertible TDL is preserved along a massive flow, and study how the TDL constrains the IR TQFT.  Consider the $\sigma'$ deformation of the tricritical Ising model (Section~\ref{Sec:Trising}), which breaks the $\mathbb{Z}_2$ invertible line $\eta$ and the $N$ line, but preserves the $W$ line, which has the fusion relation 
\begin{align}
W^2=I+W\,.
\end{align}
  This RG flow is expected to end up in a massive phase, described by a TQFT \cite{Zamolodchikov:1990xc, Ellem:1997vz}.   Modular invariance demands that
\ie
{\rm tr}\,\widehat W = {\rm dim}\, {\cal H}_W,
\fe
where the trace on the LHS is over the vacua of the IR theory. The possible eigenvalues of $\widehat W$ are $\zeta={1+\sqrt{5}\over 2}$ and $-\zeta^{-1}={1-\sqrt{5}\over 2}$, and their corresponding eigenstates must come in pairs for ${\rm tr}\,\widehat W$ to be an integer. Thus, we learn that the number of vacua must be even, and is twice the dimension of ${\cal H}_W$. Recall that the spin selection rule \eqref{spinw} on ${\cal H}_W$ is $s\in\mathbb{Z}$ or $s\in \pm{2\over 5}+\mathbb{Z}$, which indeed allows for a nonempty ${\cal H}_W$ in the TQFT.

\begin{figure}[H]
\centering
\begin{minipage}{0.16\textwidth}
\includegraphics[width=1\textwidth]{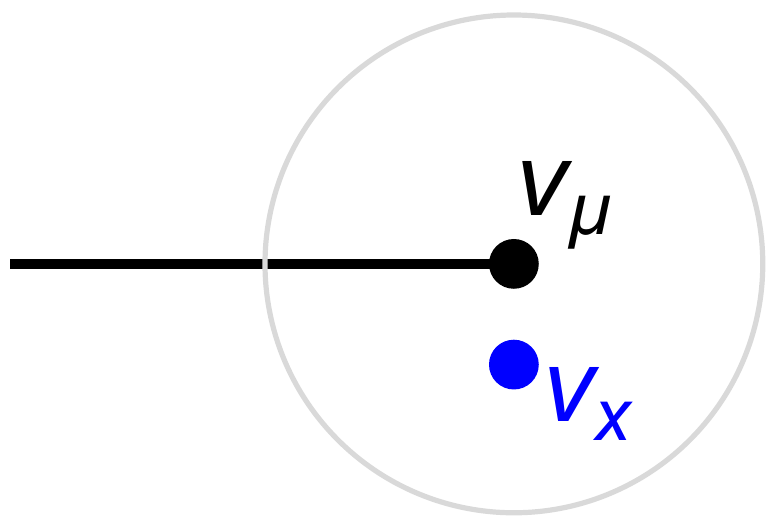}
\end{minipage}%
\begin{minipage}{0.08\textwidth}\begin{eqnarray*}~=\beta \\ \end{eqnarray*}
\end{minipage}%
\begin{minipage}{0.08\textwidth}
\includegraphics[width=1\textwidth]{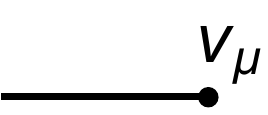}
\end{minipage}%
\begin{minipage}{0.1\textwidth}
\begin{eqnarray*}~\end{eqnarray*}
\end{minipage}%
\begin{minipage}{0.09\textwidth}
\includegraphics[width=1\textwidth]{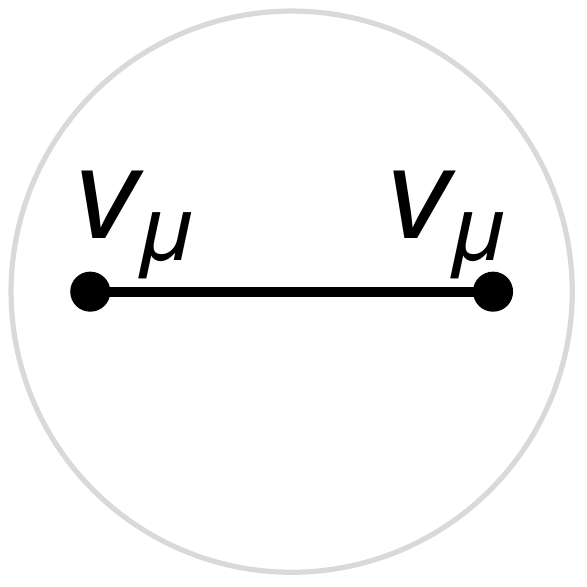}
\end{minipage}%
\begin{minipage}{0.08\textwidth}\begin{eqnarray*}~=1+\C v_x \\ \end{eqnarray*}
\end{minipage}%
\begin{minipage}{0.1\textwidth}
\begin{eqnarray*}~\end{eqnarray*}
\end{minipage}%
\begin{minipage}{0.12\textwidth}
\includegraphics[width=1\textwidth]{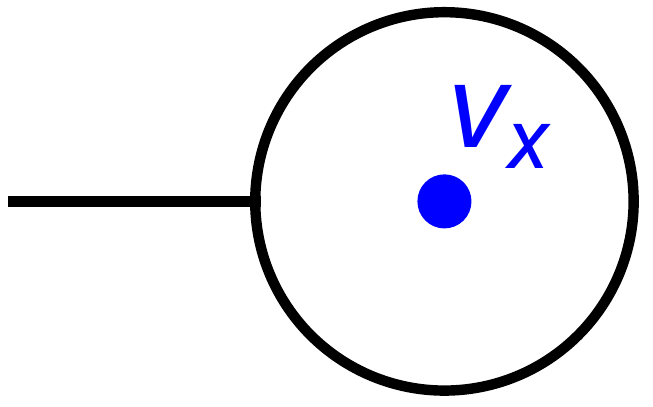}
\end{minipage}%
\begin{minipage}{0.08\textwidth}\begin{eqnarray*}~=\delta \\ \end{eqnarray*}
\end{minipage}%
\begin{minipage}{0.08\textwidth}
\includegraphics[width=1\textwidth]{figures/vmu.pdf}
\end{minipage}%
\caption{Some OPEs in the IR TQFT.}
\label{fig:TQFTOPEs}
\end{figure}

Suppose ${\rm tr}\,\widehat W=1$, namely, that there are two vacua. Let 1 be the canonical vacuum inherited from the CFT vacuum through the RG flow.  Recall that the cylinder vacuum expectation value of $W$ is $\la W\ra =\zeta$, {\it i.e.}, $\widehat W|1\ra = \zeta|1\ra$. The other vacuum $v_x$ obeys 
\ie
\widehat W |v_x\ra=-\zeta^{-1}|v_x\ra\,,
\fe
with the OPE  $v_x^2=1+\A v_x$, for some constant $\A$. Let $v_\mu$ be the unique defect operator in ${\cal H}_W$. Then it must obey OPEs of the form $v_\mu v_x = \B v_\mu\,,~\overbrace{v_\mu v_\mu} = 1+\C v_x.$ 
 Finally, we demand that a $W$ tadpole diagram enclosing $v_x$ produces $\delta v_\mu$, for some constant $\delta$. These relations are summarized in Figure~\ref{fig:TQFTOPEs}.  In particular, the rightmost lasso diagram in Figure~\ref{fig:TQFTOPEs} defines a map $\widehat W^v$  from the bulk Hilbert space $\cal H$ to the defect Hilbert space ${\cal H}_W$ (see Section~\ref{sec:action}),  with $\widehat W^v:~|v_x\rangle \mapsto \delta |v_\mu\rangle$.  The junction vector $v\in V_{WWW}$ is normalized  such that the crossing kernels are given as in Section~\ref{Sec:LeeYangAlgebra}, with $\tilde \zeta= \zeta$.

\begin{figure}[H]
\centering
\begin{minipage}{0.12\textwidth}
\includegraphics[width=1\textwidth]{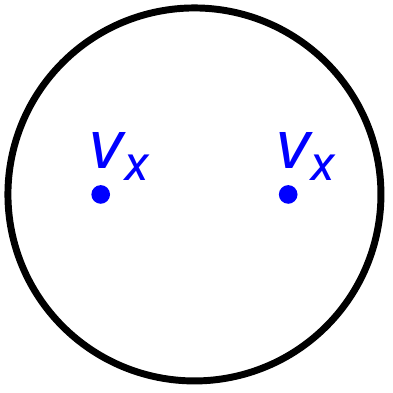}
\end{minipage}%
\begin{minipage}{0.08\textwidth}\begin{eqnarray*}~=\zeta^{-1} \\ \end{eqnarray*}
\end{minipage}%
\begin{minipage}{0.12\textwidth}
\includegraphics[width=1\textwidth]{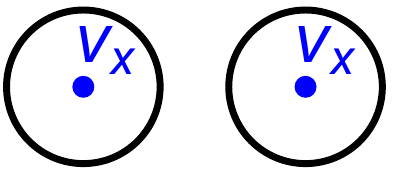}
\end{minipage}%
\begin{minipage}{0.08\textwidth}\begin{eqnarray*}~+\zeta^{-1} \\ \end{eqnarray*}
\end{minipage}%
\begin{minipage}{0.12\textwidth}
\includegraphics[width=1\textwidth]{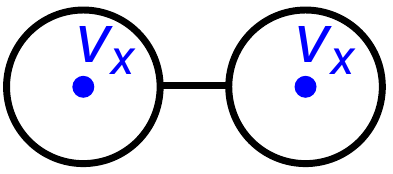}
\end{minipage}%
\caption{Applying partial fusion to $\widehat{W}(v_x^2)$.}
\label{fig:vxvx}
\end{figure}

The associativity of $\overbrace{v_\mu v_\mu} v_x$ gives
\ie\label{vxcra}
v_x + \C (1+\A v_x) = \beta (1+\C v_x).
\fe
Applying partial fusion to a $W$ loop encircling $v_x v_x$, as in Figure~\ref{fig:vxvx}, gives (see \eqref{LeeYangCK} for the crossing kernel)
\ie\label{vxcrb}
\widehat{W}(v_x^2) &= \widehat{W}(1+\A v_x) = \zeta - \zeta^{-1} \A v_x
\\
&= \zeta^{-3}(1+\A v_x) + \delta^2 \zeta^{-1} (1+\C v_x).
\fe
From (\ref{vxcra}) and (\ref{vxcrb}), we can solve (up to signs that can be absorbed into a redefinition of $v_x$ and $v_\mu$)
\ie
\A = 1,\quad  \B = \C = -\zeta^{-1}, \quad  \delta = 5^{1\over 4}.
\fe

\begin{figure}[H]
\centering
\begin{minipage}{0.08\textwidth}
\includegraphics[width=1\textwidth]{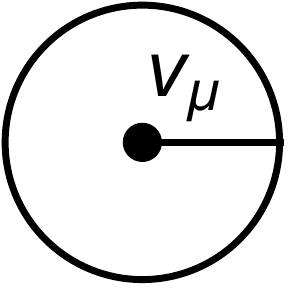}
\end{minipage}%
\begin{minipage}{0.08\textwidth}\begin{eqnarray*}\quad = \quad 5^{1\over4} v_x \,,\\ \end{eqnarray*}
\end{minipage}%
\begin{minipage}{0.1\textwidth}
\begin{eqnarray*}~~~~~~~~~~~~~~\,\,~~~~\end{eqnarray*}
\end{minipage}%
\begin{minipage}{0.12\textwidth}
\includegraphics[width=1\textwidth]{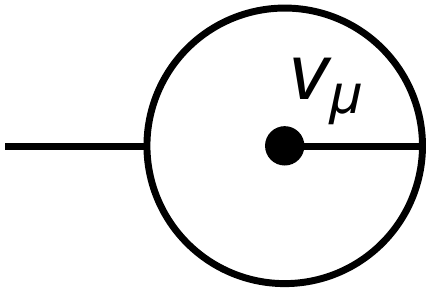}
\end{minipage}%
\begin{minipage}{0.08\textwidth}\begin{eqnarray*}\quad = \quad \zeta^{-1}~~~ \\ \end{eqnarray*}
\end{minipage}%
~~~~
\begin{minipage}{0.08\textwidth}
\includegraphics[width=1\textwidth]{figures/vmu.pdf}
\end{minipage}%
\caption{Lassoing the defect operator $v_\mu \in {\cal H}_W$.}
\label{fig:LYLasso}
\end{figure}

\begin{figure}[H]
\centering
\begin{minipage}{0.12\textwidth}
\includegraphics[width=1\textwidth]{figures/LYTQFT1.pdf}
\end{minipage}%
\begin{minipage}{0.08\textwidth}\begin{eqnarray*}~~= \\ \end{eqnarray*}
\end{minipage}%
\begin{minipage}{0.12\textwidth}
\includegraphics[width=1\textwidth]{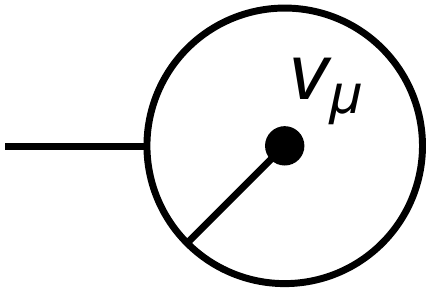}
\end{minipage}%
\begin{minipage}{0.08\textwidth}\begin{eqnarray*}~~= \\ \end{eqnarray*}
\end{minipage}%
\begin{minipage}{0.12\textwidth}
\includegraphics[width=1\textwidth]{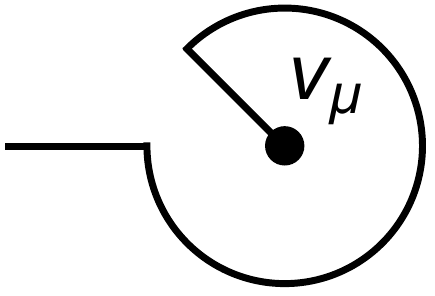}
\end{minipage}%
\begin{minipage}{0.08\textwidth}\begin{eqnarray*}~- \zeta^{-1} \\ \end{eqnarray*}
\end{minipage}%
\begin{minipage}{0.12\textwidth}
\includegraphics[width=1\textwidth]{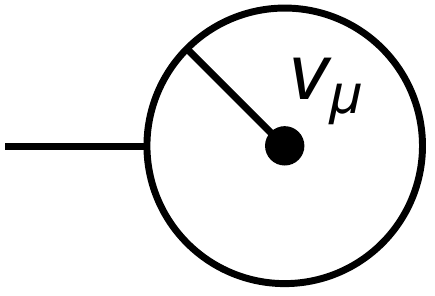}
\end{minipage}%
\caption{Determining the lassoing of the defect operator $v_\mu \in {\cal H}_W$.}
\label{fig:LYLasso2}
\end{figure}

To complete our analysis of the IR TQFT, we also need to compute the lassoing of the $W$ line on the defect operator $v_\mu$ as shown in~Figure~\ref{fig:LYLasso}.  The coefficient for the left figure is readily fixed to be the same as $\delta = 5^{1\over4}$, by considering the two-point function on the sphere of the LHS with $v_x$, and unwrapping the $W$ line to circle $v_x$. The coefficient $\zeta^{-1}$ for the right figure is fixed by the H-junction crossing relation, as illustrated in Figure~\ref{fig:LYLasso2}. Thus, we determine the Frobenius algebra of the IR TQFT of the $\sigma'$-deformed tricritical Ising model to be 
\ie\label{vxu}
v_x^2 = 1+v_x, \quad  v_\mu v_x = -\zeta^{-1} v_\mu,\quad  \overbrace{v_\mu v_\mu} = 1 - \zeta^{-1} v_x.
\fe

We emphasize that the TQFT structure constant $\la v_x v_x v_x\ra = \A$ is fixed to be 1 only through the consideration of TDLs. If a primary $\phi$ of the UV CFT flows to $v_x$, it should be possible to reproduce the structure constant $\A$ by studying the RG flow of a three-point function, say $\la \phi| \phi |\phi\ra$ on the cylinder, using the truncated conformal space approach (TCSA) \cite{Yurov:1989yu}. It would also be interesting to relate $\A$ to the S-matrix of the solitons interpolating the degenerate vacua \cite{Zamolodchikov:1990xc, Ellem:1997vz}.

\subsubsection{Tricritical Ising model deformed by $\varepsilon'$}
\label{tritoising}

\begin{figure}[H]
\centering
\begin{minipage}{0.1\textwidth}
\includegraphics[width=1\textwidth]{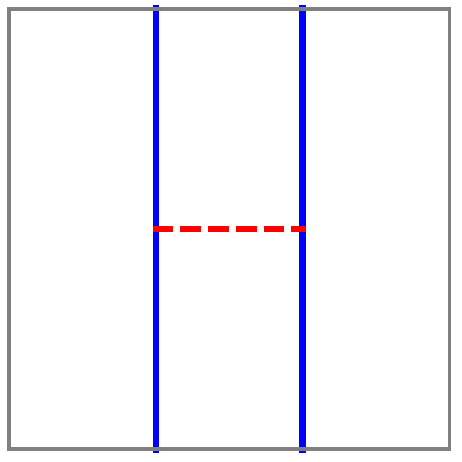}
\end{minipage}%
\begin{minipage}{0.1\textwidth}\begin{eqnarray*}~= \sqrt{2} \\ \end{eqnarray*}
\end{minipage}%
\begin{minipage}{0.1\textwidth}
\includegraphics[width=1\textwidth]{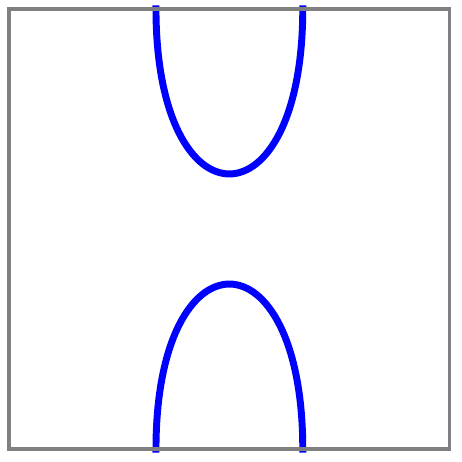}
\end{minipage}%
\begin{minipage}{0.05\textwidth}\begin{eqnarray*}~- \\ \end{eqnarray*}
\end{minipage}%
\begin{minipage}{0.1\textwidth}
\includegraphics[width=1\textwidth]{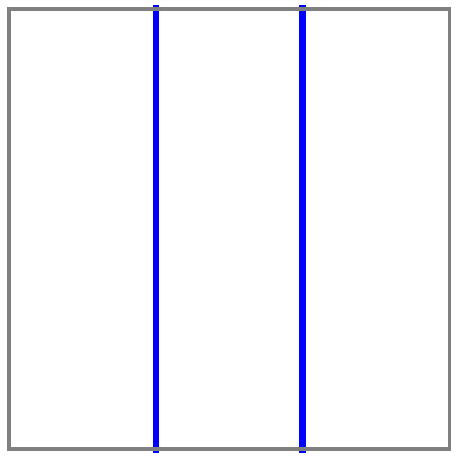}
\end{minipage}%
\begin{minipage}{0.1\textwidth}\begin{eqnarray*}~= 2\, {\rm tr}\, 1 - {\rm tr}\,\widehat N^2
= {\rm tr}\, 1 - {\rm tr}\, \widehat\eta \\ \end{eqnarray*}
\end{minipage}%
\begin{minipage}{0.37\textwidth}~
\end{minipage}%
\\
\bigskip
\begin{minipage}{0.05\textwidth}\begin{eqnarray*}~=  \\ \end{eqnarray*}
\end{minipage}%
\begin{minipage}{0.1\textwidth}
\includegraphics[width=1\textwidth]{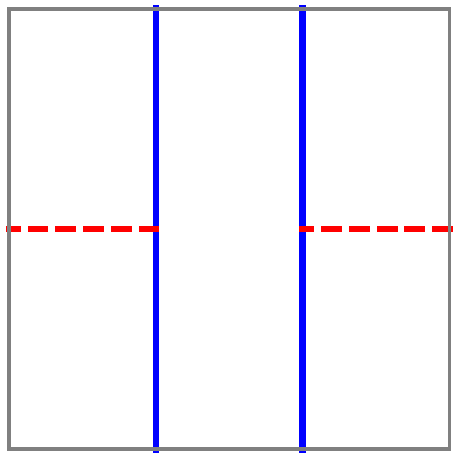}
\end{minipage}%
\begin{minipage}{0.1\textwidth}\begin{eqnarray*}~= {1\over\sqrt{2}} \\ \end{eqnarray*}
\end{minipage}%
\begin{minipage}{0.1\textwidth}
\includegraphics[width=1\textwidth]{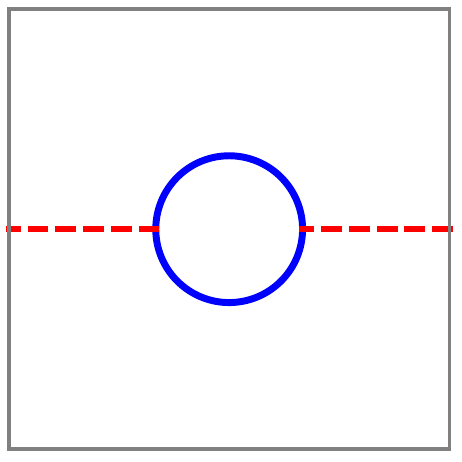}
\end{minipage}%
\begin{minipage}{0.1\textwidth}\begin{eqnarray*}~ + {1\over\sqrt{2}} \\ \end{eqnarray*}
\end{minipage}%
\begin{minipage}{0.1\textwidth}
\includegraphics[width=1\textwidth]{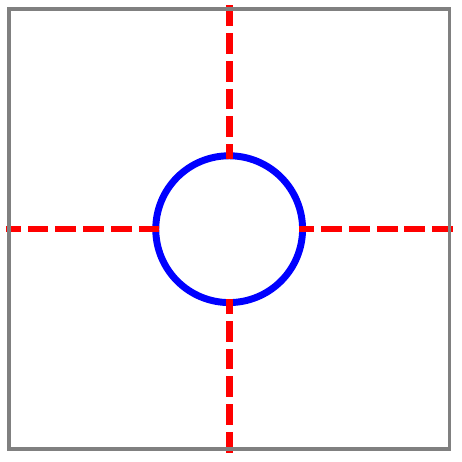}
\end{minipage}%
\begin{minipage}{0.05\textwidth}\begin{eqnarray*}~= \\ \end{eqnarray*}
\end{minipage}%
\begin{minipage}{0.1\textwidth}
\includegraphics[width=1\textwidth]{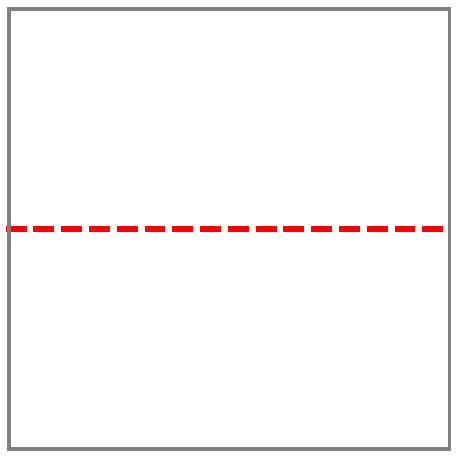}
\end{minipage}%
\begin{minipage}{0.05\textwidth}\begin{eqnarray*}~ + \\ \end{eqnarray*}
\end{minipage}%
\begin{minipage}{0.1\textwidth}
\includegraphics[width=1\textwidth]{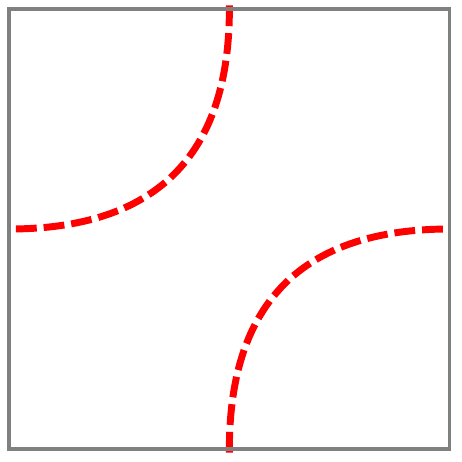}
\end{minipage}%
\begin{minipage}{0.1\textwidth}\begin{eqnarray*}~= 2\,{\rm tr}\, \widehat\eta \\ \end{eqnarray*}
\end{minipage}%
\caption{The determination of ${\rm tr}\,\widehat\eta$ in the IR TQFT by repeatedly applying the H-junction crossing relations and modular invariance.}
\label{fig:treta}
\end{figure}

Let us consider a more nontrivial example: the tricritical Ising model (Section~\ref{Sec:Trising}) deformed by $\varepsilon'$, with a negative coupling such that the RG flow ends up in a massive phase. Both the $\mathbb{Z}_2$ invertible line $\eta$ and the $N$ line are preserved under this RG flow. Since $\la N\ra=\sqrt{2}$ is not an integer, there must be degenerate vacua by the conclusion in Section~\ref{sec:degvacua}. As we demonstrate in Figure~\ref{fig:treta}, $ {\rm tr}\,1=3\,{\rm tr}\,\widehat \eta$, implying that there must be at least three degenerate vacua, one of which is $\widehat\eta$-odd.

The states in ${\cal H}_N$ obey the nontrivial spin selection rule $s\in \pm {1\over 16}+{1\over 2}\mathbb{Z}$ along the entire RG flow, so ${\cal H}_N$ is empty in the IR. Following the same arguments as in Section~\ref{sec:degvacua}, there must be degenerate vacua in the IR such that ${\rm tr}\,\widehat N=0$ by modular invariance. The fusion relation $N^2=I+\eta$ implies that $\widehat N$ takes the eigenvalues $\pm\sqrt{2}$ over a basis of $\mathbb{Z}_2$-even states ($\widehat\eta=1$), and annihilates all $\mathbb{Z}_2$-odd states ($\widehat\eta=-1$).  

Note that the $N$ line in this TQFT is an example of a TDL on which no defect operator can end, {\it i.e.}, ${\cal H}_N=\emptyset$. This is not to be confused with the $N$ line in the critical Ising model, since as discussed around \eqref{HLnonempty}, in a unitary, compact CFT with a unique vacuum, we expect the defect operator Hilbert space ${\cal H}_{\cal L}$ to be non-empty by modular invariance. The $N$ line discussed above violates this expectation because the TQFT in question has degenerate vacua.
 
Indeed, it is known that there are precisely three vacua \cite{PhysRevB.30.3908}.
From the discussions above, we deduce from this fact that there is a unique set of assignments of the $\eta$ and $N$ charges
\begin{equation}
\label{TriEpsCharges}
\begin{tabu}{ccccc}
 & & |1\ra & |v_\varepsilon\ra & |v_\sigma\ra
\\
\widehat\eta : & \quad & 1 & 1 & -1
\\
\widehat N : & \quad & \sqrt2 & -\sqrt2 & 0
\end{tabu}
\end{equation}
where we labeled the operators corresponding to the degenerate vacua in the TQFT by $1, v_\varepsilon, v_\sigma$, by analogy to the critical Ising model. Under OPE, the three degenerate vacua form a commutative Frobenius algebra, which can be fixed by the $\widehat\eta$- and $\widehat N$-charges, the emptiness of ${\cal H}_N$, and together with associativity to be
\ie
v_\varepsilon v_\varepsilon = 1,\quad  v_\sigma v_\sigma = 1 + v_\varepsilon,\quad  v_\varepsilon v_\sigma = v_\sigma.
\fe
In particular, the $v_\varepsilon v_\varepsilon$ OPE does not contain $v_\varepsilon$ because $v_\varepsilon$ anticommutes with the $N$ line. The forms of these OPEs are formally identical to the fusion rules in the critical Ising model.

\begin{figure}[H]
\centering
\begin{minipage}{0.15\textwidth}
\includegraphics[width=1\textwidth]{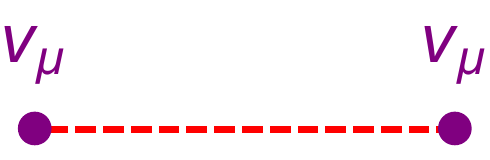}
\end{minipage}%
\caption{Two $v_\mu$'s connected by an $\eta$ segment.}
\label{fig:vmuvmured}
\end{figure}

While ${\cal H}_N$ contains no state, ${\cal H}_\eta$ should be one-dimensional since ${\rm tr}\,\widehat \eta=1$. The state in ${\cal H}_\eta$ corresponds to a topological defect operator which we denote by $v_\mu$.
We normalize $v_\mu$ such that the coefficient of 1 in the $v_\mu v_\mu$ OPE is one.  
 In a correlation function, $v_\mu$ must appear in pairs connected by $\eta$ lines. An $\eta$ line segment ending on a pair of $v_\mu$'s as in Figure~\ref{fig:vmuvmured}, which we denote by $\overbrace{v_\mu v_\mu}$, can be rewritten as a topological bulk local operator. Since $v_\sigma$ anticommutes with $\eta$, it follows that $v_\sigma v_\mu = 0$. By associativity, we determine
\ie
\overbrace{v_\mu v_\mu} = 1 - v_\varepsilon, \quad  v_\varepsilon v_\mu = - v_\mu.
\fe
This completes the description of the IR TQFT of the deformed tricritical Ising model, including the data of TDLs and defect operators.

\subsubsection{On TQFTs admitting ${R}_{\bC}(S_3)$ fusion ring}

One way to realize TDLs of the ${R}_{\bC}(S_3)$ fusion ring is to begin with a CFT with $S_3$ global symmetry, where the $\mathbb{Z}_2$ subgroups are free of an 't Hooft anomaly, and orbifold by a $\mathbb{Z}_2$. We have seen this in the example of the relation between the three-state Potts model and the tetracritical Ising model. In general, depending on whether the $S_3$ has a $\mathbb{Z}_3$ 't Hooft anomaly, the result after the $\mathbb{Z}_2$ orbifold would be either the ${\rm Rep}(S_3)$ (with trivial cyclic permutation map on $V_{Y,Y,Y}$), or one of the two twisted ${\rm Rep}(S_3)$ fusion categories (with nontrivial cyclic permutation map) \cite{Bhardwaj:2017xup}. In this sense, the twisted ${\rm Rep}(S_3)$ fusion categories are analogous to situations with 't Hooft anomalies'.

We saw in Section~\ref{repsthree} that the spin selection rules for the twisted ${\rm Rep}(S_3)$ fusion categories still allow integer spin states in ${\cal H}_Y$. One may wonder whether the existence of twisted ${\rm Rep}(S_3)$ TDLs would still in general forbid the possibility of an RG flow to a trivial massive phase, namely, a TQFT with a unique vacuum. We will show that this is indeed the case, even though it does not follow directly from the spin selection rule on defect operators.

\begin{figure}[H]
\centering
\begin{minipage}{0.19\textwidth}
\includegraphics[width=1\textwidth]{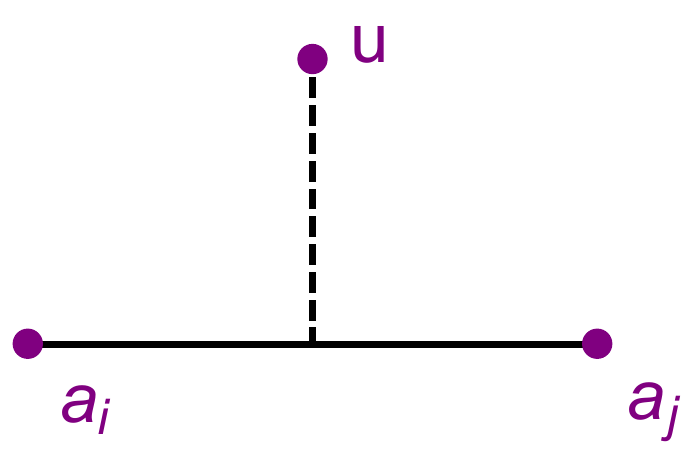}
\end{minipage}%
\begin{minipage}{0.05\textwidth}\begin{eqnarray*}~\equiv U_{ij} \\ \end{eqnarray*}
\end{minipage}%
\begin{minipage}{0.15\textwidth}
\begin{eqnarray*}~\end{eqnarray*}
\end{minipage}%
\begin{minipage}{0.16\textwidth}
\includegraphics[width=1\textwidth]{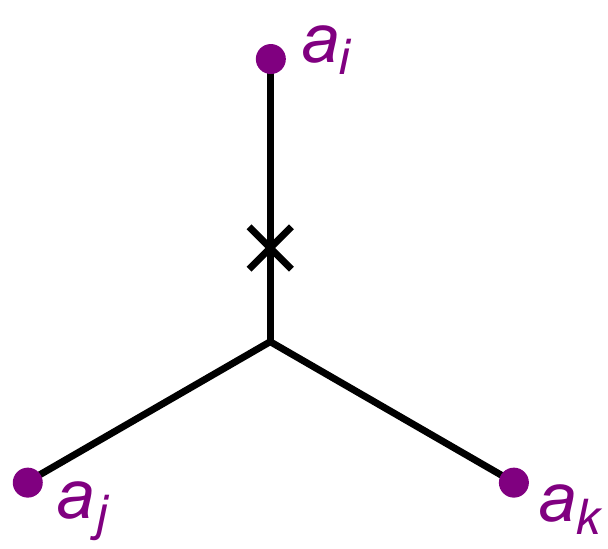}
\end{minipage}%
\begin{minipage}{0.05\textwidth}\begin{eqnarray*}~\equiv C^i_{jk} \\ \end{eqnarray*}
\end{minipage}%
\caption{Structure constants $U_{ij}$ and $C^i_{jk}$ of defect operators. Here the dotted line stands for $X$ and the solid line stands for $Y$.}
\label{fig:strconst}
\end{figure}

\begin{figure}[H]
\begin{minipage}{0.2\textwidth}~
\end{minipage}%
\begin{minipage}{0.1\textwidth}\begin{eqnarray*}{\rm tr}\,U = \\ \end{eqnarray*}
\end{minipage}%
\begin{minipage}{0.15\textwidth}
\includegraphics[width=1\textwidth]{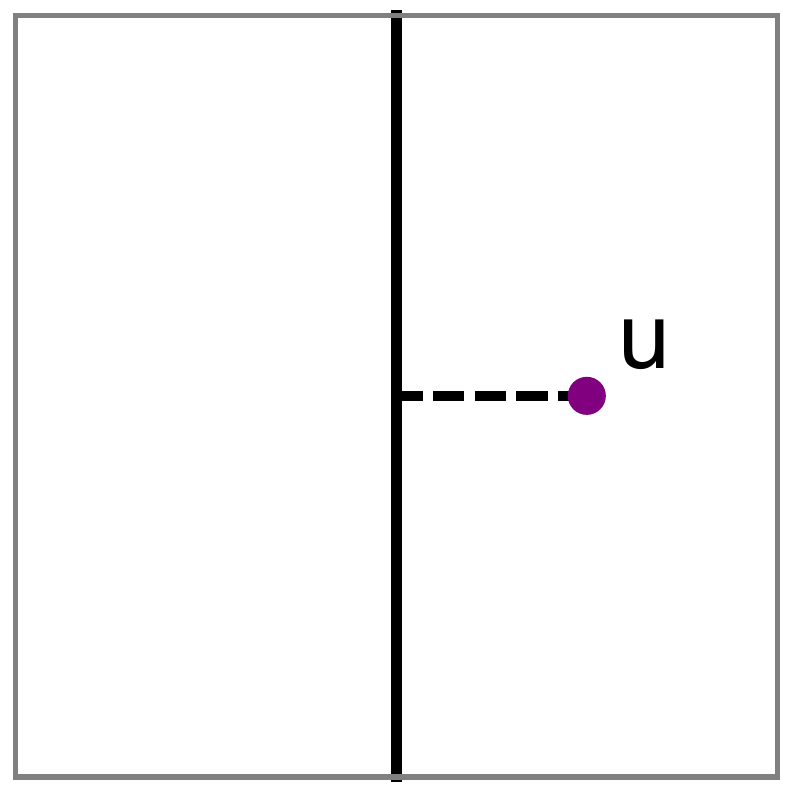}
\end{minipage}%
\begin{minipage}{0.09\textwidth}\begin{eqnarray*}\quad = \\ \end{eqnarray*}
\end{minipage}%
\begin{minipage}{0.15\textwidth}
\includegraphics[width=1\textwidth]{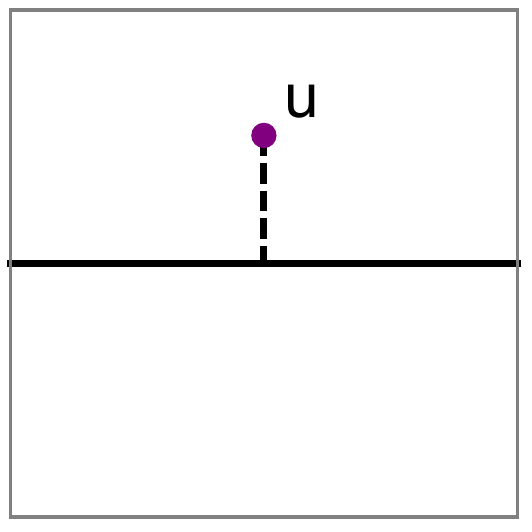}
\end{minipage}%
\begin{minipage}{0.09\textwidth}\begin{eqnarray*}\quad = \\ \end{eqnarray*}
\end{minipage}%
\begin{minipage}{0.07\textwidth}
\includegraphics[width=1\textwidth]{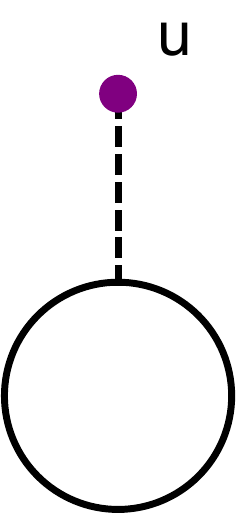}
\end{minipage}%
\begin{minipage}{0.05\textwidth}\begin{eqnarray*}\quad =0 \\ \end{eqnarray*}
\end{minipage}%
\\
\bigskip
\\
\begin{minipage}{0.2\textwidth}~
\end{minipage}%
\begin{minipage}{0.1\textwidth}\begin{eqnarray*}(U^2)_{ij} = \\ \end{eqnarray*}
\end{minipage}%
\begin{minipage}{0.2\textwidth}
\includegraphics[width=1\textwidth]{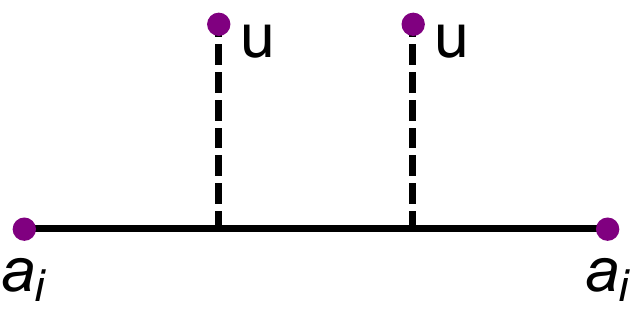}
\end{minipage}%
\begin{minipage}{0.09\textwidth}\begin{eqnarray*}\quad = \\ \end{eqnarray*}
\end{minipage}%
\begin{minipage}{0.2\textwidth}
\includegraphics[width=1\textwidth]{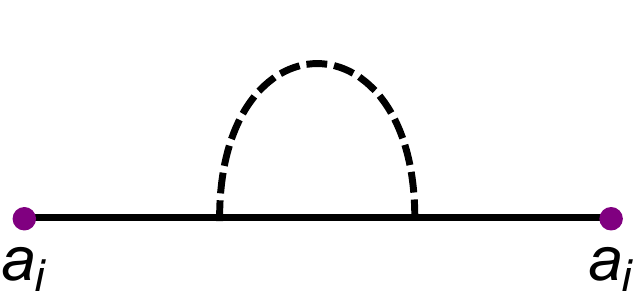}
\end{minipage}%
\begin{minipage}{0.05\textwidth}\begin{eqnarray*}\quad = \delta_{ij} \\ \end{eqnarray*}
\end{minipage}%
\caption{Some constraints on $U_{ij}$. In the first line, going from the torus one-point function of $u$ attached to a $Y$ loop to the $u$-tadpole graph on the plane, we made use of the assumption that the TQFT has a unique vacuum, {\it i.e.}, the only bulk local operator is the identity. In the second line, we used the triviality of the crossing kernels with $X$ external lines.}
\label{fig:utadpole}
\end{figure}

Suppose that there is a TQFT with TDLs obeying the crossing relations (\ref{kxyrep}), and with a unique vacuum -- the identity operator. It follows from $\la X \ra=1$ and $\la Y \ra=2$ that ${\rm dim}\,{\cal H}_X=1$, ${\rm dim}\,{\cal H}_Y=2$. Let $u\in {\cal H}_X$ and $a_i\in {\cal H}_Y$ be a basis of defect operators, normalized such that
\ie
\overbrace{uu} = 1, \quad  \overbrace{a_ia_j} = \delta_{ij}.
\fe
The nontrivial structure constants are depicted in Figure~\ref{fig:strconst}.
Unitarity demands that $U$ is a Hermitian $2\times 2$ matrix. As shown in Figure~\ref{fig:utadpole}, it follows from the torus one-point function of $u$ attached to a $Y$ loop, and the vanishing tadpole condition for the $X$ line (see Section~\ref{Sec:VanishingTadpole}) that ${\rm tr}\, U=0$. Furthermore, from the crossing relations, we can show that $U^2=1$.

\begin{figure}[H]
\centering
\begin{minipage}{0.15\textwidth}
\includegraphics[width=1\textwidth]{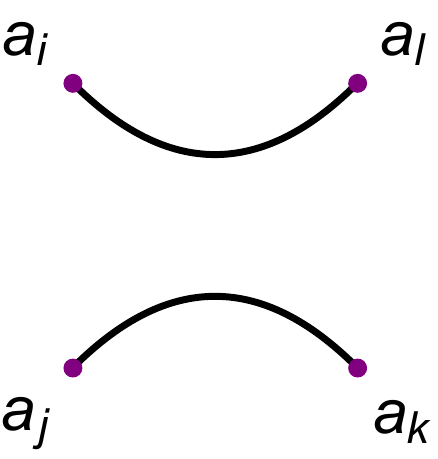}
\end{minipage}%
\begin{minipage}{0.11\textwidth}\begin{eqnarray*}\quad = ~~ {1\over 2}\\ \end{eqnarray*}
\end{minipage}%
\begin{minipage}{0.12\textwidth}
\includegraphics[width=1\textwidth]{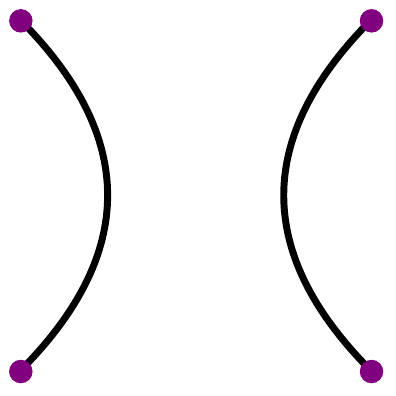}
\end{minipage}%
\begin{minipage}{0.08\textwidth}\begin{eqnarray*}\quad +~ {1\over 2} \\ \end{eqnarray*}
\end{minipage}%
\begin{minipage}{0.12\textwidth}
\includegraphics[width=1\textwidth]{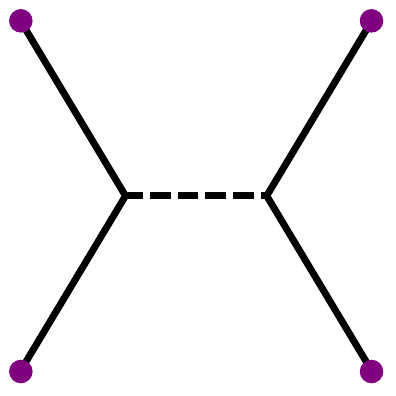}
\end{minipage}%
\begin{minipage}{0.1\textwidth}\begin{eqnarray*}\quad +~ {1\over 2}\omega^2 \\ \end{eqnarray*}
\end{minipage}%
\begin{minipage}{0.12\textwidth}
\includegraphics[width=1\textwidth]{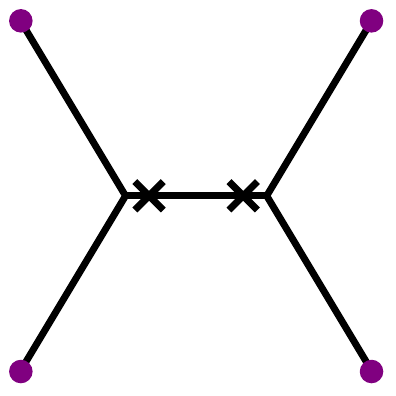}
\end{minipage}%
\caption{The crossing equation on the four-point function of the defect operator $a_i$.}
\label{fig:aaaacross}
\end{figure}

Finally, the crossing kernel ${\widetilde K}_{Y,Y}^{Y,Y}$ applied to the four-point function of $a_i$ (with a cyclic permutation map applied to one of the YYY junctions) implies the following identity among the structure constants (see Figure~\ref{fig:aaaacross}),
\ie\label{uucc}
\delta_{i\ell}\delta_{jk} = {1\over 2} \delta_{ij}\delta_{k\ell} + {1\over 2} U_{ij} U_{kl} + {\omega^2\over 2} \sum_{m=1,2} C^m_{ij} C^m_{k\ell}.
\fe
where $\omega$ is a third root of unity coming from the cyclic permutation map on the T-junction of $Y$.  Note that $\omega = 1$ for the standard Rep$(S_3)$, and $\omega = e^{\pm 2\pi i \over 3}$ for the two cases of twisted Rep$(S_3)$.

Given this restriction on $U$, one finds that only for $U_{ij}=\pm i \epsilon_{ij}$ does there exist a solution $C^m_{ij}$ compatible with {\it some} cyclic permutation map on $V_{Y,Y,Y}$. Up to a change of basis, the structure constants are
\ie
C^1_{12} = C^1_{21} = C^2_{11} = - \omega^{-1}, \quad C^2_{22}= \omega^{-1}, \quad C^1_{11} = C^1_{22} = C^2_{12}=C^2_{21}=0.
\fe
The non-vanishing $C^2_{22}$, for instance, is only possible if the cyclic permutation map acts trivially on $V_{Y,Y,Y}$, which is the case for the ${\rm Rep}(S_3)$ fusion category. This shows that the twisted ${\rm Rep}(S_3)$ fusion categories admit no solution to the crossing equations of defect operators. Thus, there must be degenerate vacua in the TQFT.

\subsubsection{$(A_{10},E_6)$ minimal model deformed by $\phi_{2,1}$}
\label{Sec:A10E6}

As already discussed, the $(A_{10},E_6)$ minimal model (Section~\ref{Sec:hE6}) admits TDLs $X$ and $Y$ that obey the ${1\over 2}E_6$ fusion ring  (which is commutative),
\ie
X^2=I, \quad  Y^2=I+X+2Y, \quad  XY=Y.
\fe
The $X$ line is associated to the $\bZ_2$ symmetry of the model. These TDLs commute with the relevant operator $\phi_{2,1}$ of weight $({7\over 22},{7\over 22})$. We expect the $(A_{10},E_6)$ minimal model perturbed by $\phi_{2,1}$ to flow to a TQFT in the IR that admits the TDLs $X$ and $Y$. It follows from the fusion ring that the possible eigenvalues of $(\widehat X, \widehat Y)$ are $(1, 1+\sqrt{3})$, $(1,1-\sqrt{3})$, and $(-1,0)$. Modular invariance of the TQFT immediately implies that the bulk vacua have to be degenerate, and the defect Hilbert space ${\cal H}_Y$ has to be even-dimensional.

	\begin{figure}[H]
	\centering
	\begin{minipage}{0.15\textwidth}
	\begin{eqnarray*}{\rm tr}_{{\cal H}_Y}
	\widehat X_- = \\ \end{eqnarray*}
	\end{minipage}%
	\begin{minipage}{0.15\textwidth}
		\includegraphics[width=1\textwidth]{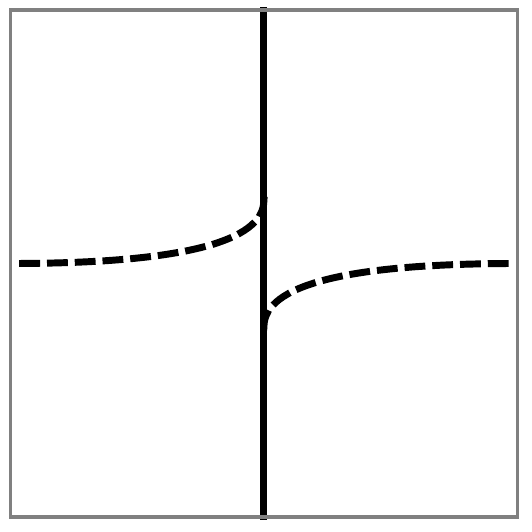}
	\end{minipage}%
	\begin{minipage}{0.09\textwidth}
	\begin{eqnarray*}\quad = \\ \end{eqnarray*}
	\end{minipage}%
	\begin{minipage}{0.15\textwidth}
		\includegraphics[width=1\textwidth]{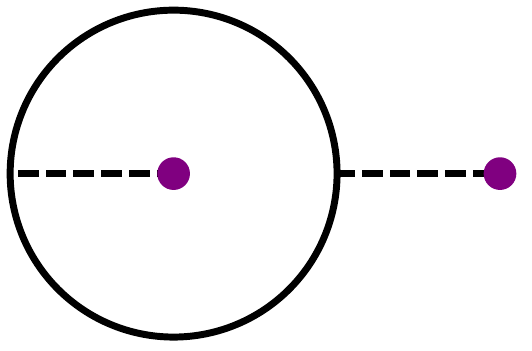}
	\end{minipage}%
	\begin{minipage}{0.05\textwidth}
	\begin{eqnarray*}\quad =0 \\ \end{eqnarray*}
	\end{minipage}%
	\caption{Vanishing ${\rm tr}_{{\cal H}_Y} \widehat X_-$. The first picture represents the torus partition function with a temporal $Y$ loop (solid), and a spatial $X$ line (dashed). The second picture represents a two-point function of the defect operator $a\in {\cal H}_X$ connected to a $Y$ loop through $X$ lines. If the operator $a$ lying outside is brought around the circle, a minus sign is acquired when the two $XYY$ junctions cross one another, due to the crossing phase ${\widetilde K}_{X,Y}^{Y,X}(Y,Y)=-1$. Therefore, the correlator vanishes.}\label{fig:torusXY}
	\end{figure}

	\begin{figure}[H]
	\centering
	\begin{minipage}{0.12\textwidth}
		\includegraphics[width=1\textwidth]{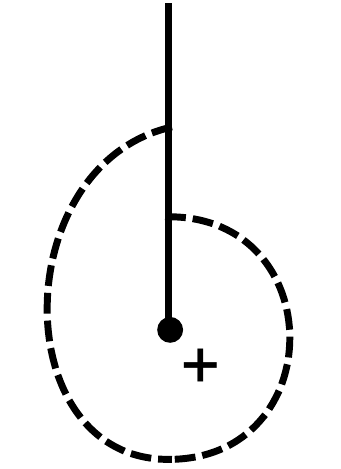}
	\end{minipage}%
	\begin{minipage}{0.09\textwidth}
	\begin{eqnarray*}=~ i \\ \end{eqnarray*}
	\end{minipage}%
	\begin{minipage}{0.025\textwidth}
		\includegraphics[width=1\textwidth]{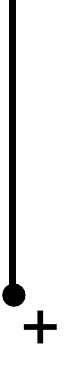}
	\end{minipage}%
	\begin{minipage}{0.15\textwidth}~
	\end{minipage}%
	\begin{minipage}{0.12\textwidth}
		\includegraphics[width=1\textwidth]{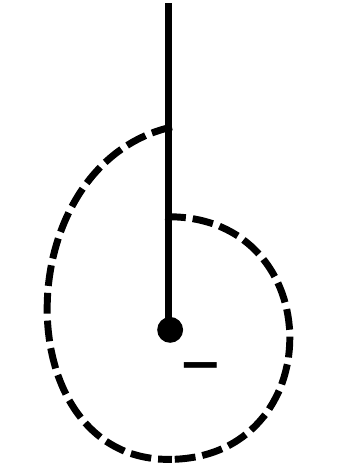}
	\end{minipage}%
	\begin{minipage}{0.09\textwidth}
	\begin{eqnarray*}=~-i \\ \end{eqnarray*}
	\end{minipage}%
	\begin{minipage}{0.025\textwidth}
		\includegraphics[width=1\textwidth]{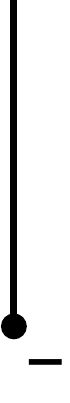}
	\end{minipage}%
	\caption{The action of $\widehat X_-$ on $b_+, b_-\in {\cal H}_Y$.}\label{fig:Wbact}
	\end{figure}

	\begin{figure}[H]
	\begin{minipage}{0.15\textwidth}~
	\end{minipage}%
	\begin{minipage}{0.15\textwidth}
		\includegraphics[width=1\textwidth]{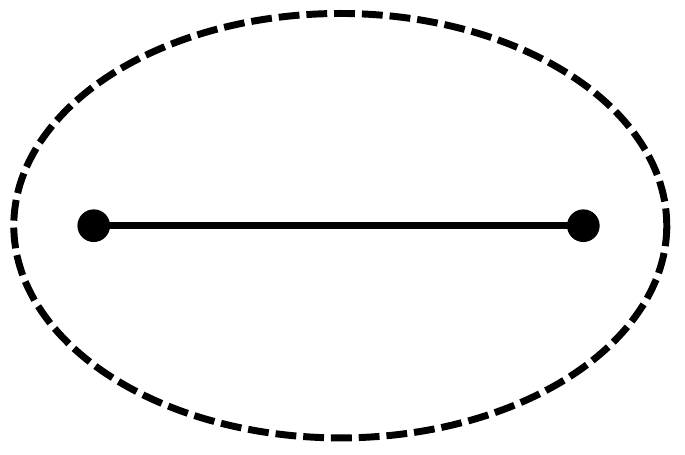}
	\end{minipage}%
	\begin{minipage}{0.08\textwidth}
	\begin{eqnarray*}~~=  \\ \end{eqnarray*}
	\end{minipage}%
	\begin{minipage}{0.2\textwidth}
		\includegraphics[width=1\textwidth]{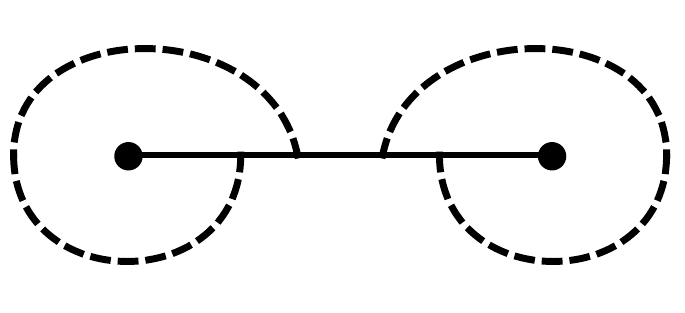}
	\end{minipage}%
	\\
	\bigskip
	\\
	\bigskip
	\begin{minipage}{0.15\textwidth}~
	\end{minipage}%
	\begin{minipage}{0.18\textwidth}
		\includegraphics[width=1\textwidth]{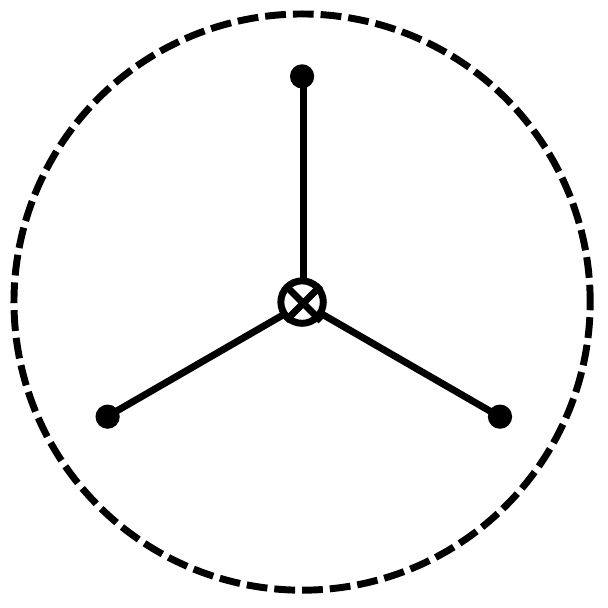}
	\end{minipage}%
	\begin{minipage}{0.06\textwidth}
	\begin{eqnarray*}\quad =  \\ \\ \end{eqnarray*}
	\end{minipage}%
	\begin{minipage}{0.19\textwidth}
		\includegraphics[width=1\textwidth]{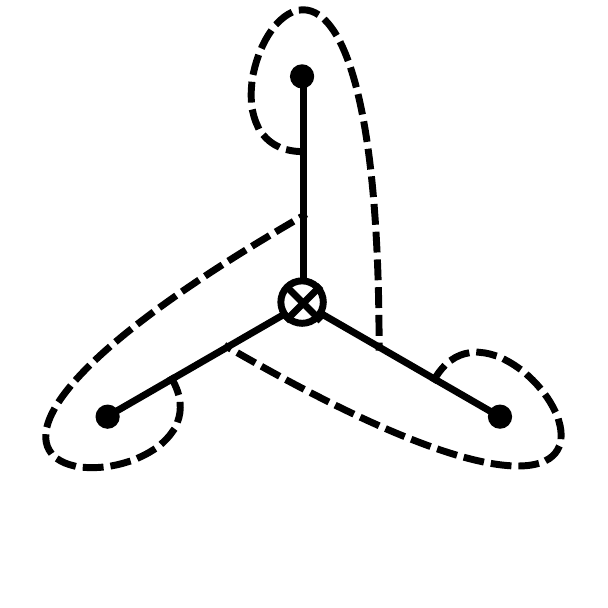}
	\end{minipage}%
	\begin{minipage}{0.06\textwidth}
	\begin{eqnarray*}~= ~~i \\ \\ \end{eqnarray*}
	\end{minipage}%
	\begin{minipage}{0.19\textwidth}
		\includegraphics[width=1\textwidth]{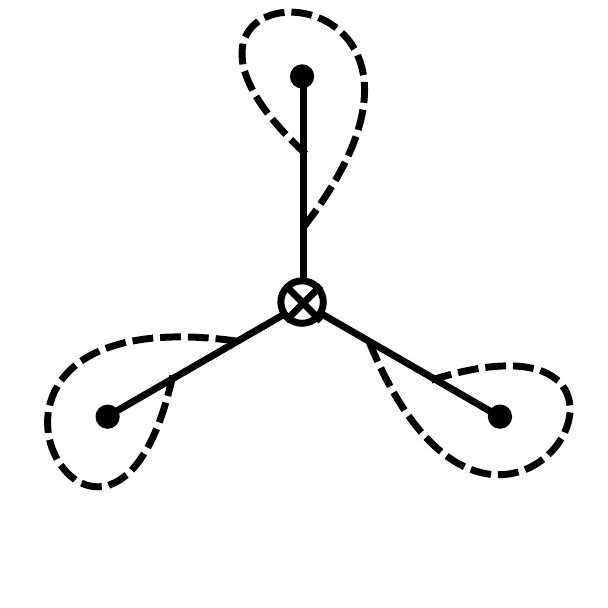}
	\end{minipage}%
	\caption{The $\mathbb{Z}_2$ charge ($\widehat X$ value) of the OPE of a pair of defect operators $b_\pm$, as well as three defect operators joined by a $YYY$ junction, can be reduced to $\widehat X_-$ acting on the defect operators by a sequence of crossing relations. In the bottom row, an arbitrary $YYY$ junction vector is assigned. In deriving the second equality, we used the identity $({\widetilde K}_{Y,Y}^{Y,X}{\widetilde K}_{Y,Y}^{Y,I})^3 = i$, which follows from one set of crossing kernels (see Appendix~\ref{1/2E6CK}) that solve the pentagon identity (an alternative set of crossing kernels gives the result $-i$, corresponding to the ``charged conjugated" ${1\over 2}E_6$ fusion category).}\label{fig:bbX}
	\end{figure}

In Section~\ref{esixselect}, we defined the operator $\widehat X_-$ acting on ${\cal H}_Y$ as a spatial $X$ line that splits off the temporal $Y$ line. It follows from the nontrivial crossing kernel between $X$ and $Y$, ${\widetilde K}_{X,Y}^{Y,X} (Y,Y)=-1$, that $\widehat X_-^2=-1$. Moreover, by modular invariance, we obtain ${\rm tr}_{{\cal H}_Y}\widehat X_- = 0$ (Figure~\ref{fig:torusXY}), and thus $\widehat X_-$ must have the same number of $\pm i$ eigenstates. We denote them collectively by $b_+$ and $b_-$, such that (Figure~\ref{fig:Wbact})
\ie
\widehat X_- |b_+\ra = i |b_+\ra, \quad  \widehat X_- |b_-\ra = -i |b_-\ra.
\fe
By crossing, we can then identify the $\widehat X$ charges of various TDL configurations ending on defect operators (see Figure~\ref{fig:bbX}). For example, a $Y$ segment connecting either $b_+$ and $b_+$, or $b_-$ and $b_-$, is even under $\bZ_2$, while one connecting $b_+$ and $b_-$ is odd under $\bZ_2$.\footnote{We emphasize here that the subscript of $b_\pm$ should not be confused with their $\widehat X$ charges.
} Similarly, a $YYY$ junction ending on either $b_+,b_+,b_+$, or $b_+,b_-,b_-$, is even under $\bZ_2$, while the other possibilities are odd. Since these TDL configurations ending on defect operators can be expanded in bulk operators (by locality), the $\bZ_2$-invariance put constraints on the structure constants of the TQFT, which we exploit in the following sections to pin down the TQFT.

	\begin{figure}[H]
 	\centering
	\hspace{-.4in}
 	\begin{minipage}{0.18\textwidth}
 		\includegraphics[width=1\textwidth]{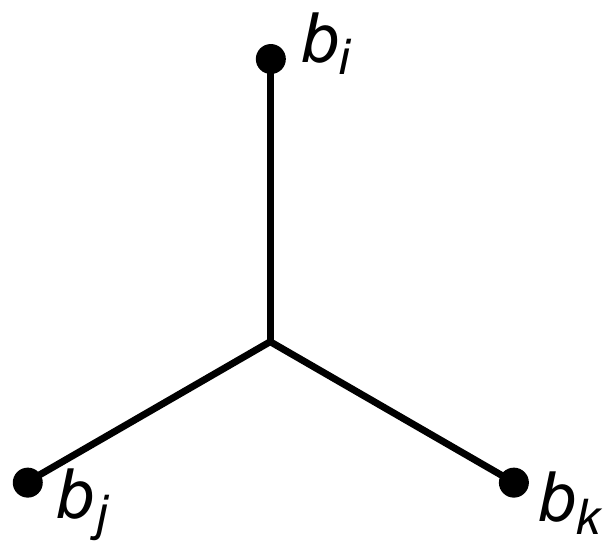}
 	\end{minipage}%
 	\begin{minipage}{0.09\textwidth}
	\begin{eqnarray*}=~ \la b_i b_j b_k \ra_{v_0}  
	\\ \end{eqnarray*}
 	\end{minipage}%
 	\begin{minipage}{0.1\textwidth}~
 	\end{minipage}%
 	\begin{minipage}{0.18\textwidth}
 		\includegraphics[width=1\textwidth]{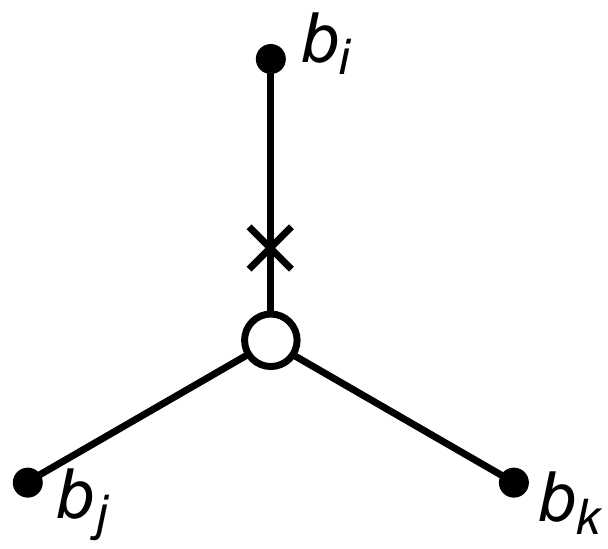}
 	\end{minipage}%
	\begin{minipage}{0.05\textwidth}
	\begin{eqnarray*}=\omega^2  \\ 
	\end{eqnarray*}
 	\end{minipage}%
	\begin{minipage}{0.18\textwidth}
 		\includegraphics[width=1\textwidth]{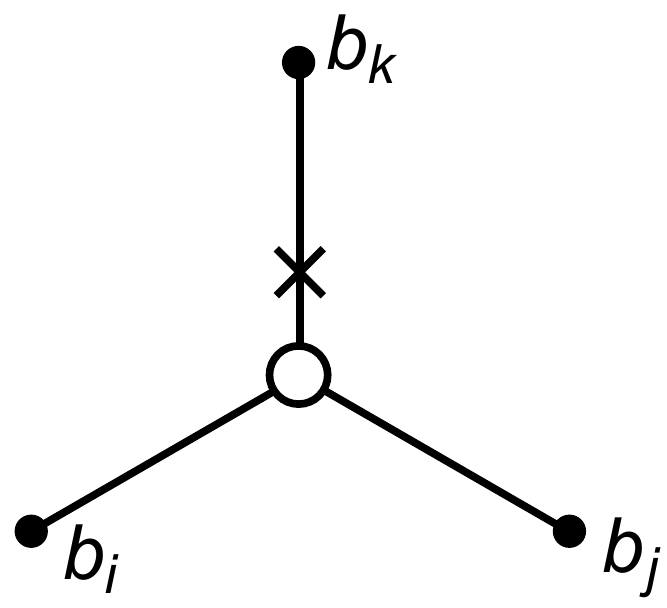}
 	\end{minipage}%
 	\begin{minipage}{0.09\textwidth}
	\begin{eqnarray*}=~\la b_i, b_j b_k \ra_{v_1}
	\\ \end{eqnarray*}
 	\end{minipage}%
 	\caption{Three-point function of defect operators $b_\pm\in{\cal H}_Y$ connected through a $YYY$ junction. The unmarked junction in the first figure stands for the junction vector $v_0 \in V_{Y,Y,Y}$, which is permutation invariant. The circle junction in the second figure stands for the junction vector $v_1$, which transforms by the phase $\omega=e^{2\pi i \over 3}$ under the cyclic permutation map.}\label{fig:bbb}
	\end{figure}

The remaining task is to identify the extended TQFT that is consistent with unitarity, crossing, and modular invariance. As explained in the beginning of this subsection, the bulk must have degenerate vacua. We will start by ruling out the two vacua possibility, and then present a consistent solution in the three vacua case.

\begin{figure}[H]
 	\begin{minipage}{0.13\textwidth}
		\includegraphics[width=1\textwidth]{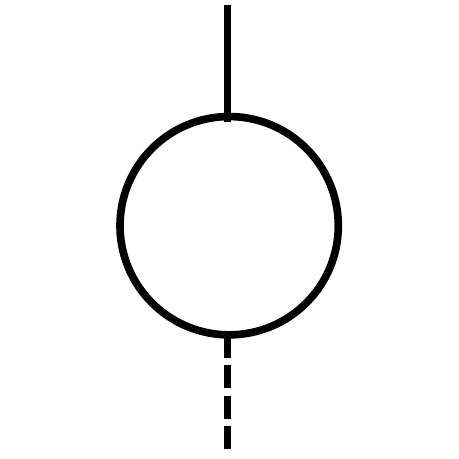}
	\end{minipage}%
	\begin{minipage}{0.01\textwidth}
	\begin{eqnarray*}  =   \\ \end{eqnarray*}
	\end{minipage}%
	\begin{minipage}{0.13\textwidth}
		\includegraphics[width=1\textwidth]{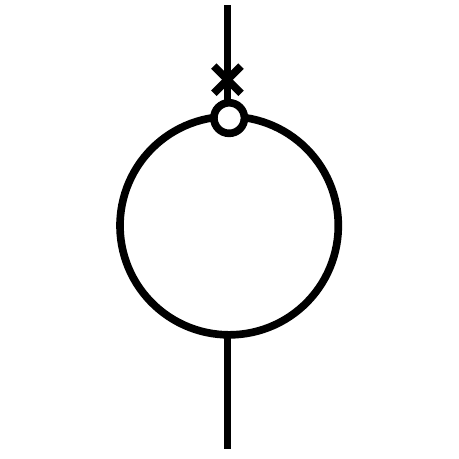}
	\end{minipage}
	\begin{minipage}{0.01\textwidth}
	\begin{eqnarray*}=0\\ \end{eqnarray*}
	\end{minipage}%
	\begin{minipage}{0.08\textwidth}
	\begin{eqnarray*} \\ \end{eqnarray*}
	\end{minipage}%
	\begin{minipage}{0.13\textwidth}
	\includegraphics[width=1\textwidth]{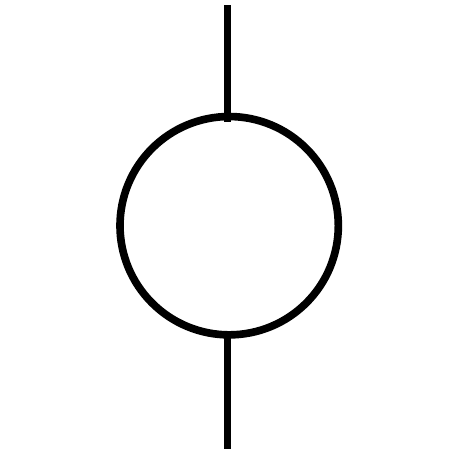}
	\end{minipage}%
\begin{minipage}{0.06\textwidth}\begin{eqnarray*}={\sqrt{3}-1\over 2}   \\ \end{eqnarray*}
\end{minipage}%
\begin{minipage}{0.13\textwidth}
	\includegraphics[width=1\textwidth]{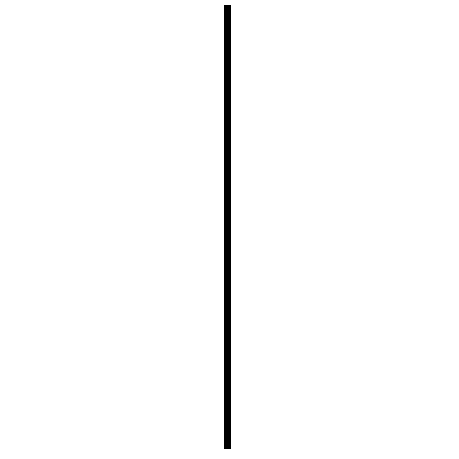}
\end{minipage}%
\begin{minipage}{0.13\textwidth}
	\includegraphics[width=1\textwidth]{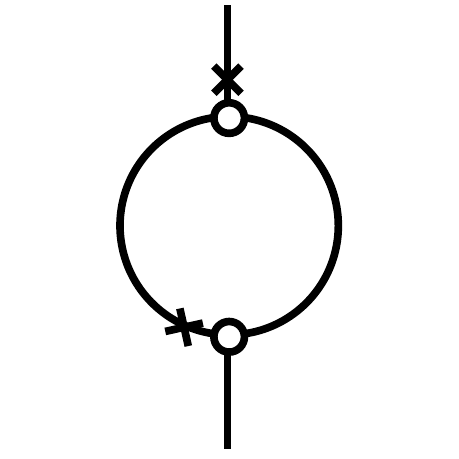}
\end{minipage}%
\begin{minipage}{0.1\textwidth}\begin{eqnarray*}~={\sqrt{3}-1\over 2}\omega^2 ~ \\ \end{eqnarray*}
\end{minipage}%
\begin{minipage}{0.13\textwidth}
	\includegraphics[width=1\textwidth]{figures/ss0.pdf}
\end{minipage}%
\caption{Identities for removing loops deduced from the crossing relations and the vanishing tadpole property.}
\label{fig:E6sunset}
\end{figure}

In the following analysis, it is convenient to work with a basis $\{v_0,v_1\}$ of the $YYY$ junction vector space that diagonalizes the cyclic permutation map $\widetilde K^{Y,I}_{Y,Y}(Y,Y)$ (see Figure~\ref{fig:bbb}) and makes $\widetilde K^{Y,Y}_{Y,Y}(I,Y)$ equal to the identity matrix. In this basis, the conjugation map acts on the junction vector space $V_{YYY}$ as
\ie
\iota(v_0)=v_0,\quad \iota(v_1)=\omega v_1,
\fe
where $\omega=e^{2\pi i\over 3}$. We record below some crossing kernels written in this basis that we will explicitly use,
\ie
&
{\widetilde K^{Y,I}_{Y,Y}(Y,Y)= \begin{pmatrix}
	1 & 0 \\0 & \omega 
\end{pmatrix}
,\quad 
\widetilde K^{Y,Y}_{Y,Y}(Y,I)={\sqrt{3}-1\over 2}\begin{pmatrix}
	1 & 0 \\0 & \omega^2\end{pmatrix},}
\\
&
{\widetilde K^{Y,Y}_{Y,Y}(Y,X)={1\over 3+\sqrt{3}}\begin{pmatrix}
		-1 & \sqrt{2} \\ \sqrt{2}\omega^2 & -\omega^{1/2}
\end{pmatrix},}
\\
&
{\widetilde K^{Y,Y}_{Y,Y}(Y,Y)(v_0,v_0)= \begin{pmatrix}
	1-{1\over \sqrt{3}} & 0 \\ 0 & -{1\over \sqrt{3}}
\end{pmatrix},\quad 
\widetilde K^{Y,Y}_{Y,Y}(Y,Y)(v_0,v_1)= \begin{pmatrix}
	0 & -{1\over \sqrt{3}} \\ -{1\over \sqrt{3}} & -{\sqrt{2}\over 3+\sqrt{3}}
\end{pmatrix}},
\label{E6CKnb}
\fe
We also make repeated use of the vanishing tadpole property discussed in Section~\ref{Sec:VanishingTadpole} to simplify TDL configurations. Some useful consequences of the vanishing tadpole property are summarized in Figure~\ref{fig:E6sunset}.

\subsubsection*{Ruling out ${1\over 2}E_6$ TQFT with two vacua}

In this case, we denote the bulk operators of the TQFT by 1 and $u$, which are even under $\widehat X$ and have eigenvalues $1+\sqrt{3}$ and $1-\sqrt{3}$ under $\widehat Y$.\footnote{In light of the nontrivial crossing kernel $\widetilde K^{Y,X}_{X,Y}(Y,Y)=-1$ between the $X$ and $Y$ lines, it is somewhat counterintuitive to have a TQFT whose bulk states are all invariant under $\widehat X$, even though we do not have a general argument against this possibility, and would have to analyze the full set of TQFT structure constants to rule it out.
}
Modular invariance implies that the defect Hilbert spaces for $X$ and $Y$ are both two-dimensional. We label the basis elements of ${\cal H}_X$ by $a_{1,2}$. For ${\cal H}_Y$, we use $b_{\pm}$ in accordance with the $\widehat X_-$ charges as explained in the previous section.
 
\begin{figure}[h]
		\centering
		\begin{minipage}{0.15\textwidth}
			\includegraphics[width=1\textwidth]{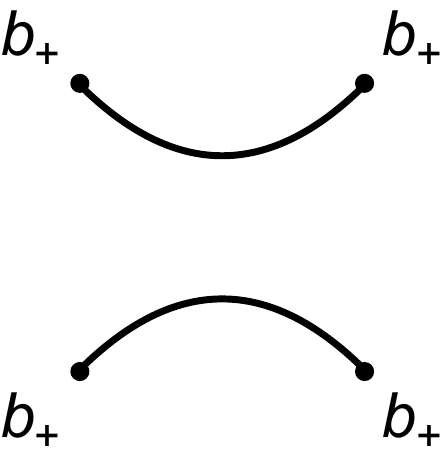}
		\end{minipage}%
		\begin{minipage}{0.15\textwidth}\begin{eqnarray*}=\quad {\sqrt{3}-1\over 2}  \\ \end{eqnarray*}
		\end{minipage}%
		\begin{minipage}{0.15\textwidth}
			\includegraphics[width=1\textwidth]{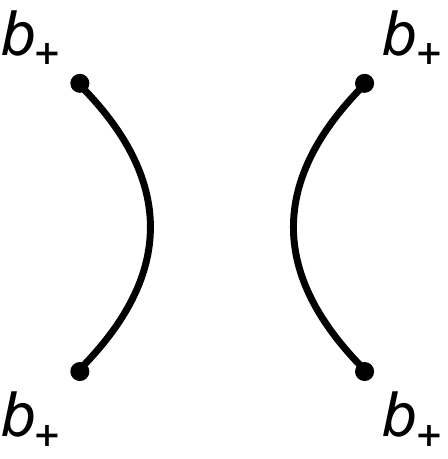}
		\end{minipage}%
		\begin{minipage}{0.09\textwidth}\begin{eqnarray*}~~+  \\ \end{eqnarray*}
		\end{minipage}%
		\begin{minipage}{0.15\textwidth}
			\includegraphics[width=1\textwidth]{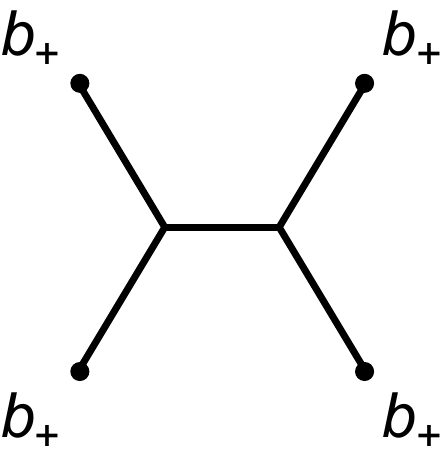}
		\end{minipage}%
		\caption{An example of a crossing equation for the four-point function of $b_+$ connected by an H-junction. In this example, the internal line on the LHS is the trivial line. On the RHS, we have used ${\widetilde K}_{Y,Y}^{Y,Y}(I,I) = {\sqrt{3}-1\over 2}$, $\la a_i b_+ b_+\ra=0$, and  $\la b_+b_+b_-\ra_{v_{0,1}}=0$, which follow from the $\mathbb{Z}_2$-invariance, and $\la b_+, b_+b_+\ra_{v_1}=0$ by the nontriviality of the cyclic permutation map when acting on $v_1$. }\label{fig:bbbb}
	\end{figure}

We will normalize all defect operators $a_i$, $b_+$, $b_-$, and $u$ to be self-conjugate with unit norm. 
	 Modular invariance of the torus one-point function of $u$ with a spatial $X$ loop implies that
	\ie
	\tr u \widehat X = \la u a_1 a_1\ra +\la u a_2 a_2\ra=\la uuu\ra.
	\label{uX1pt}
	\fe
Along with associativity and unitarity, the OPE of the bulk operator $u$ and the defect operators $a_{1,2}$ of the $X$ line are determined to be\footnote{{\it A priori}, one could also satisfy associativity with $\overbrace{a_2 a_2} = 1 +\A u$ in \eqref{2vfa}, but then the modular invariance of $\tr u\widehat X$ \eqref{uX1pt} would imply that $\A=\pm i$, violating unitarity.}
	\ie\label{2vfa}
	& u^2 = 1+(\A - \A^{-1}) u, 
	\quad u a_1= \A a_1 ,
	\\
	&u a_2=-\A^{-1}a_2,\quad \overbrace{a_1 a_1} = 1 +\A u, \quad   \overbrace{a_2 a_2} = 1 - \A^{-1} u, \ \quad   \overbrace{a_1 a_2}= 0,
	\fe
Since all the bulk operators are invariant under $\widehat X$, it follows that among the TDL configurations ending on defect operators, all the $\widehat X$-odd ones vanish, in particular,
	\ie
	\overbrace{b_+ b_-}=0. 
	\fe
The rest of the OPEs between $u$ and the defect operators $b_{\pm }$ have the following two possibilities (up to a redefinition of the operators),\footnote{Note that in \eqref{2vfa}, $a_1$ and $a_2$ can be exchanged by setting $\A\to -\A^{-1}$. We fix this ambiguity here by  our ansatz for the OPE of $u$ with $b_\pm$.}
	\ie
	& u b_{\pm}=\A b_{\pm}, 
	\quad \overbrace{b_+ b_+} = \overbrace{b_- b_-} = 1+\A u,
	\label{buope}
	\fe
	or
	\ie
	& u b_{+}=\A b_{+}, 
	\quad u b_{-}=-\A^{-1} b_{-}, 
	\quad \overbrace{b_+ b_+}  = 1+\A u,
	\quad \overbrace{b_- b_-} = 1-\A^{-1} u,
	\label{2ndbuope}
	\fe
to satisfy associativity. Unitarity requires that $\alpha$ is a real number. We further assume that it is positive, since its sign can be absorbed into a further redefinition of $u$. The possibility \eqref{2ndbuope} can be eliminated by modular invariance of the torus one-point function of $u$ with a spatial $Y$ loop.\footnote{The case of  \eqref{2ndbuope}  with  $\alpha=\pm 1$ needs special care. Although it satisfies the modular invariance of $\tr u \widehat Y$, it is ruled out by the crossing equations for the defect four-point functions, similar to those in Figure~\ref{fig:bbbb}.
}
In the case of \eqref{buope}, the same consideration determines $\alpha$ by
	\ie
	\tr u \widehat Y = \la u b_+ b_+\ra +\la u b_- b_-\ra =(1-\sqrt{3})\la uuu\ra,
	\label{uY1pt}
	\fe
which gives a unique positive real solution $\A = \sqrt{2-\sqrt{3}}$. From now on, we will proceed with the OPE \eqref{buope} with $\A$ understood to take the aforementioned value.
	
Moving on to the structure constants of the TQFT that involve nontrivial junctions, we deduce from the invariance under $\widehat X$, the consistency with the nontrivial cyclic permutation maps, and the OPEs with $u$, that the non-vanishing three-point functions are $\la b_+ b_- b_-\ra_{v_0}$, $\la b_+ b_+ b_+\ra_{v_0}$, $\la b_+, b_- b_-\ra_{v_1}$, and  $\la a_1 b_+ b_-\ra$, together with their conjugates (recall Figure~\ref{fig:iota}). In particular, our choice of junction vectors ensures that $\la b_+ b_- b_-\ra_{v_0}$ and $\la b_+ b_+ b_+\ra_{v_0}$ are self-conjugate, and
	\ie
	\la b_+, b_- b_-\ra_{v_1}^*=\omega^2 \la b_+, b_- b_-\ra_{v_1}, \quad \la a_1 b_+ b_-\ra^*= \la a_1 b_- b_+\ra .
	\label{bbbconj}
	\fe

From the crossing relations of all four-point functions with arbitrary H-junctions (see Figure~\ref{fig:bbbb} for an example), the unique solution is (up to a redefinition of $b_\pm$ and $a_1$)
 	\ie
	&\la b_+ b_+ b_+\ra_{v_0}={\sqrt{3}-3\over \sqrt{2}},\quad 
	\la b_+ b_- b_-\ra_{v_0}={\sqrt{2-\sqrt{3}}},\quad 
	\la b_+, b_- b_-\ra_{v_1}=(1-{\sqrt{3}}) \, \omega^2,\quad 
	\\
	&\la a_1 b_+ b_-\ra=-{\sqrt{3-\sqrt{3}}} \, e^{\pi  i \over 4},
	\label{e62vsc}
	\fe
with the the rest determined by conjugation.

	\begin{figure}[h]
		\begin{minipage}{0.12\textwidth}\begin{eqnarray*}\\ \end{eqnarray*}
		\end{minipage}%
		\begin{minipage}{0.15\textwidth}
			\includegraphics[width=1\textwidth]{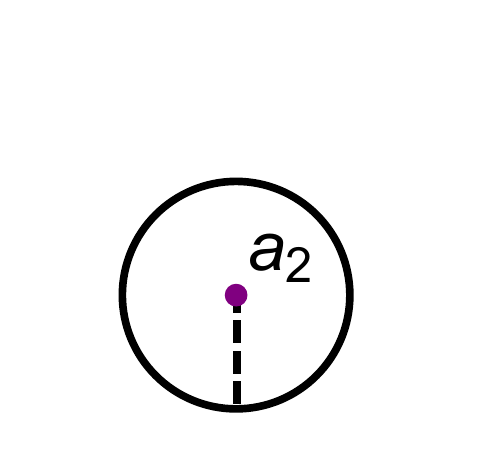}
		\end{minipage}%
		\begin{minipage}{0.01\textwidth}\begin{eqnarray*}=\\ \end{eqnarray*}
		\end{minipage}%
		\begin{minipage}{0.15\textwidth}
			\includegraphics[width=1\textwidth]{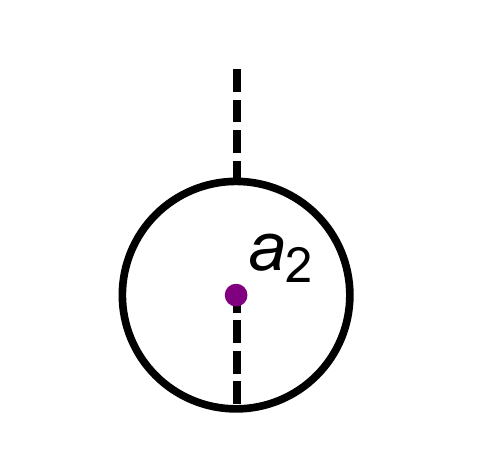}
		\end{minipage}%
		\begin{minipage}{0.01\textwidth}\begin{eqnarray*}=\\ \end{eqnarray*}
		\end{minipage}%
		\begin{minipage}{0.15\textwidth}
			\includegraphics[width=1\textwidth]{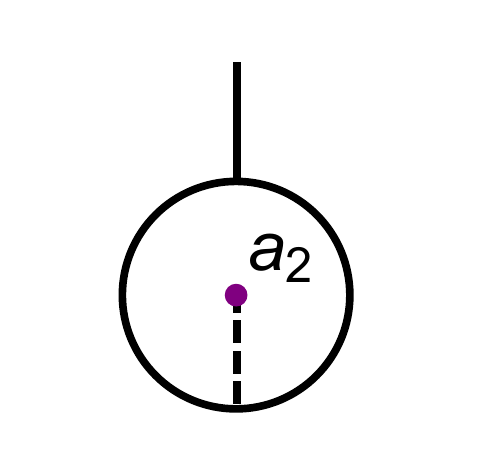}
		\end{minipage}%
		\begin{minipage}{0.01\textwidth}\begin{eqnarray*}=\\ \end{eqnarray*}
		\end{minipage}%
		\begin{minipage}{0.15\textwidth}
			\includegraphics[width=1\textwidth]{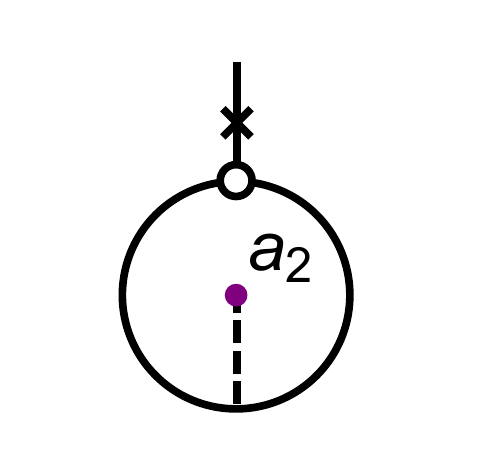}
		\end{minipage}%
		\begin{minipage}{0.01\textwidth}\begin{eqnarray*}=0\\ \end{eqnarray*}
		\end{minipage}%
				\caption{Lassoing the defect operator $a_2 \in {\cal H}_X$. The first diagram vanishes by modular invariance of the torus one point function of $a_2$ attached to a spatial $Y$ loop via an $XYY$ junction. The second diagram vanishes as a consequence of the crossing phase ${\widetilde K}_{X,Y}^{Y,X}(Y,Y)=-1$. The third and last diagrams can be shown to have vanishing correlators with the defect operators $b_\pm \in {\cal H}_Y$ using crossing and the previous vanishing results (see Figure~\ref{fig:2ve6sr}). Similar arguments ensure that the first two lasso diagrams for $a_1$ also vanish (but not the last two).}\label{fig:2vE6las}
	\end{figure}

	\begin{figure}[H]
\begin{minipage}{0.1\textwidth}\begin{eqnarray*}\\ \end{eqnarray*}
		\end{minipage}
			\begin{minipage}{0.04\textwidth}\begin{eqnarray*}0=\\ \end{eqnarray*}
		\end{minipage}%
		\begin{minipage}{0.05\textwidth}\begin{eqnarray*}\\ \end{eqnarray*}
		\end{minipage}%
		\begin{minipage}{0.2\textwidth}
			\includegraphics[width=1\textwidth]{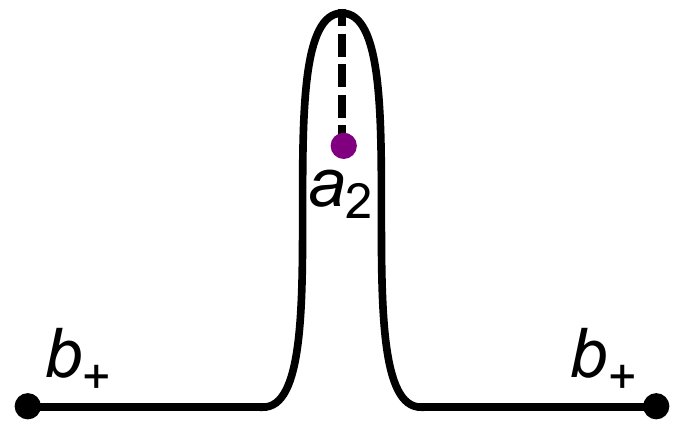}
		\end{minipage}%
		\begin{minipage}{0.05\textwidth}\begin{eqnarray*}=\\ \end{eqnarray*}
		\end{minipage}%
		\begin{minipage}{0.2\textwidth}
			\includegraphics[width=1\textwidth]{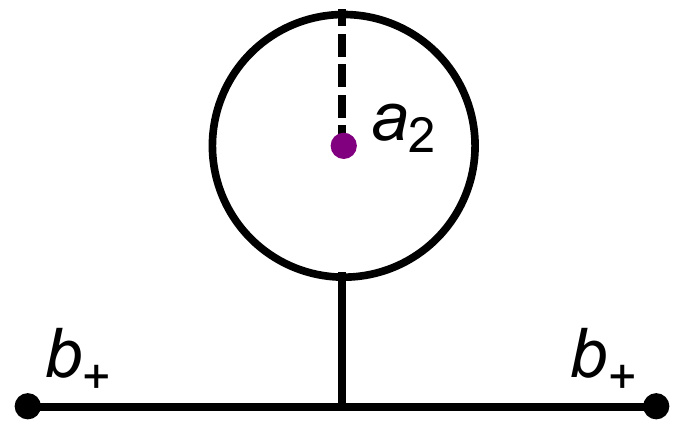}
		\end{minipage}%
\medskip\\
\begin{minipage}{0.1\textwidth}\begin{eqnarray*}\\ \end{eqnarray*}
		\end{minipage}
			\begin{minipage}{0.04\textwidth}\begin{eqnarray*}0=\\ \end{eqnarray*}
		\end{minipage}%
		\begin{minipage}{0.05\textwidth}\begin{eqnarray*}\\ \end{eqnarray*}
		\end{minipage}%
		\begin{minipage}{0.2\textwidth}
			\includegraphics[width=1\textwidth]{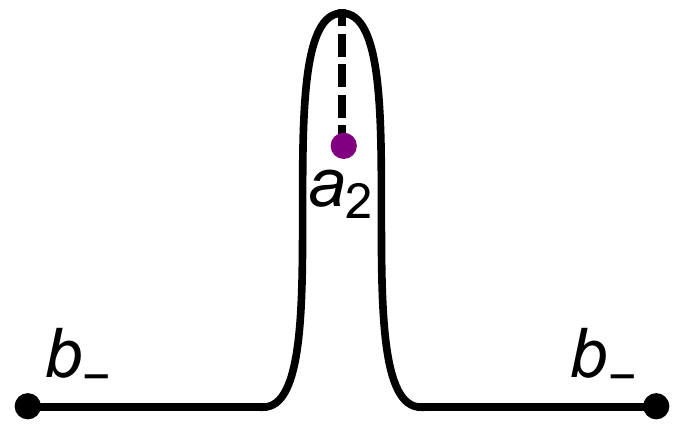}
		\end{minipage}%
		\begin{minipage}{0.05\textwidth}\begin{eqnarray*}=\\ \end{eqnarray*}
		\end{minipage}%
		\begin{minipage}{0.2\textwidth}
			\includegraphics[width=1\textwidth]{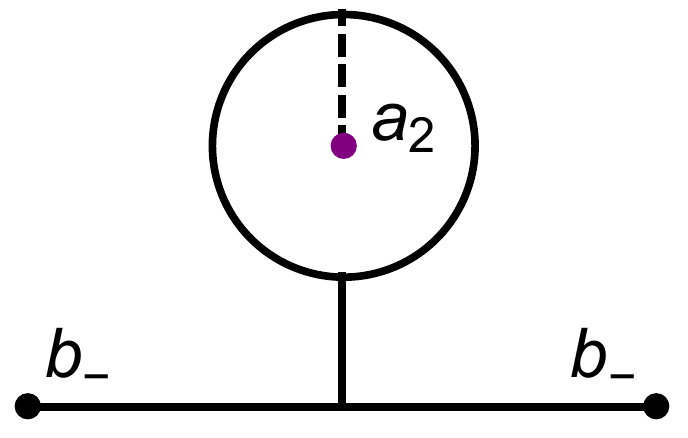}
		\end{minipage}%
		\begin{minipage}{0.05\textwidth}\begin{eqnarray*}+\\ \end{eqnarray*}
		\end{minipage}%
		\begin{minipage}{0.2\textwidth}
			\includegraphics[width=1\textwidth]{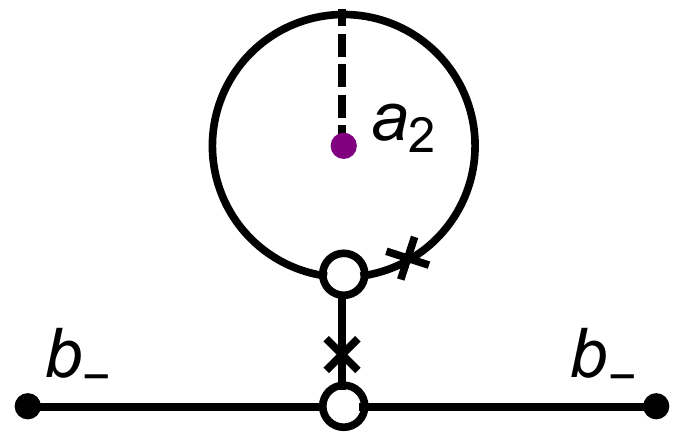}
		\end{minipage}%
\caption{Derivation of the last two lasso diagrams in Figure~\ref{fig:2vE6las}. We start with three-point functions of defect operators $a_2$ and $b_\pm$ which vanish identically, and then pass $a_2$ through the $Y$ line via crossing. We thus obtain a set of four algebraic equations involving the structure constants \eqref{e62vsc}, and the lasso diagrams of $a_2$ ending on $b_\pm$. Here, we display two of the four equations (the other two are obtained by setting the pair of defect operators in ${\cal H}_Y$ to be $b_+,\,b_-$ and $b_-,\,b_+$). The unique solution to these equations is that all lasso diagrams of $a_2$ involving $YYY$ junctions vanish.}
\label{fig:2ve6sr}
\end{figure}

	\begin{figure}[H]
		\centering
		\begin{minipage}{0.15\textwidth}
			\includegraphics[width=1\textwidth]{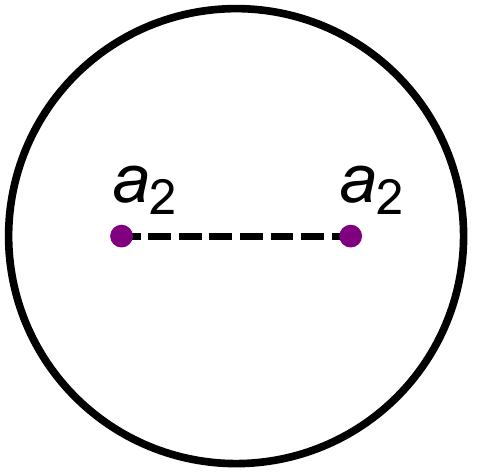}
		\end{minipage}%
		\begin{minipage}{0.05\textwidth}\begin{eqnarray*}~=~ \\ \end{eqnarray*}
		\end{minipage}%
		\begin{minipage}{0.15\textwidth}
			\includegraphics[width=1\textwidth]{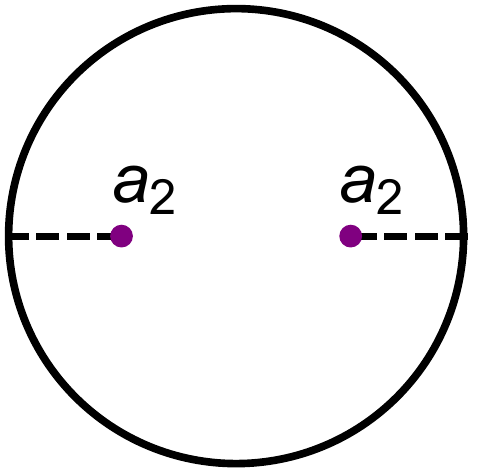}
		\end{minipage}%
		\begin{minipage}{0.16\textwidth}\begin{eqnarray*}~=~ {\sqrt{3}-1\over 2}  \\ \end{eqnarray*}
		\end{minipage}%
		\begin{minipage}{0.15\textwidth}
			\includegraphics[width=1\textwidth]{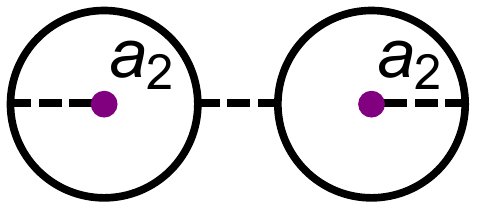}
		\end{minipage}%
		\begin{minipage}{0.05\textwidth}\begin{eqnarray*}~ +  \\ \end{eqnarray*}
		\end{minipage}%
		\begin{minipage}{0.15\textwidth}
			\includegraphics[width=1\textwidth]{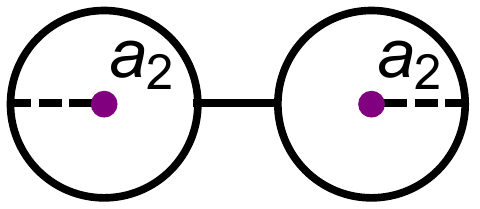}
		\end{minipage}
		\\ \hspace{3in}
		\begin{minipage}{0.05\textwidth}\begin{eqnarray*}  + \\ \end{eqnarray*}
		\end{minipage}%
		\begin{minipage}{0.15\textwidth}
			\includegraphics[width=1\textwidth]{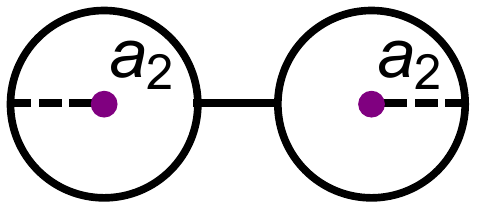}
		\end{minipage}
		\begin{minipage}{0.05\textwidth}\begin{eqnarray*}  + \\ \end{eqnarray*}
		\end{minipage}%
		\begin{minipage}{0.15\textwidth}
			\includegraphics[width=1\textwidth]{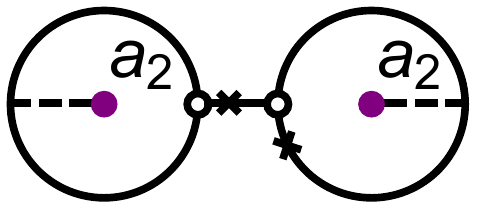}
		\end{minipage}
		\begin{minipage}{0.05\textwidth}\begin{eqnarray*}   \\ \end{eqnarray*}
		\end{minipage}%
		\caption{The above diagrams are related by partial fusions first between $X$ and $Y$ TDLs and then between two $Y$ TDLs. Since we obtain a combination of products of lasso diagrams of $a_2$, the RHS vanish identically but the LHS clearly does not.}\label{fig:2va2la}
	\end{figure}

Furthermore, the lassoing can be determined from the modular invariance of torus one-point functions together with crossing. In particular, we find that all lasso diagrams of the defect operator $a_2 \in {\cal H}_X$ vanish identically (see Figure~\ref{fig:2vE6las}). This leads to an immediate contradiction with the crossing relations in Figure~\ref{fig:2va2la}. We thus conclude that there is no ${1\over 2}E_6$ TQFT with two vacua.

\subsubsection*{${1\over 2}E_6$ TQFT with three vacua}

In this case, we label  the three degenerate vacua of the TQFT by 1, $u$, and $w$, whose eigenvalues under $(\widehat X, \widehat Y)$ are $(1, 1+\sqrt{3})$, $(1,1-\sqrt{3})$, and $(-1,0)$, respectively. By modular invariance, the defect Hilbert space ${\cal H}_X$ is one-dimensional and generated by $a$, while ${\cal H}_Y$ is two-dimensional with basis elements $b_\pm$, labeled in accordance with their charges under $\widehat X_-$.  
	
We normalize all the defect operators $a$, $b_+$, $b_-$, $u$, and $w$ to have unit norm. From the associativity (without using the $YYY$ junction) and the selection rule by the $\widehat X$-invariance, we deduce the following relations,
	\ie\label{3vwope}
	& u^2 = 1+(\A - \A^{-1}) u,\quad  w^2 = 1+ \A u,\quad  u w = \A w,
	\\
	&
	u a=-\A^{-1}a,\quad  u b_{\pm}=\A b_{\pm},\quad  w a=0,\quad w b_{\pm}=\sqrt{1+\A^2}b_{\mp}
	\\
	& \overbrace{a a} = 1 - \A^{-1} u, \quad  \overbrace{b_+ b_+} = \overbrace{b_- b_-} = 1+\A u, \quad  \overbrace{b_+ b_-} = \sqrt{1+\A^2} w.
	\fe
Here, $\A$ is a real number, which we further assume to be positive, since its sign can be absorbed by a redefinition of $u$. By the modular invariance of the torus one-point function $\tr u\hat Y$ as in \eqref{uY1pt}, we determine $\A=\sqrt{2-\sqrt{3}}$.
	
As before, we use the $\widehat X$-invariance to constrain the structure constants involving $b_\pm$, so that the potential non-vanishing structure constants involving $XYY$ and $YYY$ junctions are $\la a b_+ b_-\ra$, $\la b_+ b_- b_-\ra_{v_0}$, $\la b_+ b_+ b_+\ra_{v_0}$, $\la b_+, b_+ b_+\ra_{v_1}$, and $\la b_+, b_- b_-\ra_{v_1}$. Furthermore, the consistency of OPEs with bulk operators requires $ \la a b_+ b_-\ra=0$, and the consistency with the nontrivial cyclic permutation map requires $\la b_+, b_+ b_+\ra_{v_1}=0$.
	
To determine the rest of the structure constants, let us consider the crossing equations of all defect operator four-point functions involving nontrivial H-junctions (see Figure~\ref{fig:bbbb}). There exists a unique solution (up to a redefinition of $b_\pm$) given by 
	\ie
	\la b_+ b_- b_-\ra_{v_1}=0,\quad\la b_+ b_+ b_+\ra_{v_0} = \la b_+ b_- b_-\ra_{v_0} = \sqrt{{3-\sqrt{3}\over 2} (1+\A^2)}={3-\sqrt{3}\over \sqrt{2}}.
	\fe

\begin{figure}[H]
		\begin{minipage}{0.1\textwidth}\begin{eqnarray*}  ~\\ \end{eqnarray*}
		\end{minipage}%
		\begin{minipage}{0.13\textwidth}
			\includegraphics[width=1\textwidth]{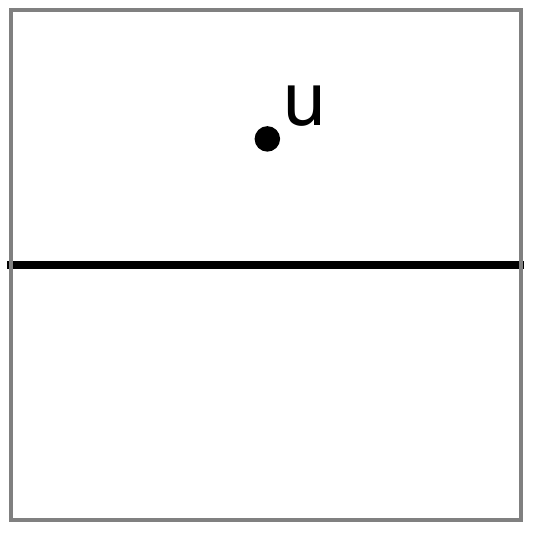}
		\end{minipage}%
		\begin{minipage}{0.15\textwidth}\begin{eqnarray*}~= (1-\sqrt{3})\la uuu\ra  \\ \end{eqnarray*}
		\end{minipage}%
		\begin{minipage}{0.15\textwidth}\begin{eqnarray*}  ~\\ \end{eqnarray*}
		\end{minipage}%
		\begin{minipage}{0.13\textwidth}
			\includegraphics[width=1\textwidth]{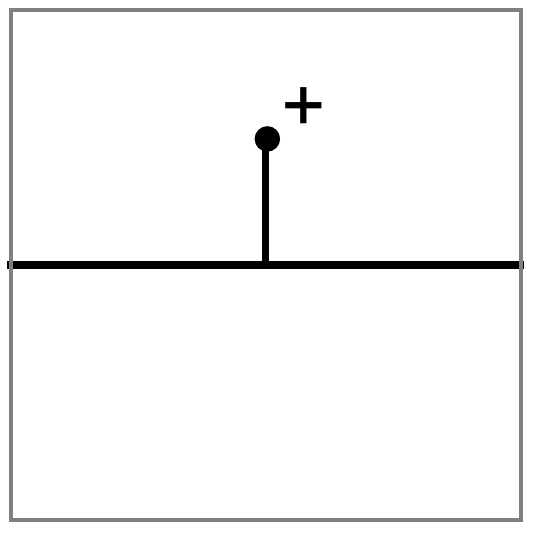}
		\end{minipage}%
		\begin{minipage}{0.15\textwidth}\begin{eqnarray*}~= 2 \la b_+b_+b_+\ra_{v_0} \\ \end{eqnarray*}
		\end{minipage}%
		\\
		\bigskip
		\\
		\begin{minipage}{0.1\textwidth}\begin{eqnarray*}  ~\\ \end{eqnarray*}
		\end{minipage}%
		\begin{minipage}{0.13\textwidth}
			\includegraphics[width=1\textwidth]{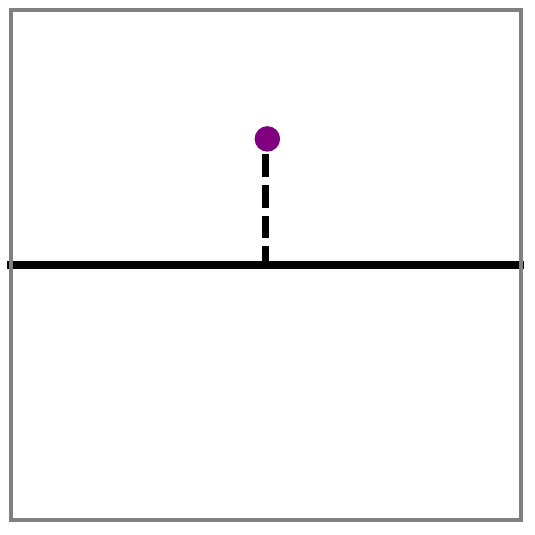}
		\end{minipage}%
		\begin{minipage}{0.05\textwidth}\begin{eqnarray*}~= 0  \\ \end{eqnarray*}
		\end{minipage}%
		\begin{minipage}{0.15\textwidth}\begin{eqnarray*}  ~\\ \end{eqnarray*}
		\end{minipage}%
		\begin{minipage}{0.13\textwidth}
			\includegraphics[width=1\textwidth]{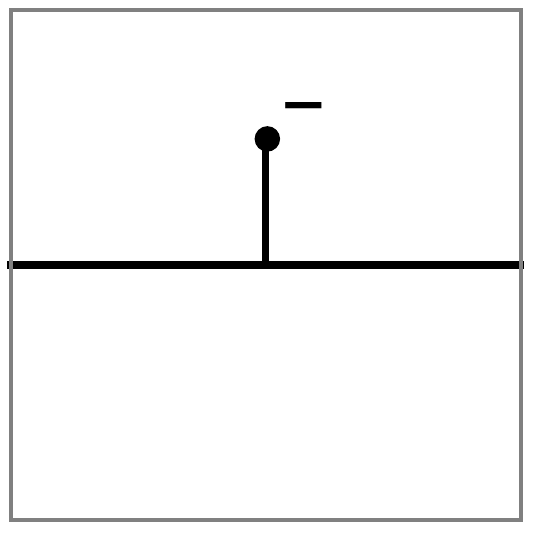}
		\end{minipage}%
		\begin{minipage}{0.05\textwidth}\begin{eqnarray*}~= 0 \\ \end{eqnarray*}
		\end{minipage}%
		\begin{minipage}{0.15\textwidth}\begin{eqnarray*}  ~\\ \end{eqnarray*}
		\end{minipage}%
		\begin{minipage}{0.13\textwidth}
			\includegraphics[width=1\textwidth]{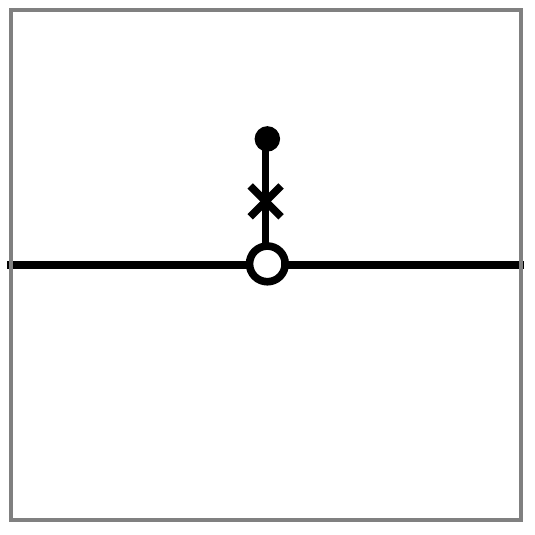}
		\end{minipage}%
		\begin{minipage}{0.05\textwidth}\begin{eqnarray*}~= 0 \\ \end{eqnarray*}
		\end{minipage}%
		\caption{Some torus one-point functions of defect operators.}\label{fig:torusone}
	\end{figure}

\begin{figure}[H]
		\centering
		\begin{minipage}{0.15\textwidth}
			\includegraphics[width=1\textwidth]{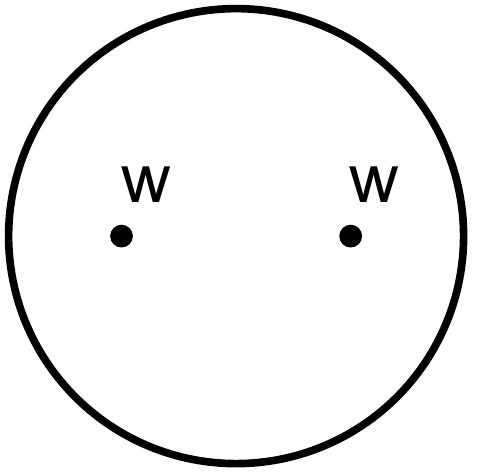}
		\end{minipage}%
		\begin{minipage}{0.16\textwidth}\begin{eqnarray*}~=\quad {\sqrt{3}-1\over 2}  \\ \end{eqnarray*}
		\end{minipage}%
		\begin{minipage}{0.15\textwidth}
			\includegraphics[width=1\textwidth]{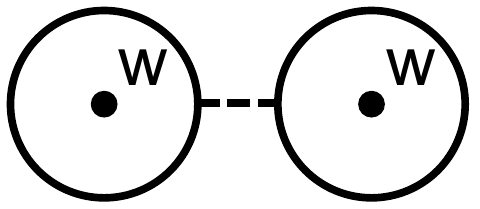}
		\end{minipage}%
		\begin{minipage}{0.04\textwidth}\begin{eqnarray*}~ +  \\ \end{eqnarray*}
		\end{minipage}%
		\begin{minipage}{0.15\textwidth}
			\includegraphics[width=1\textwidth]{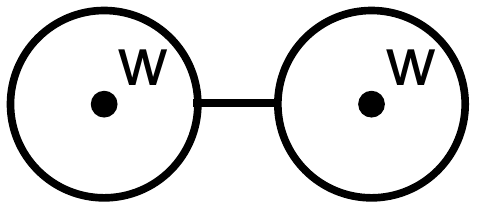}
		\end{minipage}
		\begin{minipage}{0.04\textwidth}\begin{eqnarray*}  + \\ \end{eqnarray*}
		\end{minipage}%
		\begin{minipage}{0.15\textwidth}
			\includegraphics[width=1\textwidth]{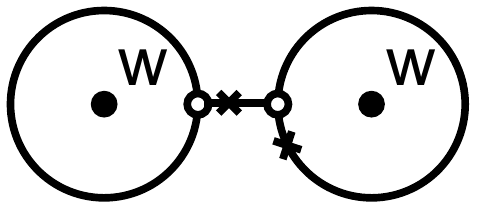}
		\end{minipage}
		\caption{An example of a crossing equation that constrains the lassoing of bulk operators. On the right, we have dropped the contribution from the H-junction with the identity line in the middle, because $\widehat Y$ annihilates the bulk operator $w$.}\label{fig:E6lassocross}
	\end{figure}

\begin{figure}[H]
		\begin{minipage}{0.12\textwidth}
			\includegraphics[width=1\textwidth]{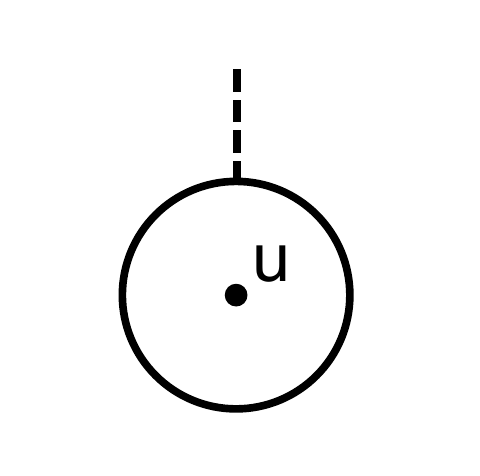}
		\end{minipage}%
		\begin{minipage}{0.01\textwidth}\begin{eqnarray*}=0\\ \end{eqnarray*}
		\end{minipage}%
		\begin{minipage}{0.175\textwidth}\begin{eqnarray*} \\ \end{eqnarray*}
		\end{minipage}%
		\begin{minipage}{0.12\textwidth}
			\includegraphics[width=1\textwidth]{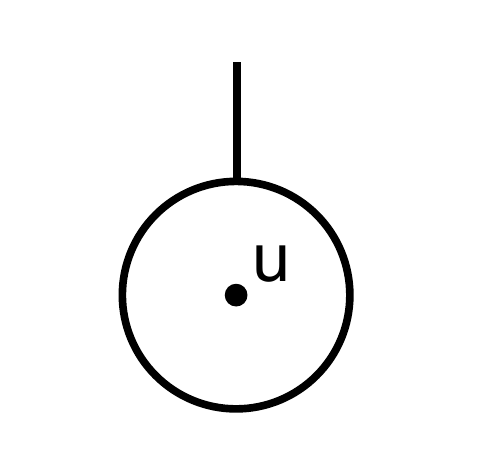}
		\end{minipage}%
		\begin{minipage}{0.06\textwidth}\begin{eqnarray*} ={\sqrt{3}-1 }  ~ \\ \end{eqnarray*}
		\end{minipage}%
		\begin{minipage}{0.12\textwidth}
			\includegraphics[width=1\textwidth]{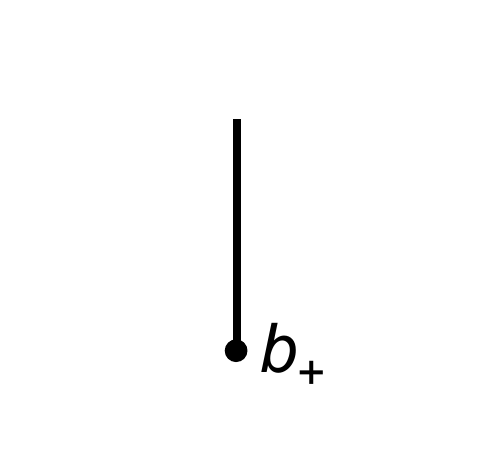}
		\end{minipage}%
		\begin{minipage}{0.027\textwidth}\begin{eqnarray*} \\ \end{eqnarray*}
		\end{minipage}%
		\begin{minipage}{0.12\textwidth}
			\includegraphics[width=1\textwidth]{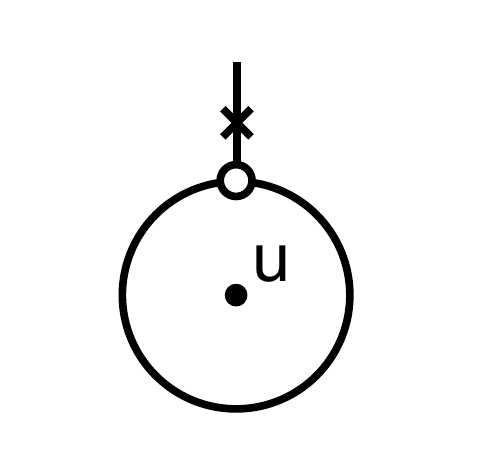}
		\end{minipage}%
		\begin{minipage}{0.04\textwidth}\begin{eqnarray*} ={\sqrt{2}\omega^2}  ~ \\ \end{eqnarray*}
		\end{minipage}%
		\begin{minipage}{0.12\textwidth}
			\includegraphics[width=1\textwidth]{figures/e6tpbp.pdf}
		\end{minipage}%
		\\
		\begin{minipage}{0.12\textwidth}
			\includegraphics[width=1\textwidth]{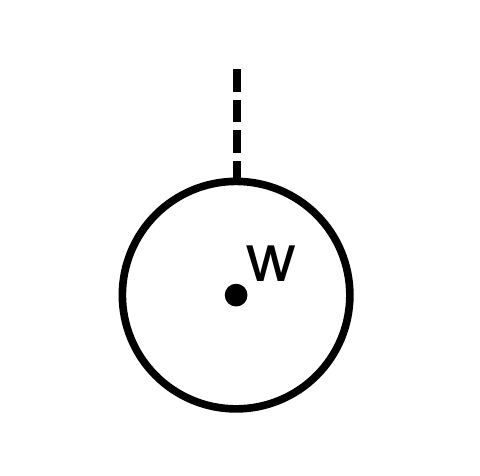}
		\end{minipage}%
		\begin{minipage}{0.01\textwidth}\begin{eqnarray*}= \sqrt{2}    \\ \end{eqnarray*}
		\end{minipage}%
		\begin{minipage}{0.12\textwidth}
			\includegraphics[width=1\textwidth]{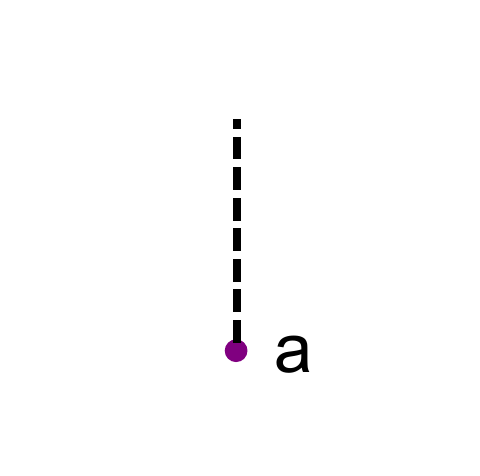}
		\end{minipage}%
			\begin{minipage}{0.052\textwidth}\begin{eqnarray*} \\ \end{eqnarray*}
		\end{minipage}%
		\begin{minipage}{0.12\textwidth}
			\includegraphics[width=1\textwidth]{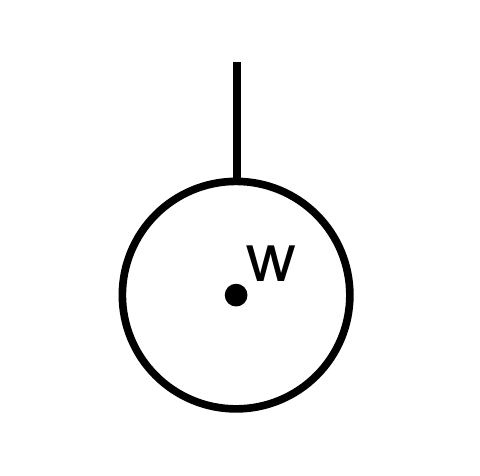}
		\end{minipage}%
		\begin{minipage}{0.09\textwidth}\begin{eqnarray*} =\sqrt{1+\frac{1}{\sqrt{3}}}  \\ \end{eqnarray*}
		\end{minipage}%
		\begin{minipage}{0.12\textwidth}
			\includegraphics[width=1\textwidth]{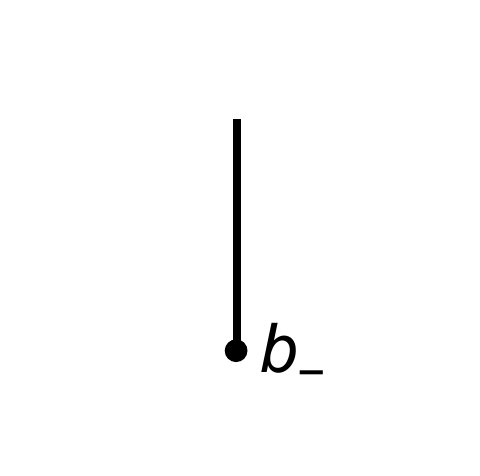}
		\end{minipage}%
		\begin{minipage}{0.12\textwidth}
			\includegraphics[width=1\textwidth]{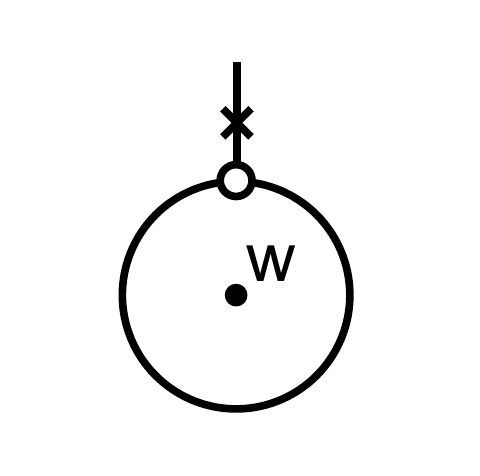}
		\end{minipage}%
		\begin{minipage}{0.14\textwidth}\begin{eqnarray*} =-\omega^2\sqrt{1-\frac{1}{\sqrt{3}}}\\ \end{eqnarray*}
		\end{minipage}%
		\begin{minipage}{0.12\textwidth}
			\includegraphics[width=1\textwidth]{figures/e6tpbm.pdf}
		\end{minipage}%
				\\
		\begin{minipage}{0.12\textwidth}
			\includegraphics[width=1\textwidth]{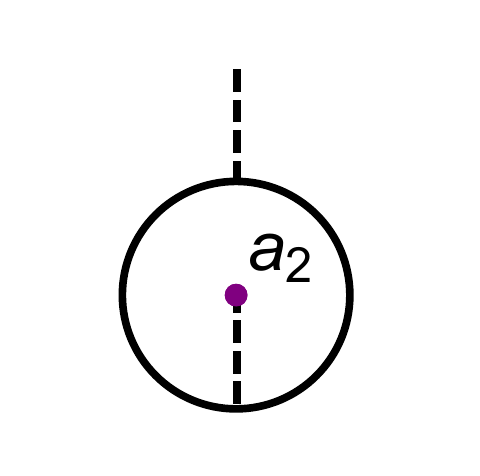}
		\end{minipage}%
		\begin{minipage}{0.01\textwidth}\begin{eqnarray*}= 0   \\ \end{eqnarray*}
		\end{minipage}%
		\begin{minipage}{0.175\textwidth}\begin{eqnarray*} \\ \end{eqnarray*}
		\end{minipage}%
		\begin{minipage}{0.12\textwidth}
			\includegraphics[width=1\textwidth]{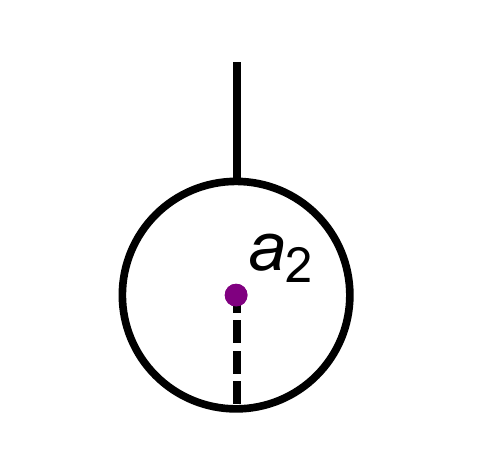}
		\end{minipage}%
		\begin{minipage}{0.09\textwidth}\begin{eqnarray*} = \sqrt{1-\frac{1}{\sqrt{3}}}  \\ \end{eqnarray*}
		\end{minipage}%
		\begin{minipage}{0.12\textwidth}
			\includegraphics[width=1\textwidth]{figures/e6tpbm.pdf}
		\end{minipage}%
		\begin{minipage}{0.12\textwidth}
			\includegraphics[width=1\textwidth]{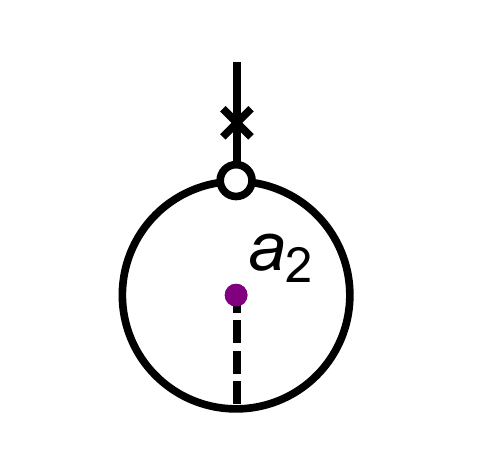}
		\end{minipage}%
		\begin{minipage}{0.14\textwidth}\begin{eqnarray*} =\omega^2\sqrt{1+\frac{1}{\sqrt{3}}}\\ \end{eqnarray*}
		\end{minipage}%
		\begin{minipage}{0.12\textwidth}
			\includegraphics[width=1\textwidth]{figures/e6tpbm.pdf}
		\end{minipage}%
			\\
		\begin{minipage}{0.12\textwidth}
			\includegraphics[width=1\textwidth]{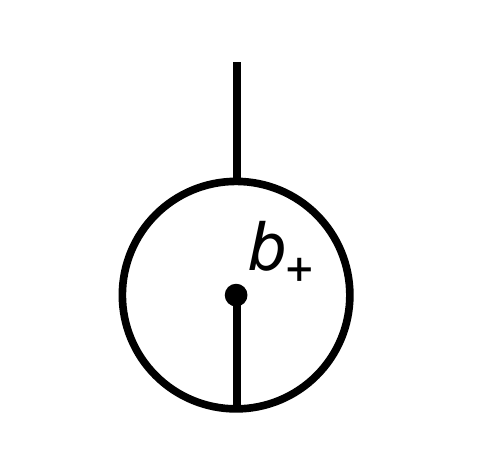}
		\end{minipage}%
		\begin{minipage}{0.06\textwidth}\begin{eqnarray*}= {2 -{2\over \sqrt{3}}}   \\ \end{eqnarray*}
		\end{minipage}%
				\begin{minipage}{0.12\textwidth}
			\includegraphics[width=1\textwidth]{figures/e6tpbp.pdf}
		\end{minipage}%
			\begin{minipage}{0.005\textwidth}\begin{eqnarray*} \\ \end{eqnarray*}
		\end{minipage}%
		\begin{minipage}{0.12\textwidth}
			\includegraphics[width=1\textwidth]{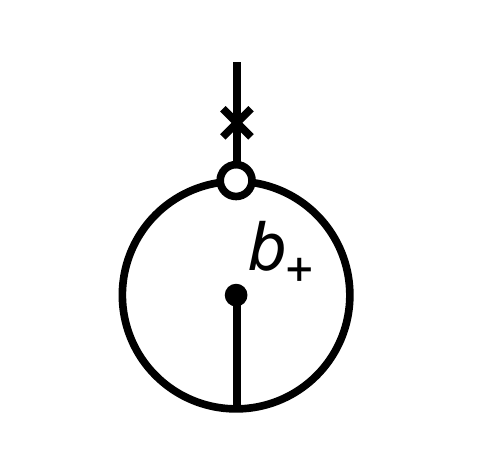}
		\end{minipage}%
		\begin{minipage}{0.09\textwidth}\begin{eqnarray*} ={1-\sqrt{3}\over \sqrt{6}}  \\ \end{eqnarray*}
		\end{minipage}%
		\begin{minipage}{0.12\textwidth}
			\includegraphics[width=1\textwidth]{figures/e6tpbp.pdf}
		\end{minipage}%
		\begin{minipage}{0.12\textwidth}
			\includegraphics[width=1\textwidth]{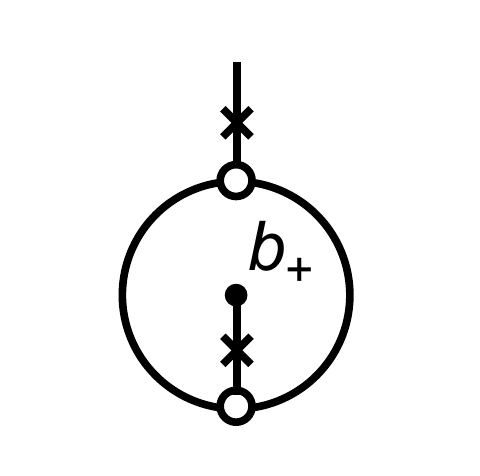}
		\end{minipage}%
		\begin{minipage}{0.14\textwidth}\begin{eqnarray*} =2\omega^2 \left(1-\frac{2}{\sqrt{3}}\right)\\ \end{eqnarray*}
		\end{minipage}%
		\begin{minipage}{0.12\textwidth}
			\includegraphics[width=1\textwidth]{figures/e6tpbp.pdf}
		\end{minipage}%
				\\
		\begin{minipage}{0.12\textwidth}
			\includegraphics[width=1\textwidth]{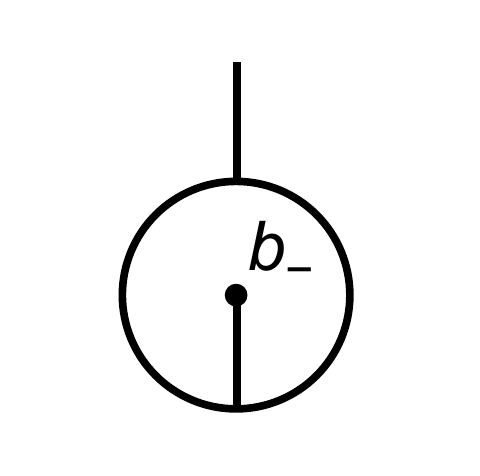}
		\end{minipage}%
		\begin{minipage}{0.06\textwidth}\begin{eqnarray*}=  1-{1\over \sqrt{3}} \\ \end{eqnarray*}
		\end{minipage}%
				\begin{minipage}{0.12\textwidth}
			\includegraphics[width=1\textwidth]{figures/e6tpbm.pdf}
		\end{minipage}%
			\begin{minipage}{0.005\textwidth}\begin{eqnarray*} \\ \end{eqnarray*}
		\end{minipage}%
		\begin{minipage}{0.12\textwidth}
			\includegraphics[width=1\textwidth]{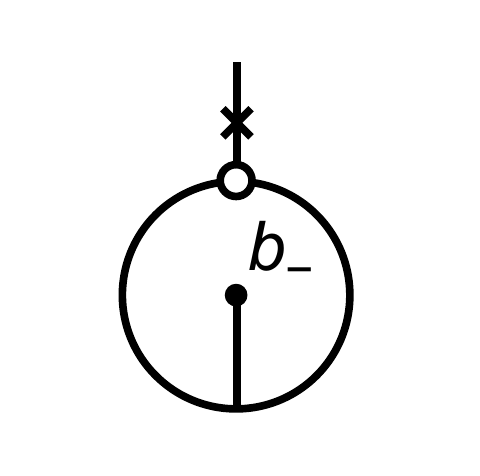}
		\end{minipage}%
		\begin{minipage}{0.09\textwidth}\begin{eqnarray*} ={\sqrt{3}-1\over \sqrt{6}}  \\ \end{eqnarray*}
		\end{minipage}%
		\begin{minipage}{0.12\textwidth}
			\includegraphics[width=1\textwidth]{figures/e6tpbm.pdf}
		\end{minipage}%
		\begin{minipage}{0.12\textwidth}
			\includegraphics[width=1\textwidth]{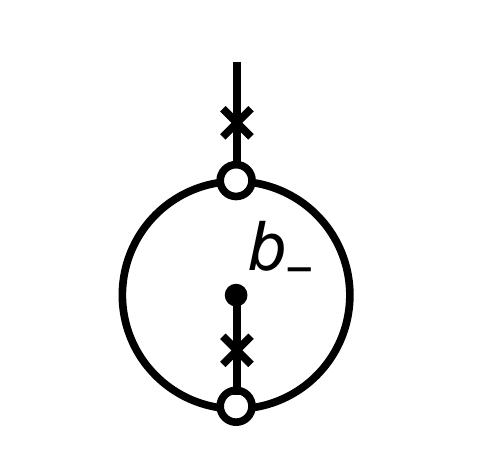}
		\end{minipage}%
		\begin{minipage}{0.14\textwidth}\begin{eqnarray*} =\omega^2 \left(\frac{1}{\sqrt{3}}-1\right)\\ \end{eqnarray*}
		\end{minipage}%
		\begin{minipage}{0.12\textwidth}
			\includegraphics[width=1\textwidth]{figures/e6tpbm.pdf}
		\end{minipage}%
		\caption{Lassoing of bulk and defect operators in the ${1\over 2}E_6$ TQFT. We only include the independent lasso diagrams here, as the rest can be obtained by unwrapping the loop on the sphere.}
		\label{fig:E6las}
	\end{figure}

Next, using modular invariance of the torus one-point functions in Figure~\ref{fig:torusone} and  crossing relations such as in Figure~\ref{fig:E6lassocross} we determine the lasso diagrams of bulk operators. By similar manipulations as in Figure~\ref{fig:2ve6sr} and Figure~\ref{fig:2va2la}, we also obtain the lasso diagrams of defect operators.  We summarize the full set of lassoing results in Figure~\ref{fig:E6las}.

\begin{figure}[H]
	\centering
	\hspace{-2in}
	\begin{minipage}{0.18\textwidth}
		\includegraphics[width=1\textwidth]{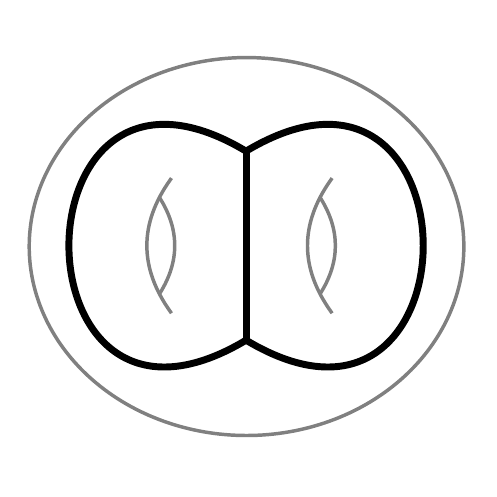}
	\end{minipage}%
	\begin{minipage}{0.218\textwidth}\begin{eqnarray*}= {\widetilde K}_{Y,Y}^{Y,Y}(Y,I)(v_0,v_0)  \\ \end{eqnarray*}
	\end{minipage}%
	\begin{minipage}{0.18\textwidth}
		\includegraphics[width=1\textwidth]{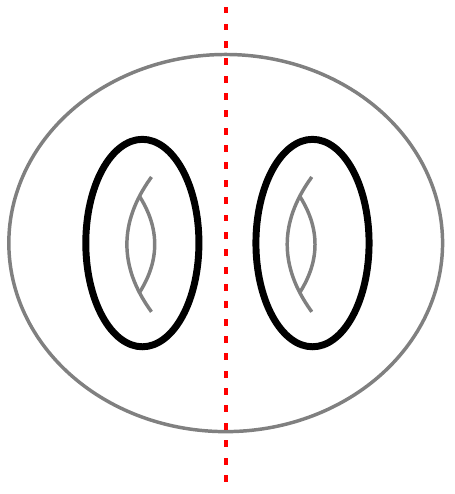}
	\end{minipage}%
	\vspace{-.25in}
	\\
	\hspace{2in}
	\begin{minipage}{0.305\textwidth}\begin{eqnarray*}~+ {\widetilde K}_{Y,Y}^{Y,Y}(Y, Y)(v_0, v_0; v_0, v_0)  \\ \end{eqnarray*}
	\end{minipage}%
	\begin{minipage}{0.18\textwidth}
		\includegraphics[width=1\textwidth]{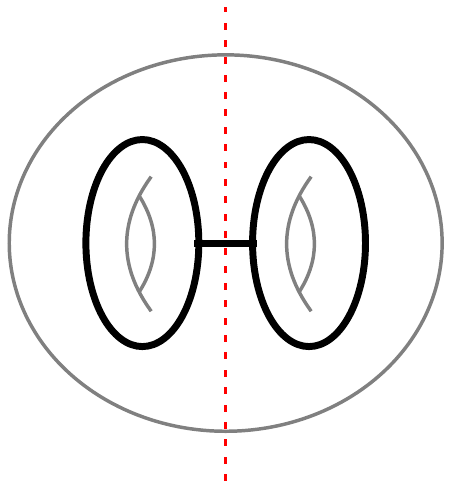}
	\end{minipage}%
	\caption{Crossing transformation on the genus-two partition function with two $Y$ loops. The $YYY$ junction only involves the junction vector $v_0$, as the contributions involving the junction vector $v_1$ vanish, due to the vanishing results of the torus one-point functions in Figure~\ref{fig:torusone}. Likewise, there is no contribution from a pair of $Y$ loops connected by an $X$ segment.}\label{fig:genustwo}
\end{figure}

We provide a nontrivial consistency check by considering the genus-two partition function of the TQFT, with three $Y$ line segments running along the three handles, joining at a pair of $YYY$ junctions, as shown in Figure~\ref{fig:genustwo}. Using the crossing kernel ${\widetilde K}_{Y,Y}^{Y,Y}$, and the vanishing of the torus one-point function of the defect operator $a$ attached to a $Y$ loop (Figure~\ref{fig:torusone}), we find the relation
\ie
4 \la b_+, b_+ b_+\ra^2 &= {\widetilde K}_{Y,Y}^{Y,Y}(Y, I)(v_0, v_0) \left[ \left( {\rm tr}\,\widehat Y \right)^2 + \left( {\rm tr}\,\widehat Y u \right)^2 \right]
\\
& \hspace{.5in} + {\widetilde K}_{Y,Y}^{Y,Y}(Y, Y)(v_0, v_0; v_0, v_0) 4 \la b_+, b_+ b_+\ra^2.
\label{g2eq}
\fe
Using ${\rm tr}\,\widehat Y=2$,  ${\rm tr}\,\widehat Y w =0$, ${\rm tr}\,\widehat Y u = (1-\sqrt{3})(\A-\A^{-1})$ (the first diagram of Figure~\ref{fig:torusone}), and the crossing kernels in \eqref{E6CKnb}, 
we find that $\A = \sqrt{2-\sqrt{3}}$ is the unique positive real solution for \eqref{g2eq},   in agreement with our previous findings.

\subsubsection{Tetracritical Ising model perturbed by $\phi_{1,3}$}

Recall that the tetracritical Ising model (Section~\ref{Sec:Tetra}) admits a TDL $W$ that obeys the fusion relation $W^2=I+W$, and commutes with the bulk local primary $\phi_{1,3}$ of weight $({2\over 3},{2\over 3})$. When perturbed by $\phi_{1,3}$, theory flows to either the tricritical Ising model or a massive phase, depending on the sign of the coupling \cite{Zamolodchikov:1987ti, Ludwig:1987gs, Zamolodchikov:1989cf, Zamolodchikov:1991vh}. The $C$ and $W$ lines, as well as their crossing relations, are preserved under this RG flow. Under the flow to the tricritical Ising model, the $C$ and $W$ lines flow to the $\mathbb{Z}_2$ invertible line $\eta$ and the $W$ line in the tricritical Ising model, respectively. 
The $N$ line in the tricritical Ising model (not to be confused with that in the tetracritical Ising model, which is broken by the $\phi_{1,3}$ flow) is emergent and is not inherited from the tetracritical Ising model. Under the flow to the massive phase, there is a nontrivial TQFT in the IR with at least two-fold degenerate vacua. Note that there is no nontrivial crossing phase between $C$ and $W$, and we cannot deduce {\it a priori} whether $\widehat C$ acts nontrivially on the vacua of the TQFT. The TCSA study of this RG flow was carried out in \cite{Martins:1991jj}, indicating four-fold degenerate vacua. Presumably, the IR TQFT is a tensor product of the one described in Section~\ref{lytqft} with  an extra $\mathbb{Z}_2$ factor.

We can also use TDLs to constrain the IR limit of this RG flow.  As discussed above, the $C$ and $W$ lines in the tetracritical Ising model are preserved along the entire  flow.  From the viewpoint of the IR tricritical Ising model deformed by irrelevant operators, this implies that the irrelevant operators should also commute with the $C$ (which becomes $\eta$ in the tricritical Ising model) and the $W$ lines.  From \eqref{triIsingaction},   the unique such irrelevant operator in the tricritical Ising model is $\phi_{3,1} = \varepsilon''$. Indeed, it was shown in \cite{Feverati:1995hy} that the leading irrelevant operator that should be turned on in the IR regime of this flow is $\phi_{3,1}$.

Similarly, the three-state Potts model perturbed by $Z+Z^*$ also flows to either the tricritical Ising model or a massive phase depending on the sign of the coupling. In this case, the flow to the massive phase again preserves the $C$ and $W$ lines, where $C$ is the charge conjugation symmetry that exchanges $Z$ with $Z^*$. We expect the IR TQFT to be the $\mathbb{Z}_2$ orbifold of the TQFT of $M(6,5)$ perturbed by $\phi_{1,3}$, namely, one that is identical to the TQFT of Section~\ref{lytqft}, with two-fold degenerate vacua. We can also consider the more general perturbation by $e^{i\A}Z + e^{-i\A} Z^*$. When $e^{i\A}$ is not a third root of unity,  this flow is not expected to be integrable \cite{Fateev:1991bv}. The perturbation  breaks the $S_3$ symmetry completely, and only the $W$ line is preserved. Since we still expect the vacuum to be two-fold degenerate, and the IR TQFT fixed by the $W$ line to be that of Section~\ref{lytqft}.

\subsubsection{Pentacritical Ising model perturbed by $\phi_{2,1}$}

The pentacritical Ising model (Section~\ref{Sec:Penta}) admits TDLs $X$ and $Y$ that generate the ${\rm Rep}(\widehat{so(3)_5})$ fusion category, and 
commute with the relevant operator $\phi_{2,1}$ of weight $({3\over 8},{3\over 8})$. Since this fusion category does not exist in minimal models of smaller central charges, we expect the pentacritical Ising model perturbed by $\phi_{2,1}$ to flow to a nontrivial TQFT.

In the IR TQFT, in order for ${\rm tr}\,\widehat X = {\rm dim}\,{\cal H}_X$ to be an integer, there must be at least three degenerate vacua. This indeed agrees with the TCSA results of \cite{Martins:1991jj}, where a three-fold vacuum degeneracy is seen numerically. Thus, we expect the IR TQFT to contain three vacua, 1, $v_1$, $v_2$, one defect operator $a\in {\cal H}_X$, and two defect operators $b_1, b_2 \in {\cal H}_Y$. We leave the determination of the full IR TQFT to future work.

\subsection{Comments on RG walls and boundary states}

Let ${\cal O}$ be a relevant scalar primary, and consider the deformation of the CFT by turning on the coupling $\lambda \int_D d^2z {\cal O}(z,\bar z)$ on a disc $D$ (see Section~\ref{Sec:IsingFlow}), with positive $\lambda$. After flowing to the IR -- which may be viewed as taking the $\lambda\to \infty$ limit -- the boundary of the disc $\partial D$ becomes a conformal interface between the original CFT outside the disc and a new phase inside the disc \cite{Gaiotto:2012np}, which is either a new CFT or a massive phase (TQFT). In this section, we focus on the latter case. The RG flow inside the disc produces an Ishibashi state on the boundary of the disc, which we denote by $|{\cal O}\ra\ra_{RG}$. However, when the massive phase inside the disc is a nontrivial TQFT, $|{\cal O}\ra\ra_{RG}$ may not be a boundary (Cardy) state. We will illustrate this phenomenon with a few examples.

To begin with, consider the critical Ising model deformed by $\varepsilon$ inside the disc, with positive coupling $\lambda$. It is well known that this flow produces the $\mathbb{Z}_2$-invariant boundary state
\ie
|\varepsilon\ra\ra_{RG}= |f\ra\ra = |1\ra\ra - |\varepsilon\ra\ra,
\fe
where $|\phi\ra\ra$ denotes the Ishibashi state corresponding to the bulk local primary $\phi$. The fusion of the TDL $N$ with $|f\ra\ra$ produces a new boundary state
\ie
|Nf\ra\ra = \widehat N |f\ra\ra = \sqrt{2} \left( |1\ra\ra + |\varepsilon\ra\ra \right) = |+\ra\ra + |-\ra\ra,
\fe
where $|\pm\ra\ra$ are two Cardy states, given by 
\ie
|\pm\ra\ra = {1\over \sqrt{2}} \left(  |1\ra\ra + |\varepsilon\ra\ra \pm 2^{1\over 4} |\sigma\ra\ra \right).
\fe
The action of $N$ on the boundary state $|f\ra\ra$ can be understood from its action on the relevant deformation. When $N$ moves past $\varepsilon$, it flips the sign of $\varepsilon$. Shrinking an $N$ loop encircling the disc by moving it inside the disc, we see that
\ie
|-\varepsilon\ra\ra_{RG} = {1\over\sqrt2} |Nf\ra\ra = |1\ra\ra + |\varepsilon\ra\ra.
\fe
So the RG wall construction based on the deformation by $-\varepsilon$ produces the Ishibashi state $|-\varepsilon\ra\ra_{RG}$, which is {\it not} a boundary state.
This can only happen when the flow ends up in a nontrivial TQFT with degenerate vacua, which is indeed the case in this example (the TQFT being that of the $\mathbb{Z}_2$ fusion category).

\begin{figure}[H]
\centering
\begin{minipage}{0.3\textwidth}
\includegraphics[width=1\textwidth]{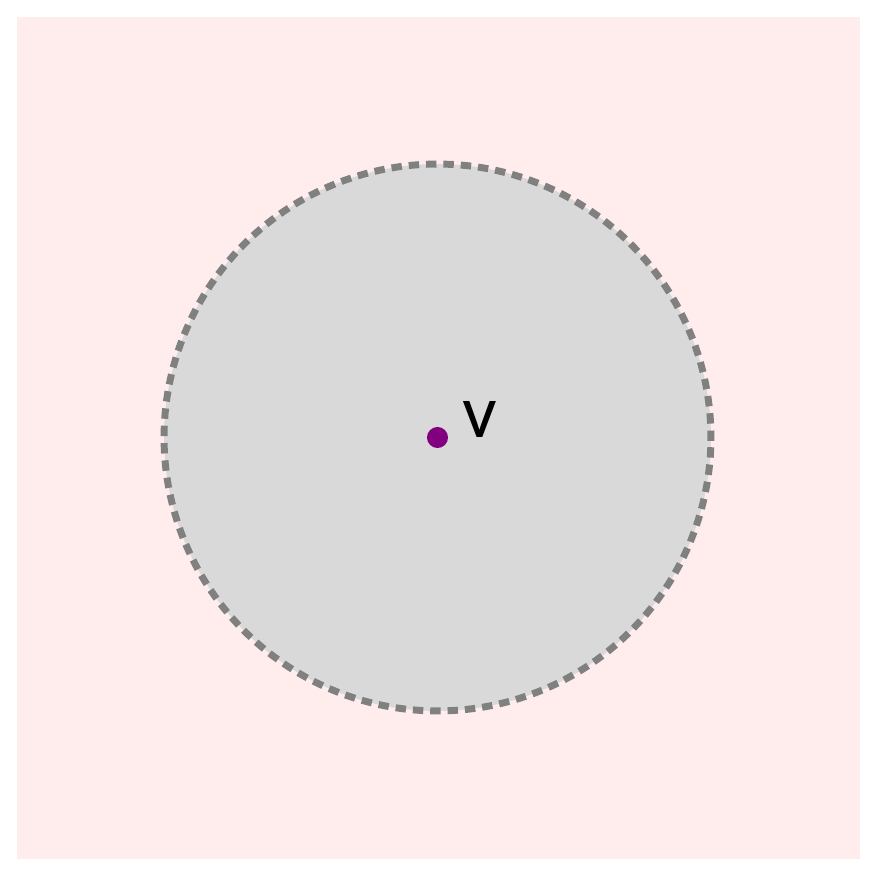}
\end{minipage}%
\begin{minipage}{0.2\textwidth}\begin{eqnarray*}\quad \\ \end{eqnarray*}
\end{minipage}%
\begin{minipage}{0.3\textwidth}
\includegraphics[width=1\textwidth]{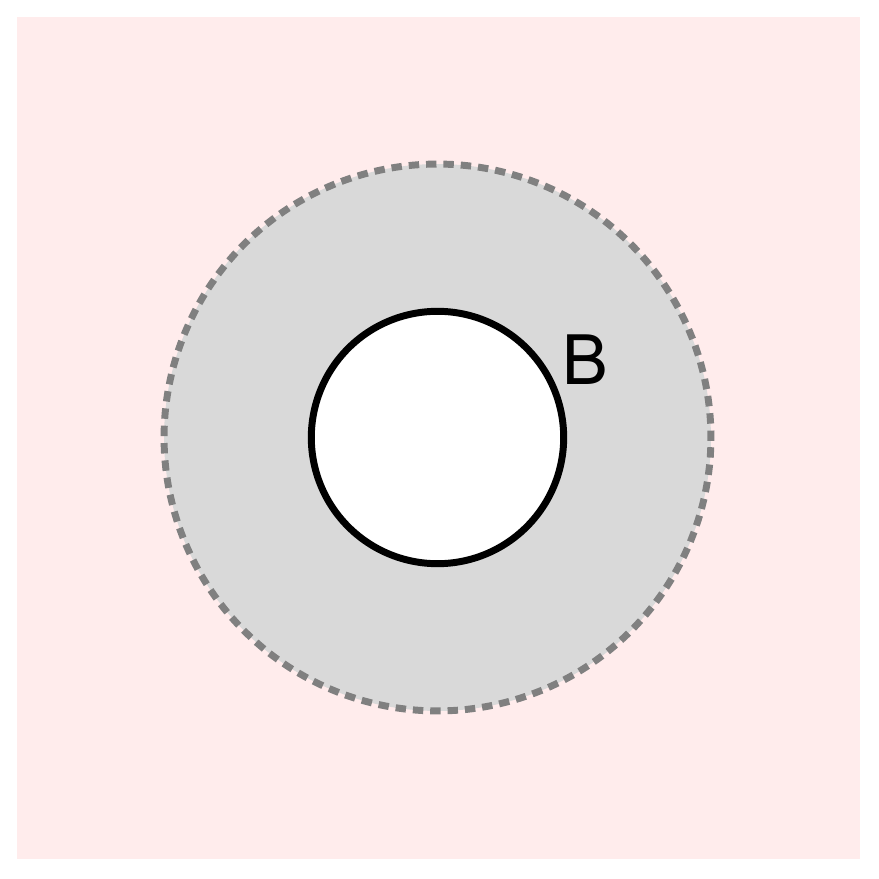}
\end{minipage}%
\caption{A relevant deformation inside the disc (gray region) triggering an RG flow to a TQFT. In the left picture, a point-like topological operator $v$ in the TQFT is inserted inside the disc. This construction produces an Ishibashi state on the dotted circle, which is generally not a boundary state. In the right picture, we insert a boundary state $B$ of the TQFT instead, which produces a boundary state of the CFT on the dotted circle.}
\label{fig:RGwall}
\end{figure}

Next, consider a relevant operator ${\cal O}$ in a CFT that {\it commutes} with a TDL ${\cal L}$ and drives the CFT to a massive phase. Now turning on the deformation ${\cal O}$ on a disc, and shrinking an ${\cal L}$ loop encircling the disc, we have
\ie
\widehat {\cal L}|{\cal O}\ra\ra_{RG} = \la {\cal L} \ra |{\cal O}\ra\ra_{RG}.
\fe
If $|{\cal O}\ra\ra_{RG}$ is a boundary state, then so must be $\widehat {\cal L}|{\cal O}\ra\ra_{RG}$, as it corresponds to fusing the TDL ${\cal L}$ onto the boundary, but this is clearly impossible when $\la {\cal L} \ra$ is not an integer.

As an example, consider the tricritical Ising model with ${\cal O}=\sigma'$, and the TDL $W$ that commutes with $\sigma'$ (see Section~\ref{lytqft}).  Consider the following two Ishibashi states: $|\sigma', 1\ra\ra_{RG}$ and $|\sigma', v_x\ra\ra_{RG}$, where 1 and $v_x$ denote the bulk operator of the IR TQFT inserted in the disc, as depicted in Figure~\ref{fig:RGwall}.
Since the two Ishibashi states have $\widehat W$ eigenvalues $\zeta = {1+\sqrt{5}\over 2}$ and $-\zeta^{-1}$, respectively, neither is a boundary state. However, an actual boundary state is produced if we insert a boundary state of the IR TQFT inside the disc. There are two irreducible boundary states
\ie
B_1 = 1+v_x,\quad  B_2 = \zeta - \zeta^{-1} v_x.
\fe
Indeed, $\la B_1 B_1 \ra=2$, $\la B_2 B_2\ra = 3$, and $\la B_1 B_2\ra = 1$ are the numbers of states in the strip Hilbert spaces ${\cal H}_{B_1 B_1}$, ${\cal H}_{B_2 B_2}$, and ${\cal H}_{B_1B_2}$ of the TQFT, respectively. Inserting these in the interior of the disc, we can produce the two boundary states of the tricritical Ising model
\ie
|\sigma', 1\ra\ra_{RG} + |\sigma', v_x\ra\ra_{RG}\quad {\rm and}\quad  \zeta |\sigma', 1\ra\ra_{RG} -\zeta^{-1} |\sigma', v_x\ra\ra_{RG}.
\fe

\subsection{Coupled minimal models}
\label{sec:coupledminmodels}

We consider an RG flow that is not known to be integrable,
starting from the tensor product of $n$ copies of the three-state Potts model, deformed by the relevant operator (see Section~\ref{potts})
\ie\label{deps}
{\cal O}=\sum_{1\leq i<j\leq n} \varepsilon_i \varepsilon_j.
\fe
The deformation ${\cal O}$ preserves the global symmetry $S_n \ltimes S_3^n$, and the TDL $N\equiv \prod_{i=1}^n N_i$.\footnote{The deformation ${\cal O}$ also preserves the analogous TDLs where an arbitrary number of the $N_i$ lines are replaced by $N_i' = C_iN_i$.
}
Note that ${\cal O}$ and $\sum_{i=1}^n\varepsilon_i$ are the only relevant operators preserving all the global symmetries, but $\sum_{i=1}^n\varepsilon_i$ anticommutes with $N$, whereas ${\cal O}$ commutes with $N$. Therefore, no new relevant operator can be generated in the RG flow generated by ${\cal O}$.

Note that in the large $n$ limit, ${\cal O}$ may be viewed as a double trace deformation, and a $1 \over n$ expansion may be employed to compute the spectrum and correlation functions at the fixed point. In particular, $\sum_i \varepsilon_i$ flows to an operator of scaling dimension $2-{4\over 5} = {6\over 5}$ in the $n\to \infty$ limit.

The spin selection rule on ${\cal H}_N$ is such that the states in ${\cal H}_N$ have spins,
\ie
\label{CoupledSpins}
s \in {1\over 2}\mathbb{Z} + \sum_{i=1}^n r_i,\quad r_i = \pm {1\over 24} ~{\rm or}~\pm{1\over 8},
\fe
derived from the single-copy selection rule \eqref{TYZ3Spins}. Let us consider the $n=3$ case. In this case, ${\cal H}_N$ contains states of spin $s\in {1\over 24} + {1\over 12}\mathbb{Z}$.

It follows that theory must either flow to an IR fixed point that admits a TDL $N$ that obeys the same spin selection rule \eqref{CoupledSpins}, or to a massive phase with degenerate vacua, such that ${\rm tr}\,\widehat N=0$. 

The flow of the three-coupled Potts model has been studied using conformal perturbation theory in \cite{Dotsenko:1999erv}. With positive coupling, the flow is expected to end at an IR fixed point with central charge $c\approx 2.38$. The existence of the TDL $N$ at the IR fixed point implies that the N-twisted character ${\rm tr}\, \widehat N q^{L_0-{c\over 24}} \bar q^{\bar L_0 - {\tilde c\over 24}}$ is related by the modular $S$ transformation to a partition sum over ${\cal H}_N$, whose states obey the spin selection rule mentioned above. The TDL $N$ also constrains the OPE of bulk local operators along the RG flow. These constraints should be useful for the conformal/modular bootstrap study of the CFT at the IR fixed point.

Another example of a similar nature is the $n$-coupled tricritical Ising model, defined as the tensor product of $n$ tricritical Ising models deformed by the relevant operator (see Section~\ref{Sec:Trising})
\ie
{\cal O} = \sum_{1\leq i<j\leq n} \sigma_i' \sigma_j'.
\fe 
The $W$ lines in all $n$ copies of the tricritical Ising model, the $S_n$ permutation symmetry, and the overall $\mathbb{Z}_2$ symmetry that flips the spin fields of all $n$ copies are preserved along the RG flow.

In the special case of $n=2$, the coupled tricritical Ising model corresponds to the deformation of the $SU(2)_8/U(1)$ coset CFT by the parafermion bilinear and flows to the $c={14\over 15}$ $A_7$ minimal model in the IR. For $n \geq 3$, the flow is not expected to be integrable. In the large $n$ limit, ${\cal O}$ can once again be viewed as a double trace deformation, and in particular $\sum_i \sigma_i'$ flows to an operator of scaling dimension $2-{7\over 8} = {9\over 8}$ to leading order in $1 \over n$.

\section{Summary and discussions}
\label{sec:conclusion}

\subsection{On IR TQFTs}

Much of this paper has been devoted to formulating the definition of TDLs, constructing them in CFTs as models of various fusion categories, and deriving properties of defect operators such as the spin selection rules from the crossing relations of TDLs. One particularly interesting set of results is the use of topological defect lines in constraining RG flows, as a generalization of the 't Hooft anomaly matching. This is particularly powerful in constraining, and sometimes determining, the TQFT in the IR of a massive RG flow. Curiously, our arguments made essential use of the modular invariance of the TQFT as well as the existence of (topological) defect operators therein, ingredients that are absent in the standard definition of fusion categories. An interesting question is whether every fusion category can be modeled by a fully extended, modular invariant TQFT. This is {\it a priori} not obvious, for instance, for the ${1\over 2}E_6$ fusion category, but as we have argued, it should be realized in the IR TQFT of the $(A_{10},E_6)$ minimal model perturbed by $\phi_{2,1}$. Assuming a minimal admissible number of vacua, this TQFT was constructed in Section~\ref{Sec:A10E6}.

In several examples, we determined the structure constants of the IR TQFT by consideration of the TDLs. We emphasize that the former is not constrained by the associativity of the OPE of bulk local operators alone. It should be possible to check these results by studying the RG flow of three-point functions of bulk local operators numerically using the truncated conformal space approach (TCSA) \cite{Yurov:1989yu}. Note that the basis of vacua we worked with in the TQFT may be nontrivial linear combinations of the ones that obey cluster decomposition. It would be interesting to understand the relation between the TQFT and the data of massive particle excitations and their S-matrix.

\subsection{TDLs in irrational CFTs}

We have seen that TDLs are ubiquitous in rational CFTs, including invertible lines associated with global symmetries, and in the case of diagonal modular invariant theories, Verlinde lines associated with the chiral vertex algebra, and there are also more general TDLs that are neither invertible lines nor Verlinde lines. A large class of non-invertible TDLs in irrational CFTs can be constructed as Wilson lines in non-Abelian orbifold theories \cite{Brunner:2014lua,Bhardwaj:2017xup}. Non-invertible TDLs are also present in irrational, unitary, compact CFTs obtained as fixed points of RG flows, such as in coupled minimal models, and would be useful in bootstrapping such theories by constraining OPEs and refining modular constraints. A natural question, to which we do not know the answer, is whether TDLs exist  in ``more generic" irrational CFTs.

To illustrate with a simple (though not necessarily typical) example, consider the CFT described by a sigma model whose target space is the rectangular torus with modulus $\tau = it$. For $t=1$, the $T^2$ admits a rotation symmetry by 90 degrees, whose corresponding invertible line is denoted by $\eta$. The sigma model with $t=p/q$, where $p,q$ are a pair of positive coprime integers, can be viewed as the $\mathbb{Z}_p\times \mathbb{Z}_q$ orbifold of a sigma model on a larger $T^2$ target space with $t=1$, and radius $R$. Denote by $T_x$ the translation symmetry (or the corresponding invertible line) of the latter CFT along the $x$ direction by $2\pi R/p$, and $T_y$ the translation along $y$ direction by $2\pi R/q$, and $\eta$ the $\mathbb{Z}_4$ invertible line corresponding to the 90 degree rotation thereof. The TDL
\ie
{\cal L}=\sum_{0\leq n\leq p-1, 0\leq m\leq q-1} T_x^n T_y^m \eta T_y^{-m} T_x^{-n}
\fe
is invariant under the $\mathbb{Z}_p\times \mathbb{Z}_q$ symmetry, and gives a simple TDL in the orbifolded theory, {\it i.e.}, the sigma model on the torus with $t=p/q$. However, in the limit where $t$ becomes irrational, the fusion relation of ${\cal L}$ would involve an infinite sum of simple TDLs, which goes beyond the class of topological defects considered in this paper. It may be interesting to relax the assumption that the fusion product of a pair of TDLs involves only the direct sum of finitely many simple lines. This possibility has already been considered in the context of Liouville and Toda CFTs \cite{Sarkissian:2009aa, Drukker:2010jp}.

\subsection{From topological to conformal defects}

The TDLs in CFTs are a special case of conformal defects, or conformal interfaces. A general conformal interface ${\cal I}$ of a CFT $M$ can be characterized by the interface state $|{\cal I}\ra\ra$ which is equivalent to a boundary state of $M\otimes \overline M$, where $\overline M$ is the parity reversal of $M$ \cite{Oshikawa:1996ww}. Let $T_L$ and $T_R$ be the (non-singular) limit of the stress tensor $T(z)$ that approaches ${\cal I}$ from the left and right, respectively. They are related to the displacement operator $D$ by
\ie
D = T_L - T_R,
\fe
where $D$ is a dimension-2 operator on ${\cal I}$ that generates transverse deformations of the interface. A TDL is a conformal interface with $D(x)\equiv0$. While the fusion of TDLs is straightforward to describe, the fusion of the general conformal interfaces is much more complicated. For instance, generically the limit of a pair of conformal interfaces approaching one another is singular; viewed as two conformal defects points on the spatial circle approaching one another, there may be a Casimir energy that diverges in the coincidence limit. Further, the fusion of a pair of conformal interfaces may involve the direct sum of infinitely many conformal interfaces, each deformed by an infinite set of relevant or irrelevant local operators on the interface. We do not know a useful formulation of the fusion of general conformal interfaces that is analogous to the OPE of local operators.

It is possible to tame the fusion of conformal interfaces if the latter is obtained by deforming TDLs. Such deformations could be either due to a relevant or marginal deformation of a TDL ${\cal L}$ by a defect operator in ${\cal H}_{{\cal L}\overline{\cal L}}$ (of scaling dimension less than or equal to 1), or due to a relevant or marginal deformation of the bulk CFT. In the latter case, for instance, one may hope that the deformed conformal interface inherit certain fusion properties of the TDL.  This strategy has been applied in studying the fusion between conformal interfaces in the  critical Ising model \cite{Bachas:2013ora}. It would be interesting to explore the deformation of TDLs under exactly marginal deformations along a conformal manifold of a family of CFTs.

\section*{Acknowledgements} 

We would like to thank Philip Argyres, Nathan Benjamin, Lakshya Bhardwaj, Aleksey Cherman, Minjae Cho, Scott Collier, Clay Cordova, Thomas Dumitrescu, Davide Gaiotto, Zohar Komargodski, Liang Kong, Kantaro Ohmori,  Hirosi Ooguri, Nathan Seiberg, David Simmons-Duffin, Yuji Tachikawa, Cumrun Vafa, Juven Wang, Edward Witten, and Cenke Xu for discussions. 
We also thank Minjae Cho, Scott Collier, Liang Kong, and Yuji Tachikawa for comments on a preliminary draft. 
This project was initiated at the working group {\it 2D Conformal and Modular Bootstrap} at the Aspen Center for Physics. CC thanks California Institute of Technology, Yau Mathematical Sciences Center at Tsinghua University, XY thanks Boston University, Rutgers University, Mitchell Institute at Texas A\&M University, CC and YW thank Superconformal Field Theories in 6 and Lower Dimensions Workshop at Tsinghua Sanya International Mathematics Forum, YW and XY thank the annual meeting of the Simons Collaboration on Non-Perturbative Bootstrap, for their hospitality during the course of this work. CC is supported in part by the U.S. Department of Energy grant DE-SC0009999. YL is supported by the Sherman Fairchild Foundation, and by the U.S. Department of Energy, Office of Science, Office of High Energy Physics, under Award Number DE-SC0011632. SHS is supported by the Zurich Insurance Company Membership and the National Science Foundation grant PHY-1314311. YW is supported in part by the US NSF under Grant No.~PHY-1620059 and by the Simons Foundation Grant No.~488653. XY is supported by a Simons Investigator Award from the Simons Foundation and by DOE grant DE-FG02-91ER40654.

\appendix

\section{Gauge conditions on crossing kernels}
\label{CKbasis}

To solve the pentagon identity
\ie
&{\widetilde K}_{{\cal L}_{j_2},{\cal L}_{k_2}}^{{\cal L}_{j_1},{\cal L}_{k_1}}({\cal L}_{j},\overline{\cal L}_{k_3}) \circ {\widetilde K}_{{\cal L}_{j_4},{\cal L}_{k_4}}^{{\cal L}_{j},{\cal L}_{k_1}}(\overline{\cal L}_{j_3},{\cal L}_{k_2})
\\
&=\sum_{j'}{\widetilde K}_{{\cal L}_{j_4},{\cal L}_{k_4}}^{{\cal L}_{j_2},{\cal L}_{k_3}}({\cal L}_{j'},{\cal L}_{k_2})\circ {\widetilde K}_{{\cal L}_{j'},{\cal L}_{k_4}}^{{\cal L}_{j_1},{\cal L}_{k_1}}(\overline{\cal L}_{j_3},\overline{\cal L}_{k_3})\circ {\widetilde K}_{{\cal L}_{j_2},{\cal L}_{j_4}}^{{\cal L}_{j_1},{\cal L}_{j_3}}({\cal L}_{j},{\cal L}_{j'}),
\fe
it is practical to first remove the redundancies in the crossing kernels due to changes in the basis junction vectors. This appendix provides a set of conditions to fix the redundancies, that are not necessarily complete.

If all the TDLs involved in the crossing kernels are trivial, the pentagon identity immediately implies that 
\ie
 {\widetilde K}_{I,I}^{I,I}(I,I)=1.
\fe 
For the more general crossing kernels, we need to pick a basis for the junction vector space, which is specified by a coordinate map 
\ie\label{ncoordmap}
N_{{\cal L}_1,{\cal L}_2,{\cal L}_3} ~:~ V_{{\cal L}_1,{\cal L}_2,{\cal L}_3}\xrightarrow{\sim} \bC^{d_{123}},
\fe
where $d_{123}={\rm dim}\,(V_{{\cal L}_1,{\cal L}_2,{\cal L}_3})$ is the dimension of the junction vector space. A change of basis in $V_{{\cal L}_1,{\cal L}_2,{\cal L}_3}$ transforms the coordinate map by
\ie
N_{{\cal L}_1,{\cal L}_2,{\cal L}_3}\sim M_{{\cal L}_1,{\cal L}_2,{\cal L}_3}N_{{\cal L}_1,{\cal L}_2,{\cal L}_3},
\fe
where $M_{{\cal L}_1,{\cal L}_2,{\cal L}_3}$ is a $d_{123}\times d_{123}$ matrix. In the basis specified by the coordinate map (\ref{ncoordmap}), the crossing kernel is explicitly written as
\ie\label{crossingWbasis}
{\widetilde {\cal K}}_{{\cal L}_2,{\cal L}_3}^{{\cal L}_1,{\cal L}_4}({\cal L}_5,{\cal L}_6)&=(N_{{\cal L}_2,{\cal L}_3,\overline{\cal L}_6}\otimes N_{{\cal L}_1,{{\cal L}_6}, {\cal L}_4})\circ{\widetilde K}_{{\cal L}_2,{\cal L}_3}^{{\cal L}_1,{\cal L}_4}({\cal L}_5,{\cal L}_6) \circ (N_{{\cal L}_1,{\cal L}_2,\overline{\cal L}_5}^{-1} \otimes N_{{{\cal L}_5} ,{\cal L}_3,{\cal L}_4}^{-1}).
\fe
A change of basis by $M$ leads to a transformation of ${\widetilde {\cal K}}$ by
\ie\label{gauge}
\hspace{-.05in}
{\widetilde {\cal K}}_{{\cal L}_2,{\cal L}_3}^{{\cal L}_1,{\cal L}_4}({\cal L}_5,{\cal L}_6)&\sim (M_{{\cal L}_2,{\cal L}_3,\overline{\cal L}_6}\otimes M_{{\cal L}_1,{{\cal L}_6} ,{\cal L}_4})\circ {\widetilde {\cal K}}_{{\cal L}_2,{\cal L}_3}^{{\cal L}_1,{\cal L}_4}({\cal L}_5,{\cal L}_6) \circ  (M_{{\cal L}_1,{\cal L}_2,\overline{\cal L}_5}^{-1} \otimes M_{{{\cal L}_5} ,{\cal L}_3,{\cal L}_4}^{-1}),
\fe
where $\circ$ represents the suitable contraction of the matrix indices. We will refer to this change of basis as a ``gauge transformation" on the crossing kernel.

To begin with, consider the crossing kernels that involve a pair of trivial external lines,
\ie\label{twoIdentities}
{\widetilde {\cal K}}_{{\cal L},\overline {\cal L}}^{I, I}({\cal L},I),\quad {\widetilde {\cal K}}^{\overline{\cal L} ,{\cal L}}_{I, I}(\overline{\cal L},I),\quad {\widetilde {\cal K}}_{I,{\cal L}}^{I,\overline{\cal L}}(I,{\cal L}),\quad {\widetilde {\cal K}}_{\overline{\cal L},I}^{{\cal L},I}(I,\overline {\cal L}),\quad {\widetilde {\cal K}}_{I,{\cal L}}^{\overline{\cal L},I}(\overline{\cal L},{\cal L}),\quad {\widetilde {\cal K}}_{{\cal L},I}^{I,\overline{\cal L}}({\cal L}, {\cal L}).
\fe
Using the gauge rotation $M_{I,{\cal L},\overline{\cal L}}$ and $M_{\overline{\cal L},I,{\cal L}}$ on the corresponding trivial junctions,
\ie
&{\widetilde {\cal K}}_{{\cal L},\overline {\cal L}}^{I ,I}({\cal L},I)\sim(M_{{\cal L},\overline {\cal L},I}\otimes M_{I, I,I})\circ{\widetilde {\cal K}}_{{\cal L},\overline {\cal L}}^{I, I}({\cal L},I) \circ (M_{I,{\cal L},\overline{\cal L}}^{-1}\otimes M_{{{\cal L}} ,\overline {\cal L}, I}^{-1}),
\\
&{\widetilde {\cal K}}^{\overline{\cal L}, {\cal L}}_{I, I}(\overline{\cal L},I)\sim (M_{I,I,I}\otimes M_{\overline{\cal L},I, {\cal L}})\circ{\widetilde {\cal K}}^{\overline{\cal L} ,{\cal L}}_{I ,I}(\overline{\cal L},I)\circ(M_{\overline{\cal L},I,{\cal L}} ^{-1} \otimes M_{{\overline{\cal L}}, I ,{\cal L} }^{-1}),
\fe
we can fix the first two crossing kernels in \eqref{twoIdentities} to be
\ie\label{SIGC}
{\widetilde {\cal K}}_{{\cal L},\overline {\cal L}}^{I ,I}({\cal L},I)= {\widetilde {\cal K}}^{\overline{\cal L}, {\cal L}}_{I, I}(\overline{\cal L},I)=1.
\fe
The pentagon identities
\ie
{\widetilde {\cal K}}_{I,I}^{I,I}(I,I)\circ {\widetilde {\cal K}}_{{\cal L},\overline{\cal L}}^{I,I}({\cal L},I)&={\widetilde {\cal K}}_{{\cal L},\overline{\cal L}}^{I,I}({\cal L},I) \circ{\widetilde {\cal K}}_{{\cal L},\overline{\cal L}}^{I,I}({\cal L},I) \circ{\widetilde {\cal K}}_{I,{\cal L}}^{I,\overline{\cal L}}(I,{\cal L}),
\\
{\widetilde {\cal K}}_{\overline {\cal L},I}^{{\cal L},I}(I,\overline{\cal L})\circ {\widetilde {\cal K}}_{I,I}^{I,I}(I,I)&={\widetilde {\cal K}}_{I,I}^{\overline {\cal L},{\cal L}}(\overline {\cal L},I) \circ{\widetilde {\cal K}}_{\overline {\cal L},I}^{{\cal L},I}(I,\overline{\cal L})\circ {\widetilde {\cal K}}_{\overline {\cal L},I}^{{\cal L},I}(I,\overline {\cal L}),
\\
{\widetilde {\cal K}}_{{\cal L},I}^{I,\overline{\cal L}}({\cal L},{\cal L})\circ {\widetilde {\cal K}}^{{\cal L},\overline{\cal L}}_{I,I}({\cal L},I)&={\widetilde {\cal K}}^{{\cal L},\overline{\cal L}}_{I,I}({\cal L},I)\circ{\widetilde {\cal K}}_{{\cal L},I}^{I,\overline{\cal L}}({\cal L},{\cal L})\circ{\widetilde {\cal K}}_{{\cal L},I}^{I,\overline{\cal L}}({\cal L},{\cal L}),
\\
{\widetilde {\cal K}}_{I,{\cal L}}^{\overline{\cal L},I}(\overline{\cal L},{\cal L}) \circ{\widetilde {\cal K}}_{I,{\cal L}}^{\overline{\cal L},I}(\overline{\cal L},{\cal L})&={\widetilde {\cal K}}_{I,{\cal L}}^{I\overline {\cal L}}(I,{\cal L}) \circ{\widetilde {\cal K}}_{I ,{\cal L}}^{\overline{\cal L},I}(\overline{\cal L},{\cal L})\circ{\widetilde {\cal K}}_{I,I}^{\overline{\cal L},{\cal L}}(\overline {\cal L},I),
\fe
then determine
\ie
 {\widetilde {\cal K}}_{I,{\cal L}}^{I,\overline{\cal L}}(I,{\cal L})= {\widetilde {\cal K}}_{\overline{\cal L},I}^{{\cal L},I}(I,\overline {\cal L})= {\widetilde {\cal K}}_{I,{\cal L}}^{\overline{\cal L},I}(\overline{\cal L},{\cal L})= {\widetilde {\cal K}}_{{\cal L},I}^{I,\overline{\cal L}}({\cal L}, {\cal L})=1.
\fe

The following crossing kernels that involve one trivial external line,
\ie\label{SICK}
{\widetilde {\cal K}}_{{\cal L}_2,{\cal L}_3}^{I,{\cal L}_4}({\cal L}_2,\overline{\cal L}_4),\quad{\widetilde {\cal K}}_{I,{\cal L}_3}^{{\cal L}_1,{\cal L}_4}({\cal L}_1,{\cal L}_3),\quad{\widetilde {\cal K}}_{{\cal L}_2,I}^{{\cal L}_1,{\cal L}_4}(\overline{\cal L}_4,{\cal L}_2)
\fe
are fixed to be identity matrices by the pentagon identities
\ie
{\widetilde {\cal K}}_{I,\overline {\cal L}_{4}}^{I,{\cal L}_4}(I,\overline{\cal L}_4)\circ {\widetilde {\cal K}}_{{\cal L}_2,{\cal L}_3}^{I,{\cal L}_4}({\cal L}_2,\overline{\cal L}_4)&={\widetilde {\cal K}}_{{\cal L}_2,{\cal L}_3}^{I,{\cal L}_4}({\cal L}_2,\overline{\cal L}_4)\circ {\widetilde {\cal K}}_{{\cal L}_2,{\cal L}_3}^{I,{\cal L}_4}({\cal L}_2,\overline{\cal L}_4)\circ {\widetilde {\cal K}}_{I,{\cal L}_2}^{I,\overline{\cal L}_2}(I,{\cal L}_2),
\\
{\widetilde {\cal K}}_{I,{\cal L}_3}^{{\cal L}_1,{\cal L}_4}({\cal L}_1,{\cal L}_3)\circ {\widetilde {\cal K}}_{I,{\cal L}_3}^{{\cal L}_1,{\cal L}_4}({\cal L}_1,{\cal L}_3)&={\widetilde {\cal K}}_{I,{\cal L}_3}^{I,\overline{\cal L}_3}(I,{\cal L}_3)\circ {\widetilde {\cal K}}_{I,{\cal L}_3}^{{\cal L}_1,{\cal L}_4}({\cal L}_1,{\cal L}_3)\circ {\widetilde {\cal K}}_{I,I}^{{\cal L}_1,\overline{\cal L}_1}({\cal L}_1,I),
\\
{\widetilde {\cal K}}_{{\cal L}_2,I}^{{\cal L}_1,{\cal L}_4}(\overline{\cal L}_4,{\cal L}_2)\circ {\widetilde {\cal K}}^{\overline{\cal L}_4,{\cal L}_4}_{I,I}(\overline{\cal L}_4,I)&={\widetilde {\cal K}}_{I,I}^{{\cal L}_2,\overline{\cal L}_2}({\cal L}_2,I)\circ {\widetilde {\cal K}}_{{\cal L}_2,I}^{{\cal L}_1,{\cal L}_4}(\overline{\cal L}_4,{\cal L}_2)\circ {\widetilde {\cal K}}_{{\cal L}_2,I}^{{\cal L}_1,{\cal L}_4}(\overline{\cal L}_4,{\cal L}_2).
\fe

Next, consider the crossing kernels ${\widetilde {\cal K}}_{{\cal L},\overline{\cal L}}^{\overline{\cal L},{\cal L}}(I,I)$, which transforms under the gauge transformation as
\ie
{\widetilde {\cal K}}_{{\cal L},\overline{\cal L}}^{\overline{\cal L},{\cal L}}(I,I)& \sim (M_{{\cal L},\overline{\cal L},I}\otimes M_{\overline{\cal L},I, {\cal L}})\circ{\widetilde {\cal K}}_{{\cal L},\overline{\cal L}}^{\overline{\cal L},{\cal L}}(I,I)  \circ(M_{\overline{\cal L},{\cal L},I}\otimes M_{I,\overline{\cal L},{\cal L}})^{-1}.
\fe
In the case $\cal L\neq\overline{\cal L}$, {\it i.e.}, ${\cal L}$ is a TDL of a different type from its orientation reversal $\overline{\cal L}$, using the gauge freedom of $M_{{\cal L},\overline {\cal L},I}$, we can fix
\ie\label{eqn:GCinterID}
{\widetilde {\cal K}}_{{\cal L},\overline{\cal L}}^{\overline{\cal L},{\cal L}}(I,I)={\widetilde {\cal K}}_{\overline{\cal L},{\cal L}}^{{\cal L},\overline{\cal L}}(I,I)=|{\widetilde {\cal K}}_{\overline{\cal L},{\cal L}}^{{\cal L},\overline{\cal L}}(I,I)|.
\fe
Note that the product ${\widetilde {\cal K}}_{{\cal L},\overline{\cal L}}^{\overline{\cal L},{\cal L}}(I,I){\widetilde {\cal K}}_{\overline{\cal L},{\cal L}}^{{\cal L},\overline{\cal L}}(I,I)$ is invariant under the gauge transformation generated by $M_{{\cal L},\overline {\cal L},I}$.  This gauge condition implies the relation between the empty loop expectation values on the plane, 
\ie
R({\cal L})=R(\overline{\cal L})=|R(\overline{\cal L})|.
\fe
It follows from the relation between $R({\cal L})$ and the vacuum expectation value $\la {\cal L}\ra$ on the cylinder that the isotopy anomaly coefficient $\A_{\cal L}$ is given by 
\ie
\alpha_{\cal L}=\begin{cases}0\quad{\rm for}\quad  \la {\cal L}\ra>0,
\\
\pi\quad{\rm for}\quad  \la {\cal L}\ra<0.
\end{cases}
\fe
Note that $\la {\cal L}\ra>0$ is required by unitarity.
In the case $\cal L=\overline{\cal L}$, ${\widetilde {\cal K}}_{{\cal L},{\cal L}}^{{\cal L},{\cal L}}(I,I)$ is invariant under the gauge transformation generated by $M_{{\cal L},\overline {\cal L},I}$, and cannot be used for gauge-fixing.

Finally, we want to fix the gauge freedom $M_{{\cal L}_i,{\cal L}_j,{\cal L}_k}$ for ${\cal L}_i$, ${\cal L}_j$, ${\cal L}_k\neq I$. Consider the crossing kernels (differing from the ones in (\ref{SICK}) by the position of the trivial external line)
\ie\label{KKL1L2L3}
{\widetilde {\cal K}}_{{\cal L}_j,{\cal L}_k}^{{\cal L}_i,I}(\overline{\cal L}_k,\overline{\cal L}_i)\quad{\rm for}\quad {\cal L}_i,{\cal L}_j,{\cal L}_k\neq I,
\fe
whose gauge transformations take the form
\ie\label{gaugeLiLjLk}
{\widetilde {\cal K}}_{{\cal L}_j,{\cal L}_k}^{{\cal L}_i,I}(\overline{\cal L}_k,\overline{\cal L}_i)&\sim (M_{{\cal L}_j,{\cal L}_k,{\cal L}_i}\otimes M_{{\cal L}_i,{\overline{\cal L}_i} ,I})\circ{\widetilde {\cal K}}_{{\cal L}_j,{\cal L}_k}^{{\cal L}_i,I}(\overline{\cal L}_k,\overline{\cal L}_i)  \circ(M_{{\cal L}_i,{\cal L}_j,{\cal L}_k}^{-1}\otimes M_{{\overline{\cal L}_k} ,{\cal L}_k,I}^{-1}).
\fe
We will fix an ordering convention on the index $i$ that labels the type of the TDL ${\cal L}_i$. The orientation reversal $\overline {\cal L}_i$ will be labeled by $\overline i$. When the indices $i,j,k$ satisfy
\ie\label{cIset}
(i,j,k)\notin {\cal I}\equiv\{(i,j,k)|i>j>k~{\rm or}~i\le j\le k\},
\fe
the crossing kernels ${\widetilde {\cal K}}_{{\cal L}_j{\cal L}_k}^{{\cal L}_iI}(\overline{\cal L}_k,\overline{\cal L}_i)$ can be gauge-fixed to identity matrices by \eqref{gaugeLiLjLk}.

We are still left with the unfixed gauge transformations $M_{{\cal L}_i,{\cal L}_j,{\cal L}_k}$ for $(i,j,k)\in {\cal I}$, and we will not attempt to fix them in the most general setting. Let us consider the special case where all the TDLs involved are of the same type as their orientation reversals, ${\cal L}_i=\overline{\cal L}_i$. In this case, the crossing kernels
\ie
\quad{\widetilde {\cal K}}_{{\cal L}_j,{\cal L}_j}^{{\cal L}_i,{\cal L}_i}({\cal L}_k,I),\quad{\widetilde {\cal K}}_{{\cal L}_i,{\cal L}_j}^{{\cal L}_i,{\cal L}_j}(I,{\cal L}_k)\quad{\rm for}\quad {\cal L}_i,{\cal L}_j,{\cal L}_k\neq I
\fe
are subject to gauge redundancies of the form
\ie\label{internalidentity}
{\widetilde {\cal K}}_{{\cal L}_j,{\cal L}_j}^{{\cal L}_i,{\cal L}_i}({\cal L}_k,I)&\sim (M_{{\cal L}_j,{\cal L}_j,I}\otimes M_{{\cal L}_i, I,{\cal L}_i})\circ{\widetilde {\cal K}}_{{\cal L}_j,{\cal L}_j}^{{\cal L}_i,{\cal L}_i}({\cal L}_k,I)  \circ(M_{{\cal L}_i,{\cal L}_j,{\cal L}_k}^{-1}\otimes M_{{{\cal L}_k} ,{\cal L}_j,{\cal L}_i}^{-1}),
\\
{\widetilde {\cal K}}_{{\cal L}_i,{\cal L}_j}^{{\cal L}_i,{\cal L}_j}(I,{\cal L}_k)& \sim (M_{{\cal L}_i,{\cal L}_j,{\cal L}_k}\otimes M_{{\cal L}_i,{{\cal L}_k}, {\cal L}_j})\circ{ K}_{{\cal L}_i,{\cal L}_j}^{{\cal L}_i,{\cal L}_j}(I,{\cal L}_k)  \circ(M_{{\cal L}_i,{\cal L}_i,I}\otimes M_{I,{\cal L}_j,{\cal L}_j})^{-1}.
\fe
For $i \le j \le k$, we can use $M_{{\cal L}_i,{\cal L}_j,{\cal L}_k}$ to gauge-fix the crossing kernel ${\widetilde {\cal K}}_{{\cal L}_j,{\cal L}_j}^{{\cal L}_i,{\cal L}_i}({\cal L}_k,I)$ in the first line above. For $ i>  j>   k$, we can use $M_{{\cal L}_i,{\cal L}_j,{\cal L}_k}$ to gauge-fix ${\widetilde {\cal K}}_{{\cal L}_i{\cal L}_j}^{{\cal L}_i{\cal L}_j}(I,{\cal L}_k)$ to gauge-fix the second line.

By the gauge conditions described here, the pentagon identity can be implemented in \texttt{Mathematica} to find the explicit solutions, for the fusing rings up to rank-three that are considered in this paper.

\section{Recovering fusion ring from crossing kernels}
\label{sec:FR}

\begin{figure}[H]
\centering
\hspace{-1.8in}
\begin{minipage}{0.1\textwidth}\begin{eqnarray*} \quad \sum_{{\cal L}_i} \\ \end{eqnarray*}
\end{minipage}%
\begin{minipage}{0.18\textwidth}
\includegraphics[width=1\textwidth]{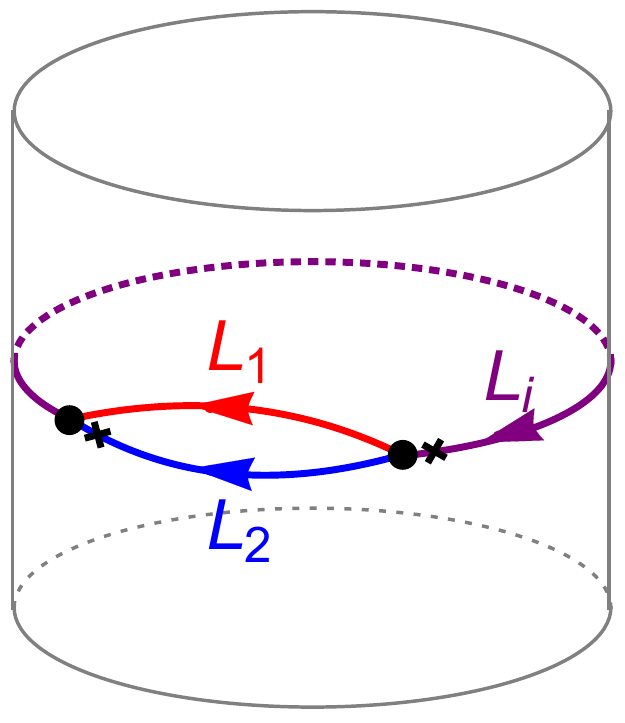}
\end{minipage}%
\hspace{-.3in}
\begin{minipage}{0.3\textwidth}\begin{eqnarray*} \circ ~ {\widetilde K}^{\overline{\cal L}_1,\overline{\cal L}_2}_{{\cal L}_1,{\cal L}_2}(I,{\cal L}_i) \\ \end{eqnarray*}
\end{minipage}%
\\
\hspace{-1.9in}
\begin{minipage}{0.3\textwidth}\begin{eqnarray*} 
=\sum_{{\cal L}_i} \\ \end{eqnarray*}
\end{minipage}%
\hspace{-.7in}
\begin{minipage}{0.18\textwidth}
\includegraphics[width=1\textwidth]{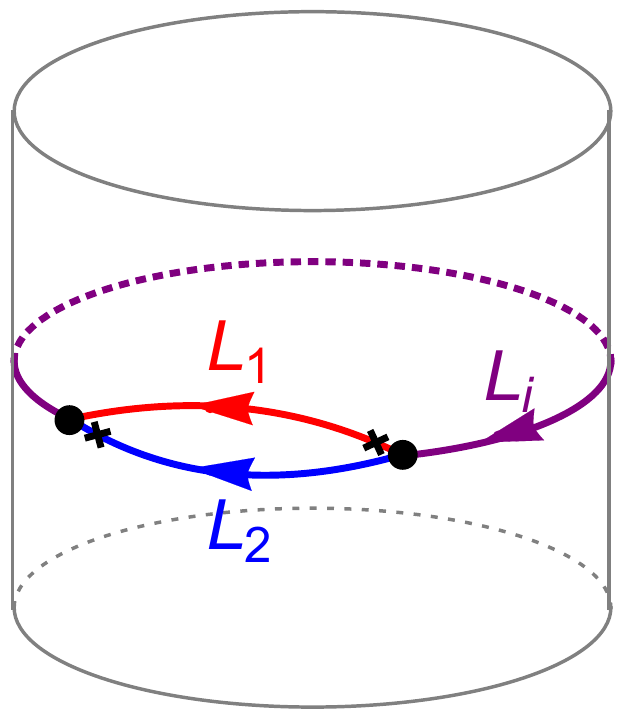}
\end{minipage}%
\begin{minipage}{0.35\textwidth}\begin{eqnarray*}  \circ ~ C_{{\cal L}_1,{\cal L}_2,\overline{\cal L}_i} \, \circ ~ {\widetilde K}^{\overline{\cal L}_1,\overline{\cal L}_2}_{{\cal L}_1,{\cal L}_2}(I,{\cal L}_i) \\ \end{eqnarray*}
\end{minipage}%
\\
\begin{minipage}{0.1\textwidth}\begin{eqnarray*}  =\sum_{{\cal L}_i}\\  \end{eqnarray*}
\end{minipage}%
\begin{minipage}{0.18\textwidth}
\includegraphics[width=1\textwidth]{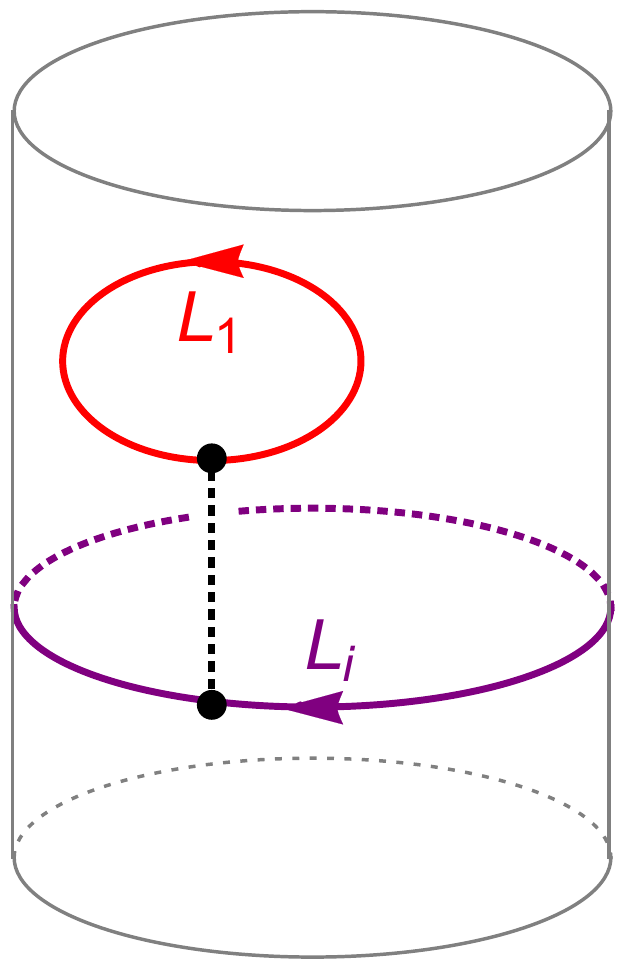}
\end{minipage}%
\begin{minipage}{0.53\textwidth}\begin{eqnarray*}{\widetilde K}^{\overline{\cal L}_1,{\cal L}_1}_{{\cal L}_i,\overline{\cal L}_i}({\cal L}_2,I)\, \circ ~ C_{{\cal L}_1,{\cal L}_2,\overline{\cal L}_i}\,\circ ~ {\widetilde K}^{\overline{\cal L}_1,\overline{\cal L}_2}_{{\cal L}_1,{\cal L}_2}(I,{\cal L}_i) ~\\  \end{eqnarray*}
\end{minipage}%
\caption{Some steps in the derivation of the fusion relations.}
\label{fig:fuseD}
\end{figure}

In this appendix, we derive a relation between the fusion coefficients and the dimensions of junction vector spaces. Let us start with two TDL loops on a cylinder, as shown on the left of Figure~\ref{fig:fuse}. The H-junction crossing gives a sum over the TDL configurations shown on the first line of Figure~\ref{fig:fuseD}, where $\circ$ represents the suitable contraction of the indices. Next, we apply a permutation on the $ {\cal L}_1,  {\cal L}_2, \overline{\cal L}_i$ junction, and obtain the second line of Figure~\ref{fig:fuseD}. Finally, the third line of Figure~\ref{fig:fuseD} is obtained by applying an H-junction crossing with the middle line being the ${\cal L}_2$. By the vanishing tadpole property, the black dotted line can only be the trivial TDL $I$, and hence, we can replace the empty ${\cal L}_1$ loop by $R(\overline{\cal L}_1)$. Using a specialization of the pentagon identity \eqref{eqn:pentagon} -- which gives the following commutative diagram,
\\

\noindent
\begin{tikzcd}
&V_{\overline {\cal L}_1, {\cal L}_1, I}\otimes V_{I, {\cal L}_2 ,\overline{\cal L}_2}\otimes V_{ {\cal L}_2,\overline {\cal L}_i ,{\cal L}_1 }  
\arrow[rd, "{{\widetilde K}_{ {\cal L}_2,\overline{\cal L}_i}^{I,{\cal L}_1}( {\cal L}_2,\overline{\cal L}_1)}"]
\arrow[ld,"{{\widetilde K}_{ {\cal L}_1, {\cal L}_2}^{\overline{\cal L}_1,\overline{\cal L}_2}(I, {\cal L}_i)}" ]
& & &
\\
V_{ {\cal L}_1, {\cal L}_2,  \overline {\cal L}_i}\otimes V_{\overline{\cal L}_1, {\cal L}_i,\overline{\cal L}_2  }\otimes V_{ {\cal L}_2 ,\overline{\cal L}_i ,{\cal L}_1 }  
\arrow[dd,"{{\widetilde K}_{ {\cal L}_i,\overline{\cal L}_i}^{\overline {\cal L}_1,{\cal L}_1}( {\cal L}_2, I)}"]
& &V_{\overline{\cal L}_1, {\cal L}_1, I}\otimes V_{ {\cal L}_2,\overline {\cal L}_i,  {\cal L}_1}\otimes V_{I,\overline{\cal L}_1 ,{\cal L}_1  } 
\arrow[dd,"{{\widetilde K}_{ {\cal L}_1,\overline{\cal L}_1}^{\overline{\cal L}_1,{\cal L}_1}(I, I)}"] 
\\
\\
V_{ {\cal L}_1, {\cal L}_2 , \overline {\cal L}_i}\otimes V_{ {\cal L}_i ,\overline{\cal L}_i , I}\otimes V_{\overline{\cal L}_1, I,{\cal L}_1  }
\arrow[rr,"{{\widetilde K}_{ {\cal L}_2,\overline{\cal L}_i}^{ {\cal L}_1,I}( {\cal L}_i,\overline{\cal L}_1)}"]
&& V_{ {\cal L}_1 ,\overline{\cal L}_1  ,I}\otimes V_{ {\cal L}_2, \overline{\cal L}_i,  {\cal L}_1}\otimes V_{\overline {\cal L}_1, I, {\cal L}_1}
\end{tikzcd}
\\

\noindent -- as well as the trivial crossing kernel \eqref{eqn:KtoR}, we obtain
\ie
{ {\rm dim}(V_{{\cal L}_1,{\cal L}_2,\overline{\cal L}_i}) \over R(\overline{\cal L}_1) } = \tr_{V_{ {\cal L}_2, \overline{\cal L}_i,  {\cal L}_1}}\left( {\widetilde K}^{{\cal L}_1,I}_{{\cal L}_2,\overline{\cal L}_i}( {\cal L}_i,\overline{\cal L}_1)\circ {\widetilde K}^{\overline{\cal L}_1,{\cal L}_1}_{{\cal L}_i,\overline{\cal L}_i}({\cal L}_2,I)\circ  {\widetilde K}^{\overline{\cal L}_1,\overline{\cal L}_2}_{{\cal L}_1,{\cal L}_2}(I,{\cal L}_i) \right).
\fe
By the relation \eqref{KToC} between the cyclic permutation maps and the crossing kernels, we have
\ie
& \tr_{V_{ {\cal L}_2, \overline{\cal L}_i,  {\cal L}_1}}\left( {\widetilde K}^{{\cal L}_1,I}_{{\cal L}_2,\overline{\cal L}_i}( {\cal L}_i,\overline{\cal L}_1)\circ {\widetilde K}^{\overline{\cal L}_1,{\cal L}_1}_{{\cal L}_i,\overline{\cal L}_i}({\cal L}_2,I)\circ  {\widetilde K}^{\overline{\cal L}_1,\overline{\cal L}_2}_{{\cal L}_1,{\cal L}_2}(I,{\cal L}_i) \right)
\\
&=\tr_{V_{ {\cal L}_2, \overline{\cal L}_i,  {\cal L}_1}}\left( C_{{\cal L}_1,{\cal L}_2,\overline{\cal L}_i}\circ {\widetilde K}^{\overline{\cal L}_1,{\cal L}_1}_{{\cal L}_i,\overline{\cal L}_i}({\cal L}_2,I)\circ  {\widetilde K}^{\overline{\cal L}_1,\overline{\cal L}_2}_{{\cal L}_1,{\cal L}_2}(I,{\cal L}_i) \right)
\\
&= {\widetilde K}^{\overline{\cal L}_1,{\cal L}_1}_{{\cal L}_i,\overline{\cal L}_i}({\cal L}_2,I) \circ C_{{\cal L}_1,{\cal L}_2,\overline{\cal L}_i}\circ  {\widetilde K}^{\overline{\cal L}_1,\overline{\cal L}_2}_{{\cal L}_1,{\cal L}_2}(I,{\cal L}_i),
\fe
where in the second equality, we used the fact that the cyclic permutation map $C_{{\cal L}_1,{\cal L}_2,\overline{\cal L}_i}$ trivially commutes with ${\widetilde K}^{\overline{\cal L}_1,{\cal L}_1}_{{\cal L}_i,\overline{\cal L}_i}({\cal L}_2,I)$.
We dropped the trace in the final line since it is a map between one-dimensional vector spaces $V_{\overline{\cal L}_1, {\cal L}_1, I}\otimes V_{I, {\cal L}_2 ,\overline{\cal L}_2}$ and $V_{ {\cal L}_1 ,\overline{\cal L}_1  ,I}\otimes V_{ \overline{\cal L}_1, I, {\cal L}_1}$. Putting everything together, we arrive at the fusion relation
\ie
{\cal L}_1{\cal L}_2 = \sum_{{\cal L}_i}{\rm dim}(V_{{\cal L}_1,{\cal L}_2,\overline{\cal L}_i}){\cal L}_i.
\fe

\section{Explicit solutions to pentagon identity}
\label{sec:pentasolutions}

This appendix presents explicit solutions to the pentagon identity for several fusion rings of rank three discussed in Section~\ref{Sec:SmallRanks}. Here, ${\widetilde {\cal K}}$ is the crossing kernel ${\widetilde K}$ in the basis specified in Appendix~\ref{CKbasis}.

\subsection{Fusion categories with ${R}_{\bC}(\widehat{so(3)}_5)$ fusion ring}
\label{app:sl2}

There are three inequivalent solutions to the pentagon identity associated to the ${R}_{\bC}(\widehat{so(3)}_5)$ fusion ring defined in Section~\ref{Sec:SmallRanks} (see also Section~\ref{Sec:Penta}). We list the nontrivial crossing kernels below, while the unlisted ones are  1 if the T-junctions involved are allowed, and 0 otherwise.
\ie\label{eqn:sl5sol}
& {\widetilde {\cal K}}^{Y,Y}_{Y,Y}(I,I)=1-\zeta^2, \quad {\widetilde {\cal K}}^{Y,Y}_{Y,Y}(I,Y)=1-\zeta^2, \quad  {\widetilde {\cal K}}^{Y,Y}_{Y,Y}(I,X)=-\zeta^4+3 \zeta^2-1,
\\ 
& {\widetilde {\cal K}}^{Y,Y}_{Y,Y}(Y,Y)=\zeta^4-2 \zeta^2+1, 
~~ {\widetilde {\cal K}}^{Y,Y}_{Y,Y}(Y,X)=-\zeta^4+3 \zeta^2-2, ~~ {\widetilde {\cal K}}^{Y,Y}_{Y,Y}(X,Y)=\zeta^4-2 \zeta^2, 
\\ 
& {\widetilde {\cal K}}^{Y,Y}_{Y,Y}(X,X)=-\zeta^4+3 \zeta^2-1, \quad {\widetilde {\cal K}}^{Y,X}_{Y,Y}(Y,Y)=\zeta^4-2 \zeta^2, \quad {\widetilde {\cal K}}^{Y,X}_{Y,Y}(Y,X)=2 \zeta^3-\zeta^5,
\\ 
& {\widetilde {\cal K}}^{Y,X}_{Y,Y}(X,Y)=2 \zeta^3-\zeta^5, 
\quad{\widetilde {\cal K}}^{Y,X}_{Y,Y}(X,X)=2 \zeta^2-\zeta^4, \quad {\widetilde {\cal K}}^{Y,Y}_{Y,X}(Y,Y)=\zeta^4-2 \zeta^2, 
\\ 
& {\widetilde {\cal K}}^{Y,Y}_{Y,X}(Y,X)=2 \zeta^3-\zeta^5, \quad {\widetilde {\cal K}}^{Y,Y}_{Y,X}(X,Y)=2 \zeta^3-\zeta^5, 
\quad {\widetilde {\cal K}}^{Y,Y}_{Y,X}(X,X)=2 \zeta^2-\zeta^4, 
\\ 
& {\widetilde {\cal K}}^{Y,X}_{Y,X}(I,Y)=1-\zeta^2, \quad {\widetilde {\cal K}}^{Y,X}_{Y,X}(I,X)=1-\zeta^2, \quad {\widetilde {\cal K}}^{Y,X}_{Y,X}(Y,Y)=2 \zeta^3-\zeta^5,
\\ 
& {\widetilde {\cal K}}^{Y,X}_{Y,X}(Y,X)=-\zeta, \quad {\widetilde {\cal K}}^{Y,Y}_{X,Y}(Y,Y)=\zeta^4-2 \zeta^2, 
\quad {\widetilde {\cal K}}^{Y,Y}_{X,Y}(Y,X)=2 \zeta^3-\zeta^5,
\\
&   {\widetilde {\cal K}}^{Y,Y}_{X,Y}(X,Y)=2 \zeta^3-\zeta^5, 
\quad {\widetilde {\cal K}}^{Y,Y}_{X,Y}(X,X)=2 \zeta^2-\zeta^4, \quad {\widetilde {\cal K}}^{Y,X}_{X,Y}(Y,Y)=1-\zeta^2, 
\\ 
& {\widetilde {\cal K}}^{Y,X}_{X,Y}(X,Y)=2 \zeta^2-\zeta^4, \quad {\widetilde {\cal K}}^{Y,X}_{X,Y}(X,X)=\zeta^2-1, 
\quad {\widetilde {\cal K}}^{Y,Y}_{X,X}(Y,I)=-\zeta^4+\zeta^2+1, 
\\
& {\widetilde {\cal K}}^{Y,Y}_{X,X}(Y,Y)=\zeta, 
\quad {\widetilde {\cal K}}^{Y,Y}_{X,X}(X,Y)=-\zeta, \quad {\widetilde {\cal K}}^{X,Y}_{Y,Y}(Y,Y)=\zeta^4-2 
\zeta^2, 
\\ 
& {\widetilde {\cal K}}^{X,Y}_{Y,Y}(Y,X)=2 \zeta^3-\zeta^5, \quad {\widetilde {\cal K}}^{X,Y}_{Y,Y}(X,Y)=2 \zeta^3-\zeta^5, 
\quad {\widetilde {\cal K}}^{X,Y}_{Y,Y}(X,X)=2 \zeta^2-\zeta^4, 
\\
& {\widetilde {\cal K}}^{X,X}_{Y,Y}(Y,I)=-\zeta^4+\zeta^2+1, 
\quad {\widetilde {\cal K}}^{X,X}_{Y,Y}(Y,Y)=\zeta, \quad {\widetilde {\cal K}}^{X,X}_{Y,Y}(X,Y)=-\zeta, 
\\ 
& {\widetilde {\cal K}}^{X,Y}_{Y,X}(Y,Y)=1-\zeta^2, \quad {\widetilde {\cal K}}^{X,Y}_{Y,X}(X,Y)=2 \zeta^2-\zeta^4, 
\quad {\widetilde {\cal K}}^{X,Y}_{Y,X}(X,X)=\zeta^2-1, 
\\
& {\widetilde {\cal K}}^{X,Y}_{X,Y}(I,Y)=1-\zeta^2, 
\quad {\widetilde {\cal K}}^{X,Y}_{X,Y}(I,X)=1-\zeta^2, \quad {\widetilde {\cal K}}^{X,Y}_{X,Y}(Y,Y)=2 
\zeta^3-\zeta^5, 
\\ 
& {\widetilde {\cal K}}^{X,Y}_{X,Y}(Y,X)=-\zeta, \quad {\widetilde {\cal K}}^{X,X}_{X,X}(I,I)=\zeta^2, 
\quad {\widetilde {\cal K}}^{X,X}_{X,X}(I,Y)=\zeta^2, 
\\
& {\widetilde {\cal K}}^{X,X}_{X,X}(Y,I)=-\zeta^4+\zeta^2+1, 
\quad {\widetilde {\cal K}}^{X,X}_{X,X}(Y,Y)=-\zeta^2,
\fe
where $\zeta=\zeta_i$ for $i=1,\cdots,6$ are the six roots of the equation $1-\zeta^2-2\zeta^4+\zeta^6=0$.  The solutions \eqref{eqn:sl5sol} with $\zeta=\zeta_i$ and $\zeta=-\zeta_i$ are gauge equivalent by a further freedom that rotates the junction vectors by $\cM_{X,X,Y}=\cM_{X,Y,X}=\cM_{Y,X,X}=-1$. 

The expectation values $\la X\ra$ and $\la Y\ra$ are
\ie
\la X\ra=\zeta^4-\zeta^2-1,\quad\la Y\ra=\zeta^2.
\fe
Out of the three solutions, only one of them give positive values of $\la X \ra $ and $\la Y \ra$, as is required by unitarity.

\newpage

\subsection{${1\over 2} E_6$ fusion category}
\label{1/2E6CK}
There are four inequivalent solutions to the pentagon identity associated to the ${1\over 2}E_6$ fusion ring defined in Section~\ref{Sec:SmallRanks} (see also Section~\ref{Sec:hE6}).  We list the nontrivial crossing kernels below, while the unlisted ones are  1 if the T-junctions involved are allowed, and 0 otherwise. The first solution is
\ie
{}
& {\widetilde {\cal K}}^{X,Y}_{Y,X}(Y,Y) = {\widetilde {\cal K}}^{Y,X}_{X,Y}(Y,Y)=-1,
\\
& {\widetilde {\cal K}}^{X,Y}_{Y,Y}(Y,Y) = \left(
\begin{array}{cc}
 0 & 1  \\
 1 & 0 \\
\end{array}
\right), \quad {\widetilde {\cal K}}^{Y,Y}_{Y,X}(Y,Y)=\left(
\begin{array}{cc}
 0 & -i  \\
 i & 0 \\
\end{array}
\right), \quad {\widetilde {\cal K}}^{Y,Y}_{X,Y}(Y,Y)=\left(
\begin{array}{cc}
 1 & 0  \\
 0 & -1 \\
\end{array}
\right),
\\
& {\widetilde {\cal K}}^{Y,I}_{Y,Y}(Y,Y) = {1\over \sqrt{2}}e^{-{\pi i\over 12}}
 \left(
\begin{array}{cc}
 1 & 1  \\
 i & -i \\
\end{array}
\right),\quad
 {\widetilde {\cal K}}^{Y,X}_{Y,Y}(Y,Y) = {1\over \sqrt{2}}e^{{5\pi i\over 12}}
 \left(
\begin{array}{cc}
 1 & -1  \\
 -i & -i \\
\end{array}
\right) ,
\\
& {\widetilde {\cal K}}^{Y,Y}_{Y,Y}(I,I) = {\widetilde {\cal K}}^{Y,Y}_{Y,Y}(I,X)={\widetilde {\cal K}}^{Y,Y}_{Y,Y}(X,I)=-{1+\sqrt{3}\over 2},\quad {\widetilde {\cal K}}^{Y,Y}_{Y,Y}(X,X)={1+\sqrt{3}\over 2},
\\
& {\widetilde {\cal K}}^{Y,Y}_{Y,Y}(Y,I) =
 \left(
\begin{array}{cc}
 -1 & -i  \\
 -i & -1 \\
\end{array}
\right), \quad
{\widetilde {\cal K}}^{Y,Y}_{Y,Y}(Y,X) =
 \left(
\begin{array}{cc}
 -1 & -i  \\
 i & 1 \\
\end{array}
\right) ,
\\
& {\widetilde {\cal K}}^{Y,Y}_{Y,Y}(I,Y) = {1\over 4}(1+\sqrt{3})e^{-{\pi i\over 6}}
 \left(
\begin{array}{cc}
 1 & 1  \\
 1 & -1 \\
\end{array}
\right),
\quad
{\widetilde {\cal K}}^{Y,Y}_{Y,Y}(X,Y) = {1\over 4}(1+\sqrt{3})e^{-{\pi i\over 6}}
 \left(
\begin{array}{cc}
 1 & 1  \\
 -1 & 1 \\
\end{array}
\right) ,
\\
& {\widetilde {\cal K}}^{Y,Y}_{Y,Y}(Y,Y)
\\
&=\left(
\begin{array}{cc}
 \left(
\begin{array}{cc}
 {\widetilde {\cal K}}^{Y,Y}_{Y,Y}(Y,Y)_{11,11} & {\widetilde {\cal K}}^{Y,Y}_{Y,Y}(Y,Y)_{11,12} \\
 {\widetilde {\cal K}}^{Y,Y}_{Y,Y}(Y,Y)_{11,21} & {\widetilde {\cal K}}^{Y,Y}_{Y,Y}(Y,Y)_{11,22} \\
\end{array}
\right) & \left(
\begin{array}{cc}
 {\widetilde {\cal K}}^{Y,Y}_{Y,Y}(Y,Y)_{12,11} & {\widetilde {\cal K}}^{Y,Y}_{Y,Y}(Y,Y)_{12,12} \\
 {\widetilde {\cal K}}^{Y,Y}_{Y,Y}(Y,Y)_{12,21} & {\widetilde {\cal K}}^{Y,Y}_{Y,Y}(Y,Y)_{12,22} \\
\end{array}
\right) \\
 \left(
\begin{array}{cc}
 {\widetilde {\cal K}}^{Y,Y}_{Y,Y}(Y,Y)_{21,11} & {\widetilde {\cal K}}^{Y,Y}_{Y,Y}(Y,Y)_{21,12} \\
 {\widetilde {\cal K}}^{Y,Y}_{Y,Y}(Y,Y)_{21,21} & {\widetilde {\cal K}}^{Y,Y}_{Y,Y}(Y,Y)_{21,22} \\
\end{array}
\right) & \left(
\begin{array}{cc}
 {\widetilde {\cal K}}^{Y,Y}_{Y,Y}(Y,Y)_{22,11} & {\widetilde {\cal K}}^{Y,Y}_{Y,Y}(Y,Y)_{22,12} \\
 {\widetilde {\cal K}}^{Y,Y}_{Y,Y}(Y,Y)_{22,21} & {\widetilde {\cal K}}^{Y,Y}_{Y,Y}(Y,Y)_{22,22} \\
\end{array}
\right) \\
\end{array}
\right)
\\
&={1\over 2}
\left(
\begin{array}{cc}
 \left(
\begin{array}{cc}
 \sqrt{2+\sqrt{3}} e^{-\frac{i \pi }{12}} & e^{-\frac{i \pi }{3}} \\
 \sqrt{2+\sqrt{3}} e^{-\frac{i \pi }{12}} & e^{\frac{2 i \pi }{3}} \\
\end{array}
\right) & \left(
\begin{array}{cc}
 e^{\frac{i \pi }{6}} & \sqrt{2+\sqrt{3}} e^{\frac{5 i \pi }{12}} \\
 e^{\frac{i \pi }{6}} & \sqrt{2+\sqrt{3}} e^{-\frac{7 i \pi }{12}} \\
\end{array}
\right) \\
 \left(
\begin{array}{cc}
 \sqrt{2+\sqrt{3}} e^{\frac{5 i \pi }{12}} & e^{\frac{i \pi }{6}} \\
 \sqrt{2+\sqrt{3}} e^{-\frac{7 i \pi }{12}} & e^{\frac{i \pi }{6}} \\
\end{array}
\right) & \left(
\begin{array}{cc}
 e^{-\frac{i \pi }{3}} & \sqrt{2+\sqrt{3}} e^{-\frac{i \pi }{12}} \\
 e^{\frac{2 i \pi }{3}} & \sqrt{2+\sqrt{3}} e^{-\frac{i \pi }{12}} \\
\end{array}
\right) \\
\end{array}
\right).
\fe
The second solution is
\ie
{} & {\widetilde {\cal K}}^{X,Y}_{Y,X}(Y,Y) = {\widetilde {\cal K}}^{Y,X}_{X,Y}(Y,Y)=-1,
\\
& {\widetilde {\cal K}}^{X,Y}_{Y,Y}(Y,Y) = \left(
\begin{array}{cc}
 0 & 1  \\
 1 & 0 \\
\end{array}
\right), \quad {\widetilde {\cal K}}^{Y,Y}_{Y,X}(Y,Y)=\left(
\begin{array}{cc}
 0 & -i  \\
 i & 0 \\
\end{array}
\right), \quad {\widetilde {\cal K}}^{Y,Y}_{X,Y}(Y,Y)=\left(
\begin{array}{cc}
 1 & 0  \\
 0 & -1 \\
\end{array}
\right),
\\
& {\widetilde {\cal K}}^{Y,I}_{Y,Y}(Y,Y) = {1\over \sqrt{2}}e^{{7\pi i\over 12}}
 \left(
\begin{array}{cc}
 1 & 1  \\
 i & -i \\
\end{array}
\right), \quad
{\widetilde {\cal K}}^{Y,X}_{Y,Y}(Y,Y) = {1\over \sqrt{2}}e^{-{11\pi i\over 6}}
 \left(
\begin{array}{cc}
 1 & -1  \\
 -i & -i \\
\end{array}
\right) ,
\\
& {\widetilde {\cal K}}^{Y,Y}_{Y,Y}(I,I) = {\widetilde {\cal K}}^{Y,Y}_{Y,Y}(I,X)={\widetilde {\cal K}}^{Y,Y}_{Y,Y}(X,I)=-{1-\sqrt{3}\over 2},\quad {\widetilde {\cal K}}^{Y,Y}_{Y,Y}(X,X)={1-\sqrt{3}\over 2},
\\
& {\widetilde {\cal K}}^{Y,Y}_{Y,Y}(Y,I) =
 \left(
\begin{array}{cc}
 -1 & -i  \\
 -i & -1 \\
\end{array}
\right), \quad
{\widetilde {\cal K}}^{Y,Y}_{Y,Y}(Y,X) =
 \left(
\begin{array}{cc}
 -1 & -i  \\
 i & 1 \\
\end{array}
\right) ,
\\
& {\widetilde {\cal K}}^{Y,Y}_{Y,Y}(I,Y) = {1\over 4}(\sqrt{3}-1)e^{{\pi i\over 6}}
 \left(
\begin{array}{cc}
 1 & 1  \\
 1 & -1 \\
\end{array}
\right),\quad
 {\widetilde {\cal K}}^{Y,Y}_{Y,Y}(X,Y) = {1\over 4}(\sqrt{3}-1)e^{\pi i\over 6}
 \left(
\begin{array}{cc}
 1 & 1  \\
 -1 & 1 \\
\end{array}
\right) ,
\\
& {\widetilde {\cal K}}^{Y,Y}_{Y,Y}(Y,Y)
\\
&= \left(
\begin{array}{cc}
 \left(
\begin{array}{cc}
 {\widetilde {\cal K}}^{Y,Y}_{Y,Y}(Y,Y)_{11,11} & {\widetilde {\cal K}}^{Y,Y}_{Y,Y}(Y,Y)_{11,12} \\
 {\widetilde {\cal K}}^{Y,Y}_{Y,Y}(Y,Y)_{11,21} & {\widetilde {\cal K}}^{Y,Y}_{Y,Y}(Y,Y)_{11,22} \\
\end{array}
\right) & \left(
\begin{array}{cc}
 {\widetilde {\cal K}}^{Y,Y}_{Y,Y}(Y,Y)_{12,11} & {\widetilde {\cal K}}^{Y,Y}_{Y,Y}(Y,Y)_{12,12} \\
 {\widetilde {\cal K}}^{Y,Y}_{Y,Y}(Y,Y)_{12,21} & {\widetilde {\cal K}}^{Y,Y}_{Y,Y}(Y,Y)_{12,22} \\
\end{array}
\right) \\
 \left(
\begin{array}{cc}
 {\widetilde {\cal K}}^{Y,Y}_{Y,Y}(Y,Y)_{21,11} & {\widetilde {\cal K}}^{Y,Y}_{Y,Y}(Y,Y)_{21,12} \\
 {\widetilde {\cal K}}^{Y,Y}_{Y,Y}(Y,Y)_{21,21} & {\widetilde {\cal K}}^{Y,Y}_{Y,Y}(Y,Y)_{21,22} \\
\end{array}
\right) & \left(
\begin{array}{cc}
 {\widetilde {\cal K}}^{Y,Y}_{Y,Y}(Y,Y)_{22,11} & {\widetilde {\cal K}}^{Y,Y}_{Y,Y}(Y,Y)_{22,12} \\
 {\widetilde {\cal K}}^{Y,Y}_{Y,Y}(Y,Y)_{22,21} & {\widetilde {\cal K}}^{Y,Y}_{Y,Y}(Y,Y)_{22,22} \\
\end{array}
\right) \\
\end{array}
\right)
\\
&={1\over 2}
\left(
\begin{array}{cc}
 \left(
\begin{array}{cc}
 \sqrt{2-\sqrt{3}} e^{-\frac{5 i \pi }{12}} & e^{\frac{i \pi }{3}} \\
 \sqrt{2-\sqrt{3}} e^{-\frac{5 i \pi }{12}} & e^{-\frac{2 i \pi }{3}} \\
\end{array}
\right) & \left(
\begin{array}{cc}
 e^{\frac{5 i \pi }{6}} & \sqrt{2-\sqrt{3}} e^{\frac{i \pi }{12}} \\
 e^{\frac{5 i \pi }{6}} & \sqrt{2-\sqrt{3}} e^{-\frac{11 i \pi }{12}} \\
\end{array}
\right) \\
 \left(
\begin{array}{cc}
 \sqrt{2-\sqrt{3}} e^{\frac{i \pi }{12}} & e^{\frac{5 i \pi }{6}} \\
 \sqrt{2-\sqrt{3}} e^{-\frac{11 i \pi }{12}} & e^{\frac{5 i \pi }{6}} \\
\end{array}
\right) & \left(
\begin{array}{cc}
 e^{\frac{i \pi }{3}} & \sqrt{2-\sqrt{3}} e^{-\frac{5 i \pi }{12}} \\
 e^{-\frac{2 i \pi }{3}} & \sqrt{2-\sqrt{3}} e^{-\frac{5 i \pi }{12}} \\
\end{array}
\right) \\
\end{array}
\right).
\fe 
The third and forth solutions are the complex conjugates of the first and second solutions. Only the second and fourth solutions give positive values of $\la X \ra $ and $\la Y \ra$, as is required by unitarity.

\bibliographystyle{JHEP}
\bibliography{TDL}

\end{document}